\documentclass[a4paper,openany, 12pt]{book}

\usepackage[text={15.5cm,23cm},centering]{geometry}
\usepackage[sectionbib]{chapterbib} 
\usepackage{mathptmx}
\usepackage{float}
\usepackage{amssymb}
\usepackage{amsmath}
\usepackage{graphicx}
\graphicspath{{Figures/}}
\usepackage{subfigure}
\usepackage[T1]{fontenc}  
\usepackage{color}
\usepackage[colorlinks=true,
            linkcolor=blue,
            urlcolor=blue,
            citecolor=blue]{hyperref} 
\newcommand{\cm}{cm$^{-1}$}
\newcommand{\etal}{\textit{et al.}}
\newcommand{\abinitio}{\textit{ab initio}}
\newcommand{\SE}{Schr\"odinger equation}

\newcommand{\tj}[6]{ \begin{pmatrix}
       #1 & #2 & #3 \\
       #4 & #5 & #6 
    \end{pmatrix}}

\newcommand{\sj}[6]{ \begin{Bmatrix}
		#1 & #2 & #3 \\
		#4 & #5 & #6 
\end{Bmatrix}}

\title{Theoretical methods for calculating rotational-vibrational-electronic transition intensities in triatomic molecules}
\author{Emil J. Zak}

\begin{document}
\frontmatter
\maketitle
\tableofcontents
\chapter*{Acknowledgments}

I would like to thank prof. Jonathan Tennyson and the ExoMol team at University College London, thanks to whom this article could be written. 
I also thank Ania for her support.
\chapter*{Preface}
\label{Preface}

This article focuses on few selected aspects of quantum theory of molecular rotations and vibrations. The field of the nuclear motion theory for small molecules is as broad as it is an old one. An excellent book by P.R. Bunker and P. Jensen "Molecular symmetry and spectroscopy" constitutes a comprehensive cross section of existing methods. In this paper, I aim at giving an introduction to a specific, yet very successful methodology originally derived by B.T. Sutcliffe and J. Tennyson, which utilizes the exact kinetic energy operator for the description of the rotational-vibrational motion in triatomic molecules. This methodology has been proven useful in high-accuracy calculations of transition intensities, when aided with high-level electronic structure theory. Only resonant transitions are considered here, that is those molecular transitions whose intensities are determined by the electronic dipole moment of the molecule.

Computation of highly accurate transition intensities for small molecules, with H$_2$O, CO$_2$ and O$_3$ as primary examples, is important for the Earth's atmospheric science and more recently for spectroscopic characterization of Exoplanets. 
High quality models for the retrieval of molecular concentrations in the atmosphere require reference transition intensities, which can be either calculated or measured in the laboratory. The latter is often challenging due to technical complications, such as limited laser stability, accurate pressure and temperature measurements.
A theoretical calculation can provide a comprehensive list of lines (line list), which can be converted to an absorption spectrum. In this work, I discuss some of the state-of-the-art procedures for calculating such spectra for the rotational-vibrational transitions, mainly located in the infrared region of the electromagnetic spectrum, as well as rotational-vibrational-electronic transitions located in the visible and ultra-violet region. 

Chapter 2 constitutes a concise tutorial on the calculations of the rotational-vibrational energy levels of triatomic molecules using the Sutcliffe-Tennyson approach. Among a number of techniques for solving the \SE\ the discrete-variable representation (DVR) method is a popular choice. For this reason, I briefly discuss the fundamentals of the DVR technique at the end of Chapter 2.

Chapter 3 outlines a procedure for the calculation of transition intensities from first principles. The theoretical machinery introduced in Chapters 2 and 3 is then applied to study the rotational-vibrational spectra for isotopologues of carbon dioxide. Finally, at the end of Chapter 3 a bullet-point guide is given, which lays out some the practical rules for calculating high accuracy transition intensities. 

Theoretically calculated transition intensities can be of similar accuracy to the measured values; they can be calculated for spectral regions that are difficult to probe by an experiment. 
This is because the accuracy of \textit{ab initio} calculated transition intensities is typically constant for the rotational transitions within a vibrational band. The only exceptions are transitions between energy levels, where one of the levels interacts strongly via the so called resonance interaction with another rotational-vibrational level. In such an instance, transition intensity borrowing may occur, leading to complicated spectral patterns. 
For this reason in Chapter 4, a method for searching such resonance-perturbed transitions is discussed. In our procedure each transition intensity receives a reliability factor, a particularly useful descriptor for detecting the resonance interactions between the rotational-vibrational energy levels.

A tentative attempt is made in Chapter 5 to generalize the Sutcliffe-Tennyson framework onto the spin-rotational-vibrational-electronic \SE\ for polytaomic molecules. This piece of theory awaits a computer implementation. 

This article is dedicated to a reader who wishes to specialize in quantum theory of  the nuclear motions in molecules. Triatomic molecules are the simplest systems, which posses qualitative characteristics of larger polyatomic molecules. Thus on the minimal example of triatomic molecules one can introduce general concepts used in the broad field of nuclear motion theory. 
The concepts and techniques used to compute transition intensities are not limited to calculating spectra of molecules. Time-dependent quantum descriptions of molecules interacting with external fields, quantum optics, and many other areas in modern physics have their grounds in resonant transitions between energy levels of the system. Similarities with the nuclear motion theory appear abundantly in other fields. The reader can view this article as a tutorial, by no means comprehensive, which still touches some of the unsolved problems in modern theoretical molecular spectroscopy, and thus shall leave the reader with a knowledge necessary to expand the field.

\paragraph{A note from the author.}
Initially, this article was meant to be published in the form of a monograph book with one of the reputable publishers in physical sciences. 
After positive peer-reviews, followed by an initial approval from the editorial board, a last-moment decision was made to refrain from production. 
I was left alone with a piece of work, which I have been invited to write. With a promise of time-cost remuneration, I have spent over one year preparing this manuscript. 
Therefore, after being abandoned by the publisher, I have decided to make this article open to everyone in the scientific community. Should \textit{bibliometry} take over completely our ambitions, and more worryingly the science funding scheme? This arXiv submission comes along with the hope that knowledge is going to be passed to a curious reader anyway. Is not it what science should be about?

\mainmatter

\chapter{Introduction}
When the electromagnetic radiation passes through a sample filled with molecules in the gas phase, depending on the chemical composition of the sample, different wavelengths of the radiation are absorbed by the molecules at different probabilities. As a result the absorption intensity of the electromagnetic radiation gains a profile which depends on the wavelength. This profile can be measured or calculated and is known as the \textit{molecular absorption spectrum}. Depending on the wavelength of the absorbed electromagnetic radiation different internal motions in the molecule are excited. Absorption of the microwave radiation usually puts molecules in states with higher rotational energy, due to the interaction of the electric field of the electromagnetic wave with the electric dipole moment vector of the molecule. Infrared radiation can excite vibrations of atoms in the molecule, whereas visible and ultraviolet light often excites electrons in the molecule to occupy molecular orbitals of higher energy. 

There are two ways of telling, given a molecule, which wavelengths of the incident electromagnetic radiation are absorbed: a) measurement; b) calculation. Both empirical and theoretical approaches rely on the \textit{Lambert-Beer law} for absorption, which connects the absorption intensity of the electromagnetic radiation with the concentration of molecules in the sample and the molecule-characteristic quantity known under the names of the \textit{attenuation coefficient} or \textit{absorption cross-section}. Lambert-beer law states that the intensity of the incident radiation $\Phi_0$ is directly related to the intensity of the attenuated radiation $\Phi$, which passed through the sample at a given wavelength $\lambda$ \cite{06BuJexx.method}, by means of the relation

\begin{equation}
\frac{\Phi}{\Phi_0}=e^{-\sigma(\lambda)\cdot N \cdot L}
\label{eq:lambert-beer}
\end{equation}
where $L$ is the optical path length, $\sigma(\lambda)$ is the absorption cross-section and $N$ is the concentration of molecules. To be able to retrieve information about the concentration of the molecules in the sample one needs to know $\frac{\Phi}{\Phi_0}$, which is found in the measurement. The optical path length $L$ is also precisely determined in the experiment. The absorption-cross section $\sigma(\lambda)$ can be either determined in the experiment by measuring $\frac{\Phi}{\Phi_0}$ for known concentration of molecules in the sample or can be obtained from calculation. The absorption cross-section, which informs about the wavelength dependence of the absorption intensity is a central object in molecular absorption spectroscopy. 
In this article we focus on theoretical methods for accurately determining the $\sigma(\lambda)$ function in triatomic molecules, for wavelengths $\lambda$ ranging from the UV (electronic excitations) to the microwave (rotational excitations). 

Most often, when the UV electromagnetic wave interacts with a molecule in the gas phase, a multiple of internal motions are activated in the molecule. In the majority of cases transitions to electronically excited states, even for a simple diatomic or triatomic molecule, are accompanied with excitation in the rotational and vibrational quantum states. The intensity of the absorbed radiation carries contributions from excitations of rotations, vibrations and the electrons. However because electrons are three orders of magnitude lighter than the atomic nuclei, it is often reasonable to separate the dynamics of the electrons from the dynamics of the rotational and vibrational motion. The energetics of the electronic motion is typically 2-3 orders of magnitude greater than the energetics of the vibrational motion of nuclei, whereas the energetics of the nuclear vibrations is several times higher than the energy of rotations of the molecule. In instances when not a very high accuracy is required, the rotational, vibrational and electronic motions can be treated independently. 
High-accuracy calculations of the absorption cross-sections require using quantum-mechanical models which account for the coupling between the rotational and vibrational motion of nuclei (centrifugal distortion and Coriolis forces) and sometimes also the coupling between the motion of the electrons and nuclei (vibronic coupling). Particularly when molecular energy levels are involved in resonance interactions it is unavoidable to work with the fully coupled rotational-vibrational-electronic models. 

In this article we introduce selected theoretical techniques for high-accuracy calculations of the rotational-vibrational-electronic molecular absorption spectra. The focus is on the rotational-vibrational nuclear motion calculations with the dynamics of the electrons treated separately. Resonance interactions of rotational-vibrational energy levels are also discussed. 

One important application of molecular spectroscopy is in studying the Earth's atmosphere. Triatomic molecules represent a major class of species in the Earth's atmosphere, with water, carbon dioxide, sulphur dioxide, nitrous oxide and ozone as examples. A key to understanding the physics and chemistry of atmospheric processes is the identification of sources, sinks and migration mechanisms of these gases.

Molecular spectroscopy delivers tools for quantifying the concentrations of molecules in the gas phase samples. In the concentration \textit{retrieval} procedure, one measures the absorption spectrum of a gas sample of interest and compares the measured absorption intensity profile (spectrum) with a reference spectrum. By means of the Lambert-Beer law given in eq. \ref{eq:lambert-beer}, fitting the measured and the reference spectrum to match as closely as possible, allows to extract information about concentrations of individual species in the sample.
 
The reference spectrum can be obtained by measuring the absorption cross-sections in samples with known molecular concentrations or it can be calculated from the principles of Quantum Mechanics. Experimental determination of high-resolution and high-accuracy absorption cross-sections is often difficult for a number of reasons:
\begin{itemize}
\item identifying individual absorption lines from raw experimental data is often a tedious task and therefore only a limited number of transition lines can be extracted from the measurement. As a result, often multiple data sources must be used in atmospheric models, which presents consistency issues in the form of systematic shifts in transition intensities and line positions. 
\item it is particularly challenging to measure infrared absorption intensities at sub-percent accuracy for atmospherically relevant molecules. This problem is particularly pronounced for less abundant species, such as rare isotopologues of molecules, which are important trackers of human activity. Experimental determination of the reference spectra requires samples with high isotopic purity and known concentration of the isotopologues.
\item high-temperature absorption measurements are particularly costly and difficult
\item hazardous chemical compounds (such as H$_2$S or HCN) add cost, risk and complexity to the absorption measurements
\end{itemize}
In recent years, absorption cross-sections for high-resolution molecular spectroscopy are more favourably obtained from the calculation. First-principles quantum-mechanical calculations guarantee completeness of the spectrum. With the advent of powerful computational techniques for high-accuracy calculations of rotational-vibrational spectra, the accuracy in the calculated transition intensities for triatomic molecules has become competitive with the accuracy in the experimentally sourced intensities. This advancement in the level of accuracy offered by first-principles calculations rendered computation as an attractive alternative to the experiment. 

One example of theoretical calculations for predicting high-accuracy rotational-vibrational spectra are recent works on carbon dioxide (CO$_2$)  \cite{15ZaTePo.CO2,17ZaTePo.CO2,Zak2017}. Carbon dioxide is one of the main greenhouse gases and has been monitored over the years by several government and private funded projects \cite{OCO-2, TCCON, NDACC}. For determination of the concentration of CO$_2$ in the Earth's atmosphere, both satellite and ground based measurements use infrared transitions between rotational-vibrational states of this molecule. 
In such measurements, it is often desired to have several isotopologues of the same molecule quantified simultaneously, to learn about the sources of this gas.  For example, the unstable $^{14}$CO$_2$ isotopologue containing radioactive $^{14}$C is a key trace species used as a marker for industrial activities \cite{15Grxxxx.CO2,13LeMiWo.CO2,08ToWaTa.CO2,11GaBaBo.CO2}. Low natural abundance of isotopologues of CO$_2$ containing $^{13}$C, $^{14}$C, $^{17}$O or $^{18}$O makes an accurate measurement challenging.

The variations in the atmospheric CO$_2$ concentration are typically of order of few parts-per-million (ppm), with preferably 1-ppm target resolution in the CO$_2$ concentration retrieval. For this to be possible the accuracy of the reference spectra must be about 0.3\% in transition intensities \cite{XOCO}, which have been, until recently, beyond the reach of the experiment \cite{17OyPaDr.CO2,jt613}.  Even for the main isotopologue $^{12}$C$^{16}$O$_2$, the very recent highly sophisticated Cavity Ring-Down measurements reached 0.3\%--1\% accuracy for only a limited number of transitions \cite{11WuViJo.CO2,jt613,Hodges2017,16BeDeSu.CO2,17OdFaMo.CO2}. Theory brings the possibility to reach the level of accuracy needed for 1-ppm resolution in the atmospheric CO$_2$ retrieval.

Detection and quantification of other very important, but less abundant molecules in the Earth's atmosphere, such as SO$_2$ or O$_3$ relies on absorption of the ultraviolet (UV) radiation \cite{Chance1997}, which is accompanied by transition between rotational-vibrational-electronic (ro-vibronic) states of the molecule. 
Rotational-vibrational-electronic transitions in the UV for the SO$_2$ molecule play a central role in monitoring volcanic activity \cite{Carn2017}. The H$_3^+$ molecular ion detected in the interstellar medium (ISM) and in planetary atmospheres (Jupiter) is another important triatomic species carrying information about the history of an early universe \cite{McCall1999}.
Last but not least, the water molecule (H$_2$O) and its isotopologues are of great importance in climate science as well as in studies of Exoplanets \cite{Tinetti2012}.

Every high-accuracy absorption intensity measurement ought to be supported with an appropriate high-level theoretical model. Such a model should aid to eliminate spurious absorption lines from contaminant species, but most of all should allow to spectrally assign the observed absorption lines. So far infrared absorption calculations have been shown very accurate, unfortunately UV ro-vibronic calculations still lag behind the progress in the ro-vibrational calculations.

A general procedure for calculating molecular absorption spectra from first principles can be summarized as follows. Prior to performing the quantum-mechanical calculations, one needs to know the masses of atoms building up the molecule as well as the approximate molecular geometry, both of which can be found in tables.
In the first step of calculations, the \SE\ for the motion of electrons is solved for a set of molecular geometries.  For doing this, one has to employ the Born-Oppenheimer approximation, in which the electron dynamics can be viewed from the clamped-nuclei perspective, as is discussed in Chapter 2. The resulting electronic energies, being functions of coordinates of the nuclei in the molecule, define the potential energy surface (PES). Because the electronic \SE\ can only be solved for a limited number of nuclear geometries, it is often necessary to fit a functional form to model the shape of the PES. The continuous representation to the PES is then used as the potential energy for the motion of the nuclei. 

By solving the \SE\ for the nuclear motion on an electronic PES one obtains rotational-vibrational energy levels and wavefunctions. There are several methods available to find solutions to the rotational-vibrational \SE. Some of them are described in Chapter 2. The calculated rotational-vibrational energy levels, with the aid of selection rules (discussed in Chapter 3), determine between which energy levels transitions can occur upon absorption of a single photon of the electromagnetic radiation interacting with the dipole moment of the molecule. The absorption intensity can be calculated knowing the \textit{transition dipole moments}, which are calculated from the rotational-vibrational wavefunctions and the electronic transition dipole moment surface (TDMS) computed with electronic structure packages. 
The end product of the procedure outlined above, that is transition frequencies, intensities and quantum numbers uniquely characterize individual transitions and compose the final spectrum, often called a \textit{line list.}

There are a number of computer codes available, which are dedicated to calculations of rotational-vibrational and rotational-vibrational-electronic spectra of triatomic molecules. To name a few: DVR3D \cite{DVR3D} (Tennyson \etal), MORBID \cite{MORBID}  (Jensen \etal), RENNER \cite{RENNER} (Odaka \etal), RVIB3 \cite{RVIB3} (Carter \etal), MULTIMODE \cite{MULTIMODE} (Bowman \etal), a code by Schwenke \cite{Schwenke1992}, DOPI3 \cite{DOPI3} (Czako \etal) and others are only selected examples of programs specializing in particular aspects of modeling of the molecular absorption spectra of triatomics. These programs are based on several different methodologies and implementations, introduced by Islamapour \cite{Islampour2005}, Schwenke \cite{Schwenke2003},
Lukka \cite{Lukka1995}, Rey, Nikitin and Tyuterev \cite{Rey2012} and others. In this article we employ a general methodology to nuclear motion calculations of Sutcliffe and Tennyson \cite{Sutcliffe1986} which is implemented in the DVR3D suite \cite{DVR3D}.

Complementary to the class of \abinitio\ variationally-derived methodologies listed above the effective Hamiltonian (EH) techniques have been very popular. In high-resolution molecular spectroscopy the EH models have been by far the most accurate in finding line positions. These models rely on input experimental transition frequencies and intensities of lines; they provide a means of interpolation and extrapolation of rotational-vibrational spectra. Notable contributions to the theory of effective Hamiltonians  include works by Tyuterev \cite{Tyuterev2004}, Perevalov \cite{Teffo1992}, Nikitin and Tashkun \cite{CassamChena2011} and others. The accuracy achieved in effective Hamiltonian calculations is typically much higher than in variational calculations. The drawback of the effective Hamiltonian techniques is their sensitivity to the quality of the experimental input data.

The article is most suitable for students and researchers with background in quantum mechanics and ideally with basic knowledge about molecular physics.  A researcher working in another field related to quantum theory should already be equipped with the conceptual and mathematical knowledge to study problems introduced in this paper. Due to space limitations, rather than giving a comprehensive and elementary introduction to molecular spectroscopy, the aim of the present article is to present a specific methodology for calculating transition intensities in triatomic molecules.

In the text, when a problem is discussed which requires a prior knowledge of an elementary concept such as the basic solutions to the \SE\ for the harmonic oscillator or the rigid rotor Hamiltonian,  the reader is referred to specific handbooks, reviews or individual publications. For a wider context in molecular spectroscopy of small molecules, which is often not discussed in the present article I strongly recommend the reader to familiarize with excellent books by Bunker and Jensen \cite{06BuJexx.method} or Bernath \cite{bernath} which cover the fundamentals of molecular spectroscopy. An ideal recipient of the present paper has some background knowledge in molecular spectroscopy, for instance from the aforementioned handbooks, and the read-through the present text serves gaining a certain level of expertise in the topic of accurate transition intensities calculations.

Chapter 2 introduces basic concepts in the quantum theory of rotational-vibrational motion of triatomic molecules. We follow a methodology originally introduced by Sutcliffe and Tennyson for calculating energy levels and wavefunctions of triatomic molecules. Other approaches, such as the effective Hamiltonian or collocation methods are also briefly discussed. At the end of the chapter we discuss techniques used to represent the \SE\ in the matrix form as well as symmetry properties of these representations, which can be utilized in making the calculations more efficient.

Chapter 3 gives a derivation of the general expression for the transition absorption intensity. A number of popular approximations are discussed and appropriate selection rules are outlined for each level of approximation. Example results of absorption intensity calculations are presented for the CO$_2$ (infrared) and SO$_2$ (UV) molecules. 

Chapter  4 introduces the concept of a resonance interaction between energy levels. An example list of known types of ro-vibrational resonances is given  and their role in spectroscopy is discussed. Next, potential issues with variational calculations are pointed out, on the example of transition intensities calculations for energy levels affected by resonance interactions. At the end of Chapter 4 a method is given for quantifying the reliability of the variationally calculated transition intensities. 

Chapter 5 briefly discusses a generalized framework for representing the spin-rotational-vibrational-electronic \SE\ for polytaomic molecules. The role of vibronic coupling in transition intensity borrowing is explained on a simple example. 
\bibliographystyle{plain}
\bibliography{References}

\chapter{Rotational-vibrational energy levels}
This chapter introduces a popular methodology used to calculate rotational-vibrational energy levels and wavefunctions of triatomic molecules. We begin with the discussion of the Born-Oppenheimer approximation, which allows to treat the dynamics of electrons separately from the dynamics of the nuclei. The rotational-vibrational Hamiltonian describing the dynamics of the nuclei is then derived in a general and elegant procedure proposed by Sutcliffe and Tennyson. This procedure, although one of few available, has been successfully utilized in high-accuracy calculations of infrared and UV spectra of triatomic molecules. Having the rotational-vibrational Hamiltonian, a number of popular computational methods for solving the \SE\ is given in section \ref{sec:methods} along with popular choices of the rotational and vibrational basis sets. Next, in section \ref{sec:additionalsymmetries} physical symmetries are discussed, application of which in the construction of the wavefunction simplifies the rotational-vibrational calculations. We close the chapter with an outline of strategies for solving the rotational-vibrational \SE.

\begin{figure}[H]
\begin{center}
  \includegraphics[width=15cm]{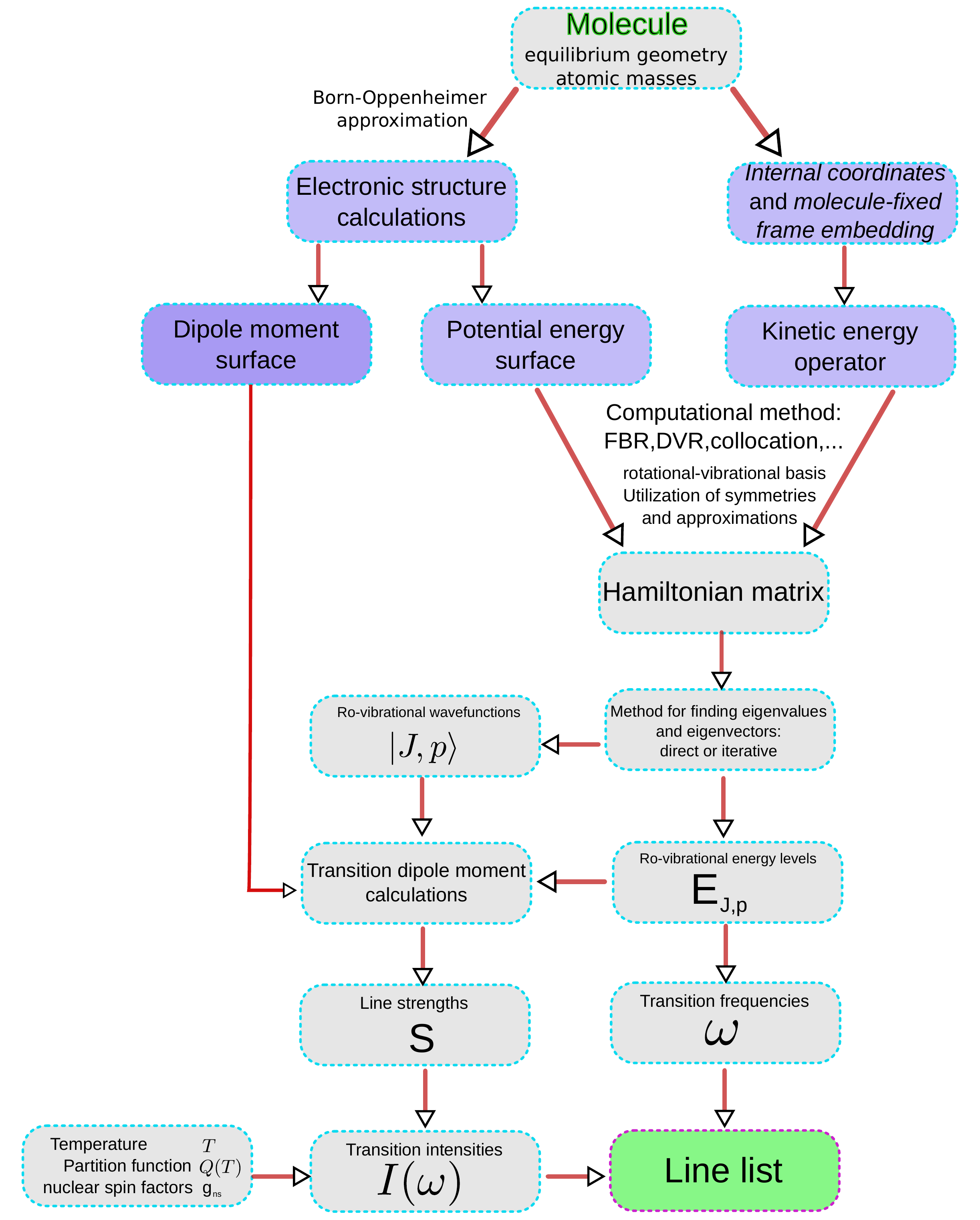}
\caption{A scheme strategy for finding solutions to the rotational-vibrational \SE\ . }
\label{fig:scheme}
\end{center}
\end{figure}

A strategy for finding solutions to the rotational-vibrational \SE\ is schematically summarized in Figure \ref{fig:scheme}. In this article, we are going to discuss individual pieces that build up the scheme given in Figure \ref{fig:scheme}, so that the reader can not only gain a bird's eye view on the methodologies used in nuclear motion theory, but also absorb some level of detail, allowing one to perform their own calculation. By no means the scheme presented in Figure \ref{fig:scheme} and the description given for its individual blocks is unique. We are only going to discuss a particular choice of methods and techniques, which although have been shown particularly successful, are still merely author's subjective selection.  

Having chosen a molecule, we are in possession of information about the masses of atoms constituting the molecule as well as usually we know an approximate molecular geometry (linear, bent, etc.). Having this knowledge, the procedure to follow, as sketched in Figure \ref{fig:scheme}, reads:
\begin{enumerate}
\item In the Born-Oppenheimer approximation: solve the electronic \SE\ given in eq. \ref{eq:SEel} for a set of molecular geometries (inter-atomic distances)
\item Fit a functional form to the calculated electronic energy levels, to obtain the potential energy surface (PES)
\item Choose: a) internal coordinates to describe the vibrational motion in the molecule; b) molecule-fixed coordinate frame (see eq.~\ref{eq:embedding}); c) basis set associated with the vibrational coordinates (see section \ref{sec:theorybasis}). Derive the KEO for the motion of nuclei (section \ref{sec:KEO}).
\item Choose a method of solution to the \SE\ and run a computer code for solving the nuclear motion problem for each $J$ value separately to obtain the rotational-vibrational energy levels and wavefunctions
\item Calculate the electronic transition dipole moments (by solving the electronic \SE) for a set of molecular geometries
\item Fit a functional form to the calculated electronic transition dipole moments, to obtain the transition dipole moment surface (TDMS) (see section \ref{sec:tdms})
\item Calculate ro-vibrational transition frequencies and transition intensities using eq. \ref{eq:Int}.
\item Put together transition frequencies, transition intensities and appropriate quantum numbers to form the final line list
\end{enumerate}

\section{The Born-Oppenheimer approximation}
\label{sec:BO}
The time-independent \SE\ for a molecular system can be written as 
\begin{equation}
\hat{H}_{mol}\Psi_{\nu}\left(\mathbf{Q},\mathbf{q}\right)=E_{\nu}\Psi_{\nu}\left(\mathbf{Q},\mathbf{q}\right)
\label{eq:SE}
\end{equation}
where the molecular wavefunction $\Psi_{\nu}\left(\mathbf{Q},\mathbf{q}\right)$ depends on the coordinates of nuclei $\mathbf{Q}=(Q_1,Q_2,...,Q_{3N_{nuc}})$ and the coordinates of electrons $\mathbf{q}=(q_1,q_2,...,q_{3N_{el}})$. The molecular Hamiltonian can be written in the following form
\begin{equation}
\hat{H}_{mol}=\hat{T}_{nuc}(\mathbf{Q})+\hat{H}_{el}(\mathbf{q},\mathbf{Q})
\label{eq:Hmol}
\end{equation}
where the kinetic energy operator of the nuclei $\hat{T}_{nuc}(\mathbf{Q})$ depends on coordinates of $N_{nuc}$  nuclei and the electronic Hamiltonian $\hat{H}_{el}(\mathbf{q},\mathbf{Q})$ depends on coordinates of $N_{el}$ electrons and as well as on all nuclear coordinates. In the Cartesian representation, the kinetic energy operator of the nuclei is given by

\begin{equation}
\hat{T}_{nuc}(\mathbf{Q})=-\frac{1}{2}\sum_{k=1}^{N_{nuc}}\frac{1}{M_k}\Delta_{k}
\label{eq:KSF}
\end{equation}
where $M_k$ is the mass of the $k$-th nucleus and $\Delta_k=\left(\frac{\partial^2}{\partial x^2_k},\frac{\partial^2}{\partial y^2_k},\frac{\partial^2}{\partial z^2_k}\right)$ represents a vector of second derivatives with respect to the Cartesian coordinates of the $ k$-th nucleus.The electronic Hamiltonian can be written as:
\begin{equation}
\hat{H}_{el}(\mathbf{q},\mathbf{Q})=\hat{T}_e(\mathbf{q})+\hat{V}_{ne}(\mathbf{q},\mathbf{Q})+\hat{V}_{ee}(\mathbf{q})+\hat{V}_{nn}(\mathbf{Q})
\label{eq:elham}
\end{equation}
where $\hat{T}_e$ is the kinetic energy operator of all electrons in the molecule, $\hat{V}_{ne}$ is the Coulomb electron-nuclei attraction operator and $\hat{V}_{ee}$, $\hat{V}_{nn}$ are Coulomb repulsion operators for the electrons and the nuclei, respectively. 

For a molecule isolated from any external fields it is possible to separate three centre-of-mass coordinates: $\mathbf{R}_{CM}$. The kinetic energy operator associated with the centre-of-mass motion is exactly separable from the remaining part of the Hamiltonian describing the internal motion of the molecule:

\begin{equation}
\hat{H}_{mol}=\hat{K}_{CM}(\mathbf{R}_{CM})+\hat{H}_{int}(\mathbf{q},\mathbf{Q},\Theta)
\label{eq:Hmolint}
\end{equation}
where $\mathbf{Q}=(Q_1,...,Q_d)$ denotes a set of internal coordinates of the molecule, which describe the vibrational motion of atoms relative to the molecular centre-of-mass. The number of internal coordinates is $d=3N_{nuc}-6$ for non-linear polyatomic molecules  and $d=3N_{nuc}-5$ for linear molecules, where $N_{nuc}$ is the number of atoms. The rotational motion of the molecule is described by three angular coordinates $\Theta=(\phi,\theta,\chi)$, which specify instantaneous orientation of the molecule with respect to a selected frame of reference fixed in 3-D space.

Upon defining the centre-of-mass coordinate system, the internal motion Hamiltonian is given as
\begin{equation}
\hat{H}_{int}(\mathbf{q},\mathbf{Q},\Theta)=\hat{K}_{nuc}(\mathbf{Q},\Theta)+\hat{K}_{cross}+\hat{H}_{el}(\mathbf{q},\mathbf{Q})
\label{eq:Ktrans}
\end{equation}
where $\hat{K}_{nuc}(\mathbf{Q},\Theta)$ is the kinetic energy operator for $3N_{nuc}-6$ internal coordinates $\mathbf{Q}$ and 3 rotational coordinates $\Theta$. In the above equation $\hat{K}_{cross}=\frac{1}{2}\sum_{k,l=2}^{N_{nuc}}\frac{1}{M}\vec{\nabla}_{k}\cdot\vec{\nabla}_{l}$ represents the mixed-derivatives operator, which is an artifact from the coordinate system transformation into the centre-of-mass system.
$\hat{K}_{cross}$ is often referred to as the \textit{nuclear mass polarization term}. $M$ stands for the total mass of all nuclei in the molecule.  The appearance of the mass polarization term can be understood as follows: because we introduced the centre-of-mass dynamical variable, the coordinates of any one particle are fully determined by the coordinates of the remaining N-1 particles. Motion of any chosen particle slightly changes the position of the centre-of-mass coordinate in the laboratory frame. But because no external forces act on the molecule, the position of the centre-of-mass must be fixed in the laboratory frame. For this reason the coordinates (and momenta) of the remaining N-1 particles must be corrected so that the centre-of-mass is unchanged. As a result the positions (and momenta) of all particles in the system become inter-dependent. The mass-polarization term is inversely proportional to the total mass of the particles (electrons and nuclei) and can be neglected, unless ultra-high accuracy calculations are performed. 

Because the total molecular Hamiltonian is separable into the centre-of-mass part $\hat{K}_{CM}(\mathbf{R}_{CM})$ and the internal motion Hamiltonian $\hat{H}_{int}(\mathbf{q},\mathbf{Q},\Theta)$, the total wavefunction is the product of the wavefunction for the centre-of-mass motion $f^{CM}_{\mathbf{k}}(\mathbf{R}_{CM})=\frac{1}{\sqrt{2\pi}}e^{i\mathbf{k}\cdot\mathbf{R}_{CM}}$ and the internal motion wavefunction $\Psi_{\nu}\left(\mathbf{q},\mathbf{Q},\Theta\right)$. It is always possible to choose the translational wavefunction which corresponds to $\mathbf{k}=0$ eigenvalue (wavevector) of $\hat{K}_{CM}$, meaning that the molecule is not moving in the laboratory reference frame and the translational wavefunction $f^{CM}_{k=0}$ is constant. The translation-free \SE\ can be then written as

\begin{equation}
\hat{H}_{int}(\mathbf{q},\mathbf{Q},\Theta)\Psi_{\nu}\left(\mathbf{q},\mathbf{Q},\Theta\right)=E^{int}_{\nu}\Psi_{\nu} \left(\mathbf{q},\mathbf{Q},\Theta\right)
\label{eq:SEint}
\end{equation}
where $\nu$ labels eigenfunctions of the total internal motion Hamiltonian and $E^{int}_{\nu}$ is the total internal energy of the molecule. The above equation contains $3N_{el}+3N_{nuc}-3$ variables (internal coordinates) which are coupled.

Even for a simple triatomic molecule, such as H$_2$O, its 10 electrons combined with 3 nuclear vibrational coordinates and 3 rotational coordinates totals 36 coupled internal coordinates. A direct and accurate solution to the \SE\ given in eq. \ref{eq:SEint} with 36 coupled coordinates remains intractable. This is because in typical variational calculations the memory requirements for storing information about the wavefunction grows exponentially with the number of coordinates, which is known as \textit{the curse of dimensionality}. So far, numerically exact solutions to eq. \ref{eq:SEint} have been obtained only for 4-5 particle systems such as $H_2$ or $Li$ \cite{Mtyus2018}. 
For this reason, approximations are needed for the \SE\ to be useful in any chemical applications. 

A particularly successful approximation to eq.~\ref{eq:SEint} is based on the observation that the proton, which is the lightest atomic nuclei, has the rest mass about 1836 times greater than the mass of the electron. As a consequence, the average kinetic energy associated with the motion of the nuclei in molecules is much lower than the kinetic energy of the electrons. The latter move much faster than the former. This results in a significant difference in the quantum mechanical energy scales associated with the motion of electrons and the motion of the nuclei. Energy of the electronic motion is typically at least 10 times greater than the energy of the vibrational motion, whereas the energy of the vibrational motion is typically about 10 times greater than the energy associated with rotations of the molecule.  Thus from the perspective of the electronic motion the nuclei are nearly stationary. Conversely, from the perspective of the nuclei the electrons form a cloud, which generates an averaged electric potential for the nuclear motion.

It is therefore reasonable to assume the following procedure for solving the \SE\ given in eq. \ref{eq:SEint}.
Assume for the moment that the nuclei are stationary, i.e. $M_{nuclei} \rightarrow \infty $ ($\hat{K}_{nuc}(\mathbf{Q},\Theta) = 0$, $\hat{K}_{cross}=0$) and solve the stationary \SE\ for the motion of the electrons first:
\begin{equation}
\hat{H}_{el}(\mathbf{q};\mathbf{Q})\Phi_{i}(\mathbf{q};\mathbf{Q})=E_i(\mathbf{Q})\Phi_{i}(\mathbf{q};\mathbf{Q})
\label{eq:SEel}
\end{equation}
where the semicolon in $\Phi_{i}(\mathbf{q};\mathbf{Q})$ means that the nuclear coordinates $\mathbf{Q}$ are treated as parameters. Such camped-nuclei treatment is possible because $\hat{H}_{el}$ does not contain any derivatives with respect to nuclear coordinates.   Also note that the electronic Hamiltonian $\hat{H}_{el}(\mathbf{q};\mathbf{Q})$ does not depend on rotational coordinates of the molecule, so does not the electronic wavefunction $\Phi_{i}(\mathbf{q};\mathbf{Q})$.\textit{ The electronic} \SE\ \ref{eq:SEel} is then solved at a number of positions $\mathbf{Q}$ of the nuclei. As a result the electronic energy $E_i(\mathbf{Q})$ becomes a function of relative positions $\mathbf{Q}$ of all nuclei in the molecule. Such function is called the electronic \textit{potential energy surface} (PES). A separate PES is calculated for every electronic eigenstate $\Phi_{i}(\mathbf{q};\mathbf{Q})$.

For each nuclear configuration $\mathbf{Q}$, the wavefunctions $\lbrace\Phi_{i}(\mathbf{q};\mathbf{Q})\rbrace_{i=1,2,...}$ obtained by solving the electronic Schrodinger equation \ref{eq:SEel} form a complete orthonormal basis of the Hilbert space, in which the total molecular wavefunction can be expanded
\begin{equation}
\Psi_{\nu}(\mathbf{q},\mathbf{Q},\Theta) =\sum_{i=0}^{+\infty} \chi_{(i,\nu)} \left( \mathbf{Q},\Theta \right)\Phi_{i}(\mathbf{q};\mathbf{Q})
\label{eq:BHexpansion}
\end{equation}
which is known as the \textit{Born-Huang} expansion of the molecular wavefunction. The summation in eq. \ref{eq:BHexpansion} goes over all electronic eigenstates of the system and the $\chi_{(i,\nu)} \left( \mathbf{Q},\Theta \right)$ coefficients represent wavefunctions of the nuclei in the $\nu$-th molecular state projected onto the $i$-th electronic state. 

In order to obtain an effective equation for $\chi_{(i,v)}\left(\mathbf{Q},\Theta\right)$, that is to describe the motion of the nuclei,  equation \ref{eq:SEint} is multiplied from the left by the $j-th$ electronic eigenstate $\langle \Phi_{j}(\mathbf{q};\mathbf{Q})|$ and integrated over all electronic coordinates. The resulting set of differential equations for the nuclear wavefunctions $\mathbf{U}_{i,v} \equiv \chi_{(i,v)}\left(\mathbf{Q}\right)$ can be written in the matrix form as:

\begin{equation}
\left(\hat{K}_{nuc}(\mathbf{Q},\Theta)+\mathbf{V}(\mathbf{Q})\right) \mathbf{U}=\mathbf{U}\mathbf{E}^{int}
\label{eq:vibronic}
\end{equation}
where 
\begin{equation}
\mathbf{V}_{ij}(\mathbf{Q})=E_i(\mathbf{Q})\delta_{ij}+\mathbf{C}_{ij}(\mathbf{Q})
\end{equation}
Here $E_i(\mathbf{Q})$ stands for the potential energy surface of the $i-th$ electronic state and the \textit{vibronic coupling} elements $\mathbf{C}_{ij}(\mathbf{Q})$ are given as

\begin{equation}
\mathbf{C}_{ij}(\mathbf{Q})=\sum_{\alpha=1}^{N_{nuc}-1} \left(\vec{\mathbf{F}}^{\alpha}_{ij}\cdot \vec{\nabla}_{\alpha}+G^{\alpha}_{ij}\delta_{ij}\right)
\label{eq:vibcouplel}
\end{equation}
where summation goes over $N_{nuc}-1$ nuclei. The off-diagonal terms in eq. \ref{eq:vibcouplel} are referred to as the vibronic coupling elements or non-adiabatic coupling terms (NACT) written explicitly as

\begin{equation}
\vec{\mathbf{F}}^{\alpha}_{ij}(\mathbf{Q})=\frac{1}{M_{\alpha}}\langle \Phi_{i}(\mathbf{q};\mathbf{Q})|\left( \vec{\nabla}_{\alpha} \Phi_{j}\right)(\mathbf{q};\mathbf{Q})\rangle
\label{eq:NACT}
\end{equation}
and the diagonal terms

\begin{equation}
G^{\alpha}_{ii}(\mathbf{Q})=\frac{1}{M_{\alpha}}\langle \Phi_{i}(\mathbf{q};\mathbf{Q})|\left(\Delta_{\alpha} \Phi_{i}\right)(\mathbf{q};\mathbf{Q})\rangle
\label{eq:DBOC}
\end{equation}
represent corrections to potential energy surfaces resulting from the coupling of nuclear and electronic motions, which it is often called diagonal Born-Oppenheimer correction (DBOC).

Solutions to the system of differential equations given in eq. \ref{eq:vibronic} represent full solutions to the molecular \SE\ from eq. \ref{eq:SEint}. To obtain these solutions however one needs to solve eq. \ref{eq:SEel} for many electronic states $\Phi_{i}(\mathbf{q};\mathbf{Q})$ followed by solution to the coupled set of equations \ref{eq:vibronic} for the nuclear wavefunctions. Unfortunately, this is often difficult, and in some cases impossible in the present representation.

The big difference in masses of the nuclei and the electrons in molecules allows to assume that the electron cloud follows the motion of the nuclei nearly instantaneously from the perspective of the dynamics of the nuclei. Quantum-mechanically it means that the nuclear motion does not mix eigenstates of the electrons. 
Consistently with this assumption one can neglect to a good approximation the $\vec{\mathbf{F}}^{\alpha}_{ij}\cdot \vec{\nabla}_{\alpha}$  terms in eq. \ref{eq:vibronic}, which couple different electronic states. The $G^{\alpha}_{ii}$ elements are also inversely proportional to nuclear masses and are dropped in the Born-Oppenheimer approximation. $\mathbf{F}^{\alpha}_{ij}$ contains the derivative of the electronic wavefunction with respect to nuclear coordinates $\left( \vec{\nabla}_{\alpha} \Phi_{j}\right)(\mathbf{q};\mathbf{Q})$. Because the dynamics of nuclei is expected not to affect the electronic eigenstates, the electronic wavefunctions, as functions of nuclear coordinates, are very smooth, so that the dynamics (momentum operator) of the nuclei does not generate significant change in the electronic wavefunction. For this reason the vibronic coupling elements $\vec{\mathbf{F}}^{\alpha}_{ij}$ and $\mathbf{G}^{\alpha}_{ii}$  are to a good approximation negligible. As a result of neglecting the non-adiabatic coupling terms and the DBOC term in eq. \ref{eq:vibronic}, the system of differential equations \ref{eq:vibronic} becomes separable. The separability of eq. \ref{eq:vibronic} leads to separability of the resulting molecular wavefunction, which is equivalent to truncating the Born-Huang expansion \ref{eq:BHexpansion} at the first term:
\begin{equation}
\Psi_{i,\nu}^{BO}(\mathbf{q},\mathbf{Q},\Theta) = \chi_{(i,\nu)} \left( \mathbf{Q},\Theta \right)\Phi_{i}(\mathbf{q};\mathbf{Q})
\label{eq:BOwf}
\end{equation}
where $i$ labels isolated electronic states and $\nu$ labels rotational-vibrational states. Then the full system of rotational-vibrational-electronic Schrodinger equations given in \ref{eq:vibronic} becomes a set of rotational-vibrational Schrodinger equations for each electronic state separately:
\begin{equation}
\left(\hat{K}_{nuc}(\mathbf{Q},\Theta)+E_i(\mathbf{Q})\right) \mathbf{U}^{BO}=\mathbf{U}^{BO}\mathbf{E}^{int}_i \qquad i=0,1,2,...
\label{eq:BO}
\end{equation}
where $\mathbf{U}^{BO}_{\nu}=\chi_{(i,\nu)} \left( \mathbf{Q},\theta \right)$ is a vector of rotational-vibrational states for a given electronic state $i$.
As a result of the Born-Oppenheimer approximation presented above, the partial differential equation for  $3N_{el}+3N_{nuc}-3$  coupled coordinates has been decomposed into the electronic \SE\ with  $3N_{el}$ coordinates and a decoupled nuclear \SE\ with  $3N_{nuc}-3$ coordinates. This decoupling often represent a significant reduction in the computational complexity and enables to apply first-principles quantum mechanics in making predictions of chemical properties of molecules.

Decoupling of the nuclear and electronic motion is not always a good approximation. For instance, when the electronic energy as a function of the nuclear coordinates (the PES) is approaching the energy of another electronic state, it is more and more likely that some vibrations of atoms in molecule can cause change in the electronic state. This means that the motion of the nuclei and electrons can no longer be treated separately. The vibrational motion of the nuclei becomes correlated with the shape of the electron cloud in the molecule and \textit{vice versa}. The magnitude of mixing of two electronic states by means of the change in an internal nuclear coordinate is a measure of the vibronic coupling. Qualitatively the mixing of electronic states by the dynamics of nuclei can be described by the expression
\begin{equation}
\vec{\mathbf{F}}^{\alpha}_{ij}=\frac{\langle \Phi_i|( \vec{\nabla}_{\alpha}\hat{H}_{el})|\Phi_j\rangle}{E_j(\mathbf{Q})-E_i(\mathbf{Q})}
\label{eq:BOcommut}
\end{equation}
Analysis of eq. \ref{eq:BOcommut} suggests that the smaller the separation of potential energy surfaces of two electronic states $E_i(\mathbf{Q})$ and $E_j(\mathbf{Q})$, the stronger the mixing between the electronic states due to the nuclear motion. In such cases the Born-Oppenheimer approximation breaks down. In practice, for example in calculations of infrared spectra of molecules we are usually interested in the electronic ground state, which is often energetically well isolated from other electronic states, hence the Born-Oppenheimer approximation holds well.

By finding solutions to the rotational-vibrational \SE\ \ref{eq:BO} a great amount of chemically relevant information about the molecule can be obtained. Molecular line lists are the end product of solving eq. \ref{eq:BO}.
Molecular line lists and derived quantities, such as heat capacity, partition functions or radiative lifetimes are widely used in atmospheric science, astrophysics, monitoring of volcanic activity, analysis of industrial combustion gases and many others areas of physics and engineering. 
Because many triatomic molecules are important infrared absorbing atmospheric species (CO$_2$, SO$_2$, H$_2$O, NO$_2$, N$_2$O etc.) the next section describes a procedure for solving  eq. \ref{eq:BO} in the electronic ground state, which is relevant for the absorption in the infrared. 

\section{General rotational-vibrational Hamiltonian for a triatomic molecule}
\label{sec:KEO}

In this section we discuss a particularly successful procedure introduced by Sutcliffe and Tennyson for deriving the molecular Hamiltonian in the Born-Oppenheimer approximation. The procedure will be applied to triatomic molecules, which beside their simplicity, possess many characteristics of larger polyatomic molecules. 9 Cartesian coordinates are required to describe the positions of the nulcie in a triatomic molecule; three of them describe the motion of the centre-of-mass, three parametrize rotations of the molecule in 3-D space and the remaining three correspond to internal coordinates describing vibrations of the nuclei around the molecular centre-of-mass.

Because the quantization rules of the canonical position and momentum observables in the Cartesian representation are straightforward, it is historically, practically and perhaps pedagogically justified to begin with the nuclear Hamiltonian written in Cartesian coordinates. Therefore beginning with the Cartesian representation, we will proceed with deriving the Born-Oppenheimer nuclear Hamiltonian given in eq. \ref{eq:BO}: $\hat{H}_{nuc}(\mathbf{Q},\Theta)=\hat{K}_{nuc}(\mathbf{Q},\Theta)+E(\mathbf{Q})$ expressed in terms of chosen internal coordinates.  

\begin{figure}[H]
\begin{center}
  \includegraphics[width=10cm]{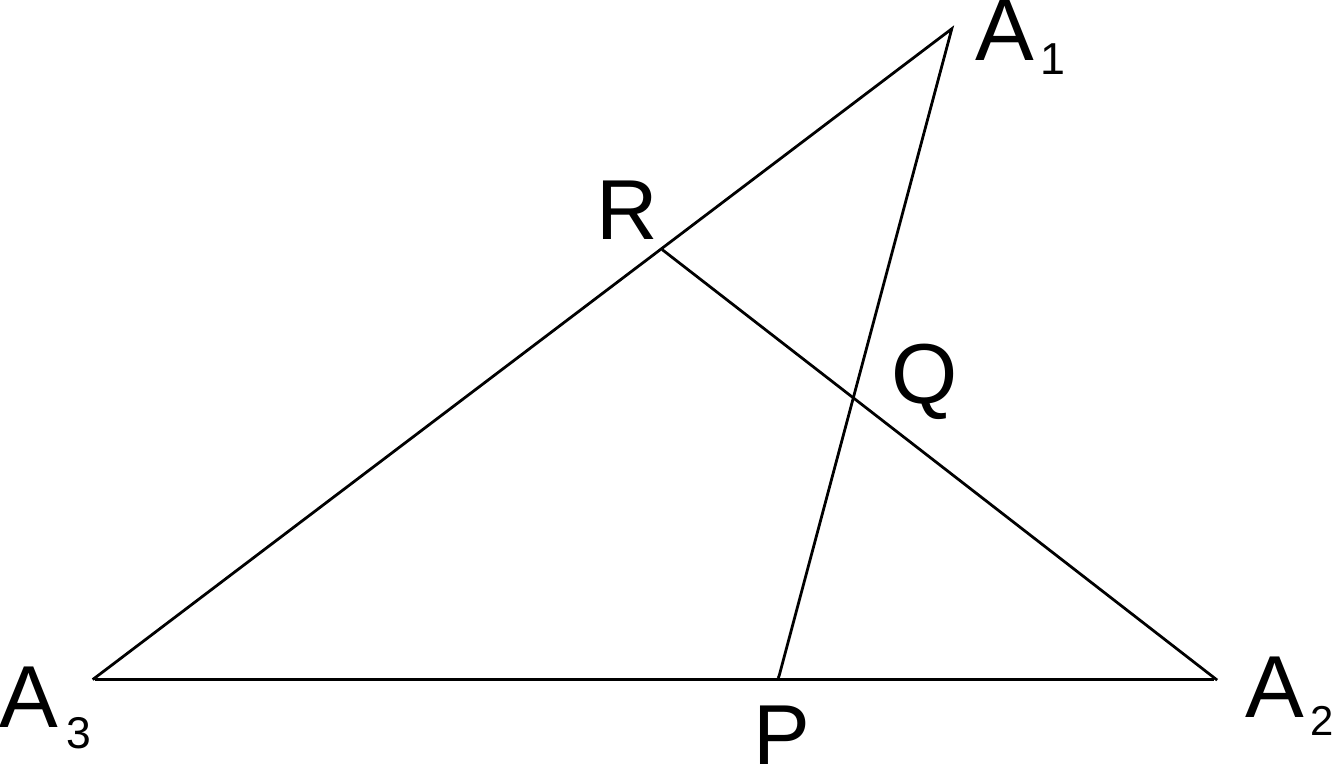}
\caption{A generalized coordinate system for a triatomic system introduced by Sutcliffe and Tennyson \cite{Sutcliffe1986}. $A_i$ represents the point at which atom \textit{i} is positioned.}
\label{fig:coordinates}
\end{center}
\end{figure}

There is a number of ways of deriving the nuclear Hamiltonian for a given choice of internal curvlinear (non-Cartesian) coordinates. An elegant and general framework has been proposed by Sutcliffe and Tennyson \cite{Sutcliffe1986,Sutcliffe2007}.
In their approach the $3N$ molecular Cartesian coordinates $\lbrace\textbf{x}\rbrace_{j=1,...,9}$ are transformed into three centre-of-mass coordinates and a set of six translationally invariant coordinates, by means of the following relations 

\begin{equation}
\textbf{t}_i=\sum_{j=1}^3 \textbf{x}_j V_{ji} \qquad i=1,2
\label{eq:tcoord}
\end{equation}
where  $V_{ji}$ is the transformation matrix element satisfying the condition: $\sum_{j=1}^3 V_{ji}=0$ for $i=1,2$. This condition ensures the translational invariance in the set of the $\textbf{t}_i$ coordinates, which can now be called the \textit{space-fixed coordinate system}, i.e. the coordinate system which moves along with the centre-of-mass of the molecule. In this new frame of reference, the three translational degrees of freedom of the centre-of-mass are fixed and the corresponding kinetic energy operator can be removed, thus one can focus on the remaining six internal degrees of freedom only.
The nuclear Hamiltonian given in the translationally-invariant coordinates takes a general form
 \begin{equation}
\hat{H}_{nuc}(\textbf{t}_1,\textbf{t}_2)=\hat{K}_{nuc}(\textbf{t}_1,\textbf{t}_2)+E(\textbf{t}_1,\textbf{t}_2)
\label{eq:BOt}
\end{equation}
where $\textbf{t}_1,\textbf{t}_2$ are three-dimensional vectors describing the six internal degrees of freedom of a triatomic molecule. The nuclear Hamiltonian expressed in terms of the $\textbf{t}_i$ coordinates reads:
\begin{equation}
\hat{H}_{nuc}(\textbf{t}_1,\textbf{t}_2)=-\frac{1}{2}\sum_{i,j=1}^2\frac{1}{\mu_{ij}}\vec{\nabla}(\textbf{t}_i)\cdot \vec{\nabla}(\textbf{t}_j)+E(\textbf{t}_1,\textbf{t}_2)
\label{eq:Htcoord}
\end{equation}
where $\vec{\nabla}(\textbf{t}_i)$ is the Nabla operator in $\textbf{t}_i$ coordinate and $\mu^{-1}_{ij}=\sum_{k=1}^3 m^{-1}_k V_{ki}V_{kj}$. 
Following Sutcliffe and Tennyson \cite{Sutcliffe1986,Sutcliffe2007} the transformation given in eq. \ref{eq:tcoord} can be uniquely characterized by two independent parameters, $g_1$ and $g_2$ in the following way:

\begin{equation}
\textbf{V}=\begin{pmatrix}
-g_2 & 1 \\
1   & -g_1 \\
g_2-1 & g_1-1 
\end{pmatrix}
\label{eq:Vdef}
\end{equation}
which are determined from the geometric definition of the internal coordinate system, as displayed in Figure \ref{fig:coordinates}. The choice of values for $g_1$ and $g_2$ defines two translationally-invariant vectors, from which a desired set of three internal coordinates of the molecule can be derived.  

With the definition presented in Figure \ref{fig:coordinates}, the geometric parameters $g_1$ and $g_2$ can be written as:
\begin{equation}
g_1=\frac{A_3P}{A_3A_2} \qquad g_2=\frac{A_3R}{A_3A_1}
\label{eq:g1g2}
\end{equation}
and $0 \leq g_1,g_2 \leq 1 $. $A_3P$ stands for the length of the segment connecting points $A_3$ and $P$. Different choices of $g_1$ and $g_2$ yield different popular types of internal coordinates: $g_1=\frac{m_1}{m_2+m_3}, g_2=0$ gives Jacobi (scattering) coordinates \cite{DVR3D}, $g_1=0, g_2=0$ gives bond-length-bond-angle coordinates and $g_1=1-\frac{\alpha}{\alpha+\beta-\alpha\beta}, g_2=1-\frac{\alpha}{1-\beta+\alpha\beta}$ defines Radau coordinates  \cite{Johnson1986}, with $\alpha = \left(\frac{m_3}{m_1+m_2+m_3}\right)^{\frac{1}{2}}$ and $\beta = \frac{m_2}{m_1+m_2}$.  

In the computational practice,  direct solution to the stationary Schr{\"o}dinger equation \ref{eq:Htcoord} is inconvenient, especially when a large number of energy levels is needed. This is because the Cartesian representation of the Hamiltonian does not make use of any symmetries of the system. For this reason, it is more convenient to pick a coordinate system in which the internal energy of the system can be at least approximately decomposed into contributions from vibrational and rotational degrees of freedom. A choice of such coordinates not only provides an intuitive picture of the quantum state of the system but also allows to make very efficient approximations. Which of the internal nuclear coordinates are selected is primarily dictated by the geometry and mass distribution in the molecule. The second element which should be correlated with the choice of internal coordinates is the quantum-mechanical basis set used to represent the wavefunction. As a rule of thumb, a good choice of internal coordinates corresponds to a matrix representation of the nuclear Hamiltonian, which is possibly close to diagonal. Appropriately chosen internal coordinates can significantly reduce the computational cost of calculations, by ensuring fast convergence of the variational procedure \cite{jt312}. Particularly welcome are sparse matrix representations of the Hamiltonian, as they can be efficiently used with an iterative eigensolver.

\begin{figure}[H]
\centering
\includegraphics[width=10cm]{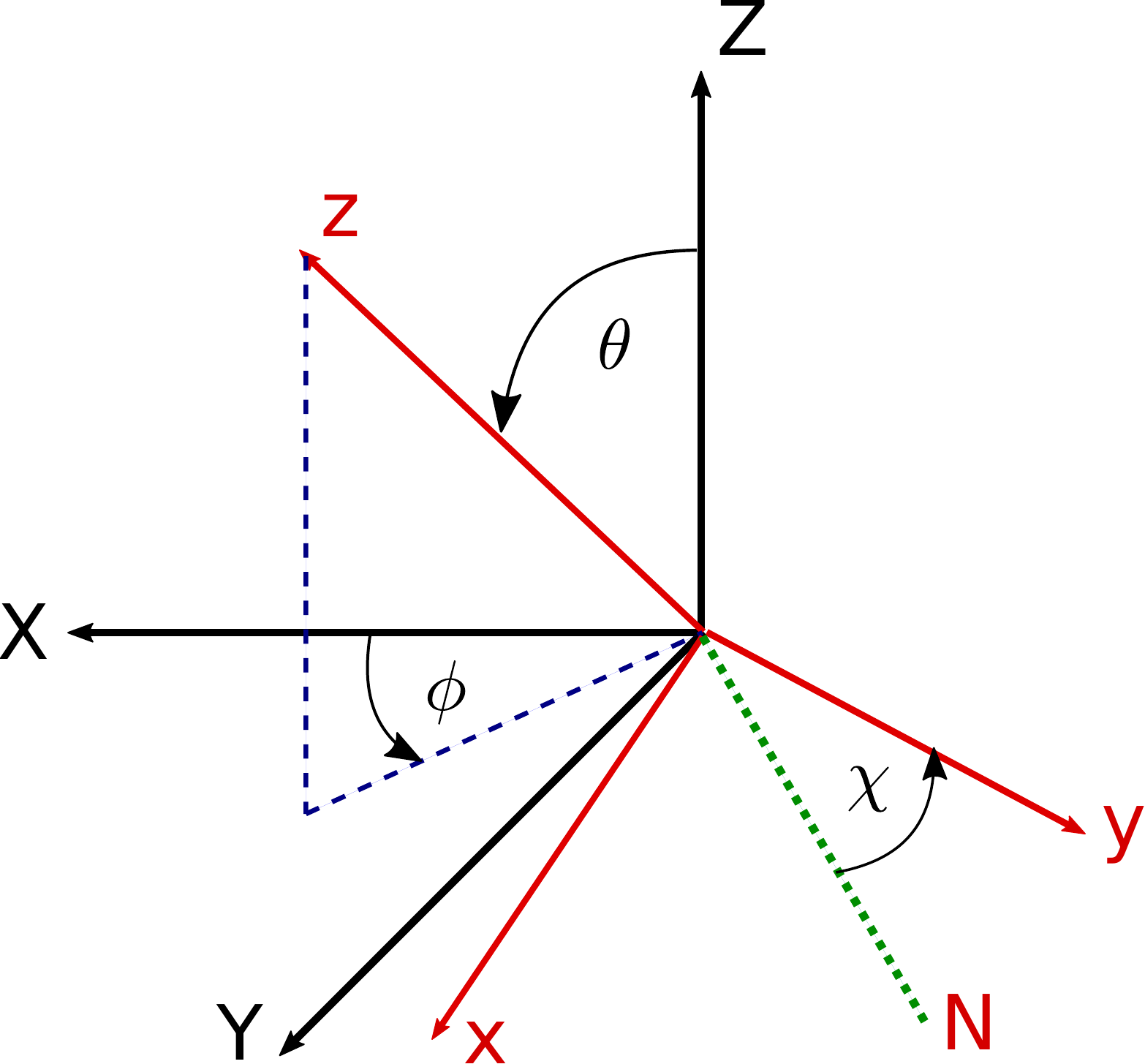}
\caption[Euler Angles in ZYZ' convention]{Euler angles in the ZYZ' convention.}
\label{fig:euler}
\end{figure}

The length and the mutual orientation of the $\textbf{t}_1$ and $\textbf{t}_2$ vectors determine the three internal coordinates of the system, whereas directions of $\textbf{t}_1,\textbf{t}_2$ in space define the orientation of the molecule with respect to the space-fixed frame, as displayed in Figure \ref{fig:euler}. The internal coordinates will be denoted as follows: $r_1=|\textbf{t}_1|$,  $r_2=|\textbf{t}_2|$ and $\gamma=arccos\left(\frac{\textbf{t}_1 \cdot \textbf{t}_2}{|\textbf{t}_1|\cdot |\textbf{t}_2|}\right)$. The translation-free Hamiltonian from eq. \ref{eq:Htcoord} can be now transformed into a Hamiltonian, which is a function of $r_1,r_2,\gamma$ and three angles $(\theta,\phi,\chi)$ determining the orientation of the $\textbf{t}_1,\textbf{t}_2$ vectors with respect to the space-fixed coordinate system. 

Relating the orientation of $\textbf{t}_1,\textbf{t}_2$ vectors to coordinates of nuclei in molecule is called embedding of the molecule-fixed coordinate system. Embedding of the molecule-fixed coordinate frame is mathematically realised by an orthogonal transformation (rotation) $\mathbf{C}$ of vectors $\textbf{t}_1,\textbf{t}_2$:

\begin{equation}
\textbf{t}_i= \mathbf{C} \cdot \textbf{z}_i \qquad i=1,2.
\label{eq:Crotation}
\end{equation}

The orientation of the new vectors $\textbf{z}_1, \textbf{z}_2$ defined by the above transformation depends on the orientation of the nuclei in space. We have full freedom in choosing how these body-fixed vectors are related to nuclear coordinates.
Atoms in triatomic molecules always lay in a plane. It is thus reasonable to choose the molecule-fixed frame so that one of the axis is perpendicular to the molecular plane. In such case we can write a general condition for $\textbf{z}_1, \textbf{z}_2$  expressed in terms of three internal coordinates $r_1,r_2,\gamma$:
\begin{equation}
\mathbf{z}=\left(\begin{array}{cc}
z_{x_1} &z_{x_2} \\
0 & 0 \\
z_{z_1}  & z_{z_2}  
 \end{array} \right) 
 \label{eq:embedding}
\end{equation}
where $\mathbf{z}=\textbf{C}^T \mathbf{t}$. Here we assumed that the \textit{y}-axis of the molecule-fixed frame is perpendicular to the molecular plane. The conditions for the embedding of the molecule-fixed frame can be written in terms of a single parameter $a$:

\begin{equation}
\mathbf{z}=\left(\begin{array}{cc}
-r_1\sin(a\gamma) & r_2\sin((1-a)\gamma)\\
0 & 0 \\
r_1\cos(a\gamma) & r_2\cos((1-a)\gamma) 
 \end{array} \right) 
\end{equation}
where $a\in [0,1]$ determines the orientation of the molecule-fixed \textit{z}-axis with respect to the direction of $r_1$ (more specifically $r_1$ makes an $a\gamma$ angle with the \textit{z}-axis). For instance, for the embedding of the molecule-fixed frame in which the $x$-axis bisects the angle between the directions of $r_1$ and $r_2$ we get 
\begin{equation}
\mathbf{z}=\left(\begin{array}{cc}
r_1\cos(\gamma/2) & r_2\cos(\gamma/2)\\
0 & 0 \\
-r_1\sin(\gamma/2) & -r_2\sin(\gamma/2) 
 \end{array} \right)
\end{equation}
and for the embedding in which the $z$-axis of the molecule-fixed frame is placed along the direction of $\mathrm{a}=r_1$ or $\mathrm{a}=r_2$ we have
\begin{equation}
\mathbf{z}=\left(\begin{array}{cc}
0 &  \mathrm{a}\sin\gamma\\
0 & 0 \\
 \mathrm{a} &  \mathrm{a}\cos\gamma 
 \end{array} \right) 
 \label{eq:bond}
\end{equation}
which is often referred to as the \textit{bond embedding}. 

The choice of embedding together with a particular parametrization of the $\textbf{C}$ rotation matrix fully defines the relation between the space-fixed set of translationally-invariant coordinates $\mathbf{t}_1, \mathbf{t}_2$ and the set of coordinates in the molecule-fixed frame (internal coordinates). 
Note that because the rotation matrix $\mathbf{C}$ is orthogonal, that is $|\det \mathbf{C}| = 1$ the lengths of $\mathbf{t}$ and $\mathbf{z}$ are identical, so are the scalar products $\mathbf{t}_1\cdot \mathbf{t}_2$ and  $\mathbf{z}_1\cdot \mathbf{z}_2$. This means that the internal coordinates $r_1,r_2,\gamma$ expressed as lengths and the angle between $\mathbf{t}_1$ and $\mathbf{t}_2$, respectively, are unchanged by the transformation to the molecule-fixed frame. As a result these new coordinates $\mathbf{z}_1, \mathbf{z}_2$ are invariant with respect to rotations in three-dimensional space.  The transformation matrix $\textbf{C}$ can be parametrized in terms of the \textit{Euler angles} $(\phi,\theta,\chi)$ \cite{06BuJexx.method,Wilson1955}.

Euler angles define the position of the rotated reference frame with respect to the space-fixed frame. There are several conventions for defining the Euler angles. For the present purposes the  \textit{ZYZ'} convention is most suitable \cite{Wilson1955, 06BuJexx.method}, as shown in Figure \ref{fig:euler}. In the  \textit{ZYZ'} convention a sequence of three rotations to the space-fixed frame are applied in the following order: $\phi$ angle rotation around the $Z$-axis, followed by $\theta$ angle rotation around the $Y$-axis, and finally $\chi$ angle rotation around the $z$-axis. It is also convenient to introduce a \textit{node line} $N$ which represents the positive sense of rotation from the $Z$-axis to the $z$-axis and lays in the intersection of $xy$ and $XY$ planes. Ranges for the Euler angles are $\phi,\chi \in [0,2\pi]$ and $\theta \in [0,\pi]$, so that all possible orientations in 3-D space are covered.
The $\textbf{C}$ transformation matrix given in eq.  \ref{eq:Crotation} can be found as the product of rotation matrices along the respective  \textit{Z}, \textit{Y} and \textit{Z}' directions: 
\small
\begin{align}
 &  R_{y}(\alpha)=\left(\begin{array}{ccc}
\cos\alpha& 0 & \sin\alpha \\
0 & 1 & 0\\
-\sin\alpha & 0  & \cos\alpha \end{array} \right) && \\ \nonumber
 & R_{z}(\alpha)=\left(\begin{array}{ccc}
\cos\alpha & -\sin\alpha & 0\\
 \sin\alpha & \cos\alpha & 0 \\
 0 & 0 & 1
 \end{array} \right) &&
 \label{rotation}
\end{align} 
Therefore the general rotation matrix in the $ZYZ'$ scheme is given by
\scriptsize
\begin{equation}
\textbf{C}(\theta,\phi,\chi)=R_{z'}(\chi)R_{y}(\theta)R_{z}(\phi)=\left(\begin{array}{ccc}
\cos\theta\cos\phi\cos\chi-\sin\phi\sin\chi & \cos\theta\sin\phi\cos\chi+\cos\phi\sin\chi & -\sin\theta\cos\chi \\
-\cos\theta\cos\phi\sin\chi-\sin\phi\cos\chi & -\cos\theta\sin\phi\sin\chi+\cos\phi\cos\chi & \sin\theta\sin\chi\\
\sin\theta\cos\phi & \sin\theta\sin\phi & \cos\theta \end{array} \right)
\label{eq:euler}
\end{equation} 
\normalsize
where the Euler angles $\phi,\theta,\chi$ become rotational coordinates of the system. With the transformation given in eq. \ref{eq:Crotation}, it is possible with the use of the \textit{chain rule} to express the Hamiltonian given in eq. \ref{eq:Htcoord} in a form which depends on the internal coordinates $r_1,r_2,\gamma$ and the total angular momentum operators $\hat{J}_{x}, \hat{J}_{y}, \hat{J}_{z}$. These angular momentum operators generate rotations associated with the $x,y$ and $z$ axis of the molecule-fixed coordinate frame, respectively, and can be expressed analytically in terms of the three Euler angles $ \phi, \chi$ and $\theta$ \cite{06BuJexx.method,Wilson1955}. Further on we assume that the molecule is located in the x--z plane of the right-handed ($\det(\textbf{C})=+1$) molecule-fixed coordinate system.

Now, with the use of the \textit{chain-rule} the space-fixed Hamiltonian of eq. \ref{eq:Htcoord} can be transformed into the  Hamiltonian in molecule-fixed coordinate frame. After a rather tedious algebra, details of which are presented in ref. \cite{Sutcliffe2007}, the final molecule-fixed Hamiltonian is given by:

\begin{equation}
\small
\hat{H}(r_1,r_2,\gamma,\phi,\theta,\chi)= \hat{K}^{(1)}_V(r_1,r_2,\gamma)+\mu_{12}^{-1}\hat{K}^{(2)}_V(r_1,r_2,\gamma)+\hat{K}_{VR}(r_1,r_2,\gamma,\phi,\theta,\chi)+E(r_1,r_2,\gamma)
\label{eq:Hbody}
\end{equation}

where:

\begin{equation}
\begin{split}
\hat{K}^{(1)}_V(r_1,r_2,\gamma)=-\frac{1}{2}\left[\frac{1}{\mu_1r_1^2}\left(\frac{\partial}{\partial r_1}r_1^2\frac{\partial}{\partial r_1}+\frac{1}{\sin\gamma}\frac{\partial}{\partial \gamma}\sin\gamma\frac{\partial}{\partial \gamma}\right)\right.+ \\
\left. + \frac{1}{\mu_2r_2^2}\left(\frac{\partial}{\partial r_2}r_2^2\frac{\partial}{\partial r_2}+\frac{1}{\sin\gamma}\frac{\partial}{\partial \gamma}\sin\gamma\frac{\partial}{\partial \gamma}\right)\right]
\end{split}
\label{eq:Kv1}
\end{equation}

\begin{equation}
\begin{split}
\hat{K}^{(2)}_V(r_1,r_2,\gamma)= - \cos \gamma \frac{\partial^2}{\partial r_1 \partial r_2} + \frac{\cos \gamma}{r_1r_2}\left(\frac{1}{\sin\gamma}\frac{\partial}{\partial \gamma}\sin\gamma\frac{\partial}{\partial \gamma}\right)+\\
+\sin\gamma\left(\frac{1}{r_1}\frac{\partial}{\partial r_2}+\frac{1}{r_2}\frac{\partial}{\partial r_1}+\frac{1}{r_1r_2}\right)\frac{\partial}{\partial \gamma}
\end{split}
\label{eq:Kv2}
\end{equation}

\begin{equation}
\begin{split}
\hat{K}_{VR}(r_1,r_2,\gamma,\phi,\theta,\chi)=\frac{1}{2}\left[M_{xx}\hat{J}_x^2+M_{yy}\hat{J}_y^2+M_{zz}\hat{J}_z^2+M_{xz}\left(\hat{J}_x\hat{J}_z+\hat{J}_z\hat{J}_x\right)\right]+\\
+\frac{1}{i}\left[\left(\frac{1-a}{\mu_1r_1^2}-\frac{a}{\mu_2r_2^2}\right)\left(\frac{\partial}{\partial\gamma}+\frac{\cot}{2}\right)+\frac{2a-1}{\mu_{12}r_1r_2}\left(\cos\gamma\frac{\partial}{\partial\gamma}+\frac{1}{2\sin\gamma}\right)\right.+\\
\left.+\frac{\sin\gamma}{\mu_{12}}\left(\frac{a}{r_2}\frac{\partial}{\partial r_1}-\frac{(1-a)}{r_1}\frac{\partial}{\partial r_2} \right)\right]\hat{J}_y
\end{split}
\label{eq:Kvr}
\end{equation}
where $\hat{J}_i$ are the molecule-fixed angular momentum operators, obeying the standard commutation relations $\left[ \hat{J}_i,\hat{J}_j\right] =i\epsilon_{ijk}\hat{J}_k$ (where $\epsilon_{ijk}$ is the totally antisymmetric Levi-Civita tensor). $M_{\alpha\beta}$ are the elements of the inverse generalized moment of inertia tensor, given explicitly in ref. \cite{jt95}.
$\mu_{12}^{-1}$, $\mu_{1}^{-1}$ and $\mu_{2}^{-1}$ are reduced massed depending on the choice of internal coordinates, as defined in eq. \ref{eq:Htcoord}.  The Jacobian associated with the space-fixed to molecule-fixed transformation is given by $r^2_1r^2_2\sin\gamma$. 

In this article, we restrict ourselves to using \textit{orthogonal coordinates} only, which by definition give $\mu_{12}^{-1}=0$ \cite{Sutcliffe1986}, so that the second term in eq. \ref{eq:Hbody} vanishes. Jacobi(scattering coordinates), Radau and bond coordinates are examples of orthogonal internal  coordinates. 
The last term in eq. \ref{eq:Hbody}, $E(r_1,r_2,\gamma)$ is the potential energy surface (PES) for a given electronic state, defined as the total electronic energy in this state for a given configuration of clamped-nuclei. Note that, as discussed in section \ref{sec:BO}, the PES depends only on internal nuclear coordinates of the system, hence is invariant to transformations between the molecule-fixed and space-fixed frames.  

To summarize, we arrived at an explicit and exact form of the molecular Hamiltonian in the Born-Oppenheimer approximation for a triatomic molecule for a given choice of internal coordinates. The Hamiltonian written in eq. \ref{eq:Hbody} is exactly the Hamiltonian from eq. \ref{eq:BO} where $\hat{K}_{nuc}(\mathbf{Q},\Theta)=\hat{K}^{(2)}_V(r_1,r_2,\gamma)+\hat{K}^{(1)}_V(r_1,r_2,\gamma)+\hat{K}_{VR}(r_1,r_2,\gamma,\phi,\theta,\chi)$.

Any choice of the internal curvilinear coordinates inevitably generates singularities in the ro-vibrational Hamiltonian. Singularity is a set of internal coordinates for which the Hamiltonian becomes infinite. By inspecting eqs. \ref{eq:Kv1}-\ref{eq:Kvr} we can see that the Hamiltonian becomes infinite at $r_1=0$ or $r_2=0$ or when $\gamma=0,\pi$. These points in the configuration space are called the singular points of the Hamiltonian. Due to the presence of singularities, expectation values of observables and scalar products can be obscured by a large magnitude numerical noise. It is therefore reasonable to choose coordinates for which these singularities can be appropriately handled by the choice of the basis set. Sometimes, when it is impossible to remove singularities by a choice of the basis set, it is sensible to choose internal coordinates so that when they occur, singularities unveil at geometries with very high electronic potential energy, so that the wavefunction is nearly vanishing in these regions.

\section{Methods for solving the rotational-vibrational \SE}
\label{sec:methods}
\subsection{Variational basis representation}
A choice of the rotational-vibrational basis set introduces a representation to the time-independent \SE\ in the form of an algebraic matrix equation $\mathbf{H}\mathbf{U}=\mathbf{S}\mathbf{UE}$, where $\mathbf{H}$ is the \textit{Hamiltonian matrix},  $\mathbf{S}$ is the \textit{overlap matrix}, $\mathbf{U}$ is the matrix of eigenvectors and $\mathbf{E}$ is the diagonal matrix of energies.
Once the basis set has been chosen, the calculation can
proceed with building (exactly) and diagonalizing the Hamiltonian matrix $\mathbf{H}$. This way of solving the \SE\ is called the variational basis representation (VBR) approach, which is by far the most popular computational method in nuclear motion theory.

The time-independent \SE\ can be written as
\begin{equation}
\hat{H}|\psi \rangle = E |\psi \rangle
\label{eq:hpsiepsi}
\end{equation}
where typically the wavefunction is represented as a linear combination of basis functions
\begin{equation}
|\psi \rangle = \sum_{n=1}^N u_n |\phi_n\rangle
\label{eq:linexp}
\end{equation}
with constant coefficients $u_n$ which need to be determined. $|\phi_n\rangle$ is the $n$-th basis function used to model the wavefunction $|\psi \rangle $. $N$ denotes the size of the variational basis set. Eq. \ref{eq:linexp} can be inserted into eq. \ref{eq:hpsiepsi} to give 
\begin{equation}
\sum_{n=1}^N u_n \hat{H}|\phi_n\rangle = E \sum_{m=1}^N u_m |\phi_m\rangle
\label{eq:linexp2}
\end{equation}
At this point, one possibility for proceeding with the representation of the \SE\  is to multiply eq. \ref{eq:linexp2} from the \textit{left} by a basis function $\langle \phi_i |$, which results in the following matrix equation:
\begin{equation}
\mathbf{H}^{VBR}\mathbf{U}=\mathbf{S}\mathbf{UE}
\label{eq:galerkin}
\end{equation}
where $H^{VBR}_{ij}=\langle \phi_i | \hat{H} | \phi_j \rangle$ is the matrix element of the Hamiltonian operator, $\mathbf{U}$ is the vector of coefficients $u_j$, where $j=1,...,N$ and $S_{ij}=\langle \phi_i | \phi_j \rangle$ is the overlap matrix. $VBR$ stands for \textit{variational basis representation}. This method of solving the \SE\ is called \textit{variational} or \textit{Galerkin} method. The second order differential \SE\ is replaced by a set of $N$ algebraic equations. The task is to solve the generalized eigenvalue problem given in eq. \ref{eq:galerkin} to determine $\mathbf{E}$ and $\mathbf{U}$.

In the Born-Oppenheimer approximation, when an exact kinetic energy operator is used, such as the one given in eq. \ref{eq:Hbody}, the only source of errors in the variational calculations are the error in the potential energy surface and the error associated with the finite variational basis set size. The former error is associated with imperfect representation to the electronic wavefunction of the molecule whereas the basis set size error indicates that the variational basis is not complete. Whenever a complete variational basis set is used, the solutions to the \SE\ can be represented exactly. Such a complete representation is given by the wavefunctions of the symmetric top model, for which analytic solutions to the \SE\ are known. Symmetric top eigenfunctions serve as a complete, finite-size basis set for the rotational motion of any molecule. 

The variational basis representation or \textbf{VBR} assumes that all matrix elements of the Hamiltonian $H^{VBR}_{ij}=\langle \phi_i | \hat{H} | \phi_j \rangle$ are calculated exactly. This is often very difficult if not impossible to achieve, especially when several internal coordinates of the molecule are coupled in the PES, meaning that a multidimensional integral has to be calculated.
For this reason, it is often necessary to employ approximate methods of evaluating matrix elements of the Hamiltonian. The following sections discuss three popular approximate methodologies for finding solutions to the \SE .

\subsection{Finite basis representation and discrete variable representation}
\label{sec:dvr}

\subsubsection{Finite basis representation}
The \textbf{VBR} Hamiltonian given in eq. \ref{eq:galerkin} can be expressed as a sum of the kinetic energy operator matrix and the potential energy matrix. Matrix elements of this Hamiltonian are integrals over all internal degrees of freedom of the nuclei (rotational and vibrational) $H^{VBR}_{ij}=\langle \phi_i | \hat{H} | \phi_j \rangle=\int_{V,\Omega}d\Omega d\vec{Q} \phi_i(\vec{Q})\hat{H}(\vec{Q},\Omega)\phi_j(\vec{Q})$. Integration over the rotational degrees of freedom $\Omega=(\theta,\phi,\chi)$ can be done analytically in the complete representation of the symmetric-top Hamiltonian, which will be discussed in section \ref{sec:rotbasis}. However for the integration over vibrational degrees of freedom $\vec{Q}=(Q_1,...,Q_D)$ one typically must use a quadrature. A quadrature is defined by its grid points $\lbrace x_k \rbrace_{k=1,...,M}$ and quadrature weights $\lbrace w_k \rbrace_{k=1,...,M}$. 

\begin{figure}[H]
\begin{center}
  \includegraphics[width=14cm]{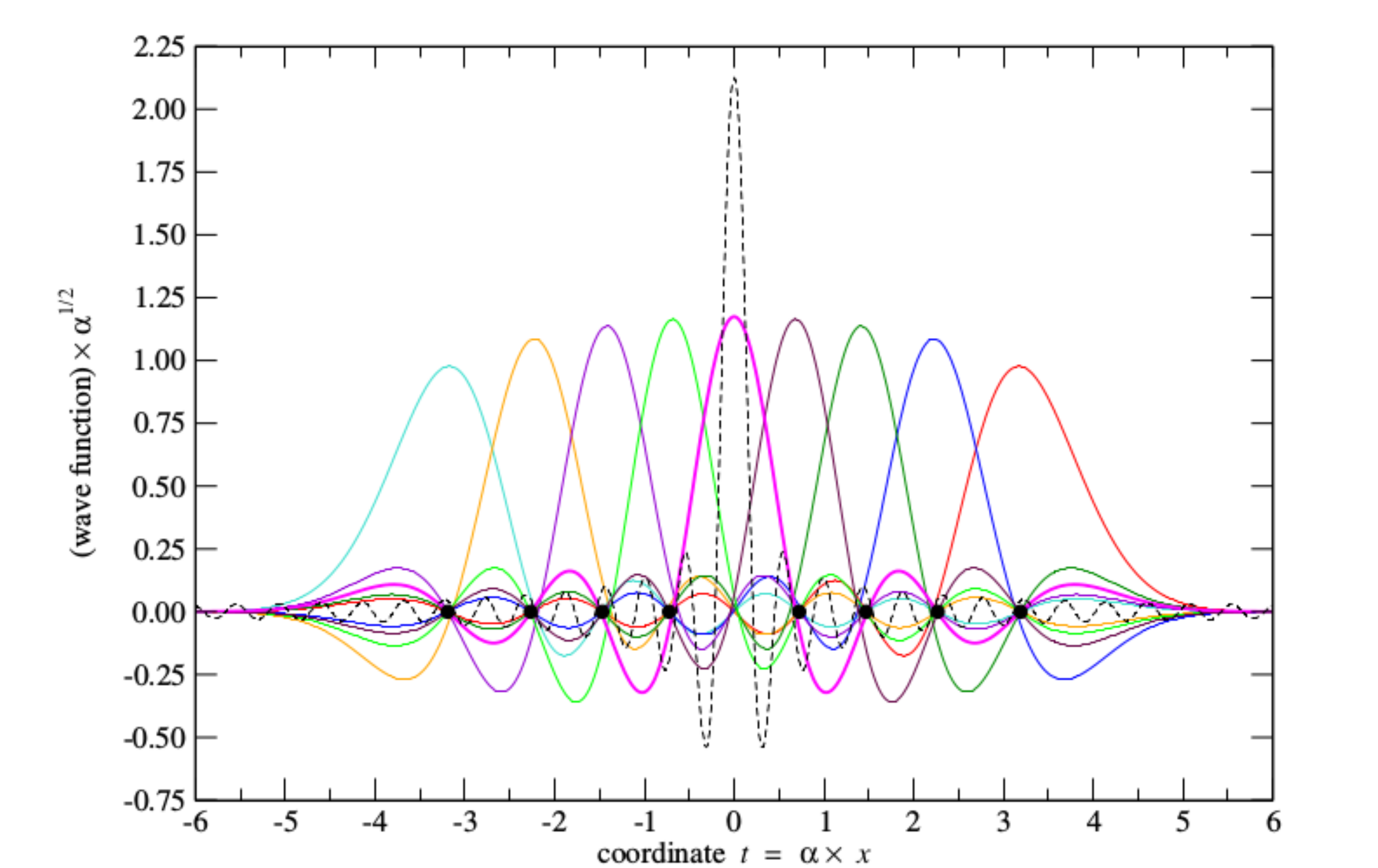}
  \caption{Example DVR basis function (dashed line) constructed from Harmonic oscillator wavefunctions (colours). Gauss-Hermite quadrature points are shown. Reproduced from PhD thesis of L.Lodi, UCL, 2008.}
  \label{Fig:DVRbasis}
  \end{center}
\end{figure}

In 1-dimensional problems the VBR Hamiltonian can be approximately written as
\begin{equation}
\mathbf{H}^{VBR}\approx \mathbf{H}^{FBR}=\mathbf{K}^{VBR}+\mathbf{T}^{T}\mathbf{V}^{diag}\mathbf{T}
\label{eq:FBR1D}
\end{equation}
 where the kinetic energy operator $\mathbf{K}^{VBR}$ remains in its VBR representation and a quadrature approximation was applied to the potential energy matrix elements $\mathbf{V}^{VBR} \approx \mathbf{T}^{T}\mathbf{V}^{diag}\mathbf{T}$ with $\mathbf{T}=\sqrt{\mathbf{w}/\omega(x)}\mathbf{B}$, where $\mathbf{w}_{kk'}=w_k\delta_{kk'}$ is the diagonal matrix of quadrature weights and $\mathbf{B}_{kn}=\phi_n(x_k)$ is the matrix of values of the $n$-th basis function $\phi_n$ at $k$-th grid point $x_k$. $\omega(x)$ is positive weight function associated with the quadrature. It is easily seen that $ \mathbf{V}^{VBR}_{ij}=\int \phi_i(x)V(x)\phi_j(x) dx \approx \sum_{k=1}^M \phi_i(x_k)w_{kk}V(x_k)\phi_j(x_k)\omega^{-1}(x_k)=\left(\mathbf{T}^{T}\mathbf{V}^{diag}\mathbf{T}\right)_{ij}$ is the quadrature approximation to $\mathbf{V}^{VBR}_{ij}$. Note that the matrix of quadrature weights $\mathbf{w}$ commutes with the diagonal matrix of values of the potential evaluated at grid points $\mathbf{V}^{diag}$. This approximate representation of the Hamiltonian matrix is called \textit{Finite Basis Representation} or \textbf{FBR}. In 1-D the \textbf{FBR} \SE\ is given by

\begin{equation}
\left(\mathbf{K}^{VBR}+\mathbf{T}^{T}\mathbf{V}^{diag}\mathbf{T}\right)\mathbf{U}=\mathbf{T}^T\mathbf{T}\mathbf{U}\mathbf{E}
\label{eq:FBR1DSE}
\end{equation}
In \textbf{FBR} it is necessary to evaluate $N(N+1)/2$ matrix elements in the potential energy part of the Hamiltonian matrix. When many internal coordinates are coupled in the PES the resulting \textbf{FBR} potential energy matrix can be a dense matrix, which is inconvenient for use with iterative eigensolvers. To circumvent this one can multiply eq. \ref{eq:FBR1DSE} from the left by $\mathbf{T}^{-T}$ and define a new basis by $\mathbf{Z}=\mathbf{T}\mathbf{U}$ which gives
\begin{equation}
\left(\mathbf{T}^{-T}\mathbf{K}^{VBR}\mathbf{T}^{-1}+\mathbf{V}^{diag}\right)\mathbf{Z}=\mathbf{Z}\mathbf{E}
\label{eq:DVR1DSE}
\end{equation}
In this new representation, called the \textit{discrete variable representation} or \textbf{DVR} the potential energy matrix is diagonal. Both the Hamiltonian matrix elements and the wavefunctions $\mathbf{Z}$ are labelled by grid indices, unlike in FBR. The \textbf{FBR}-to-\textbf{DVR} transformation $\mathbf{Z}=\mathbf{T}\mathbf{U}$ is a basis change from the basis labeled by functions to the basis labeled by grid points. The characteristic of DVR basis functions is their localization around grid points which label them, as shown in Figure \ref{Fig:DVRbasis}. The elements of the kinetic energy operator matrix $\mathbf{K}^{VBR}$ can usually be calculated analytically. The general form of the KEO is a sum of first and second derivative operators multiplied by functions of the internal coordinates. With a wise choice of internal coordinates the KEO can be represented by a sparse matrix which is convenient to use with iterative eigensolvers. 

Whenever the FBR or DVR matrix elements are non-exact, it is possible that the variational principle (see MacDonald's theorem \cite{Helgaker2000}) is not satisfied. In that case the FBR/DVR calculated energy levels can lie lower than the variationally converged energy levels. An FBR usually uses extra quadrature points to ensure variational behaviour.

In multi-dimensional problems the standard and easiest approach to define an FBR is to use a direct product basis set and a direct product quadrature grid. Direct product basis means that the indices of basis functions describing the internal coordinates are independent: $\lbrace \phi_{n_1}(Q_1)\phi_{n_2}(Q_2)...\phi_{n_D}(Q_D)\rbrace_{n_c=1,2,...,N_c; c=1,..,D}$. Direct product quadrature grid is a grid composed as a simple sum of 1-dimensional grids so that the indices of the multidimensional grid are independent. The multi-dimensional FBR \SE\ then takes the form
\begin{equation}
\left(\mathbf{K}^{VBR}+\mathbf{T}^{T}\mathbf{V}^{diag}\mathbf{T}\right)\mathbf{U}=\mathbf{T}^T\mathbf{T}\mathbf{U}\mathbf{E}
\label{eq:FBRDDSE}
\end{equation}
where $\mathbf{T}=\mathbf{^{(1)}T}\mathbf{^{(2)}T}...\mathbf{^{(D)}T}$ with $\mathbf{^{(c)}T}_{k_cn_c}=\sqrt{\frac{^{(c)}\mathbf{w}_{k_ck_c}}{\omega(Q_{k_c})}}\mathbf{^{(c)}B}_{k_cn_c}$. Here $\mathbf{^{(c)}B}_{k_cn_c}=\phi_{n_c}(Q_{k_c})$. Note that the size of the wavefunction $\mathbf{U}$ is $N^D$ which quickly becomes a prohibitively large number with increasing $D$. For instance, if 10 basis functions per dimension are needed  to converge the variational wavefunction, then for a 5-atom molecule ($D=9$) e.g. CH$_4$, as many as $10^9$ basis functions are required. For molecules with more than 5 atoms, the amount of memory needed to store even a single wavefunction vector becomes problematic. To circumvent this \textit{curse of dimensionality} several approaches have been developed to define FBR and DVR with non-direct product basis sets and non-direct product quadrature grids (see Refs. \cite{gaborig,AvCaV,Zak2019,Avila2017ch2nh}). For triatomic molecules ($D=3$) direct product basis can be used without concerns about memory.

\subsubsection{Gaussian DVRs}
The discrete variable representation (DVR) is a technique originally developed  by Harris \etal\ \cite{Harris1965} which was later implemented by Light \etal\ \cite{Bacic1989,Light2007} to solve quantum-mechanical problems in the nuclear motion theory.  Historically, the first DVRs used basis sets of orthogonal polynomials which represented solutions to the \SE\ for some model systems, e.g. Harmonic oscillator (Hermite polynomials), Hydrogen atom (Laguerre polynomials), Morse oscillator (associated Laguerre polynomials), particle in the square potential well (Chebyszev polynomials) or Legendre Polynomials for problems with spherical symmetry. The DVR grid points which label the DVR wavefunctions are then zeros of these orthogonal polynomials. The square $\mathbf{T}=\sqrt{\mathbf{w}/\omega(x)}\mathbf{B}$ matrix in eq. \ref{eq:FBR1DSE} is then defined over the zeros of these orthogonal polynomials with the diagonal matrix of weights ensuring that $\mathbf{T}^{\dag}\mathbf{T}=\mathbf{1}$. Quadrature schemes like those described above are referred to as the Gaussian quadratures. The associated DVR is called Gaussian DVR. For the past two decades this type of DVR has been very popular in computational quantum physics \cite{Light2007}, mainly due to their fast convergence even for complicated potential energy functions.

A $N$-point Gaussian quadrature is by its definition exact for polynomial functions up to degree $2N-1$.  The unitarity condition on the DVR-to-FBR transformation $\mathbf{T}$ represents the Gaussian quadrature approximation to the overlap integral: $\int_{-\infty}^{+\infty}\phi_i(x)\phi_j(x)dx \approx (\mathbf{T}^{\dag}\mathbf{T})_{ij}=\sum_{k=1}^N w_k \phi_i(x_k)\phi_j(x_k)\omega^{-1}(x_k)$ and because $i$ and $j$ are both smaller or equal $N-1$, their combined degree is maximally $2N-2$. This scheme works if the basis function remains a product of an orthogonal polynomial $p_n(x)$ and a positive weight function $\omega(x)$.

The DVR basis functions are localized around appropriate quadrature points. The type of quadrature is dictated by the choice of the orthogonal polynomial $p(x)$ and the exact location of points is given by the order of the quadrature $N$. The DVR basis functions become more localized as the number of points in the quadrature is increased and in the limit of infinite basis the DVR functions become \textit{Dirac delta} distributions:

\begin{equation} 
\begin{split}
\left(\mathbf{V^{DVR}}\right)_{kl}=\int d_k(x)V(x)d_l(x)dx  \longrightarrow \int \delta(x-x_k)V(x)\delta(x-x_l)dx=\\
=\int V(x_k)\delta(x_k-x_l)dx=V(x_k)\delta_{kl}
\end{split}
\label{delta}
\end{equation}
where $d_k(x)$ is a DVR basis function, localized around the $k$-th grid point. Finite size of the variational basis set does not allow perfect localization of the DVR functions. For this reason, in computational practice, when a finite set of variational basis functions is used, eq. \ref{delta} is satisfied only approximately: $\left(\mathbf{V^{DVR}}\right)_{ij}\approx V(x_i)\delta_{ij}$. 
Therefore when the potential energy operator matrix is calculated directly in the DVR basis, it is \textit{not diagonal}, although the off-diagonal elements are typically small. Nonetheless, the assumption that the potential matrix is diagonal in the DVR basis leads surprisingly to more accurate energy levels than if all accurate off-diagonal elements were kept in the calculation. This observation is discussed in Szalay \etal \cite{Szalay2011}. The diagonal DVR potential matrix $\mathbf{V}^{diag}$ should be rather viewed as a direct consequence of the quadrature approximation, where the quadrature error was transferred to the KEO part, as shown in eq. \ref{eq:DVR1DSE}. Calculating the potential energy operator matrix directly in the DVR basis typically gives a non-diagonal matrix, when the DVR basis is incomplete. But by making the quadrature approximation to the potential energy matrix elements in FBR followed by a unitary FBR-to-DVR transformation washes these off-diagonal elements to the KEO, hence one can expect more accurate results when the potential matrix is assumed diagonal, even in an incomplete DVR basis.

One interesting property of the DVR functions is that they vanish at all quadrature grid points except one. To show this let us take the FBR-to-DVR transformation:
 \begin{equation}
d_l(x)=\sum_{j=0}^{N-1}T_{lj}\phi_j(x)
\end{equation}
The new DVR function, when evaluated at the $k$-th grid point takes the value
\begin{equation}
d_l(x_k)=\sum_{j=0}^{N-1}\sqrt{w_l\omega(x_k)}p_j(x_l)p_j(x_k)
\end{equation}
and from orthogonality of $p_j(x)$ we find
 \begin{equation}
d_l(x_k)=\sqrt{\frac{\omega(x_k)}{w_l}}\delta_{lk}
\end{equation}

The DVR basis functions generally take non-zero values between quadrature points, however they tend to be smaller with increasing distance from the point of localization of the DVR function, as shown in Figure \ref{Fig:DVRbasis}. This observation intuitively supports the earlier discussed diagonal approximation to the potential matrix:
  \begin{equation}
  \begin{split}\label{eq:dvrkey}
V^{DVR}_{ij}=\langle d_i(x)|V(x)|d_j(x)\rangle=\int_a^bd_i(x)V(x)d_j(x)dx\approx\\
\approx \sum_{k=0}^{N-1}\frac{w_k}{\omega(x_k)}d_i(x_k)V(x_k)d_j(x_k)
=\sum_{k=0}^{N-1}\frac{w_k}{\omega(x_k)}\sqrt{\frac{\omega(x_k)}{w_i}}\delta_{ik}V(x_k)\sqrt{\frac{\omega(x_k)}{w_j}}\delta_{jk}=V(x_i)\delta_{ij}
\end{split}
\end{equation}
To summarize, Gaussian-quadrature based DVRs use the same number of points as basis functions, which are products of orthogonal polynomials and a weight function. For such basis functions the Gaussian quadrature integration is exact.  When this condition is fulfilled the DVR can be considered variational. For the potential energy operators represented as polynomial functions of coordinates (quadratic or higher degrees), the Gaussian quadrature rule is not exact and the FBR/DVR approximation can no longer be considered variational. One of the advantages of using DVR over FBR is that in the DVR the Hamiltonian matrix is often represented by a sparse matrix, which allows for the use of iterative eigensolvers and in a consequence solve problems, which require large basis sets. The potential energy matrix is diagonal in DVR, meaning that one only needs to know the values of the PES at quadrature grid points. The off-diagonal elements are transferred to the KEO part, in which the $\mathbf{K}^{VBR}$ matrix has often a banded structure.  

\begin{figure}[H]
\begin{center}
  \includegraphics[width=15cm]{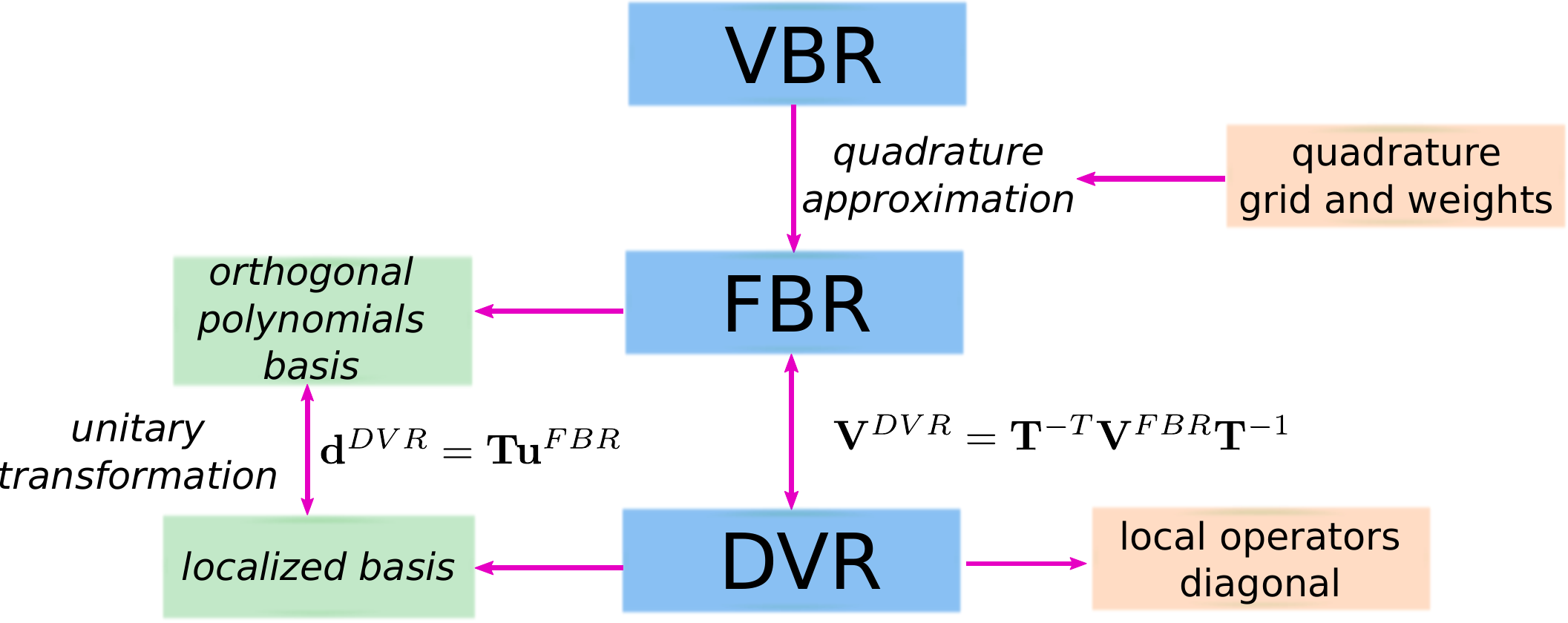}\\
  \caption{A general scheme for the VBR-FBR-DVR transformation.}\label{fig:dvrscheme}
  \end{center}
\end{figure}

\subsection{Product approximation}
The DVR-FBR transformation can be derived from yet another, more practical point of view. From the procedure described in the previous section it is not clear how to obtain quadrature points and weights necessary to define a DVR. Of course, many popular quadrature rules are tabulated, but for some applications new quadrature schemes or quadratures with many nodes are needed, which are not tabulated. A practical way of obtaining the DVR can be achieved through the so called \textit{product approximation} \cite{Light2007}. It provides a procedure for generating the DVR-FBR transformation matrices by means of diagonalisation of the position operator matrix. 

In general, the position operator matrix is of infinite size, as the position operator $X$ is unbounded and acts irreducibly on the Hilbert space (in other words the largest invariant subspace of the Hilbert space for the position operator is the full space). In the VBR the matrix elements of $X$ are written as:
\begin{equation}\label{X}
 \left(\mathbf{X^{VBR}}\right)_{ij}=\int_a^b\phi_i(x)x\phi_j(x)dx\approx \sum_{k=0}^{N-1}\frac{w_k}{\omega(x_k)}\phi_i(x_k)x_k\phi_j(x_k)=\sum_{k=0}^{N-1}\mathbf{T}_{ik}^{T}\mathbf{X}_k\mathbf{T}_{kj}
\end{equation}
As long the $ \left(\mathbf{X^{VBR}}\right)_{ij}$ matrix is truncated to size $N \times N$ and the basis functions are orthogonal polynomials of degree $N-1$, the integrated function is of degree $2N-1$, and can be evaluated exactly by a Gaussian quadrature. It means that the $\mathbf{T}$ matrix diagonalizes the position operator matrix in the orthogonal polynomials basis. As a result,  eigenvalues of $\mathbf{X}$ in this basis correspond to quadrature points and the diagonalising  transformation matrix is related to the quadrature weights. Diagonalisation of $\mathbf{X^{VBR}}$ unambiguously defines the DVR:
\begin{equation}\label{X1}
 \mathbf{X^{DVR}}=\mathbf{T}\mathbf{X^{VBR}}\mathbf{T}^{T}
\end{equation}
where we used the fact that for the Gaussian quadratures the FBR-DVR transformation matrix is unitary: $\mathbf{T}^{-1}=\mathbf{T}^{T}$.
The VBR representation of the position matrix in the orthogonal polynomials basis is straightforward to derive, thanks to the three-term recurrence relations for orthogonal polynomials: $p_{n+1}(x)=(A_nx+B_n)p_n(x)+C_np_{n-1}(x)$, where $A_n,B_n,C_n$ are constants characteristic for a given class of polynomials. As a result the position operator matrix is tridiagonal. In practice, diagonalising the position operator matrix is the most efficient way of finding a DVR.
To show how the DVR approximation relates to Gaussian quadrature (FBR) approximation we follow Harris \etal\ \cite{Harris1965} and write $\mathbf{V^{VBR}}$ approximately as:
\begin{equation}
\mathbf{V^{VBR}}\approx \mathbf{V\left(X^{VBR}\right)}
\end{equation}
Let us now postulate that such approximation is equivalent to the FBR approximation:
\begin{equation}
\mathbf{V\left(X^{VBR}\right)}=\mathbf{V^{FBR}}
\end{equation}
In other words, we replaced \textit{the matrix element of a function of the position operator with the same function of the matrix element of the position operator}:
\begin{equation}\label{prodapprox}
\langle\psi_i(x)|V(x)|\psi_j(x)\rangle \approx V\left(\langle\psi_i(x)|X|\psi_j(x)\rangle\right)
\end{equation}
If we assume that the potential energy function is expandable in a power series, then within a given radius of convergence we may write:
\begin{equation}
V(x)=\sum_{n=0}^{+\infty}c_nx^{n}
\end{equation}
In the postulated FBR approximation the matrix elements of the potential energy are given as
\begin{equation}
V(\mathbf{X^{VBR}})=\sum_{n=0}^{+\infty}c_n\left(\mathbf{X^{VBR}}\right)^{n}
\end{equation}
On account of relation \ref{X1} the VBR monomials of the position operator $\mathbf{X}$ can be written in the form:
\begin{equation}
\left(\mathbf{X^{VBR}}\right)^{n}=\left(\mathbf{T}^{T}\mathbf{X^{VBR}}\mathbf{T}\right)^{n}=\mathbf{T}^{T}\left(\mathbf{X^{DVR}}\right)^{n}\mathbf{T}
\end{equation}
where orthogonality of the $\mathbf{T}$ matrix was used. As long as $\mathbf{X^{DVR}}$ is diagonal, so is its $n-$th power. Finally it is possible to write
\begin{equation} \label{equiv}
\mathbf{V\left(X^{VBR}\right)}=\mathbf{T}^{T}\left(\sum_{n=0}^{+\infty}c_n\left(\mathbf{X^{VBR}}\right)^{n}\right)\mathbf{T}
\end{equation}
proving the equivalence of the product approximation with the Gaussian quadrature FBR. The inherent error of the DVR(FBR) approximation lies within the approximation given in eq. \ref{prodapprox}. Another way of showing this is to notice that the matrix elements of the potential energy function are built from terms containing powers of the position operator. But because the \textit{Hilbert space} is complete (as well as the orthogonal polynomial basis set) it is possible to exactly decompose the matrix elements of $n$-th power of the position operator into a sum of products of $n$ matrices:
\begin{equation} \label{sums}
\begin{split}
\langle\psi_i(x)|x^{n}|\psi_j(x)\rangle=\sum_{k=0}^{+\infty}\langle\psi_i(x)|x^{n-1}|\alpha_k(x)\rangle\langle\alpha_k(x)|x|\psi_j(x)\rangle=...\\
\sum_{k_{1}=0}^{+\infty}...\sum_{k_{n}=0}^{+\infty}\langle\psi_i(x)|x|\alpha_{k_{n}}(x)\rangle...\langle\alpha_{k_{s}}(x)|x|\alpha_{k_{s}}(x)\rangle...\langle\alpha_{k_{1}}(x)|x|\psi_j(x)\rangle
\end{split}
\end{equation}
Now by truncating the resolutions of identity inserted in between the position operators we formally conduct an approximation equivalent to the Gaussian quadrature approximation. We reduce the number of orthogonal polynomials used, hence also the number of quadrature points.  Infinite matrices in eq. \ref{sums} are replaced with truncated matrices, and the product approximation is retrieved:
\begin{equation} \label{sums1}
\begin{split}
\sum_{k_{s}=0}^{+\infty}|\alpha_{k_{s}}(x)\rangle\langle\alpha_{k_{s}}(x)|\approx \sum_{k_{s}=0}^{n}|\alpha_{k_{s}}(x)\rangle\langle\alpha_{k_{s}}(x)|\\
\langle\psi_i(x)|x^{n}|\psi_j(x)\rangle\approx \Pi_{l=1}^n\mathbf{X^{VBR}}=\left(\mathbf{X^{VBR}}\right)^n
\end{split}
\end{equation}
Eq. \ref{sums1} is in close relation to eq. \ref{eq:dvrkey}, where the infinite dimensional scalar product (integral) is being replaced by a finite sum.

To conclude, diagonalization of the position operator matrix is equivalent to applying a Gaussian quadrature approximation to the VBR integrals. In other words Gaussian quadratures and the product approach are operating at exactly the same level of approximation and can be used interchangeably. 

\subsubsection{Other DVRs}
The idea of the Discrete Variable Representation can be used with any quadrature scheme, yet Gaussian-quadratures are very efficient for low-dimensional problems. For this reason Gaussian based DVRs have become popular in nuclear motion and electronic structure calculations, but, of course the are many more types of DVRs and their use is currently much broader than molecular spectroscopy. See reviews by Light and Bacic \cite{Bacic1989}, Light and Carrington \cite{Light2007} or Szalay \cite{Szalay1996}. For instance, in the \textit{Lobbato-DVR} one can arbitrarily set the values of the basis functions at two boundary grid points, which is often of use in scattering calculations and finite-element DVR methods. 

\subsection{Collocation method}
An alternative approach to proceeding from eq. \ref{eq:linexp2} is to multiply from the left by an eigenstate of the position operator $\langle x_k|$ giving:
\begin{equation}
\mathbf{H}^{COL}\mathbf{U}=\mathbf{B}\mathbf{UE}
\label{eq:collocation}
\end{equation}
where $B_{kj}= \langle x_k | \phi_j \rangle = \phi_j(x_k)$ is the \textit{collocation matrix} of values of basis functions at points $x_k$. The number of points $\lbrace x_k\rbrace_{k=1,2,...,M}$ at which we require the \SE\ to be satisfied exactly can be different from the number of basis functions $\lbrace \phi_j\rbrace_{j=1,2,...,N}$ used to represent the wavefunction. As a result the collocation matrix can be rectangular. The form of the collocation equations given below
\begin{equation}
\mathbf{H}^{COL}_{kj}=\langle x_k |\hat{H}| \phi_j \rangle = \left(\hat{\mathbf{K}}\mathbf{B}\right)_{kj}+(\mathbf{V}^{diag}\mathbf{B})_{kj}
\label{eq:collocation2}
\end{equation}
 suggests that one only needs to know the action of the Hamiltonian on the basis functions and the value of the resulting function at grid points $\lbrace x \rbrace _{k=1,...,M}$. Despite the collocation Hamiltonian being non-symmetric, there are several advantages of using collocation such as: no need for calculating integrals, no requirement for a carefully chosen quadrature grid to ensure that the overlap matrix is calculated accurately, no real disadvantage of using non-orthogonal basis sets over the orthogonal basis sets, diagonal potential energy matrix and no need for careful choice of internal coordinates to ensure a simple form of the KEO. In collocation one only needs to calculate the KEO at grid points. For further reading see Ref. \cite{Carrington2017}.

The collocation approach, although very efficient is still much less popular than the variational method. A significant progress in adapting the collocation method to nuclear motion theory was made in the past decade by Carrington Jr. \etal\  \cite{Avila2017ch2nh}.

\subsection{Effective Hamiltonian methods}
A widely used alternative to variational calculations is based on effective operators approximation to the
Hamiltonian and the spectroscopic dipole moment \cite{Sulakshina1989,Teffo1992}.
Currently, the effective Hamiltonian approach achieves at least one order of
magnitude better accuracy in rotational-vibrational transition frequencies for triatomic molecules, than
the best-available PES \cite{Schwenke2012}. Within the effective operators framework the
calculation of intensities requires: a) eigenfunctions of an effective
Hamiltonian whose parameters were fitted to observed positions of
rotation-vibration lines; b) dipole moment operators tuned to
observed transition intensities.  Effective Hamiltonian models strongly 
depend on the quality of the input data, thus the accuracy and completeness of 
this technique are limited by the experiment.  For a more detailed discussion of the effective Hamiltonian methodology see for example refs. \cite{Teffo1992,98TaPeTe.CO2}. 

In a nutshell, an effective Hamiltonian for a triatomic molecule is commonly represented by its matrix elements in the harmonic oscillator + rigid rotor basis:

\begin{equation}
\langle \nu_1,\nu_2,l,\nu_3, J | H_{eff} |  \nu'_1,\nu'_2,l',\nu'_3, J' \rangle 
\label{eq:effham}
\end{equation}
where $|  \nu_1,\nu_2,l,\nu_3, J \rangle = |\nu_1\rangle |\nu_2,l\rangle |\nu_3\rangle |J,k=l\rangle$ is the product basis of harmonic oscillator and rigid rotor (see appendix). $\nu_1,\nu_2,l,\nu_3$ stand for vibrational quantum numbers of the harmonic oscillator basis functions; $(\nu_2,l)$ is a pair of quantum numbers defining the bending motion in the triatomic molecule, $\nu_1$ is the symmetric stretching quantum number and $\nu_3$ labels the asymmetric stretching vibrational mode. $J$ is the total angular momentum quantum number. In triatomic molecules, the $k$ quantum number, which is the projection of the total angular momentum on the molecule-fixed $z$ axis, can be associated with the $l$ quantum number which describes the \textit{vibrational angular momentum}. Further discussion of these quantum numbers is given in the reminder of this chapter.
Now, depending on which matrix elements are included in the calculations the effective Hamiltonian can model a variety of 'effects', such as \textit{Fermi resonance interaction }($\langle \nu_1,\nu_2,l,\nu_3, J | H_{eff} |  \nu_1-1,\nu_2+2,l,\nu_3, J \rangle $), \textit{l-doubling} ($\langle \nu_1,\nu_2,l,\nu_3, J | H_{eff} |  \nu_1,\nu_2,l\pm 2,\nu_3, J \rangle $) or \textit{Coriolis resonance interaction } ($\langle \nu_1,\nu_2,l,\nu_3, J | H_{eff} |  \nu_1-1,\nu_2-1,l\pm 1,\nu_3+1, J \rangle $). These 'effects' however are merely artifacts of the chosen harmonic basis set, yet they are commonly cited throughout the literature.

\section{Basis set representation to the rotational-vibrational wavefunction}
\subsection{Rotational basis functions}
\label{sec:rotbasis}
Having discussed popular methods of solving the rotational-vibrational \SE\ the next step is to choose appropriate basis set. We begin with outlining the most commonly used basis set for the description of the rotational motion. As discussed in section \ref{sec:KEO}, the transition form the space-fixed to the molecule-fixed coordinate frame carries one conspicuous advantage: the motion associated with the rotational degrees of freedom (Euler angles) of the molecule is fully described by the angular momentum operators $\hat{J}_x,\hat{J}_y,\hat{J}_z$. In eq. \ref{eq:Kvr} the most relevant terms are: $\frac{1}{2}\left[M_{xx}\hat{J}_x^2+M_{yy}\hat{J}_y^2+M_{zz}\hat{J}_z^2\right]$, taken at equilibrium geometry, which represent the quantum-mechanical operators for the rotational motion of a rigid (non-vibrating) molecule around the $x$,$y$ and $z$ molecule-fixed axis, respectively. It is sometimes advantageous to choose the molecule-fixed axis system so that it coalesces with the principle axes of inertia of the molecule, that is with the coordinate frame in which the moment of inertia tensor of the molecule is a diagonal 3-by-3 matrix. The principle axes of inertia are labelled as $a,b$ and $c$ and the corresponding pure rotational energy operator can be written as: $\frac{1}{2}\left[M_{aa}\hat{J}_a^2+M_{bb}\hat{J}_b^2+M_{cc}\hat{J}_c^2\right]$. In the principle axis system $M_{xz},M_{xy}$ and $M_{yz}$ are zero. The inverse moments of inertia in the principle axis system define so called \textit{rotational constants}: $A \equiv \frac{1}{2} M_{aa}=\frac{1}{2I_{aa}}$,  $B \equiv \frac{1}{2} M_{bb}=\frac{1}{2I_{bb}}$,  $C \equiv \frac{1}{2} M_{cc}=\frac{1}{2I_{cc}}$, with the convention $A\geq B \geq C$. The rotational constants can be straightforwardly calculated if the masses of the atoms and the equilibrium geometry of the molecule are known. The purely rotational Hamiltonian of any molecule can be now written in the form

\begin{equation}
\hat{H}_{rot}=\left(A\hat{J}_{a}^{2}+B\hat{J}_{b}^{2}+C\hat{J}_{c}^{2}\right)
\label{eq:Hrot}
\end{equation}

The Hamiltonian in eq. \ref{eq:Hrot} is called the \textit{asymmetric top Hamiltonian} \cite{06BuJexx.method}, since all rotational constants take different values. Molecules are categorized by their values of rotational constants as follows:
\begin{itemize}
\item Symmetric top molecules, for which two rotational constants are equal, dividing into two groups:
\subitem Prolate symmetric top when $A>B=C$ (e.g. $CH_{3}Cl, trans-Ni(H_{2}O)_{4}Cl_{2}$) (cigar or rugby ball shaped molecules)
\subitem Oblate symmetric top when $A=B>C$ (e.g. $SO_{3}, BF_{3}, NO_{3}^{-}, C_{6}H_{6}, NH_{3})$ (disk or doughnut shaped molecules)
\item Spherical top molecules when $A=B=C$ (e.g. $SF_{6}, CH_{4}, SiH_{4})$
\item Asymmetric top molecules when $A\ne B \ne C \ne A$ (e.g. $H_{2}O, NO_{2}$)
\item linear molecules ($N_2, CO_2, C_2H_2$, etc.)
\end{itemize}
Symmetric top molecules must contain three-fold or higher-fold symmetry axis, as an element of the point group. Molecules possessing only two-fold axis of symmetry or no symmetry are asymmetric tops (majority of polyatomic molecules are). A common convention for prolate symmetric top molecules is to choose the molecule-fixed $x,y, z$ axes as $b,c,a$, respectively which gives the following \SE

 \begin{equation}
\left(A\hat{J}_{z}^{2}+B(\hat{J}_{x}^{2}+\hat{J}_{y}^{2})\right)\Phi_{rot}(\theta, \phi, \chi)=E_{rot}\Phi_{rot}(\theta, \phi, \chi)
\label{eq:prolate}
\end{equation}
But because $\hat{J}^2= \hat{J}_x^2+\hat{J}_y^2+\hat{J}_z^2$ the equation above can be expressed with $\hat{J}^2$ and $\hat{J}_z$ operators for which exact eigenvalues and eigenfunctions are known: $\hat{J}_{z}|J,k,m\rangle=\hslash k|J,k,m\rangle$, $\hat{J}_{z}|J,k,m\rangle=\hslash^2 J(J+1)|J,k,m\rangle$. The rotational energy then reads
  \begin{equation}
    E_{rot}=BJ(J+1)+(A-B)k^{2}
    \label{eq:prolate_energy}
  \end{equation}
and the eigenfunctions are
\begin{equation}
\Phi_{rot}(\theta, \phi, \chi) \equiv | J,k,m\rangle = \sqrt{\frac{2J+1}{8\pi^2}}(-1)^k\mathcal{D}^{J*}_{m,-k}(\theta, \phi, \chi)
\label{eq:rotbasis}
\end{equation}
where $\mathcal{D}^J_{m,k}(\theta, \phi, \chi)$ is the \textit{Wigner rotation matrix} \cite{Brink}. Here $J$ stands for the total angular momentum quantum number, $k=-J,-J+1,...,J-1,J$ is the projection of the total angular momentum on the $z$-axis in the molecule-fixed coordinate frame and $m = -J, -J+1, ... , J-1, J$ is the projection of the total angular momentum on the $z$-axis in the space-fixed coordinate frame. For asymmetric top molecules the rotational Hamiltonian given in eq. \ref{eq:Hrot} commutes with $J^2$ but not with $J_z$ meaning that $k$ is no longer a good quantum number. As a result the rotational wavefunctions of asymmetric top molecules are linear combinations of symmetric top basis functions.

In the absence of external electric or magnetic fields the energy of the molecule does not depend on its orientation in space, meaning that rotational energy levels of the molecule do not depend on the $m$ quantum number ($SO(3)$ symmetry of the rotational Hamiltonian). For this reason, a shorthand notation for the rotational basis function will be used $|J,k,m\rangle\equiv |J,k\rangle$.
The symmetric top basis functions form a complete basis of the Hilbert space for a given $J$ quantum number (see appendix). As a consequence, the rotational wavefunction of any asymmetric top molecule can be exactly represented as a linear combination of the symmetric top basis functions. This fact suggests a procedure, practised in section \ref{sec:strategies}, in which the rotational degrees of freedom in the Hamiltonian are integrated-out by calculating Hamiltonian matrix elements in the symmetric top basis. This introduces $J$ and $k$ quantum numbers as labels for effective vibrational Hamiltonians.

\subsection{Rotational-vibrational basis set}
All variational, DVR, FBR and collocation methods use basis functions to represent the molecular wavefunction. An ideal basis set consists of functions which diagonalize the Hamiltonian matrix, i.e. are eigenfunctions of the Hamiltonian. Therefore given a Hamiltonian, a good basis set is such for which off-diagonal matrix elements (couplings) are possibly small. Small off-diagonal elements of the Hamiltonian matrix enable one to use perturbation theory for finding approximate molecular energy levels, but more importantly promise fast convergence of the rotational-vibrational energy levels in variational calculations. Fast convergence means that doing calculations with $N$ and subsequently with $N+1$ basis functions do not improve the values of the energy levels above a given threshold value, and the value of $N$ is small. 

In this article we focus on rotational-vibrational wavefunctions of triatomic molecules. In the Born-Oppehmeimer approximation the rotational-vibrational wavefunction for $i$-th electronic state of a triatomic molecule has the general form:

\begin{equation}
|\Psi_{rv}^{(J,h,i)}\rangle= \sum_{k=-J}^J |\Phi_{vib,J,k,h,i}\rangle |J,k\rangle
\label{eq:anzatzrv}
\end{equation}
where $h$ enumerates rotational-vibrational states, whereas $J$ and $i$ are considered good quantum numbers, labeling the total angular momentum and the electronic state, respectively. The vibrational wavefunctions $|\Phi_{vib,J,k,h,i}\rangle $ in general depend on the $k$ rotational quantum number. In the \textit{anzatz} given in eq. \ref{eq:anzatzrv} the vibrational wavefunctions are directly coupled with symmetric top rotational wavefuncitons through \textit{k}.

The vibrational wavefunction can be expressed as a function of two radial coordinates $r_1$, $r_2$ and one angle $\gamma$, as introduced in section \ref{sec:theorybasis}. The full vibrational wavefunction can be expanded as sum-of-products of single-coordinate basis functions as follows
\begin{equation}
|\Phi_{vib,J,k,h,i}\rangle = \sum_{m,n,j} c_{k,m,n,j}^{J,h,i}|m\rangle|n\rangle|jk\rangle
\label{eq:anzatzvib}
\end{equation}
although this choice is not the only one possible. There are methods \cite{Bubin2005} available, which utilize basis functions depending explicitly on more than one vibrational coordinate. Such \textit{explicitly correlated} basis functions have been used in high-accuracy rotational-vibrational calculations of simple diatomic molecules, such as H$_2$. The use of explicitly correlated basis functions in nuclear motion theory of bigger molecules is difficult because calculation (analytic or numerical) of VBR matrix elements of the Hamiltonian present serious computational challenge of evaluating multi-dimensional integrals.
 
In the most popular approach to the vibrational problem each vibrational basis function is factorized into three independent sets of primitives, each depending on a single vibrational coordinate: $|m\rangle|n\rangle|jk\rangle$, where $|m\rangle, |n\rangle$ correspond to radial coordinates $r_1$ and $r_2$, respectively, and $|jk\rangle$ is \textit{k}-dependent basis function labelled by $j$ quantum number for the bending motion associated with the angular coordinate $\gamma$. Figure \ref{fig:embeddings} depicts the internal (vibrational) coordinates along with different molecule-fixed embedding types. 

With the rotational-vibrational wavefunction defined in eq. \ref{eq:anzatzrv}, the next step in almost all nuclear motion calculations is to remove the explicit dependence of the Hamiltonian on the rotational degrees of freedom (Euler angles). Only the KEO contains rotational degrees of freedom, the potential part of the Hamiltonian depends solely on internal coordinates. The integration over Euler angles can be done exactly and analytically by calculating matrix elements of rotational-vibrational Hamiltonian in the complete symmetric top basis:
\begin{equation}
\hat{H}^{J'J}_{k'k}(r_1,r_2,\lambda)\delta_{J'J}=\langle J',k' | \hat{H}(r_1,r_2,\lambda,\theta, \phi, \chi) | J,k\rangle_{\theta, \phi, \chi}
\label{eq:Heff0}
\end{equation}
yielding a series of effective vibrational Hamiltonians labelled by $J$ and $k$. The Hamiltonian matrix formed from the Hamiltonian given in eq. \ref{eq:Heff0} in the vibrational basis defined in \ref{eq:anzatzvib} can be diagonalized to obtain eigenvalues (rotational-vibrational energy levels) and eigenvectors (rotational-vibrational wavefunctions). 

A popular and perhaps less accurate, but simpler \textit{anzatz} for the rotational-vibrational wavefunction can be written as

\begin{equation}
|\Psi_{rv}^{(J,h,i)}\rangle= |\Phi_{vib,J,h,i}\rangle |\Psi_{rot,J,h,i}\rangle
\label{eq:anzatzrv2}
\end{equation}
where the rotational wavefunction $|\Psi_{rot,J,h,i}\rangle$ is a linear combination of the symmetric top basis functions
\begin{equation}
|\Psi_{rot,J,h,i}\rangle = \sum_{k=-J}^J c_k^{J,h,i} |J,k\rangle
\label{eq:rotwf}
\end{equation} 
With the \textit{ anzatz} given in eq. \ref{eq:anzatzrv2} the rotational-vibrational wavefunction is separable into the rotational part and the vibrational part, meaning that rotational-vibrational coupling is not accounted for in the form of the wavefunction. $|\Psi_{rot,J,h,i}\rangle $ can be for instance the asymmetric top wavefunction. Note that in general the rotational wavefunction depends on the electronic state $i$ of the molecule. This is because the equilibrium molecular geometry can be different in different electronic states. As a consequence, the molecule-fixed axis system for the electronic state $j$ must be rotated when one wishes to represent the wavefunctions of both $i$-th and $j$-th electronic states in a single set of rotational coordinates. This means that in the $j$-th electronic state the definition of the Euler angles is different than in the $i$-th electronic state. This \textit{axis switching} effect is discussed in more detail in Chapter 4.

\subsection{Vibrational basis sets}
\label{sec:theorybasis}
The effective vibrational Hamiltonians given in eq. \ref{eq:Heff0} are operators which are labeled by $J$, which is a good quantum number and by $k$ and $k'$, which are usually not good quantum numbers. Matrix representation of the vibrational Hamiltonian $\hat{H}^{J}_{k'k}(r_1,r_2,\gamma)$ is obtained upon a choice of the vibrational basis set. In a triatomic molecule, as shown in eq. \ref{eq:anzatzvib}, two primitive vibrational basis functions are associated with the radial or stretching motion  and the third primitive basis function describes the bending motion. For the stretching motion, among the most popular basis functions used are the harmonic oscillator eigenfunctions (see appendix) and the Morse oscillator eigenfunctions.

The role of vibrational basis functions is to possibly best reproduce the vibrational wavefunction of the molecule in a given rotational state $(J,k)$. The choice of the vibrational basis function largely depends on the shape of the electronic potential energy surface along the vibrational coordinate. For small amplitude stretching motion of atoms it is reasonable to use basis functions, which are solutions to the Harmonic oscillator or Morse oscillator models. This is because these model basis functions are probably generating only small couplings between coordinates and the resulting matrix representation of the Hamiltonian is nearly diagonal. 

The 1-D Harmonic oscillator basis functions $|n\rangle$ are eigenfunctions of the Hamiltonian

\begin{equation}
\hat{H}_{HO}=\frac{1}{2}\left(\hat{P}^2+\hat{Q}^2\right)
\label{eq:1dho}
\end{equation}
where $\hat{P}$ is the momentum operator and $\hat{Q}$ is the dimensionless position operator. The dimensionless coordinates are useful because using them does not invoke writing molecular parameters explicitly. Dimensionless coordinate is related to dimensional position coordinate as: $x=\sqrt{\frac{\hslash}{\mu \omega}}Q$, where $\mu$ is the reduced mass of the Harmonic oscillator and $\omega$ is the frequency of the oscillator. The orthonormal \textit{harmonic oscillator basis functions} are

\begin{equation}
|n\rangle = N_n H_n(Q)e^{-\frac{1}{2}Q^2}
\label{eq:1dhobasis}
\end{equation}
where $N_n$ is the normalisation constant and $H_n(Q)$ is the n-th \textit{Hermite Polynomial}; and the corresponding vibrational energies are
\begin{equation}
E_n = n+\frac{1}{2}
\label{eq:1dhoenergy}
\end{equation}
An often exploited property of the Harmonic basis functions is their symmetry under the parity transformation $E^*|n\rangle=(-1)^n|n\rangle$. Parity transformation changes sign of all arguments of the function: $E^*f(x)=f(-x)$. For this reason we call functions with even $n$ \textit{even} and functions with odd $n$ \textit{odd}.
Harmonic basis is very efficient in describing the wavefunction of molecules with small-amplitude vibrations, which are typically near-harmonic. In some applications the radial part of the 3-D isotropic\footnote{isotropic means that all three harmonic frequencies $\omega_x, \omega_y, \omega_z$ are equal} Harmonic oscillator basis functions are useful. They are often referred to as the \textit{spherical oscillator basis functions} and are defined as

\begin{equation}
|n\rangle = N_{n \alpha+\frac{1}{2}}2^{\frac{1}{2}}\beta^{\frac{3}{4}}L_n^{\alpha+\frac{1}{2}}(x(r))x^{\frac{\alpha+1}{2}}e^{-\frac{x(r)}{2}}
\label{eq:sphericalbasis}
\end{equation}
with $x(r)=\beta r^2$ and $\beta=\left(\mu\omega_0\right)^{\frac{1}{2}}$.

Harmonic oscillator basis functions do not model correctly the long-range profile of the vibrational wavefunction. For long atom-atom distances, the electronic potential energy does not grow rapidly, and is not described by simple quadratic function of the distance. The Morse Oscillator potential accounts for this \textit{dissociation limit}:
\begin{equation}
\hat{H}_{Morse}=-\frac{1}{Q^2}\frac{d}{dQ}Q^2\frac{d}{dQ}+V_{Morse}(Q)
\label{eq:morseham}
\end{equation}
with 
\begin{equation}
V_{Morse}(Q)=D_e\left(1-e^{-a(Q-Q_0)}\right)^2
\label{eq:morsepot}
\end{equation}
where $D_e$ is the \textit{dissociation energy}, $a$ determines the width of the potential well and $Q_0$ is the equilibrium length.  Near the equilibrium ($Q\approx Q_0$) the Morse potential is similar to the Harmonic potential. Indeed, by taking a Taylor expansion of $V_{Morse}(Q)$ near $Q=Q_0$ one obtains:

\begin{equation}
V_{Morse}(Q)=D_ea^2(Q-Q_0)^2 + o((Q-Q_0)^3) \equiv \frac{1}{2}\mu\omega_0^2(Q-Q_0)^2 + o((Q-Q_0)^3) 
\label{eq:morsepottaylor}
\end{equation}
through which the characteristic 'harmonic frequency' of the Morse oscillator can be defined as $\omega_0= a\sqrt{\frac{2D_e}{\mu}}$.
The eigenvalues of the Hamiltonian in eq. \ref{eq:morseham} are given by:
\begin{equation}
E_n=\hslash \omega_0 \left(n+\frac{1}{2}\right)\left[1-x_e\left(n+\frac{1}{2}\right)\right]
\label{eq:morseenergies}
\end{equation}
and the eigenfunctions are
\begin{equation}
|n\rangle = N_{n}y^lL_n^{2l}(\frac{y}{x_e})e^{-\frac{2D_e}{\hslash \omega_0}}
\label{eq:morsebasis}
\end{equation}
where $y=e^{-(Q-aQ_0)}$, $l=\frac{2D}{\hslash \omega_0}-n-\frac{1}{2}$ and $L_n^{2l}(\frac{y}{x_e})$ are the \textit{Associated Laguerre polynomials}. There is only a finite number of bound states for the Morse potential, namely the integer part of $\frac{D_e-\hslash \omega_0}{\hslash \omega_0}$. Also, the Morse oscillator basis set does not constitute a complete basis set. This drawback of the Morse basis can be repaired by modifying the Morse basis functions in the following way
\begin{equation}
|n\rangle = N_{n \alpha}\beta^{\frac{1}{2}}L_n^{\alpha}(x(Q))x(Q)^{\frac{\alpha+1}{2}}e^{-\frac{x(Q)}{2}}
\label{eq:morselikebasis}
\end{equation}
with 
\begin{equation}
x(Q)=A e^{-\beta(Q-Q_0)}
\end{equation}
where
\begin{equation}
A=\frac{4D_e}{\beta} \qquad \beta=\omega_0\left(\frac{\mu}{2D}\right)^{\frac{1}{2}};
\end{equation}
$\alpha$ has been defined by Tennyson and Sutcliffe as the integer part of $A$ \cite{DVR3D}. With this definition, the \textit{Morse oscillator-like} functions $|n\rangle $ constitute the complete orthonormal basis set. Here $N_{n \alpha}L_n^{\alpha}(x(Q))$ is the normalized associated Laguerre polynomial and $\mu$ is the reduced mass. $\omega_0$ and $D_e$ are standard the Morse potential parameters, related to the width and the depth of the potential well, respectively. The number of bound states for the Morse (and Morse-like) oscillator is the integer part of $\frac{A-1}{2}$. 

Note that the Morse oscillator and Morse oscillator-like basis do not vanish at $Q = 0$ ($x(0)=Ae^{\beta Q_0}$), as one may expect. Nonetheless, in the $Q \approx 0$ region the Morse potential increases rapidly, making it almost un-explorable for the wavefuncion. The value $Ae^{\beta Q_e}$ is typically large enough for the exponent factor in eq.\ref{eq:morsebasis} to damp other factors almost to 0. Hence, without significant loss in accuracy it is possible to replace the finite  boundary value $Ae^{\beta Q_e}$  with $+\infty$ in integrations. With such a trick it is possible to calculate integrals over the Morse potential analytically. One has to keep in mind that, the Morse oscillator-like basis is applicable for $Q \neq 0$. Whenever the vibrational coordinate is presumed to have significant amplitude near $Q=0$, the Morse basis set can give inaccurate results; is it then more suitable to use the spherical oscillator basis functions.

From the computational perspective it is important to note that the Morse-like oscillator basis set depends on three parameters: $D_e, \omega_0$ and $r_e$ and the Spherical oscillator basis set depends on two parameters: $\alpha, \omega_0$; they can be considered as non-linear variational parameters, and should be optimized.

The bending motion of atoms in triatomic molecules is quantum-mechanically well described by \textit{associated Legendre basis functions} $P_j^k(x)\equiv |jk\rangle$, which are solutions to the differential equation:

\begin{equation}
\frac{d}{dx}\left[(1-x^2)\frac{d}{dx}\right]P_j^k(x)+\left[j(j+1)-\frac{k^2}{1-x^2}\right]P_j^k(x)
\label{eq:legendrebasis}
\end{equation}
where $0 \leq k \leq j$ are two numbers labeling the solutions. In the description of the bending motion in a rotating molecule, $k$ can be associated with the projection of the total angular momentum on molecule-fixed $z$-axis whereas $j$ can be associated with the quantum number for the bending motion excitations. Usually $x=\cos \gamma$, where $\gamma$ is the angle between the bonds, as shown in Figure \ref{fig:embeddings}. The associated Legendre functions form an orthonormal set with respect to $j$:

\begin{equation}
\int_{-1}^{+1}P_{j'}^{k}(x)P_j^k(x)dx = \frac{2(k+j)!}{(2j+1)(j-k)!}\delta_{jj'}
\label{eq:legendreort}
\end{equation}
but also they are orthogonal with respect to  $k$ with weighting function $(1-x^2)^{-1}$:
\begin{equation}
\int_{-1}^{+1}P_j^{k'}(x)P_j^k(x)(1-x^2)^{-1}dx = \frac{(k+j)!}{k(j-k)!}\delta_{kk'}
\label{eq:legendreort2}
\end{equation}
Associate Legendre functions are even or odd upon parity transformation $E^*|jk\rangle=P_j^k(-x)=(-1)^{j+k}|jk\rangle$.

\section{Parity and permutation symmetries}
\label{sec:additionalsymmetries}
From  the computational efficiency perspective is it advantageous to utilize all relevant physical symmetries of the molecule, so that the matrix representation of the molecular Hamiltonian can be factorized into smaller blocks, which are less costly to diagonalize. There are three fundamental symmetries important for nuclear motion theory:  parity, rotational symmetry in 3-D space and permutation symmetry of identical nuclei. 

Symmetries can be used in construction of basis sets, in which the Hamiltonian matrix is block-diagonal. By applying symmetry operations to rotational-vibrational basis functions it is possible to build an eigenbasis of the symmetry operators. Such basis transforms irreducibly in a given molecular symmetry group, meaning the symmetry-adapted Hamiltonian matrix acts irreducibly in the \textit{Hilbert space} spanned by the ro-vibrational basis functions. Each symmetry block is associated with a particular symmetry label (irreducible representation), which are useful in making spectroscopic assignments of energy levels. For instance the total angular momentum quantum number $J$ is associated with the 3-D rotational symmetry. The symmetric top basis functions $|J,k\rangle$ transform irreducibly in the group of 3-D rotations. Because the symmetric top basis is already an eigenbasis of rotation operators, the Hamiltonian in the $|J,k\rangle$ is block-diagonal in the $J$ quantum number, which is also a popular symmetry label for rotational states of molecules. 

Below, we give an example construction of a symmetry-adapted basis for a triatomic molecule utilizing the parity symmetry and the permutation symmetry of identical nuclei.  First, let us recall the variational basis set for a triatomic molecule defined in the previous section:

\begin{equation}
|J,h\rangle= \sum_{k=-J}^J\sum_{m,n,j} c_{k,m,n,j}^{J(h)}|J,k\rangle|m\rangle|n\rangle|jk\rangle
\label{eq:anzatzrv3}
\end{equation}
where the label for the electronic state has been dropped for clarity. 

\begin{figure}[H]
  \includegraphics[width=14cm]{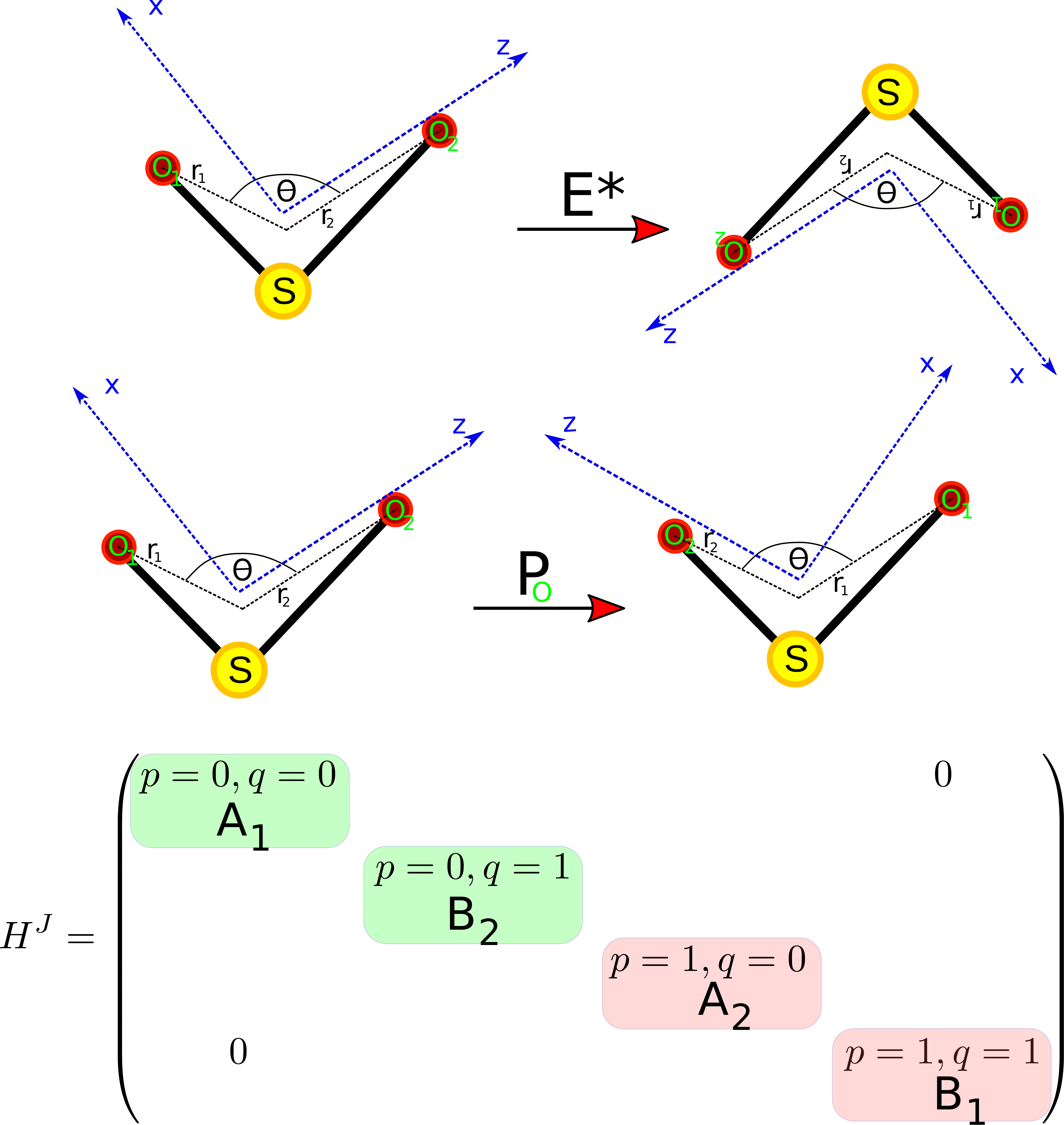}
\caption{The effect of parity operation $E^*$ and permutation $P_O$ on the example SO$_2$ molecule drawn along with the molecule-fixed frame. Symmetry adaptation of the ro-vibrational basis set leads to block-factorization of the Hamiltonian matrix into blocks which transform as $A_1,B_2,A_2$ and $B_1$ irreducible representation of the $C_{2v}(M)$ molecular symmetry group. }
\centering
\label{fig:symmetry}
\end{figure}

\paragraph{Parity.} The parity operation  $E^*$, which is a feasible symmetry operation for all molecules \footnote{parity is a symmetry for all molecules except enantiomers of chiral molecules and when the nuclear weak force is neglected}, can be used to construct a symmetry-adapted basis, which factorizes the Hamiltonian matrix and introduces spectroscopically an important symmetry label: \textit{parity} $p$.
The parity symmetry operation transforms states with $k$ quantum number into states with $-k$, and \textit{vice versa}:
which in the classical picture means that the clockwise and anti-clockwise rotation of the molecule around the molecule-fixed $z$-axis is energetically equivalent. For this reason, it is convenient to symmetry-adapt the rotational-bending basis(which depends on $k$), by means of the following unitary transformation: 

\begin{equation}
|J,K,j,p\rangle =\frac{1}{\sqrt{2(1+\delta_{k0})}}|jk\rangle \left(|J,k\rangle+(-1)^{p}|J,-k\rangle\right)^*
\label{eq:rotbendsym}
\end{equation}
where the new quantum number $p=0,1$ is associated with the \textit{parity} symmetry of the ro-vibrational state and determines the $e/f$ \textit{Wang labels} ($p=0$ for the $e$ state and $p=1$ for the $f$ state); now $K=|k|$  takes integer values from $p$ to $J$. In such a basis, only positive values of $k$ need to be considered and the full Hamiltonian is factorized into independent blocks with $p=0$ and $p=1$ of dimension $J+1$ and $J$, respectively, as shown in Figure \ref{fig:symmetry}. In spectroscopic literature molecular states are often dubbed as \textit{parity even} or \textit{parity odd}, depending on the value of the $p$ quantum number.

\paragraph{Permutation of identical nuclei.}
There are three types of triatomic molecules: $X_3$, XY$_2$ and XYZ. The XYZ type has no permutation symmetry due to lack of identical nuclei. For triatomic symmetric XY$_2$ molecules for which the molecule-fixed axis system is chosen so that the x-axis (or z-axis) bisects the Y--X--Y angle \cite{jt114} the  permutation symmetry of identical nuclei $P_{Y}$ can be used. The permutation operator interchanges $r_1$ and $r_2$ internal coordinates of the molecule, hence effectively acts on the radial vibrational basis functions, as shown in Figure \ref{fig:symmetry}. For this reason, the radial-vibrational basis set can be unitarily transformed into its symmetry-adapted form: 
\begin{equation}
|m,n,q\rangle =\frac{1}{\sqrt{2(1+\delta_{mn})}}\left(|m\rangle |n\rangle + (-1)^q |n\rangle  |m\rangle \right), \qquad m\geq n
\label{eq:vibvib}
\end{equation}
where $|m\rangle |n\rangle$ stands for the product of vibrational basis states associated with the first $r_1$ and the second $r_2$ stretching coordinate, respectively. $m$ and $n$ label the 1-D basis states. The new vibrational parity quantum number takes two values: $q=0$ for 'even' vibrational states and $q=1$ for 'odd' vibrational states. Note that the character of the permutation P$_{Y}$ of identical nuclei acting on the basis state in eq. (\ref{eq:vibvib})  is $(-1)^{q+k}$ and the character of the parity $E^*$ operation acting on the basis state in eq. (\ref{eq:rotbendsym}) is $(-1)^{p+J}$.

Utilization of the rotational parity and the vibrational parity decomposes the rotational-vibrational \textit{Hilbert space} into the simple sum of four independent subspaces for each $J$: 
\begin{equation}
\mathcal{H}^{J}=\mathcal{H}^{J}_{p,q} \bigoplus \mathcal{H}^{J}_{p,1-q}  \bigoplus \mathcal{H}^{J}_{1-p,q}  \bigoplus \mathcal{H}^{J}_{1-p,1-q}
\end{equation}
This decomposition of the \textit{Hilbert space}  allows to run calculations independently in each subspace. Mixing of these reduced \textit{Hilbert spaces} occurs through the electronic dipole moment operator $\hat{\mu}_{el}$ or the electronic polarizability in transition intensity calculations. Then for instance, $\hat{\mu}_{el}$ mixes subspaces according to rigorous selection rules:  $\left|\Delta q\right|=1$ and for $\Delta J=0$: $\left| \Delta p\right|=1$ and $\Delta J=\pm 1$: $\left|\Delta p\right|=0$ . 

The final symmetry-adapted ro-vibrational basis is given by the expression:

\begin{equation}
 |J,h,p,q\rangle =\sum_{K=p}^J  \sum_{m,n,j} C_{mnjK}^{J,(h),p,q} |m,n,q\rangle   |J,K,j,p\rangle
\label{eq:rovibfinal}
\end{equation} 
The new $ |J,h,p,q\rangle$ basis functions transform irreducibly in the complete nuclear permutation-inversion (CNPI) group for XY$_2$ molecules. This CNPI group contains the following elements: $G_2=\lbrace E,E^*,P_Y,P_Y^* \rbrace$ where $E$ is identity operation and $P_Y^*$ is combined permutation of $Y$ atoms and parity operation. The $G_2$ CNPI group corresponds to $C_{2v}(M)$ molecular symmetry group or simply $C_{2v}$ point group. The wavefunction given in eq. \ref{eq:rovibfinal} is the eigenfunction of elements of $C_{2v}(M)$ molecular symmetry group with eigenvalues given in Table \ref{table:C2v}. The irreducible representations for $ |J,h,p,q\rangle$ are determined by combination of $p$ and $q$ quantum numbers, as displayed in Figure \ref{fig:symmetry}. For comprehensive discussion of molecular symmetry see book by \textit{Bunker} and \textit{Jensen} \cite{06BuJexx.method}.
\begin{table}[h]
\begin{center}
\setlength{\tabcolsep}{4pt}
\caption{The character table for the $C_{2v}(M)$ molecular symmetry group associated with symmetry-adapted ro-vibrational basis for XY$_2$ type molecules.}
\begin{tabular}{c r r r r}
\hline\hline
$C_{2v}(M)$ &	E & (12) & E* & (12)* \\ [0.3ex] 
\hline 
$A_1$	& 1	& 1	& 1	& 1 	 	\\
$A_2$	& 1	& 1	& -1	& -1 	 	\\
$B_1$	& 1	& -1	& -1	& 1 	 	\\
$B_2$	& 1	& -1	& 1	& -1 	 	\\
\hline\hline
\end{tabular}
\label{table:C2v}
\end{center}
\end{table}

To summarize, we introduced a transformation of the rotational-vibrational basis $|J,h\rangle \rightarrow  |J,h,p,q\rangle$. This new basis transforms irreducibly in the group of permutations of identical nuclei and parity. Such an adapted basis is useful in making spectroscopic assignments as well as brings the benefit of cheaper computation of ro-vibrational energy levels and wavefunctions.

\section{Strategies for solving the matrix representation of the \SE}
\label{sec:strategies}
In the symmetry-adapted rotational-vibrational basis set given in eq. \ref{eq:rovibfinal} the full nuclear Hamiltonian matrix is labelled by the following indices: $J,K,m,n,j,p,(q)$, where $q$ is a label only for XY$_2$ type molecules. For a given $J$, without invoking the $p$ and $q$ quantum numbers, the Hamiltonian has the dimension of $(2J+1)\times N_{r_1}\times N_{r_2}\times N_{\gamma}$, where $N_{r_1}, N_{r_2}$ and $N_{\gamma}$ are sizes of 1-D vibrational basis sets associated with coordinates $r_1, r_2$ and $\gamma$, respectively. The size of the primitive vibrational basis set is system-specific and is chosen to ensure satisfactory level of convergence of energy levels. Parity symmetry allows to split the Hamiltonian into parts with $p=1$ of size $(J)\times N_{r_1}\times N_{r_2}\times N_{\gamma}$ and $p=0$ of size $(J+1)\times N_{r_1}\times N_{r_2}\times N_{\gamma}$, whereas the permutation symmetry associated with the $q$ quantum number splits each of these Hamiltonians into two, with $q=0$ and $q=1$. The size of the vibrational basis set depends on how many energy levels are requested, the quality of the basis set is and what accuracy is required.  

For example, in the author's rotational-vibrational energy levels calculations for the CO$_2$ molecule \cite{17ZaTePo.CO2} up to $J=130$ and up to 11500\cm ,  the vibrational basis consisted of $ N_{r_1}= N_{r_2}=20$ Morse-like oscillator functions and  $N_{r_{\gamma}}=120$ associated Legendre polynomials. It is important to remember that depending on the temperature and on the frequency range of the spectrum one wants to simulate, appropriate range for the $J$ quantum number must be selected along with a cut-off value for the energy of vibrational levels. These prerequisites often determine the strategy followed in finding energy levels and wavefunctions. 

If, for instance, one is only interested in few lowest rotational-vibrational energy levels an iterative eigensolver is suitable, as it provides lowest (or highest) eigenvalues and eigenvectors at much lower cost than direct diagonalization. An additional factor in favour of using iterative eigensolvers is when the Hamiltonian matrix is sparse. The computational cost of the most popular Lanczos iterative procedure \cite{Carrington2018} for finding eigenvalues is $mnd + m^3$, where $m$ is the number of requested eigenvalues, $d$ is the number of non-zero elements in a row of the matrix and $n$ is the size of the matrix. If $m$ and $d$ are small, even though $n$ is large, the cost is tiny in comparison with the $n^3$ computational cost of a direct eigensolver. 
However one often requires many energy levels and some clever ways of obtaining energy levels and wavefunctions are necessary. 

In the calculation of  CO$_2$'s $J=130$ energy levels, the total size of the Hamiltonian matrix is about 12 million - impossible to diagonalize. With the use of the symmetry-adapted basis, it is possible to reduce this number down to about 3 million, which is still too big.
In such cases it is advantageous to divide the solution plan into simpler steps. For example, one can start with finding wavefunctions for reduced-dimensionality problems. This can be done by freezing all but one coordinates at their equilibrium values, leave one active coordinate to solve the associated 1-D \SE. The wavefunctions obtained this way contain information about the shape of the PES along the active coordinate.  Such \textit{potential-optimized basis} is then used in solving a higher dimensional problem, and the procedure gradually incorporates more active degrees of freedom. A popular computer code MULTIMODE by Bowman \etal\ implements this methodology. The MULTIMODE methodology has been very successful in computing ro-vibrational spectra of small and medium sized molecules such as H$_7^+$ \cite{Qu2013} or water timer \cite{Wang2008}.
Basis sets constructed from eigenfunctions of some simpler Hamiltonian are often called \textit{contracted basis sets}. Usually, less contracted basis functions are needed than the original, e.g. harmonic oscillator functions in representing the wavefunction of the molecule.

In ro-vibrational calculations of triatomic molecules a very successful methodology utilizing the contracted basis set approach has been presented by Tennyson \etal\ \cite{DVR3D}. The problem of finding rotational-vibrational wavefunctions and energy levels is divided into two essential steps: 1) solving the rotational-vibrational \SE\ assuming that $k$ is a good quantum number (i.e. neglecting all elements in the Hamiltonian which are not diagonal in $k$); 2) using the basis from step 1 to represent solutions to the full Hamiltonian. It turns out that for many triatomic molecules, the \textit{Coriolis-decoupled} basis obtained in step 1 is excellent, as it captures most of the information about the vibrational motion, only neglecting couplings between different $k$ states.

The first step in the Tennyson strategy is to represent the rotational motion in the complete symmetric top basis, by analytic integration over rotational degrees of freedom of the molecule. Next the two-step procedure described above is utilized to find the ro-vibrational levels for every $J$ value.

In general, the rotational degrees of freedom cannot be separated out from the internal degrees of freedom of the molecule, as shown in eq. \ref{eq:Kvr}. But because the angular momentum operators $\hat{J}_i$, where $i=x,y,z$ depend on the Euler angles alone, it is possible to use the spectral representation of the symmetric-top model Hamiltonian \cite{06BuJexx.method} for the rotational degrees of freedom (eigenfunctions of $\hat{J}^2$ and $\hat{J}_z$). The complete basis set of $2J+1$ functions: $\lbrace|J,k\rangle\rbrace_{k=-J,...,J}$ can be now utilized to perform analytical integration over the rotational degrees of freedom $\theta, \phi, \chi$ in the ro-vibrational Hamiltonian given in eq. \ref{eq:Hbody}, yielding a set of effective Hamiltonians which depend only on three internal coordinates describing the vibrational degrees of freedom, and are parametrized by the $J$ and $k$ numbers:

\begin{equation}
\hat{H}^{J'J}_{k'k}(r_1,r_2,\gamma)\delta_{J'J}=\langle J',k' | \hat{H}(r_1,r_2,\gamma,\theta, \phi, \chi) | J,k\rangle_{\alpha,\beta,\gamma}
\label{eq:Heff2}
\end{equation}
where $J$ is a good quantum number associated with the invariance of the ro-vibrational Hamiltonian to 3D-space rotations, but $k$ in general is not a good quantum number for a triatomic molecule. $k$ only becomes a good quantum number, associated with the $\hat{J}_z$ operator, when the molecule becomes linear (symmetric-top) or all Coriolis-couplings ($\hat{K}_{VR}$) are neglected. 
The general form of the effective vibrational Hamiltonian is given by:

\begin{equation}
\hat{H}^{J}_{k'k}(r_1,r_2,\gamma)=\delta_{k'k}\left(\hat{K}_{V}(r_1,r_2,\gamma)+V(r_1,r_2,\gamma)\right)+\hat{K}_{VR}(r_1,r_2,\gamma)
\label{eq:Heff9}
\end{equation}

where
\begin{equation}
\hat{K}_{V}(r_1,r_2,\gamma)=-\frac{1}{2}\left[\frac{1}{\mu_1r_1^2}\frac{\partial^2}{\partial r_1^2}+\frac{1}{\mu_2r_2^2}\frac{\partial^2}{\partial r_2^2}+\left(\frac{1}{\mu_1r_1^2}+\frac{1}{\mu_2r_2^2}\right)\frac{1}{\sin\gamma}\frac{\partial}{\partial\gamma}\sin\gamma\frac{\partial}{\partial\gamma}\right],
\label{eq:Kvibv}
\end{equation}

\begin{equation}
\hat{K}_{VR}(r_1,r_2,\gamma)=\delta_{k'k\pm 2}\frac{1}{4}C^{\pm}_{Jk\pm 1}C^{\pm}_{Jk}b_-+\delta_{k'k\pm 1}\frac{1}{2}C^{\pm}_{Jk}\lambda^{\pm}+\delta_{k'k}\frac{1}{2}\left(b_+(J(J+1)-k^2)+b_0k^2\right)
\label{eq:Kvibvr}
\end{equation}

with 
 \begin{equation}
 \begin{split}
\lambda^{\pm}=\frac{1}{\mu_1r_1^2}\left[\mp (1-a)\left(\frac{\partial}{\partial\gamma}+\frac{\cot\gamma}{2}\right)+\left(k\pm \frac{1}{2}\right)\frac{z_{x2}z_{z2}}{r_2^2\sin^2\gamma}\right]+\\
+\frac{1}{\mu_2r_2^2}\left[\pm a\left(\frac{\partial}{\partial\gamma}+\frac{\cot\gamma}{2}\right)+\left(k\pm \frac{1}{2}\right)\frac{z_{x1}z_{z1}}{r_1^2\sin^2\gamma}\right]
\end{split}
\label{eq:lambda}
\end{equation}

and 
 \begin{equation}
C_{Jk}^{\pm}=\left[J(J+1)-k(k\pm 1)\right]^{\frac{1}{2}},
\label{eq:c}
\end{equation}

\begin{equation*}
b_{\pm}=\frac{M_{xx}\pm M_{yy}}{2}, \qquad b_0=M_{zz}
\label{eq:b}
\end{equation*}
where the $z_{x1}$-type terms are elements of the molecule-fixed coordinate matrix defined in eq. \ref{eq:Crotation}.
The ro-vibrational Hamiltonian in the present form takes infinite values for $\gamma =0, \pi$, due to  $\frac{1}{\sin\gamma}$ terms appearing in both its vibrational and ro-vibrational part. Sutcliffe and Tennyson suggested \cite{jt96} that these singularities can be, at least partially, eliminated with the use of the associated Legendre polynomial basis $|jk\rangle=P^{(|k|)}_{j}(\cos\gamma)$ for the bending motion. Even though the Hamiltonian is singular at $\gamma =0, \pi$, when an appropriate form of basis functions is chosen, the product of the Hamiltonian and the basis function can be non-singular. Singularities which are not cancelled out by the choice of the basis set sometimes pose serious numerical problem. For this reason one should be careful to choose basis functions which have proper behaviour (nearly vanish) near singular points of the Hamiltonian. 

The  $|jk\rangle$ bending basis couples to the rotational motion through the $k$ quantum number. Integration over the bending coordinate $\gamma$, with the chosen basis in the phase convention of \textit{Condon and Shortley} \cite{06BuJexx.method} further simplifies the effective operators to the radial-vibrational form:

\begin{equation}
\hat{H}(r_1,r_2) = \delta_{k'k}\left(\hat{K}_{V_{j'j}}(r_1,r_2)+V_{j'j}^k(r_1,r_2)\right)+\hat{K}_{VR}(r_1,r_2)
\label{eq:Hrad}
\end{equation}

\begin{equation}
\begin{split}
\hat{K}_V(r_1,r_2)=-\delta_{j'j}\left[-\frac{1}{2\mu_1}\frac{\partial^2}{\partial r_1^2}-\frac{1}{2\mu_2}\frac{\partial^2}{\partial r_2^2}+\frac{1}{2}j(j+1)\left(\frac{1}{\mu_1r_1^2}+\frac{1}{\mu_2r_2^2}\right)\right]-\\
-\frac{k^2}{2}\left(\frac{1}{\mu_1r_1^2}+\frac{1}{\mu_2r_2^2}\right)\langle j'k'|\frac{1}{\sin^2\gamma}|jk\rangle.
\end{split}
\label{eq:KVrad}
\end{equation}

\begin{figure}[H]
  \includegraphics[width=14cm]{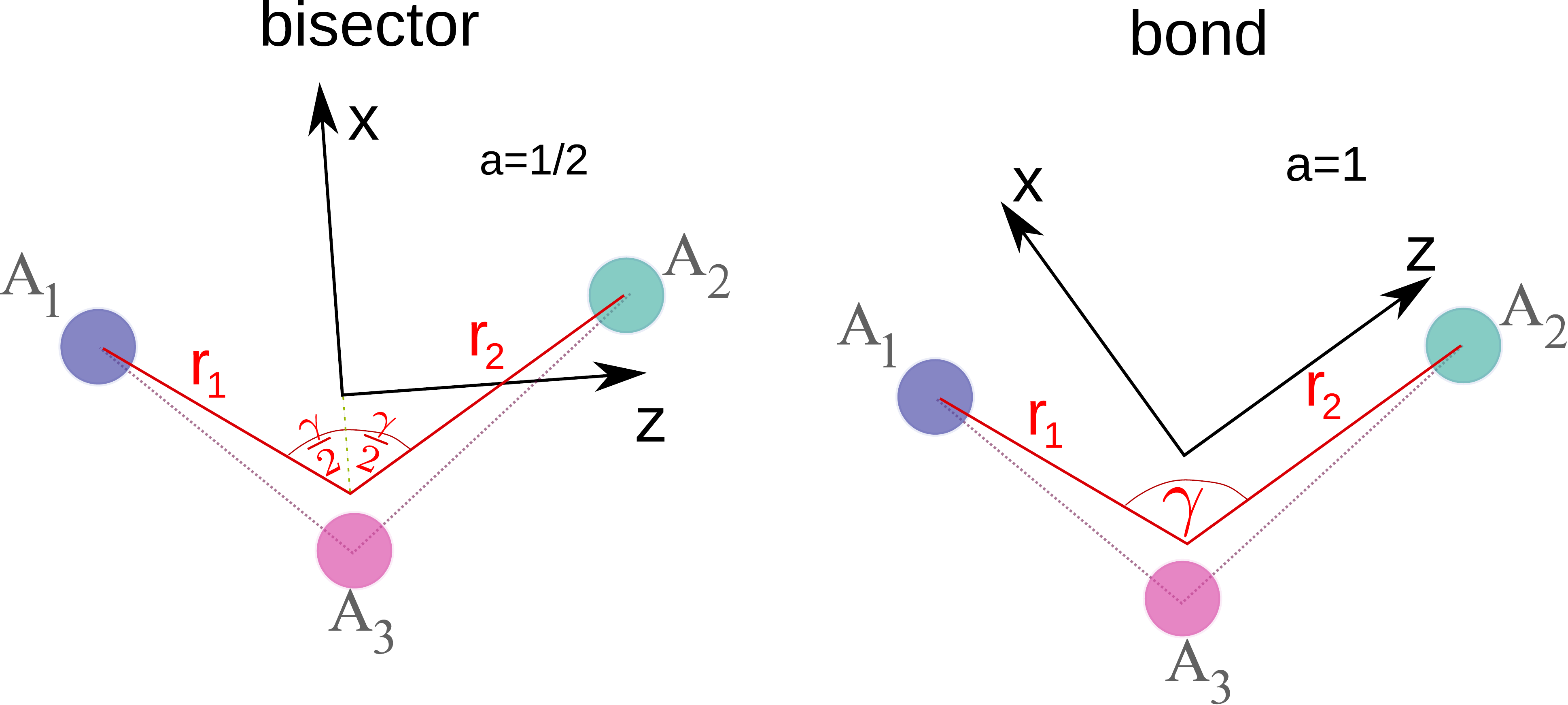}
\caption{Two types of embedding of the molecule-fixed coordinate frame used in calculations ro-vibrational spectra of triatomic molecules. $A_1,A_2,A_3$ stand for labels of atom 1, 2 and 3, respectively. $r_1,r_2,\gamma$ are Radau coordinates. The molecule-fixed axis system is centred at the nuclear centre of mass of the triatomic system. However the point at which Radau coordinates originate is a geometric mean between the distance from atom $A_3$ and the centre of mass for atoms $A_1$ and $A_2$ and the distance from the nuclear centre of mass to the centre of mass for atoms $A_1$ and $A_2$. In the bisector embedding the $x$-axis bisects the $\gamma$ angle, whereas in bond embeddings the $z$-axis is parallel to the $r_1$ or $r_2$ coordinate.}
\centering
\label{fig:embeddings}
\end{figure}

The form of the rotation-vibration operator $\hat{K}_{VR}(r_1,r_2)$ in eq. \ref{eq:Hrad} depends on the embedding of the molecule-fixed coordinate frame. Here two cases are considered: \textit{bond embedding }($a=0,1$) and \textit{bisector embedding} ($a=\frac{1}{2}$). Both types of embeddings are depicted in Figure \ref{fig:embeddings}. From now on we are going to use Radau internal coordinates \cite{Johnson1986}, which are also displayed in Figure \ref{fig:embeddings}. The bisector embedding of the molecule-fixed frame is often useful in calculations of ro-vibrational line lists for the symmetric molecules, such as CO$_2$, H$_2$O or SO$_2$ whereas the bond embedding is used for the asymmetric molecules, such as asymmetric isotopologues of CO$_2$ or HDO. 

\paragraph{Bond embedding.}
In the bond embedding the $z$-axis in the molecule-fixed coordinate frame lies along $\textbf{t}_1$ ($a=0$) or $\textbf{t}_2$ ($a=1$). The $\textbf{t}_i$ vectors are defined in eq. \ref{eq:tcoord}.   The rotation-vibration part of the KEO is then given by

 \begin{equation}
\hat{K}_{VR}(r_1,r_2)=-\delta_{j'j}\delta_{k'k}\frac{1}{2\mu_1r_1^2}(J(J+1)-2k^2)-\delta_{j'j}\delta_{k'k\pm 1}\frac{1}{2\mu_1r_1^2}C_{Jk}^{\pm}C_{jk}^{\pm}
\label{eq:KVRradbond}
\end{equation}

\paragraph{Bisector embedding.}
In the bisector embedding the $x$-axis of the molecule-fixed coordinate frame bisects the angle between $\textbf{t}_1$ and $\textbf{t}_2$ ($a=\frac{1}{2}$). In such case, the rotation-vibration KEO couples states with $k'=k\pm 1$ and $k'=k\pm 2$ and can be split into three parts:
 \begin{equation}
\begin{split}
\hat{K}_{VR}^{(1)}(r_1,r_2)=\delta_{k'k}\frac{1}{8}(J(J+1)-3k^2)\left(\frac{1}{\mu_1r_1^2}+\frac{1}{\mu_2r_2^2}\right)I^{(1)}_{j'k'jk}+\\
+\delta_{k'k}\delta_{j'j}\frac{1}{16}(J(J+1)-k^2)\left(\frac{1}{\mu_1r_1^2}+\frac{1}{\mu_2r_2^2}\right)
\end{split}
\label{eq:KVRradbisc1}
\end{equation}

 \begin{equation}
\hat{K}_{VR}^{(2)}(r_1,r_2)=\delta_{k'k\pm 1}\frac{C_{jk}^{\pm}}{4}\left(\frac{1}{\mu_2r_2^2}-\frac{1}{\mu_1r_1^2}\right)\left(\delta_{j'j}C_{jk}^{\pm}+\left(k\pm \frac{1}{2}\right)I^{(2)}_{j'k'jk}\right)
\label{eq:KVRradbisc2}
\end{equation}

 \begin{equation}
\hat{K}_{VR}^{(3)}(r_1,r_2)=\delta_{k'k\pm 2}C_{Jk\pm 1}^{\pm}C_{Jk}\left(\frac{1}{\mu_2r_2^2}+\frac{1}{\mu_1r_1^2}\right)\left(2I^{(1)}_{j'k'jk}-I^{(3)}_{j'k'jk} \right)
\label{eq:KVRradbisc3}
\end{equation}

where

 \begin{equation}
I^{(1)}_{j'k'jk}=\langle j'k'|\frac{1}{1-\cos\gamma}|jk\rangle
\label{eq:I1}
\end{equation}

 \begin{equation}
I^{(2)}_{j'k'jk}=\langle j'k'|\frac{1+\cos\gamma}{\sin\gamma}|jk\rangle
\label{eq:I2}
\end{equation}

 \begin{equation}
I^{(3)}_{j'k'jk}=\langle j'k'|jk\rangle
\label{eq:I3}
\end{equation}

Coming back to the general embedding, both operator matrices \ref{eq:Kvibv} and \ref{eq:Kvibvr} are diagonal in $J$, and the rotation-vibration coupling operator in eq. \ref{eq:Kvibvr} have a strip-pentadiagonal structure in $k$, meaning that it can have non-zero elements only two rows or two columns away from the diagonal. The final form of the rotation-vibration operator depends on the embedding chosen, which is reasonable, because individual moments of inertia and the magnitude of the rotation-vibration coupling strongly depend on where the molecule-fixed axis is placed. Another, not discussed here common choice for the molecule-fixed embedding is the so called \textit{Eckart frame} \cite{06BuJexx.method}, which by definition minimizes the coupling between rotations and vibrations. Here however, only embeddings fixed to a chosen set of internal coordinates are discussed; these have been shown to be sufficient for highly accurate nuclear motion calculations \cite{jt519,15ZaTePo.CO2,17ZaTePo.CO2,Zak2017} and give much simpler representation of the Hamiltonian operator. Of course, one could be concerned that a fixed embedding of the molecule-fixed frame will result in poor convergence of energy level calculations. The Eckart embedding comes to mind naturally. However, first of all, with the Eckart embedding the form of the kinetic energy operator becomes complicated \cite{Szalay2014,Lauvergnat2016}. Secondly, the Eckart frame is not suitable for very floppy systems (with large amplitude vibrational motion) and no advantage could be gained over the fixed embeddings for such systems \cite{Csszr2012}. Finally, an efficient algorithm by Tennyson \etal\ discussed broadly in ref. \cite{DVR3D}, for solution to the ro-vibrational \SE, diminishes the drawback of the non-minimal rotation-vibration coupling. All in all, a good level of convergence can be achieved with the fixed embeddings too.

Having the effective vibrational Hamiltonian from eq. \ref{eq:Hrad} derived, the next step is to choose a vibrational radial basis, for the calculation of the matrix elements of the operators in eqs. \ref{eq:KVRradbond}--\ref{eq:KVRradbisc3}. After the choice of the vibrational radial basis, the resulting Hamiltonian matrix has to be diagonalised, to obtain expansion coefficients in eq. \ref{eq:anzatzrv} and ro-vibrational energy levels. 

From the computational perspective, the reduction in the number of degrees of freedom outlined in section \ref{sec:KEO}, from 9 Cartesian coordinates to three effective vibrational degrees of freedom labeled by rotational quantum numbers is a huge improvement. This improvement however comes at a cost of singularities arising in the ro-vibrational Hamiltonian and the complicated form of the kinetic energy operator. On the other hand, as we expected, in the internal coordinates framework it is easy to identify vibrational, rotational and rotational-vibrational parts of the Hamiltonian, thus physically motivated approximations can be made at hand. 

As we could see in this section, the total ro-vibrational Hamiltonian matrix is constructed gradually by integrating over rotational, bending and radial stretching degrees of freedom. 
Direct calculation of all matrix elements of the KEO and the PES in the basis presented in eq. \ref{eq:rovibfinal} is impractical, and becomes prohibitive for higher $J$ values. For this reason Sutcliffe and Tennyson proposed a two-step procedure of diagonalizing the ro-vibrational Hamiltonian matrix \cite{jt114,jt46,jt66}. In the first step a Coriolis-decoupled Hamiltonian is considered:

\begin{equation}
\hat{H}_{K}=\delta_{K,K'}\hat{K}_V+\delta_{K,K'}\hat{K}_{VR}+\delta_{K,K'}V
\label{eq:ham1}
\end{equation} 
and the respective \SE\ is solved with $K=|k|=p,p+1,...,J$ as a good quantum number for each $J$ separately. This approximation is valid for any system with negligible Coriolis coupling, which mixes states with different $K$. With these assumptions, the solutions to the \SE\ in the first step can be written as:
\begin{equation}
|J,p,q,K,h\rangle= \sum_{m,n,j} d_{mnj}^{J,(h),K,p,q} |m,n,q\rangle  |J,K,j,p\rangle
\label{eq:firststep}
\end{equation}
with the corresponding energy levels $\epsilon^{J,(h),K,p,q}$. For a chosen $J$ it is necessary to solve only $J+p$ nuclear motion problems for $K=p,1,...,J$.

In the second step, solutions to the first step are used as the variational basis. Typically, good convergence is achieved with even very contracted basis set from the first step, because the Coriolis-decoupled basis captures the majority of physical information about the vibrational motion, hence serves as an excellent basis. For this reason, one can usually choose only a small percentage of solutions from the first step to achieve a good convergence level. 
The final wavefunction mixes states with different $K$'s, as suggested by the form of the $K_{VR}$ operators given in eq. \ref{eq:KVRradbisc1}-\ref{eq:KVRradbisc3}. The final symmetry-adapted wavefunction is expressed as
\begin{equation}
 |J,h,p,q\rangle = \sum_{K=p}^J f_{K}^{J,h,p,q} |J,p,q,K,h\rangle =  \sum_{K=p}^J \sum_{m,n,j} C_{mnjK}^{J,(h),p,q} |m,n,q\rangle |J,K,j,p\rangle
\label{eq:secondstep}
\end{equation} 
Diagonalization of the Hamiltonian matrix given in eq. \ref{eq:Hrad} can be done with direct methods, implemented for instance in the LAPACK computational library (netlib.org/lapack).

\section{Appendix A: Quantum harmonic oscillator}
In this appendix we give a detailed solution to the Quantum harmonic oscillator model. The Hamiltonian for the 1-D harmonic oscillator can be written as
\begin{equation}
\hat{H}_{1D}=\frac{\hat{p}_{x}^{2}}{2\mu}+\frac{1}{2}\mu\omega^{2}\hat{x}^{2}
\label{Harm:Ham1}
\end{equation}
where $\hat{p}_{x}$ denotes the $x$-component of the momentum operator and $\hat{x}$ stands for the $x$-component of the position operator. $\mu$ represents the reduced mass of the system. 
We want to solve the stationary \SE\ :
\begin{equation}
\hat{H}_{1D}|\psi\rangle=E|\psi\rangle
\label{harm:SE}
\end{equation}
For wavefunctions belonging to  $L^{2}(R,dx)$, that is to the set of square-integrable functions, the Hamiltonian given in eq. \ref{Harm:Ham1} is Hermitian and has discrete and positive spectrum.
Eq. \ref{harm:SE} can be solved analytically using the \textit{Frobenius method}
Here, we are going to follow the details of a considerably more elegant algebraic approach for solving eq. \ref{harm:SE}.

In the first place, let us introduce dimensionless operators by scaling the position and momentum operators. By doing so, one obtains convenient units of length, momentum and energy, which are natural for the harmonic oscillator. The new operators are defined as
\begin{equation} \label{Harm:alpha}
\hat{x}=\alpha\hat{X}, \; \alpha \in R \qquad \hat{p}_{x}=\frac{\hslash}{\alpha}\hat{P}
\end{equation}
where $\alpha$ has the dimension of length. The Hamiltonian in new dimensionless representation reads
\begin{equation}
\hat{H}_{1D}=\frac{\hslash^{2}}{2\mu\alpha^{2}}\hat{P}_{x}^{2}+\frac{1}{2}\mu\omega^{2}\alpha^{2}\hat{X}^{2}
\label{harm:ham2}
\end{equation}
At this point one has the freedom to choose the value for the scaling constant $\alpha$. Requiring the constants standing by $\hat{P}_{x}^{2}$ and $\hat{X}^{2}$ in eq.\ref{harm:ham2} to be equal sets $\alpha=\sqrt{\frac{\hslash}{\mu\omega}}$. Then the dimensionless position and momentum operators take the form
\begin{equation}
\hat{X}=\sqrt{\frac{\mu\omega}{\hslash}}\hat{x},  \qquad \hat{P}=\sqrt{\frac{1}{\hslash\omega\mu}}\hat{p}_{x}
\end{equation}
and the Hamiltonian simplifies to
\begin{equation}
\hat{H}_{1D}=\frac{\hslash\omega}{2}\left(\hat{P}^{2}+\hat{X}^{2}\right)
\label{harm:ham3}
\end{equation}
The Hamiltonian given in eq. \ref{harm:ham3} can be put into a more compact form by means of the following unitary transformation:
\begin{equation} \label{creationdefinition}
a=\frac{1}{\sqrt{2}}\left(\hat{X}+i\hat{P}\right), \qquad a^{\dag}=\frac{1}{\sqrt{2}}\left(\hat{X}-i\hat{P}\right)
\end{equation}

The new operators are called the \textit{annihilation} and \textit{creation} operators, respectively. Creation and annhilation operators are each other's Hermitian conjugates and together they constitute \textit{ladder operators} for the 1-D harmonic oscillator. The Hamiltonian operator expressed in these new operators reads
\begin{equation}\label{harm:ladderHamiltonian}
\hat{H}_{1D}=\hslash\omega\left(a^{\dag}a+\frac{1}{2}\right)
\end{equation}
The form of the above representation to the 1-D harmonic oscillator Hamiltonian indicates that it is a sum of a single operator $\hat{N}:=a^{\dag}a$ and a constant term $\frac{1}{2}\hslash\omega$. The former is called \textit{number operator} and the latter is \textit{zero-point energy} of the 1-D harmonic oscillator.
It is evident that if function $f$ is an eigenfunction of $\hat{N}$, so it is an eigenfunction of $\hat{H}_{1D}$. Assume that $\hat{N}f=\lambda f$. Then
\begin{equation}
\lambda=\langle f|\lambda f\rangle=\langle f|\hat{N}f\rangle=\langle\hat{N}\rangle_{f}=\langle af|af\rangle=\|af\|^{2}\ge 0 \Rightarrow \lambda \ge 0
\end{equation}
We can see that $\hat{N}$ as well as $\hat{H}_{1D}$ have a non-negative spectrum. Does there exist an eigenfunction which corresponds to the lowest possible $\lambda=0$? Assuming that $\lambda=0$ exists we can write
\begin{equation}
\hat{N}f_{0}=0
\end{equation}
for some $f_0$. We then find that $\langle f_{0}|\hat{N}f_{0}\rangle=0 \Rightarrow  \langle f_{0}|\hat{N}f_{0}\rangle=\|af_{0}\|^{2}=0 \Rightarrow \|af_{0}\|=0$. In consequence, in order to obtain the generic solution to the \SE\ for $\lambda=0$ we need to solve the equation $af_{0}=0$, which can be readily written as
\begin{equation}
\left(X+\frac{d}{dX}\right)f_{0}\left(X\right)=0
\end{equation}
with standard textbook solution:
\begin{equation}
f_{0}\left(X\right)=\pi^{-\frac{1}{4}}e^{-\frac{1}{2}X^{2}}
\end{equation}
Our goal now is to derive a general formula for the $n$-th eigenfunction of $\hat{N}$, that is for $f_{n}$, utilizing only creation and annihilation operators. This can be done with the aid of the following commutation relations:
\begin{equation}
\left[a,a^{\dag}\right]=\frac{1}{2}\left[\hat{X}+i\hat{P},\hat{X}-i\hat{P}\right]=\mathrm{1}
\end{equation}
where we made use of the \textit{canonical commutation relations} in the dimensionless representation $\left[\hat{X},\hat{P}\right]=i$. The commutation relation between the creation and annihilation operator is helpful in calculating the following commutator
\begin{equation} \label{harm:laddercommutation}
\left[a,\left(a^{\dag}\right)^{n}\right]=a^{\dag}\left[a,\left(a^{\dag}\right)^{n-1}\right]+\left[a,a^{\dag}\right]\left(a^{\dag}\right)^{n-1}=...=
n\left(a^{\dag}\right)^{n-1}.
\end{equation}
We see that for n=2: $\left[a,\left(a^{\dag}\right)^{n}\right]=2a^{\dag}$. By induction, assuming that for some arbitrary $n$ eq.\ref{harm:laddercommutation} holds, we find that:
\begin{equation}
\left[a,\left(a^{\dag}\right)^{n+1}\right]=a^{\dag}\left[a,\left(a^{\dag}\right)^{n}\right]+\left[a,a^{\dag}\right]\left(a^{\dag}\right)^{n}=n\left(a^{\dag}\right)^{n-1}+
\left(a^{\dag}\right)^{n}=\left(n+1\right)\left(a^{\dag}\right)^{n}.
\end{equation}
But at the same time we have
\begin{equation}
\left[\hat{N},\left(a^{\dag}\right)^{n}\right]=n\left(a^{\dag}\right)^{n}
\end{equation}
which gives
\begin{equation}
\left[\hat{N},\left(a^{\dag}\right)^{n}\right]f_{0}=n\left(a^{\dag}\right)^{n}f_{0} 
\end{equation}

To summarize, the eigenvalue problem for $\hat{N}$ takes the form
\begin{equation}
\hat{N}\left(a^{\dag}\right)^{n}f_{0}=n\left(a^{\dag}\right)^{n}f_{0}
\end{equation}
which means that $\left(a^{\dag}\right)^{n}f_{0}$ is the eigenfunction of $\hat{N}$ with eigenvalue $n$, hence in fact must be proportional to $f_{n}$. For this reason $\hat{N}$ is called \textit{number operator}. Normalization of $f_{n}$ is straightforward:
\begin{equation}
\|\left(a^{\dag}\right)^{n}f_{0}\|^{2}=\left(\left(a^{\dag}\right)^{n}f_{0},\left(a^{\dag}\right)^{n}f_{0}\right)=
\left(\left(a^{\dag}\right)^{n-1}f_{0},a\left(a^{\dag}\right)^{n}f_{0}\right)=n\left(\left(a^{\dag}\right)^{n-1}f_{0},\left(a^{\dag}\right)^{n-1}f_{0}\right)=...=
n!
\end{equation}
where we used of eq.~\ref{harm:laddercommutation} and the fact that $af_{0}=0$. Finally, the normalized eigenfunctions are
\begin{equation}\label{harm:fn}
f_{n}=\frac{1}{\sqrt{n!}}\left(a^{\dag}\right)^{n}f_{0}
\end{equation}
and the \SE\ can be written as:
\begin{equation}
\hat{H}_{1D}f_{n}=\hslash\omega\left(\hat{N}+\frac{1}{2}\right)f_{0}=\hslash\omega\left(n+\frac{1}{2}\right)f_{n}, \quad n=0,1,2,...
\end{equation}
Functions \ref{harm:fn} an orthonormal set in $L^{2}\left(\mathbb{R},dx\right)$ as they are eigenfunctions of the Hermitian operator. For more explicit form of the 1-D harmonic oscillator wavefunctions let us inspect eq. \ref{harm:fn}:
\begin{equation}\label{harm:explicit}
f_{n}=\frac{1}{\sqrt{n!}}\left(a^{\dag}\right)^{n}f_{0}=\frac{1}{\sqrt{2^{n}\pi^{\frac{1}{2}}n!}}\left(X-\frac{d}{dX}\right)^{n}e^{-\frac{1}{2}X^{2}}
\end{equation}
According to the identity
\begin{equation}\label{harm:explicit2}
-\frac{d}{dX}\left(g(X)e^{-\frac{1}{2}X^{2}}\right)=e^{-\frac{1}{2}X^{2}}\left(X-\frac{d}{dX}\right)g(X)
\end{equation}
for any differentiable function $g(X)$ we can write
\begin{equation}\label{harm:explicit3}
(-1)^{n}\frac{d^{n}}{dX^{n}}\left(g(X)e^{-\frac{1}{2}X^{2}}\right)=e^{-\frac{1}{2}X^{2}}\left(X-\frac{d}{dX}\right)^{n}g(X)
\end{equation}
By pasting $g(X)=e^{-\frac{1}{2}X^{2}}$ we finally get
\begin{equation}\label{harm:explicit4}
(-1)^{n}e^{\frac{1}{2}X^{2}}\frac{d^{n}}{dX^{n}}\left(e^{-X^{2}}\right)=\left(X-\frac{d}{dX}\right)^{n}e^{-\frac{1}{2}X^{2}}
\end{equation}
Hence,
\begin{equation}\label{harm:harm:1Dsolution}
f_{n}\left(X\right)=\frac{1}{\sqrt{2^{n}\pi^{\frac{1}{2}}n!}}(-1)^{n}e^{\frac{1}{2}X^{2}}\frac{d^{n}}{dX^{n}}e^{-X^{2}}:=
\frac{1}{\sqrt{2^{n}\pi^{\frac{1}{2}}n!}}H_{n}\left(X\right)e^{-\frac{1}{2}X^{2}}
\end{equation}
where $H_{n}\left(X\right)$ is $n-th$ \textit{Hermite polynomial} defined as $H_{n}\left(X\right):=(-1)^{n}e^{\frac{1}{2}X^{2}}\frac{d^{n}}{dX^{n}}e^{-\frac{1}{2}X^{2}}$. In the original representation of the position operator the 1-D harmonic oscillator wavefunction is given by
\begin{equation}\label{harm:explicit5}
\psi_{n}\left(x\right)=\frac{1}{\sqrt{2^{n}\pi^{\frac{1}{2}}n!}}\left(\frac{\mu\omega}{\hslash}\right)^{\frac{1}{4}}H_{n}\left(\sqrt{\frac{\mu\omega}{\hslash}}x\right)e^{-\frac{1}{2}\frac{\mu\omega}{\hslash}x^{2}}
\end{equation}

Lastly, let us derive here the matrix elements of the creation and annihilation operators in the harmonic oscillator eigenbasis. The action of the annihilation operator $a$ on the $n-th$ eigenstate is given by
\begin{equation}\label{harm:explicit6}
\begin{split}
af_{n}=\frac{1}{\sqrt{n!}}a\left(a^{\dag}\right)^{n}f_{0}=\frac{1}{\sqrt{n!}}\left[a,\left(a^{\dag}\right)^{n}\right]f_{0}=\\
=\frac{n}{\sqrt{n!}}\left(a^{\dag}\right)^{n-1}f_{0}=\sqrt{n}
\frac{1}{\sqrt{(n-1)!}}\left(a^{\dag}\right)^{n-1}f_{0}=\sqrt{n}f_{n-1}
\end{split}
\end{equation}
For the creation operator we can write
\begin{equation}\label{harm:explicit7}
a^{\dag}f_{n}=\frac{1}{\sqrt{n!}}\left(a^{\dag}\right)^{n+1}f_{0}=\frac{\sqrt{n+1}}{\sqrt{(n+1)!}}\left(a^{\dag}\right)^{n+1}f_{0}=\sqrt{n+1}f_{n+1}
\end{equation}
The above relations can be used to calculate matrix elements of the position and momentum operator in the eigenbasis $\{f_{n}\}_{n=0,1,2,...}$. In this basis the matrices for the creation and annihilation operators are given respectively as
\begin{equation}
\left[a^{\dag}\right]=
\left(\begin{array}{cccccc}
0 & 0 & 0 & 0 & 0 &\ldots\\
\sqrt{1} & 0 & 0 & 0 & 0 & \ldots\\
0 & \sqrt{2} &  0 & 0 & 0 &\ldots\\
0 & 0 & \sqrt{3} & 0 & 0 & \ldots\\
\vdots
\end{array}\right)
\end{equation}
\begin{equation}
\left[a\right]=
\left(\begin{array}{cccccc}
0 &\sqrt{1}& 0 & 0 & 0 &\ldots\\
0 & 0&\sqrt{2} & 0 & 0 & \ldots\\
0 & 0 &  0 & \sqrt{3} & 0 &\ldots\\
0 & 0 & 0 & 0 & \sqrt{4} & \ldots\\
\vdots
\end{array}\right)
\end{equation}

\subsection{3D isotropic harmonic oscillator in spherical coordinates}
In Cartesian coordinates the Hamiltonian of the isotropic three-dimensional harmonic oscillator has the form
\begin{equation}\label{harm:3Dcart}
\hat{H}_{3D}(X,Y,Z)=-\frac{\hslash^{2}}{2\mu}\Delta+\frac{1}{2}\mu\omega^{2}\left(\hat{x}^{2}+y^{2}+z^{2}\right)
\end{equation}
where $\mu$ is the reduced mass of the system and $\omega$ is the harmonic frequency.
The form of the above Hamiltonian suggests that it is spherically symmetric. Therefore a reasonable choice for a coordinate system is spherical coordinates
\begin{equation}
\hat{H}_{3D}(r,\theta,\phi)=-\frac{\hslash^{2}}{2\mu r^{2}}\frac{\partial}{\partial r}\left(r^{2}\frac{\partial}{\partial r}\right)+\frac{1}{2\mu r^{2}}\hat{L}^{2}+\frac{1}{2}\mu\omega^{2}r^{2}
\end{equation}
where, $\hat{L}$ stands for total angular momentum operator. Because $\hat{L}$ depends only on spherical angles, it commutes with the quadratic potential and the radial kinetic energy operator $-\frac{\hslash^{2}}{2\mu r^{2}}\frac{\partial}{\partial r}\left(r^{2}\frac{\partial}{\partial r}\right)$ \footnote{the result can be proved using the following commutation relations: $\left[\hat{L}_{i},\hat{p}_{j}^{2}\right]=\epsilon_{ikl}\left[\hat{x}_{k}\hat{p}_{l},\hat{p}_{j}^{2}\right]= \epsilon_{ikl}\left[\hat{x}_{k},\hat{p}_{j}^{2}\right]\hat{p}_{l}=2i\hslash\epsilon_{ikl}\delta_{kj}\hat{p}_{j}\hat{p}_{l}=0$, since we sum in default over $k,l$ and the expression is a product of totally symmetric $\hat{p}_{j}\hat{p}_{l}$ and totally antisymmetric species $\epsilon_{ijl}$.}.
For this reason there must exist a common eigenbasis for the total Hamiltonian and the angular part of the Hamiltonian. This leads to factorization of the total wavefunction for 3-D harmonic oscillator
\begin{equation}
\psi(r,\theta,\phi)=R_{nl}(r)Y_{lm}(\theta,\phi)
\end{equation}
After some manipulations the \SE\ reads
\begin{equation}
\left(-\frac{\hslash^{2}}{2\mu r^{2}R(r)_{nl}}\frac{\partial}{\partial r}\left(r^{2}\frac{\partial}{\partial r}\right)+\frac{1}{2}\mu\omega^{2}r^{2}\right)R_{nl}(r)+\frac{1}{2\mu r^{2}Y_{lm}(\theta,\phi)}\hat{L}^{2}Y_{lm}(\theta,\phi)=E
\end{equation}
The angular part of the above equation is a \textit{3-D homogeneous Laplace equation}, with spherical harmonics as solutions:
\begin{equation}
\frac{1}{2\mu r^{2}}\hat{L}^{2}Y_{lm}(\theta,\phi)=\frac{\hslash^{2}l(l+1)}{2\mu r^{2}}{Y_{lm}(\theta,\phi)}\equiv E_{rot}Y_{lm}(\theta,\phi)
\end{equation}
and because the angular part depends also on $r$ it must be incorporated in the radial equation, providing a coupling between the rotational and vibrational motion.
\begin{equation}\label{harm:radial}
\left(-\frac{\hslash^{2}}{2\mu r^{2}}\frac{\partial}{\partial r}\left(r^{2}\frac{\partial}{\partial r}\right)+\frac{1}{2}\mu\omega^{2}r^{2}\right)R_{nl}(r)+\frac{\hslash^{2}l(l+1)}{2\mu r^{2}}R_{nl}(r)=E R_{nl}(r)
\end{equation}
Analogically to the one-dimensional case we introduce dimensionless quantities: $\alpha=\sqrt{\frac{\hslash}{\mu\omega}}$, which yield a representation of the dimensionless position operator:
\begin{equation}
\hat{\rho}=\alpha^{-1}\hat{r}
\end{equation}
so that eq. \ref{harm:radial} takes the form
\begin{equation}\label{harm:radial1}
\frac{d^{2}R(\rho)}{d\rho^{2}}+\frac{2}{\rho}\frac{dR(\rho)}{d\rho}-\rho^{2}R(\rho)-\frac{l(l+1)}{\rho^{2}}R(\rho)+\frac{2E}{\hslash\omega}R(\rho)=0
\end{equation}
noting that $R_{nl}(r)$ and $R(\rho)$ are different functions.
In order to predetermine an appropriate anzatz for the radial wavefunction it is a common practice to investigate the asymptotic behaviour of the solutions. Let us focus on eq. \ref{harm:radial1}. In the limit of large $\rho$ the equation reduces to
\begin{equation}
\frac{d^{2}R(\rho)}{d\rho^{2}}-\rho^{2}R(\rho)=0
\end{equation}
providing Gaussian-type solutions:
\begin{equation}
R(\rho)=e^{-\frac{1}{2}\rho^{2}}
\end{equation}
In the opposite limit of small $\rho$ the equation
\begin{equation}
\frac{d^{2}R(\rho)}{d\rho^{2}}+\frac{2}{\rho}\frac{dR(\rho)}{d\rho}-\frac{l(l+1)}{\rho^{2}}R(\rho)=0
\end{equation}
suggests the following anzatz $R(\rho)=\rho^{t}$, which generates the algebraic condition
\begin{equation}
t(t-1)+2t-l(l+1)=0
\end{equation}
setting value to the exponent at $t=l$. Finally it is possible to propose a reasonable anzatz for the radial wavefunction  in the following form
\begin{equation}
R(\rho)=\rho^{l}\sum_{k=0}^{+\infty}a_{k}\rho^{k}e^{-\frac{1}{2}\rho^{2}}
\end{equation}
First we compute derivatives
\begin{equation}
\frac{dR(\rho)}{d\rho}=\sum_{k=0}^{+\infty}a_{k}\left[(l+k)\rho^{k+l-1}-\rho^{l+k+1}\right]e^{-\frac{1}{2}\rho^{2}}
\end{equation}
\begin{equation}
\frac{d^{2}R(\rho)}{d\rho^{2}}=\sum_{k=0}^{+\infty}a_{k}\left[(l+k)(l+k-1)\rho^{k+l-2}-(2l+2k+1)\rho^{l+k}+\rho^{l+k+2}\right]e^{-\frac{1}{2}\rho^{2}}
\end{equation}
Next we can place the above expressions into \ref{harm:radial1}, divide by the exponent factor and obtain
\begin{equation}
\begin{split}
\sum_{k=0}^{+\infty}a_{k}\left[(l+k)(l+k-1)\rho^{k+l-2}-(2l+2k+1)\rho^{l+k}+\rho^{l+k+2}+2(l+k)\rho^{k+l-2}-\right.\\
\left.-2\rho^{l+k}-\rho^{l+k+2}-l(l+1)\rho^{l+k-2}+\frac{2E}{\hslash\omega}\rho^{l+k}\right]=0
\label{eq:3dhofrob}
\end{split}
\end{equation}
Note that powers in above equation differ by two. After some rearrangement we get
\begin{equation}
\sum_{k=0}^{+\infty}a_{k}\left[k(2l+k+1)\rho^{k+l-2}+\left(\frac{2E}{\hslash\omega}-(2l+2k+3)\right)\rho^{l+k}\right]=0
\end{equation}
First two terms for $k=0,1$  in the above equation determine coefficients $a_{0},a_{1}$. For $k=0$ the expression standing by $a_{0}$ vanishes so $a_{0}$ is allowed in the solution and is undetermined (may be chosen arbitrarily). For $k=1$:
\begin{equation}
a_{1}\left[(2l+2)\rho^{l-1}+\left(\frac{2E}{\hslash\omega}-(2l+5)\right)\rho^{l+1}\right]=0
\end{equation}
we can see that the function in brackets do not vanish for any $l$ value, hence $a_{1}$ must be zero. By shifting the summation in eq. \ref{eq:3dhofrob} to obtain equal powers in the monomials we get
\begin{equation}
\sum_{k=2}^{+\infty}\left[a_{k+2}(k+2)(2l+k+3)+a_{k}\left(\frac{2E}{\hslash\omega}-(2l+2k+3)\right)\right]\rho^{l+k}=0
\end{equation}
Forcing all coefficients standing by the powers of $\rho$  to vanish results in the following recurrence relation:
\begin{equation}
a_{k+2}=-\frac{\frac{2E}{\hslash\omega}-(2l+2k+3)}{(k+2)(2l+k+3)}a_{k}
\label{eq:recurrence}
\end{equation}
In the limit of large $k's$ this series behaves as a harmonic-like, i.e. $a_{k+2}=\frac{2}{k}a_{k}$, hence is divergent. Therefore there must exist such $s>0$ for which $a_{s}=0$, which implies all subsequent coefficients to be also zero. From the above recurrence relation is it possible to read a condition on allowed energy levels for the 3-D harmonic oscillator
\begin{equation}
\frac{2E}{\hslash\omega}-(2l+2s+3)=0
\end{equation}
and after some rearrangement we arrive at the familiar expression for eigenvalues of the 3-D harmonic oscillator:
\begin{equation} \label{eigenenergy}
E=\hslash\omega(l+s+\frac{3}{2})
\end{equation}
Note that here the total energy  depends explicitly on the $l$-angular momentum quantum number, meaning that there exists a rotational-vibrational coupling, which manifests itself in the dependence of the height of the potential barrier on the rotational state.
Inserting the expression for the total energy into the recurrence relation given in eq. \ref{eq:recurrence} gives
\begin{equation}
a_{k+1}=\frac{k-s}{(k+1)(l+k+\frac{3}{2})}a_{k}
\end{equation}
and the final form of the solution is given by
\begin{equation}
R_{nl}(\rho)=\sum_{k=0}^{n}a_{k}(s)\rho^{l+2k}e^{-\frac{1}{2}\rho^{2}}
\end{equation}
The total energy can be written as
\begin{equation}
E=\hslash\omega(l+2s+\frac{3}{2})=\hslash\omega(n+\frac{3}{2}),\qquad n=0,1,2,...
\end{equation}
The total wavefunction of the 3-D isotropic harmonic oscillator is given by
\begin{equation}\label{harm:ultimatesolution}
\psi(r,\theta,\phi)=R_{nl}(r)Y_{lm}(\theta,\phi)
\end{equation}
The radial part of this wavefunction can be expressed by \textit{associated Laguerre polynomials}:
\begin{equation}\label{sphericaloscillator}
R_{nl}(r)=N_{nl}r^{l}L^{l+\frac{1}{2}}_{n}\left(\frac{\mu\omega}{\hslash}r\right)e^{-\frac{1}{2}\frac{\mu\omega}{\hslash}r^{2}}
\end{equation}
where the normalization constant is given by $N_{nl}=\left(\frac{\mu\omega}{\hslash}\right)^{\frac{1}{2}l+\frac{3}{4}}\sqrt{\frac{2^{k+2l+\frac{3}{2}}k!}{\pi(2k+2l+1)!!}}$. Functions from eq \ref{sphericaloscillator} are called \textit{spherical oscillator functions} and are commonly used as basis functions in solving nuclear motion problems. The energy levels of 3-D isotropic harmonic oscillator are $\frac{1}{2}(n+1)(n+2)$-degenerate. This result can be derived in a straightforward way when one considers a 3-D harmonic oscillator in Cartesian representation and counts how many possible combinations of quantum numbers $(n_x,n_y,n_z)$ associated with the $x$,$y$ and $z$ dimension, respectively, add up to the same value.  

\section{Appendix B: symmetric top eigenfunctions}
\subsection{Symmetric top}
The \SE\ for the symmetric top model of the rigid rotor is given by:
\begin{equation}
\left(A_{e}\hat{J}_{z}^{2}+B_{e}(\hat{J}_{b}^{2}+\hat{J}_{c}^{2})\right)\Phi_{rot}(\theta, \phi, \chi)=E_{rot}\Phi_{rot}(\theta, \phi, \chi)
\label{rigidSE}
\end{equation}
where $A_e,B_e$ are the rotational constants and $\hat{J}_{z},\hat{J}_{b},\hat{J}_{c}$ are the body-fixed angular momentum operators along the molecule-fixed $z$-axis, and the principal axes: $b$, $c$, respectively.

The explicit form of the angular momentum operators expressed in terms of $\theta, \phi, \chi$ Euler angles can be straightforwardly found from the definition of the angular momentum operator $\hat{\mathbf{J}}=\hat{\mathbf{r}} \times \hat{\mathbf{p}}$, the relation between the Cartesian coordinates and Euler angles and the \textit{chain rule} for differentiation. As a result we can write the \SE\ from eq. \ref{rigidSE} as

\begin{equation}
\begin{split}
\left(\frac{1}{\sin\theta}\frac{\partial}{\partial \theta}\left(\sin\theta\frac{\partial}{\partial\theta}\right)+\frac{1}{\sin^{2}\theta}\frac{\partial^{2}}{\partial \phi^{2}}
+\left(\cot^{2}\theta+\frac{A_{e}}{B_{e}}\right)\frac{\partial^{2}}{\partial \chi^{2}}-2\frac{\cos\theta}{\sin^{2}\theta}\frac{\partial^{2}}{\partial \phi \partial \chi}+\frac{E_{rot}}{B_{e}}\right)\Phi_{rot}(\theta, \phi, \chi)=0 
\end{split}
\label{eq:SEsymtop}
\end{equation}
Note that the Euler angles $\chi, \phi$ occur only in derivatives in eq. \ref{eq:SEsymtop}, they are cyclic coordinates, therefore we can make the following \textit{anzatz} for the rotational wavefunction:
\begin{equation}
\Phi_{rot}(\theta, \phi, \chi)=\Theta(\theta)e^{im\phi}e^{ik\chi}
\end{equation}
where $m$ and $k$ are some real numbers. 
The cyclic boundary conditions $\Phi_{rot}(\theta, \phi, \chi+2\pi)=\Phi_{rot}(\theta, \phi+2\pi, \chi)=\Phi_{rot}(\theta, \phi, \chi)$ imposed on the rotational wavefunction imply that $m,k$ are integers. After some manipulations on eq. \ref{eq:SEsymtop}  we obtain the following equation for $\Theta(\theta)$:
\begin{equation}
\left(\frac{1}{\sin\theta}\frac{d}{d \theta}\left(\sin\theta\frac{d}{d\theta}\right)+\left[\Delta-\frac{m^{2}-2mk\cos\theta+k^{2}}{\sin^{2}\theta}\right]\right)\Theta(\theta)=0
\label{theta}
\end{equation}
where,
\begin{equation}\label{Rigid:delta}
\Delta=\frac{\left(E_{rot}-(A_{E}-B_{e})k^{2}\right)}{B_{e}}
\end{equation}
This is already a one-dimensional second-order linear differential equation, which can be solved analytically by putting
\begin{equation}
\Theta(\theta)=x^{\frac{|k-m|}{2}}(1-x)^{\frac{|k+m|}{2}}G(x)
\end{equation}
where $x=\frac{(1-\cos\theta)}{2}$. Note that also
\begin{equation}
\Theta(\theta)=\left(\sin\frac{\theta}{2}\right)^{|k-m|}\left(\cos\frac{\theta}{2}\right)^{|k+m|}G(\sin^{2}\frac{\theta}{2})
\end{equation}
After some lengthy differentiation and algebra eq. \ref{theta} reduces to
\begin{equation}
x(1-x)\frac{d^{2}G}{dx^{2}}+(\alpha-\beta x)\frac{dG}{dx}+\gamma F=0
\label{Hypergeometric}
\end{equation}
with
\begin{equation}\label{Rigid:alpha}
\alpha=1+|k-m|
\end{equation}
\begin{equation}\label{Rigid:beta}
\beta=\alpha+1+|k+m|
\end{equation}
and
\begin{equation}\label{Rigid:gamma}
\gamma=\Delta-\frac{\beta(\beta-2)}{4}
\end{equation}
Eq. \ref{Hypergeometric} can be solved with the Frobenius method, in which the solution is found in the form of the power expansion:
\begin{equation}\label{Frobenius}
G(x)=\sum_{n=0}^{\infty}a_{n}x^{n}
\end{equation}
With the Frobenius anzatz eq. \ref{Hypergeometric} takes the form:
\begin{align}\nonumber
& \gamma a_{0}+\alpha a_{1} +\left[\left(\gamma -\beta\right)a_{1}+2\left(1+\alpha)\right)a_{2}\right]x+ &&\\
& +\sum_{n=2}^{\infty}\left[\left(\gamma - n(n-1)-\beta n\right)a_{n}+(n+1)(n+\alpha)a_{n+1}\right]x^{n}=0
\end{align}
Because subsequent powers of $x$ form a linearly independent set, the above equation requires that coefficients standing next to powers of $x$ must vanish:
\begin{align}\nonumber
& a_{1}=-\frac{\gamma}{\alpha}a_{0}&&\\\nonumber
& a_{2}=a_{1}\frac{\beta-\gamma}{2+2\alpha} &&\\
& a_{n+1}=\frac{-\gamma +\beta n +n(n-1)}{(n+1)(n+\alpha)}a_{n} &&
\label{coeffs}
\end{align}
The $a_{0}$ coefficient is chosen so that the rotational wavefunction is normalized.
The $\Theta(\theta)$ function must be finite, so that the series expansion given in eq. \ref{Frobenius} must truncate at a finite term labeled by $n_{max}$:

\begin{equation}\label{max}
a_{n_{max}+1}=0
\end{equation}
giving the condition, which provides the eigenvalues of the rigid rotor \SE\ (cf.\ref{rigidSE})

\begin{equation}\label{Rigid:condition}
\beta n_{max} +n_{max}(n_{max}-1)-\gamma=0
\end{equation}
Substituting eqs. \ref{Rigid:delta}, \ref{Rigid:beta} and \ref{Rigid:gamma} into eq. \ref{Rigid:condition} gives the rotational energies as functions of integer $J$ and $k$:

\begin{equation}\label{Rigid:eigenenergy}
E_{rot}=B_{e}J(J+1)+(A_{e}-B_{e})k^{2}
\end{equation}
where the abbreviation $J=n_{max}+\frac{|k+m|+|k-m|}{2}$ was made. The ranges for $J,k,m$ follow directly from the condition $n_{max}\geq 0$:
\begin{align}\nonumber
& J=0,1,2,... \; & k=0,\pm 1,\pm 2,...,\pm J \; and &\\
& & m=0,\pm 1,\pm 2,...,\pm J &
\label{Rigid:numbers}
\end{align} 
It is possible to identify the Frobenius expansion given in eq. \ref{Frobenius} and coefficients given in eq. \ref{coeffs} with the \textit{hypergeometric function} $F(\frac{1}{2}\beta-J-1,\frac{1}{2}\beta+J;\alpha,x)$, which enables us to write the rotational wavefunction in the compact form:

\begin{equation}\label{Rigid:Solution}
\resizebox{\hsize}{!}{$\Phi_{Jkm}(\theta,\phi,\chi)=N_{Jkm}x^{\frac{|k-m|}{2}}(1-x)^{\frac{|k+m|}{2}}F(\frac{1}{2}\beta-J-1,\frac{1}{2}\beta+J;\alpha,x)e^{im\phi}e^{ik\chi}$}
\end{equation}
The normalization constant $N_{Jkm}$ is defined by the condition:

\begin{equation}\label{Rigid:normalization}
\int_{0}^{2\pi}\int_{0}^{2\pi}\int_{0}^{\pi}\Phi_{Jkm}^{*}\Phi_{Jkm}\sin\theta d\theta d\phi d\chi =1
\end{equation}
and a choice of the phase factor. 
The symmetric-top rotational wavefunction can be compactly written as
\begin{equation}\label{Rigid:Solution2}
\Phi_{Jkm}(\theta,\phi,\chi)=\sqrt{\frac{2J+1}{8\pi^{2}}}D^{(J)}_{km}(\theta,\phi,\chi)
\end{equation}
where $D^{(J)}_{km}(\theta,\phi,\chi)$ is the \textit{Wigner D-matrix} \cite{06BuJexx.method}. Often the abbreviation $ \Phi_{Jkm}(\theta,\phi,\chi)\equiv |J,k,m\rangle$ is made. It should be however treated with caution, because the \textit{ket} $|J,k,m\rangle$  represents a \textit{state vector} in the \textit{Hilbert space} whereas  $\Phi_{Jkm}$ is  $|J,k,m\rangle$  expressed in the spectral representation of the position operator. There are other representations possible. $\Phi_{Jkm}$ belongs to the $L^{2}$ space of square-integrable functions, which is isomorphic with the original \textit{Hilbert space}. 

After expanding the hypergeometric function in eq. \ref{Rigid:Solution}, the explicit form of the wavefunction $\Phi_{Jkm}(\theta,\phi,\chi)$ is
\begin{equation}\label{Rigid:Solution3}
 N\left[\sum_{\sigma}(-1)^{\sigma}\frac{(\cos\frac{1}{2}\theta)^{2J+k-m-2\sigma}(-\sin\frac{1}{2}\theta)^{m-k+2\sigma}}{\sigma!(J-m-\sigma)!(m-k+\sigma)!(J+k-\sigma)!}\right]
	e^{im\phi}e^{ik\chi}
\end{equation}
where,
\begin{equation}\label{Rigid:normalization2}
N=\left[\frac{(J+m)!(J-m)!(J+k)!(J-k)!(2J+1)}{8\pi^{2}}\right]^{\frac{1}{2}}
\end{equation}
and $\sigma$ starts from 0 or $(k-m)$, whichever is the larger, and goes up to $(J-m)$ or $(J+k)$, whichever is smaller \cite{06BuJexx.method}.

Note that the symmetric-top Hamiltonian commutes with the square of the total angular momentum: $\hat{J}^{2}$, the $z$-component of the space-fixed angular momentum $\hat{J}_{\rho_{3}}$, and with the $z$-component of the molecule-fixed angular momentum $\hat{J}_{z}$. For these reason all these operators have a common set of eigenfunctions. 
\subsection{Asymmetric top}
A different situation occurs for the asymmetric top (all three rotational constants are different). Here the Hamiltonian \ref{rigidSE} commutes with $\hat{J}^{2}, \hat{J}_{\rho_{3}}$ but not with $\hat{J}_{z}$:
\begin{equation}
\left[\hat{H},\hat{J}_{z}\right]=\left[\hat{J}_{x},\hat{J}_{y}\right]_{+}\left(B_{e}-A_{e}\right)
\end{equation}
where the molecule-fixed angular momentum operators $\hat{J}_{x},\hat{J}_{y},\hat{J}_{z}$ satisfy the commutation relations:
\begin{equation}\label{anomalcommutation}
\left[\hat{J}_{i},\hat{J}_{j}\right]=-i\hslash \epsilon_{ijk} \hat{J}_{k}
\end{equation}
$\hat{H}$ has a common eigenbasis with $\hat{J}^{2}$ and $\hat{J}_{\rho_{3}}$ but not with $\hat{J}_{z}$. In order to obtain eigenfunctions of $\hat{H}$, a good starting point is to express the Hamiltonian matrix in the symmetric-top eigenbasis. As a result, the eigenfunctions of the asymmetric top can be represented as a linear combination of \textit{Wigner} functions (symmetric-top eigenfunctions) with different $k$ values:

\[  \left( \begin{array}{ccc}
\Psi_{1} \\
\vdots \\
\Psi_{2J+1} \end{array} \right)=\textbf{C}_{J,m} \left( \begin{array}{ccc}
|J,-J,m\rangle \\
\vdots \\
|J,J,m\rangle \end{array} \right).\] 
where $\textbf{C}$ is a $(2J+1) \times (2J+1)$ matrix which diagonalizes the block of the asymmetric top Hamiltonian for a given $(J,m)$. To find  $\textbf{C}$ we must first express the Hamiltonian matrix elements in the symmetric-top basis. The asymmetric-top Hamiltonian can be written as
\begin{align}\nonumber
\hat{H}= &\hslash^{-2}\left[\frac{1}{2}\left(B_{e}+C_{e}\right)\hat{J}^{2}+\left[A_{e}-\frac{1}{2}\left(B_{e}+C_{e}\right)\right]\hat{J}_{z}^{2}\right.+&&\\
&\left.+\frac{1}{4}\left(B_{e}-
C_{e}\right)\left((\hat{J}_{m}^{+})^{2} + (\hat{J}_{m}^{-})^{2}\right)\right] &&
\label{asymetricham}
\end{align}
\normalsize
where we introduced the molecule-fixed\textit{ ladder operators}: $\hat{J}_{m}^{\pm}:=\hat{J}_{x}\pm i\hat{J}_{y}$. The choice of the ladder operators is not accidental. Due to their commutation relation 
\begin{equation}\label{laddercommutation}
\left[\hat{J}_{z},\hat{J}_{m}^{\pm}\right]=\mp\hslash\hat{J}_{m}^{\pm}
\end{equation}
ladder operators act on the symmetric-top basis functions by lowering or raising the value of the $k$ quantum number.

Because $|J,k,m\rangle$ are eigenfunctions of $\hat{J}_{z}$: $\hat{J}_{z}|J,k,m\rangle=\hslash k|J,k,m\rangle$ the 'laddered' function $\hat{J}_{m}^{\pm}|J,k,m\rangle$ is also an eigenfunction of $\hat{J}_{z}$ to the $\hslash(k \mp 1)$ eigenvalue. 
To show this, let us act on the standard basis element with the product of operators: $\hat{J}_{z}\hat{J}_{m}^{\pm}|J,k,m\rangle$. This is equal, on account of eq. \ref{laddercommutation}, to $\hat{J}_{m}^{\pm}\hat{J}_{z}|J,k,m\rangle+\hat{J}_{m}^{\pm}|J,k,m\rangle$. Next, utilizing the eigenvalue equation for $\hat{J}_{z}$ we find
\begin{equation}
\hat{J}_{z}\hat{J}_{m}^{\pm}|J,k,m\rangle=\hslash (k\mp1)\hat{J}_{m}^{\pm}|J,k,m\rangle
\label{eq:ladder2}
\end{equation}
i.e. the 'laddered' function is also an eigenfunction of $\hat{J}_{z}$ but to the eigenvalue shifted by $1$. This is the general property of operators obeying commutation relations as given in eq. \ref{laddercommutation}.

As a consequence of eq. \ref{eq:ladder2} we can postulate that the 'laddered function' is proportional to its 'neighbour' function in the $k$ quantum number:
\begin{equation}
\hat{J}_{m}^{\pm}|J,k,m\rangle=N_{J,k}|J,k\mp 1,m\rangle
\end{equation}
Normalization of both sides ob the above equation allows to write:
\begin{equation}
\hat{J}_{m}^{\pm}|J,k,m\rangle=\hslash \sqrt{J(J+1)-k(k\mp 1)}|J,k\mp 1,m\rangle
\end{equation}
where we made use of the equality: $\hat{J}_{m}^{\mp}\hat{J}_{m}^{\pm}=\hat{J}^{2}-\hat{J}_{z}(\hat{J}_{z}\mp \hslash)$.

It is possible to construct ladder operators which change the value of the $m$ quantum number, i.e. the projection of angular momentum vector on the space-fixed $z$-axis. The space-fixed angular momentum operators satisfy standard commutation relations:
\begin{equation}\label{anomalcommutation2}
\left[\hat{J}_{\rho i},\hat{J}_{\rho j}\right]=i\hslash \epsilon_{ijk} \hat{J}_{\rho k}
\end{equation}
Space-fixed ladder operators defined as follows $\hat{J}_{s}^{\pm}:=\hat{J}_{\rho1}\pm i\hat{J}_{\rho2}$ satisfy analogical equations as the molecule-fixed ladder operators:
\begin{equation}
\hat{J}_{\rho3}\hat{J}_{s}^{\pm}|J,k,m\rangle=\hslash( m\pm1)\hat{J}_{s}^{\pm}|J,k,m\rangle
\end{equation}
and
\begin{equation}
\hat{J}_{s}^{\pm}|J,k,m\rangle=\hslash \sqrt{J(J+1)-m(m\pm 1)}|J,k,m\pm 1\rangle
\end{equation}
allowing to derive the following important relation:
\begin{equation}
|J,\pm|k|,\pm|m|>=N\left(\hat{J}_{m}^{\mp}\right)^{k}\left(\hat{J}_{s}^{\pm}\right)^{m}|J,0,0\rangle
\end{equation} 
The normalization factor is given by
\begin{equation}
N=\hslash^{-(|k|+|m|)}\sqrt{\frac{(J-|m|)!(J-|k|)!}{(J+|m|)!(J+|k|)!}}
\end{equation}
Finally, we can write down matrix elements of all relevant angular momentum operators in the symmetric-top basis:
\begin{equation}
\begin{split}
\langle J,k,m|\hat{J}^{2}|J,k,m\rangle =\hslash^{2}J(J+1)\\
\langle J,k,m|\hat{J}_{z}|J,k,m\rangle =k\hslash\\
\langle J,k,m|\hat{J}_{\rho3}|J,k,m\rangle =m\hslash\\
\langle J,k,m\pm1|\hat{J}^{\pm}_{s}|J,k,m\rangle =\hslash\sqrt{J(J+1)-m(m\pm1)}\\
\langle J,k\mp1,m|\hat{J}^{\pm}_{m}|J,k,m\rangle =\hslash\sqrt{J(J+1)-k(k\mp1)}\\
\end{split}
\end{equation}
\bibliographystyle{plain}
\bibliography{References}

\chapter{Rotational-vibrational transition intensities}

In this section we discuss the problem of calculating rotational-vibrational-electronic transition intensities at different levels of approximation. In high-accuracy quantum-mechanical calculations of transition intensities three key quantities need to be determined: the total internal wavefunction of the initial and the final state of the molecule, the electric dipole moment surface and the molecule-fixed coordinate frame. We begin with the derivation of a general expression for the transition intensity in a polyatomic molecule. Next we are going to gradually step down the ladder of approximate expressions for the transition intensity.

The Born-Oppenheimer approximation represents a standard in the calculations of infra-red spectra of molecules and is assumed from here on in. Within this approximation we first analyze the exact expression for the rotational-vibrational-electronic transition intensity in a triatomic molecule for certain choices of the molecule-fixed coordinate frame. From the general expression the well-known Frank-Condon approximation to electronic transitions is subsequently derived. Finally, an approximate separation of the rotational and vibrational motion is considered, giving the simplest quantum-mechanical view at transition intensities in molecules. The ladder of different theoretical levels represents a menu, from which the researcher can choose when attempting to calculate absorption spectrum of a molecule. The level of theory varies depending on the size of the molecule, the complexity of its electronic and geometric structure, but most of all the level of accuracy one wishes to operate at. Here we focus in detail only on triatomic molecules.

\section{Transition line strength}
\label{sec:int}
The probability of a single-photon transition between two quantum states $|i\rangle \rightarrow |f\rangle$ is given by the  square modulus of the transition dipole moment 
\begin{equation}
|\langle f |\mu |i \rangle|^2, 
\end{equation}
as stated by the Fermi golden rule.  The electric dipole moment operator of the molecule is defined as $\mu=-\sum_{i,elec.}\vec{r}_i+\sum_{\alpha,nucl.} Z_{\alpha}\vec{R}_{\alpha}$, where $Z_{\alpha}$ is the electric charge of the $\alpha$-th nuclei and $\vec{r}$, $\vec{R}$ denote position vectors of electrons and nuclei, respectively. The definition of the electric dipole moment tells us that if the molecule is neutral, the dipole moment remains unchanged upon any choice of reference frame with respect to which the position vectors $\vec{r}$, $\vec{R}$ are defined.

In this work, we assume that transitions occur via the interaction of the incident electromagnetic wave only with the electric dipole moment of the molecule. In principle, interactions with quadrupole electric moment and magnetic dipole moment of the molecule also contribute to the overall transition probability, however for most of the allowed rotational-vibrational transitions their relative contribution is about six orders of magnitude lower than the contribution from the electric dipole moment. For this reason, in spectroscopy of small molecules, when no extremely weak lines are measured and not very strong fields are used, the \textit{dipole approximation} to the transition probability is excellent.

To find an explicit expression for the transition intensity between two Born-Oppenheimer rotational-vibrational-electronic states of the molecule in the dipole approximation three quantities are required: the total internal wavefunction for the initial and the final state, a choice of the molecule-fixed coordinate frame and the electronic dipole moment surfaces for respective electronic states of the molecule. For this reason, before proceeding to the derivation of the exact quantum-mechanical expression for the dipole transition intensity, first we need to discuss a general form of the total internal wavefunction for a polyatomic molecule.

The total internal wavefunction of the molecule must contain information about the electronic, nuclear and all spin degrees of freedom. For the majority of infrared and UV spectroscopic purposes it is sufficient to assume independence of the spin degrees of freedom from other dynamical degrees of freedom. In other words the interaction between the spins of the nuclei  and other degrees of freedom  of the molecule (\textit{hyperfine interaction}) are neglected. Such an assumption is possible due to very low energy of the interaction between nuclear spins and spins of the electrons. From the perspective of the energetic separation in the rotational levels of the molecule, all energy hyperfine energy levels are split by tiny amount and can be treated as degenerate. Typical hyperfine splitting of rotational states is in the range of 1-100 kHz, whereas the rotational constants, which determine the energetic separation between the rotational levels can be anywhere between 1MHz to a hundreds GHz's.
In the Born-Oppenheimer approximation the total internal wavefunction of the molecule is separable into the nuclear spin part, the electronic part and the ro-vibrational part, as written below:
\begin{equation}
|\Phi_{int}\rangle= |\Phi_{nspin}\rangle |\Phi_{elec,(i)}\rangle |\Phi_{rv,J,h,i}\rangle 
\label{eq:wfint}
\end{equation}
where $|\Phi_{elec,(i)}\rangle$ is the  wavefunction for the $i$-th electronic state and can be obtained from quantum chemistry calculations, $|\Phi_{nspin}\rangle $ is the spin function of the nuclei. The ro-vibrational part of the wavefunction $|\Phi_{rv,J,h,i}\rangle $ is obtained from nuclear motion calculations. Here $h$ enumerates non-degenerate eigenvectors of the rotational-vibrational Hamiltonian given in eq. \ref{eq:Hbody}. Strategies for finding eigenvectors of the Hamiltonian of eq. \ref{eq:Hbody} are discussed in Chapter 2. Because the spin part of the wavefunction can be treated separately in quite a straightforward way (see section~\ref{sec:spinstat}), from now on, we are going to focus on the ro-vibronic part of the total internal wavefunction: 
$|\Phi_{elec,(i')}\rangle |\Phi_{rv,J',h',i'}\rangle  \equiv |J',h',i',\vec{D}'\rangle $, where an additional label $\vec{D}'$ has been given to denote all quantum numbers which do not affect the energy of the state (degeneracy labels). These quantum numbers are associated with symmetries of the general Hamiltonian for a polyatomic molecule in free space, e.g. $m$ - the projection of the total angular momentum of the molecule on the space-fixed $Z$-axis. Depending on the electronic state there can be additional quantum numbers for which the energy levels are degenerate.

\begin{figure}[H]
\includegraphics[width=9cm]{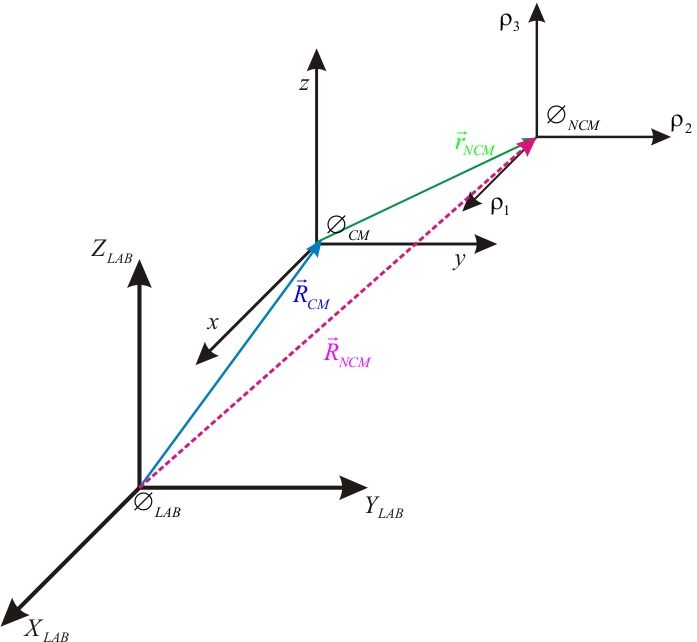}
\caption{Schematic representation of the LAB frame, space-fixed frame centred at the nuclear centre-of-mass and a space-fixed frame with the origin at the molecular centre-of-mass.}
\label{fig:coordsys}
\end{figure}

The quantum probability for the $|i\rangle \equiv  |J'',h'',i'',\vec{D}''\rangle \rightarrow |f\rangle \equiv    |J',h',i',\vec{D}'\rangle $ ro-vibronic transition is given in the dipole approximation by the square modulus of the electric transition dipole moment vector $\sum_{A=X,Y,Z}\left|T_{if}^{A,\vec{D}'',\vec{D}'}\right|^2$ \cite{06BuJexx.method}, where the summation is carried out over three Cartesian components of the electric dipole moment of the molecule in the laboratory frame  $A=X,Y,Z$. Individual transition probabilities are then summed over all degenerate states, labelled by vector $\vec{D}$:
\begin{equation}
S_{if}=\sum_{A=X,Y,Z}\sum_{\vec{D}'',\vec{D}'}\left|T_{if}^{A,\vec{D}'',\vec{D}'}\right|^2
\label{trans:transdip}
\end{equation}
giving a quantity called the \textit{line strength}. One can think of summation over the degeneracy labels $\vec{D}'$  as adding up all quantum-mechanically possible (i.e. existing) initial states which have identical energies, but which could perhaps be distinguished by interaction with an external electric or magnetic field. All such \textit{degenerate states} exist and in the combinatorial sense contribute to the overall probability for a transition between two \textit{energy levels}. Line strength can be directly related to experimentally measured integral line intensity \cite{06BuJexx.method}:

\begin{equation}
 I(\tilde{v}_{if})=\frac{8\pi^2N_A}{12\epsilon_0hc}~\frac{\tilde{v}_{if}}{Q(T)}g_{ns}\exp\left(\frac{-E_i}{k_bT}\right)\left[1-\exp\left(-\frac{\tilde{v}_{if}}{k_bT}\right)\right] S_{if}
\label{eq:Int}
\end{equation}
where $\tilde{v}_{if}$ is the transition wavenumber ($
\tilde{\nu} =c/ \lambda$ ) between the $i$-th and
$f$-th ro-vibronic state and $Q(T)$ is the
partition function at temperature $T$. $N_A$ is the Avogadro number, $k_b$ is the Boltzmann constant, $h$ is the Planck constant, $c$ is the speed of light in vacuum and $\epsilon_0$ is the permittivity of vacuum.  $g_{ns}$ is the spin statistical weight for the initial state $i$ resulting from summation over degenerate nuclear spin functions. The values of the spin statistical weights depend on the ro-vibronic symmetry of the state and on all individual spins of the nuclei. Thus, the measured transition line strength is directly related to the total internal state of the molecule, including nuclear spin states. Further discussion of spin statistical weights is given in section \ref{sec:spinstat}. Units for the integral line intensity given in eq. \ref{eq:Int} are cm/molecule and quantify the amount of absorption of the electromagnetic radiation per molecule at a given wavelength. Details of derivation of eq. \ref{eq:Int} are given further below.

Coming back to eq. \ref{trans:transdip}, the electric transition dipole moment is defined as
\begin{equation}
T_{if}^{A,\vec{D}'',\vec{D}'}=\langle J'',h'',i'',\vec{D}''\left| \hat{\mu}_{el}^{A,space}\right|J',h',i',\vec{D}' \rangle 
\end{equation}
and takes identical value for all components of $\vec{D}$ except the projection $M$ of the total angular momentum on the space-fixed Z-axis. In order to relate the transition intensity to molecular properties it is convenient to rewrite the expression for the transition dipole moment into a form which is unchanged by translating the molecule in free space and 3-D rotating the molecule. Henceforth we are going to make use of the two symmetries of an isolated molecule: the translational symmetry in free 3-D space and the rotational symmetry in 3-D space. The plan is to write the expression for line strength \ref{trans:transdip} in the form which is unchanged upon translations and rotations in 3D space. 

Cartesian representation of vector quantities, such as $T_{if}^{A,\vec{D}'',\vec{D}'}, A=X,Y,Z$ is not particularly convenient when discussing rotations in three dimensional space. Much simpler expressions arise when the space-fixed transition dipole moment is transformed into the so called \textit{spherical tensor form} \cite{06BuJexx.method}, which transforms irreducibly in the 3-D rotations group (SO(3)):
\begin{equation}
\vec{\mu}^{ space}_{el,sph}=\mathbf{K}\vec{\mu}^{space}_{el}
\label{mu_sph}
\end{equation}
where, 
\begin{equation}
\textbf{K}=\begin{pmatrix}
-\frac{1}{\sqrt{2}} & \frac{i}{\sqrt{2}} & 0 \\
	    \frac{1}{\sqrt{2}} & \frac{i}{\sqrt{2}} & 0 \\
	          0 & 0 & 1 \\
\end{pmatrix}
\label{K_matrix}
\end{equation}
is a unitary ($|\det(\mathbf{K})|=1$, $\mathbf{K}^{\dagger}\mathbf{K}=\mathbf{1}$) transformation matrix between the Cartesian operator and the \textit{rank 1 spherical tensor operator} \cite{Kutzelnigg1989}. The electric dipole moment for a neutral molecule must be invariant under translations in free space, so the 'LAB' Cartesian components ($X,Y,Z$) of the transition dipole moment can be rewritten in terms of Cartesian components in the space-fixed coordinate system ($\xi,\eta,\zeta$) with the origin at the nuclear center of mass \cite{06BuJexx.method}. Different coordinate systems are depicted in Figure \ref{fig:coordsys}. In what follows, the transition dipole moment can be expressed as

\begin{equation}
T_{if}^A=\sum_{A=\xi,\eta,\zeta}\sum_{\sigma=-1}^{1}\mathbf{K}^{\dagger}_{A\sigma}\sum_{\vec{D},\vec{D}'}\tilde{T}_{if}^{\sigma,M,M'}
\label{T_if_spherical}
\end{equation}
Upon the choice of the molecule-fixed coordinate frame, in which the rotational-vibrational wavefunction is typically expressed, one has to relate the space-fixed components of the transition dipole moment vector with the components in the molecule-fixed coordinate frame. The transformation from the \textit{Cartesian} to the \textit{spherical tensor} form of the transition dipole moment given in eq. \ref{T_if_spherical} is particularly simple. In order to relate the components of the transition dipole moment vector written in a coordinate frame which does not rotate with the molecule (space-fixed frame) to components of this vector in the molecule-fixed frame, one must know the position of the latter frame with respect to the former. This relation can be conveniently expressed by three Euler angles $\phi,\theta,\chi$ as displayed in Figure \ref{fig:euler}.

As a consequence of the chosen \textit{spherical tensor} form of the transition dipole vector $\tilde{T}_{if}^{\sigma,M,M'}$, a straightforward transformation to the molecule-fixed coordinate system can be expressed with \textit{Wigner D-matrices}:

 \begin{equation}
\vec{\mu}^{ space}_{\sigma}=\sum_{\sigma'=-1}^{1}\mathbf{D}^{(1)}_{\sigma \sigma'}(\phi,\theta,\chi)\vec{\mu}^{ mol}_{\sigma'}
\label{mu_molecule_space}
\end{equation}
where $\phi,\theta,\chi$ denote the Euler angles and subscripts '$el$' and '$sph$' have been dropped for clarity of presentation. 

The rotational-vibrational wavefunction can be written in the form:

\begin{equation}
|\Psi^{J,h,p,i}_{rv}\rangle = \sum_{K=p}^{J} |\Psi^{J,h,p,i}_{vib,K}\rangle |J,K\rangle
\label{eq:rvwf}
\end{equation}
where $K=|k|$ is the absolute value of the projection of the total angular momentum on the molecule-fixed $z$-axis, $p=0,1$ denotes the quantum number associated with the parity symmetry of the wavefunction,  $|\Psi^{J,h,p}_{vib,K}\rangle$ is the vibrational wavefunction and $|J,K\rangle$ is the symmetric-top eigenfunction for the total angular momentum $J$. Here $h$ labels the rotational-vibrational states. Note that the vibrational wavefunction depends on $K$, which adds coupling between the rotation of the molecule and the vibrational motion.

The vibrational wavefunction for a given $K$ can be expanded in a chosen primitive basis as follows:
\begin{equation}
|\Psi^{J,h,p}_{vib,K}\rangle = \sum_n c_{\textbf{n},k}^{J,h,p} |\textbf{n},k\rangle
\label{eq:vwf}
\end{equation}
where $\textbf{n}=(n_1,n_2,...,n_D)$ represents the vector of indices, which label the $D$-dimensional vibrational wavefunction.
The matrix elements of the Wigner operator in the symmetric top basis can be expressed as:
\begin{equation}
D^{J,1,J'}_{k,M,\sigma,\sigma',k',M'}=\langle J,-k,M|D^{(1)}_{\sigma\sigma'}|J',-k',M'\rangle=N_{JJ'}(-1)^{M'+k'}\tj{1}{J}{J'}{\sigma}{M}{ -M'}\tj{1}{J}{J'}{\sigma'}{k}{-k'}
\end{equation}
where $\tj{1}{J}{J'}{\sigma}{M}{ -M'}\tj{1}{J}{J'}{\sigma'}{k}{-k'}$ is the $3-j$ symbol.
Putting the anzatz from eq. \ref{eq:rvwf} into the expression for the transition dipole moment from eq. \ref{trans:transdip} followed by some algebra and with the use of symmetry properties of $3-j$ symbols, one gets \cite{06BuJexx.method,jt78,jt94,jt121}:

\begin{equation}
\begin{split}
S_{if}=\frac{1}{4}\left(2S''+1\right)\left(2S'+1\right)\left(2J''+1\right)\left(2J'+1\right)\left[(-1)^{J''+J'+1}+(-1)^{p''+p'}\right]^2\\
\times \left| \sum_{\sigma=-1}^{+1} \sum_{\substack{K'=p' \\ K''=p''}}^{J',J''}(-1)^{K''} \tj{1}{J'}{J''}{\sigma}{K'}{K''}\sum_{\textbf{n}',\textbf{n}''} c_{\textbf{n}',K'}^{J',h',p',i'} c_{\textbf{n}'',K''}^{J'',h'',p'',i''}   M_{\textbf{n}',\textbf{n}'',K',K''}^{\sigma,i',i''} \right|^2
\end{split}
\label{trans:strength}
\end{equation}
The summation over $M$ and $M'$ was carried out by utilizing the normalisation property of the $3-j$ symbol: $\sum_{\sigma=-1}^{1}\sum_{M,M'=-J,-J'}^{J,J'}\tj{1}{J}{J'}{\sigma}{M}{-M'}^2=1$.
The $(2S'+1)(2S''+1)$ prefactor in eq. (\ref{trans:strength}) comes from summation over all combinations of degenerate electron spin functions. We assume that the electronic spin does not change in the transition, that is  $ S'' = S'$. It means that we neglect, for instance, singlet to triplet transitions. Similarly, the  $(2J'+1)(2J''+1)$ prefactor comes from summation over all combinations of degenerate rotational basis functions, characterized by the $M$ quantum number. In eq.~\ref{trans:strength} $M_{\textbf{n}',\textbf{n}'',K',K''}^{\sigma,i',i''}$ is the matrix element of the electric dipole moment operator in the primitive vibrational-electronic basis:

\begin{equation}
M_{h'h''}^{\sigma,i'',i'}= \langle \Psi^{J',h',p',i'}_{vib,K'}| \mu^{i',i''}_{\sigma}(Q_1,...,Q_D) |\Psi^{J'',h'',p'',i''}_{vib,K''} \rangle  
\label{trans:transdip1}
\end{equation}
where
\begin{equation}
\mu^{i'i''}_{\sigma}(Q_1,...,Q_D)=\langle \Phi_{elec,(i')}\left| \hat{\mu}_{\sigma}^{mol}\right| \Phi_{elec,(i'')} \rangle _{elec.}
\label{eltrans0}
\end{equation}
is the $\sigma$-th spherical tensor component of the molecule-fixed electronic transition dipole moment surface between electronic states $i'$ and $i''$. In eq. (\ref{eltrans0}) the integration is carried over electronic coordinates only, leaving the dependence on vibrational coordinates $(Q_1,...,Q_D)$. This function is called the \textit{electronic transition dipole moment surface} (TDMS) and can be obtained, for example, by fitting a predefined functional form to points calculated from a quantum chemistry package. The contribution to the overall line strength from the rotational basis functions appears in the $3-j$ symbols and the prefactors in eq. \ref{trans:strength}.  

The form of the primitive basis functions $|\textbf{n},k\rangle$ for vibrations strongly depends on the choice of internal coordinates. For instance, for the H$_2$O molecule, when bond coordinates are used $(r_1,r_2,\gamma)$, a reasonable primitive vibrational basis is given as

\begin{equation}
|\textbf{n},k\rangle = |n_1\rangle |n_2\rangle |j,K\rangle
\label{eq:vprim}
\end{equation}
where $|n_1\rangle = \phi_{n_1}(r_1)$, $|n_2\rangle = \phi_{n_2}(r_2)$, $|j,K\rangle = P_K^j(\gamma)$ - the first two primitive functions labeled by $n_1,n_2$ can be Harmonic oscillator wavefunctions or Morse oscillator wavefuncitons, whereas the primitive functions for the bending coordinate $\gamma$ are typically associated Legendre polynomials $P_K^j(\gamma)$. 
With the choice of the vibrational basis given above for the water molecule, the equation for the line strength takes the following form:
\begin{equation}
\begin{split}
S_{if}=\frac{1}{4}\left(2S''+1\right)\left(2S'+1\right)\left(2J''+1\right)\left(2J'+1\right)\left[(-1)^{J''+J'+1}+(-1)^{p''+p'}\right]^2\\
\times \left| \sum_{\sigma=-1}^{+1} \sum_{\substack{K'=p' \\ K''=p''}}^{J',J''}(-1)^{K''}\tj{1}{J'}{J''}{\sigma}{K'}{K''}\sum_{\substack{m',n',j' \\ m'',n'',j''}} C_{m'n'j'K'}^{J',i',h',p'} C_{m''n''j''K''}^{J'',i'',h'',p''} M_{m'm''n'n''j'j''K'K''}^{\sigma,i'',i'} \right|^2
\end{split}
\label{trans:strength2}
\end{equation}
Line strength given in eq. \ref{trans:strength2} is directly related to the \textit{Einstein absorption coefficient}:

\begin{equation}
S_{if}=A_{if}\frac{3\epsilon_0h}{2(2*\pi)\tilde{\nu}_{if}^3}	
\label{eq:einstein}
\end{equation}
where $\epsilon_0$ is the vacuum permittivity. 
Einstein A-coefficients have the dimension of $s^{-1}$ and their inverse yield mean lifetime of a given molecular state $\tau_i=\left(\sum_f A_{if}\right)^{-1}$, where the sum is carried over all final states connected to the initial state $i$.
Eq. \ref{trans:strength2} is the most general expression for transition line strength given in the Born-Oppenheimer approximation and when interactions with the nuclear spin are neglected. From this point a series of approximations can be made.

\subsection{Condon approximation}
In many molecules, TDMS depends weakly on values of internal coordinates $(Q_1,...,Q_D)$ and it is often replaced by a constant value of the transition dipole at equilibrium geometry $\mu^{eq}_{\sigma}\equiv \mu^{ge}_{\sigma}\left(Q^{eq}_1,...,Q^{eq}_D\right)$. In such case we talk about the \textit{Condon approximation} \cite{06BuJexx.method} to the electronic transition dipole moment:

\begin{equation}
M_{\textbf{n}',\textbf{n}'','K',K''}^{\sigma,i',i''}= \mu^{i'i''}_{\sigma}\left(Q^{eq}_1,Q^{eq}_2,...,Q^{eq}_D\right) \delta_{\textbf{n}'\textbf{n}''}
\label{trans:FC2}
\end{equation}
where we assumed identical orthonormal vibrational basis sets in electronic states $i'$ and $i''$. Condon approximation is typically a reasonable one, when the molecule is rigid, that is the amplitudes of vibrational motions in the molecule are small, and when the change of vibrational coordinates within these amplitudes does not change the electric dipole moment of the molecule significantly.
The line strength is then given by
\begin{equation}
\begin{split}
S^{condon}_{if}=\frac{1}{4}\left(2S''+1\right)\left(2S'+1\right)\left(2J''+1\right)\left(2J'+1\right)\left[(-1)^{J''+J'+1}+(-1)^{p''+p'}\right]^2\\
\times \left| \sum_{\sigma=-1}^{+1} \mu^{eq}_{\sigma} \sum_{\substack{K'=p' \\ K''=p''}}^{J',J''}(-1)^{K''} \tj{1}{J'}{J''}{\sigma}{K'}{K''}\sum_{\textbf{n}',\textbf{n}''} c_{\textbf{n}',K'}^{J',h',p',i'} c_{\textbf{n}'',K''}^{J'',h'',p'',i''} \right|^2
\end{split}
\label{trans:FC3}
\end{equation}
Expression given in eq. \ref{trans:FC3} contains contributions from both the rotational and the vibrational degrees of freedom. 
The traditional vibrational \textit{Franck-Condon approximation} only refers to vibrational wavefunctions, with \textit{the Frank-Condon factors} defined as the square modulus of the overlap integral between two vibrational wavefunctions $F_{ij}=\left|\langle \phi_{vib}^{i}|\phi_{vib}^{j}\rangle \right|^2$. Then the vibrational band strength is proportional to the vibrational Frank-Condon factor. Here in eq. \ref{trans:FC3} the Frank-Condon approximation is generalized onto ro-vibrational transitions, that is the above expression accounts for the rotational-vibrational coupling, but neglects the dependence of the electronic dipole moment on nuclear coordinates.

\subsection{H{\"o}nl-London factors}
In some problems one can convincingly consider the ro-vibrational wavefunction as being separable to the rotational and vibrational wavefunctions:
\begin{equation}
|\Psi^{J,h,p,i,K}_{rv}\rangle = |\Psi^{h,i}_{vib}\rangle |J,K,p\rangle
\label{eq:rvwfsep}
\end{equation}
meaning that $K$ is treated as good quantum number. Such separation is good when no strong vibrational-rotational coupling is present or when not very high accuracy is needed. With the separable ro-vibrational wavefunction the line strength from the general expression given in eq. \ref{trans:strength} becomes
 
\begin{equation}
\begin{split}
S^{condon,vib}_{if}= |\langle \Psi^{h',i'}_{vib}|  \mu^{i',i''}_{\Delta K}(Q_1,...,Q_D)  |\Psi^{h'',i''}_{vib}\rangle|^2\cdot L(J',K',J'',K'')
\end{split}
\label{trans:FCvib}
\end{equation}
where $L(J',K',J'',K'')=\frac{1}{4}\left(2S''+1\right)\left(2S'+1\right)\left(2J''+1\right)\left(2J'+1\right)\left[(-1)^{J''+J'+1}+(-1)^{p''+p'}\right]^2
\times|\tj{1}{J'}{J''}{\Delta K}{K'}{K''}|^2$ are \textit{H{\"o}nl-London factors}, which represent the purely rotational contribution to line strength and have been precalculated and tabularized \cite{bernath}. Here $\Delta K = K' - K''$.

Sometimes, when calculation of an accurate coupled rotational-vibrational wavefunction is difficult, one corrects for the rotational-vibrational coupling by adding an extra factor to eq. \ref{trans:FCvib}:
 \begin{equation}
\begin{split}
S_{if}= \left|\langle\Psi^{h',i'}_{vib}|  \mu^{i',i''}_{\sigma}(Q_1,...,Q_D)  |\Psi^{h'',i''}_{vib}\rangle\right|^2\cdot L(J',K',J'',K'')\cdot G_{J'J''}
\end{split}
\label{trans:FCvib2}
\end{equation}
 where $G_{J'J''}$ is the so called \textit{Herman-Wallis factor} \cite{bernath}. Although the original \textit{Herman-Wallis factors} were postulated based on perturbation theory, their functional form depends on many parameters of the molecule. There is no clear recipe for the choice of the functional form of \textit{Herman-Wallis factors}, which can vary between different vibrational bands. Usually the simplest form for the Herman-Wallis factors is a low order polynomial function in $J''$ depending on the value of $\Delta J = J''-J'$. 
 
\subsection{Frank-Condon approximation}
\begin{figure}[H]
\begin{center}
  \includegraphics[width=8cm]{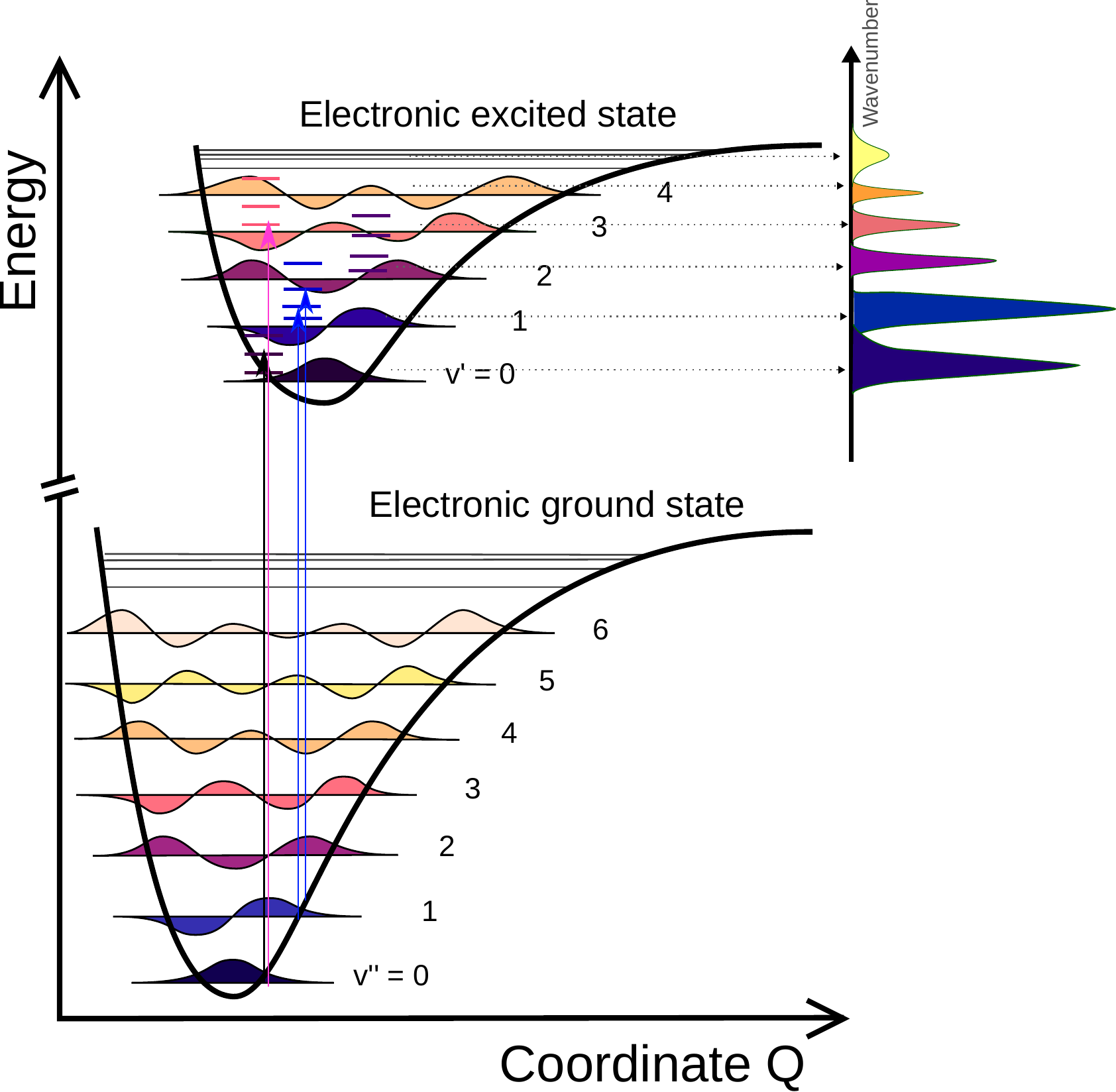}
\caption{Schematic representation of the ro-vibronic excitation. In the Frank-Condon approximation the absorption line strength is proportional to square modulus of the overlap integral between vibrational states belonging to different electronic states.}
\label{fig:FC}
\end{center}
\end{figure}

The Condon approximation applied to the case of a transition between different electronic states $i'\neq i''$ is often referred to as the \textit{Frank-Condon approximation} (cf. eq. \ref{trans:FC3}). A schematic representation of the ro-vibronic excitation in the Frank-Condon approximation is displayed in Figure \ref{fig:FC}. Because the geometry of the molecule in electronic state $i'$ is usually different from the geometry in state $i''$, the vibrational states in both electronic states are not orthogonal. As a result  
$F_{h'h''}=\left|\langle \Psi_{vib}^{h',i'}|\Psi_{vib}^{h'',i''}\rangle \right|^2 \neq 0$ and the line strength can be expressed as 	
 \begin{equation}
S^{FC}_{if}= |\mu^{i',i''}_{eq}|^2F_{h'h''}\cdot L(J',K',J'',K'')\cdot F_{J'J''}
\label{trans:FC}
\end{equation}
With a decoupled rotational-vibrational wavefunction, an additional \textit{axis switching} effect can occur. This effect is associated with the fact that the molecule-fixed coordinate frame rotate upon the geometry change in the electronic transition.

\subsection{Selection rules}
\label{par:selrules}
The $3-j$ symbol appearing in eq. (\ref{trans:strength}) and (\ref{trans:FC3}) and the \\  $\left[(-1)^{J''+J'+1}+(-1)^{p''+p'}\right]^2$ factor are responsible for the selection rules for rotational transitions. 
Within a given quantum-mechanical model selection rules are rigorous laws stating which combinations of quantum numbers for the initial and final state yield zero line strength. Such transitions are called \textit{forbidden}. It should be pointed out that \textit{forbidden} transitions in an approximate quantum-mechanical model can become allowed, when a higher level of approximation is used. This is known as \textit{relaxation of selection rules}, which are common in all quantum theory, often when a more detailed interaction picture is introduced or more accurate structure of the wavefunction is assumed. When the complete rotational basis set is used the rotational motion in the expression for the line strength in eq. \ref{eq:Int1} is modelled exactly and the selection rules for the rotational transitions given below can be considered as exact. If however one includes additional light-matter interaction mechanisms, such as the interaction with the electric quadrupole moment, magnetic moment or multi-photon absorption, the selection rules can change, and in general they become less rigorous, that is they allow for more transitions.

 From the $3-j$ symbol appearing in eq. (\ref{trans:strength}), it follows that in order for the line strength not to vanish the following conditions must be satisfied: $|J''-J'|=0,1$, $J''+J'\ge 1$ and $\Delta K=0,\pm 1$. This is because the $3-j$ symbol is zero unless the bottom row elements sum up to 0. Selection rules for the $J$ quantum number define so called $P,Q$ and $R$ branches for $\Delta J=J''-J'=+1,0,-1$, respectively. Selection rules for the $K=|k|$ quantum number allow transitions of type $k \rightarrow -k$, which reflects the time-reversal symmetry of the system. If we reverse time (which represents a transformation upon which the \SE\ remains unchanged) the direction of rotation along the well defined $z$-axis of the molecule-fixed frame, reverses. This corresponds to the change from a rotational-vibrational state with positive $k$ to a state with negative $k$, or \textit{vice versa}.  For $\Delta K=0$ only the $z$-component of the molecule-fixed electronic transition dipole moment contributes to the overall line intensity, and because the $z$-axis is chosen here as the axis of quantization for the angular momentum, we call these transitions \textit{parallel}. Accordingly $\Delta K=\pm 1$ corresponds to \textit{perpendicular} transitions, as both the $x$ and $y$ components of the electronic transition dipole moment contribute to the total intensity. For the Q branch ($\Delta J=0$) only transitions which change the $p$ quantum number are allowed, i.e. $e \leftrightarrow f, e \nleftrightarrow e, f \nleftrightarrow f$ (see section \ref{sec:additionalsymmetries}). Conversely, the $P$ and $R$ branches allow transitions conserving $p$, i.e.  $e \nleftrightarrow f, e \leftrightarrow e, f \leftrightarrow f$. Neglecting the dependence of the \textit{TDMS} on the nuclear coordinates does not affect the rotational selection rules, but it can make the vibrational selection rules more rigorous for example by forbidding vibrational overtone transitions.

\subsection{Partition function, line shapes and cross sections}
The formula for the spectral line intensity given in eq. \ref{eq:Int} is derived below. The total absorption intensity can be constructed from Einstein absorption coefficients in the following way

\begin{equation}
 I(\tilde{v}_{if})=\frac{h\nu_{if}}{c}\frac{n_i}{N_{tot}}\left(1-\frac{g_in_f}{g_fn_i}\right) B_{if}
\label{eq:Int1}
\end{equation}
where $\tilde{v}_{if}$ is the transition wavenumber between the $i$-th and
$f$-th ro-vibronic state, $h$ is the Planck constant, $c$ is the speed of light in vacuum.  The spin statistical weights $g_{i/f}$ (discussed in the next section) for the initial and the final state result from the summation over all degenerate nuclear spin functions. $B_{if}$ is the induced absorption Einstein coefficient, $\frac{n_i}{N_{tot}}$ is the fraction of the occupation of the molecular initial state at temperature $T$ and $n_f$ is the population of the final state. The Einstein $B_{if}$ coefficient  stays in a simple relation to the quantum-mechanical absorption line strength $S_{if}$ defined in eq. \ref{trans:transdip}:

\begin{equation}
B_{if} =\frac{8\pi^3}{3h^2} S_{if}
\label{eq:Int2}
\end{equation}
For systems staying in a local thermal equilibrium, such as gas chambers in absorption measurements, the population of rotational-vibrational levels of molecules is determined by the Boltzmann statistics:

\begin{equation}
\frac{g_in_f}{g_fn_i}=\exp\left(-\frac{c_2\nu_{if}}{T}\right)
\label{eq:boltzmann}
\end{equation}
where $c_2=hc/k_b=1.4388 cm\cdot K$  is the second radiation constant, so that  $\tilde{\nu}_{if}$ is expressed in wavenumbers \cm .
Similarly the fraction of the total population of molecules in the ensemble is given by the Boltzmann formula
\begin{equation}
\frac{n_i}{N_{tot}}=\frac{g_i\exp\left(-\frac{c_2 E_{i}}{T}\right)}{Q(T)}
\label{eq:boltzmann2}
\end{equation}
where $E_{i}$ is the lower state energy and $Q(T)$ is the partition function, which normalizes the Boltzmann thermal distribution of molecular populations:

\begin{equation}
Q(T)=\sum_{i}g_{i}e^{-c_2\frac{E_{i}}{T}}
\label{eq:partition}
\end{equation}
where index $i$ which represents all good quantum numbers: $J,p$ and $h$ - the index numbering non-degenerate ro-vibrational states; $g_i=(2J+1)g_{ns}$ is the state dependent degeneracy factor discussed in section \ref{sec:spinstat}. For a given temperature $T$ the contribution to the partition function $Q(T)$ decreases with increasing energy of ro-vibrational states. In practice, the partition function is calculated with a finite number of energy levels. When no significant change in the partition function is visible upon adding energy levels with higher and higher energy, the partition function is said to be \textit{converged}. The value of the partition function is not very sensitive to the accuracy of ro-vibrational energy levels obtained from variational calculations. This is because in variational calculations the energy levels calculated with highest accuracy have the lowest energy, and these levels contribute most to the value of $Q(T)$.  Sometimes, when obtaining accurate ro-vibrational energy levels is problematic the partition function is approximated as a product of the partition function for all vibrational levels $Q_{vib}(T)$ and the partition function for rotational levels $Q_{rot}(T)$ separately. Then, for instance, the rigid rotor model can be used to find rotational energy levels.

Partition function given in eq. \ref{eq:partition} can be interpreted as the total number of possible occupations of ro-vibrational states, each weighted by the exponential factor.
Combining equations \ref{eq:Int1}-\ref{eq:boltzmann2} and assuming that the isotopic abundance (number fraction) of the molecule of interest is $a$ we get

\begin{equation}
 I(\tilde{v}_{if})=\frac{8\pi^3N_A}{3\cdot 4\cdot \pi \epsilon_0hc}~\frac{ \tilde{v}_{if}}{Q(T)}a\cdot g_{ns}\exp\left(\frac{-E_i}{k_bT}\right)\left[1-\exp\left(-\frac{\tilde{v}_{if}}{k_bT}\right)\right] S_{if}
\label{eq:Int3}
\end{equation}
where $N_A$ is the Avogadro number and $\epsilon_0$ is the permittivity of the vacuum, which appear in the equation to convert to the standard cm/molecule units for integral line intensity. These units are used throughout spectroscopic databases. 
The expression for the integral absorption intensity given in eq. \ref{eq:Int3} can be used to determine the integral absorption intensity at temperature $T$ when $I$ is known for a reference temperature $T_{ref}$ and the partition function is known at both temperatures:
\begin{equation}
 I(\tilde{v}_{if};T)= I(\tilde{v}_{if};T_{ref})\frac{Q(T_{ref})}{Q(T)}\left[\frac{1-\exp\left(-\frac{c_2\tilde{v}_{if}}{T}\right)}{\exp\left(-\frac{c_2\tilde{v}_{if}}{T_{ref}}\right)}\right]\exp\left(-c_2E_i\left(\frac{1}{T}-\frac{1}{T_{ref}}\right)\right)
\label{eq:Int4}
\end{equation}

The integral transition intensity $I$ given in eq. \ref{eq:Int3} is directly related to the intensity of the attenuated incident radiation for a given wavelength in the Beer-Lambert law \cite{06BuJexx.method}

\begin{equation}
\frac{\Phi}{\Phi_0}=e^{-\sigma(\tilde{\nu})\cdot N \cdot L}
\label{eq:lambert-beer2}
\end{equation} 
which relates the intensity of the absorbed radiation $\Phi$, the reference radiation intensity $\Phi_0$, the path length $L$ and the the absorption cross-section $\sigma(\tilde{\nu})$ to $N$ - the concentration of molecules. The absorption cross section depends on the transition wavenumber $\tilde{\nu}$ and is defines as the convolution of the integral line intensity and a line shape function:
\begin{equation}
\sigma(\tilde{\nu})=\int I(\tilde{\nu}'-\tilde{\nu})f(\tilde{\nu}')d\tilde{\nu}
\label{eq:cross-section}
\end{equation} 
 This aspect of producing molecular spectra is particularly relevant for atmospheric science in determining concentrations of molecules from remote sensing and ground based telescope measurements. Each absorption line measured by the gas phase spectroscopy has a characteristic shape and width. There are three main effects responsible for broadening of spectral absorption lines: \textit{the Doppler effect}, \textit{finite lifetime of quantum states} and \textit{collision induced emission}. Each of these effects is briefly explained below.
\begin{enumerate}
\item \textbf{Doppler Broadening} \\
Molecules in the gas phase sample at temperature $T$ have a distribution of velocities described by the Maxwell-Boltzmann distribution. On average half of the molecules will move away from the direction of the propagation of the incident electromagnetic wave. This causes the molecules to experience lower frequency of the electromagnetic wave than if they were at rest. According to the\textit{ Doppler law} this shift in frequency is $\nu_0 \frac{v}{c}$, where $\nu_0$ is the frequency of the electromagnetic wave and $v$ is the speed at which the molecule moves away from the source of radiation in the direction of its propagation. The\textit{ Fermi golden rule} states that a single-photon transition between two quantum states is possible only if the energy difference between the final and the initial state matches the energy of the incident radiation. If the molecule experiences slightly lower frequency, it will need to absorb light with a bit higher frequency (in the frame of reference of the light source/detector). Molecules moving towards the source of radiation will experience elevated frequency of incident radiation hence will absorb the part of the incoming beam at slightly lower frequency. Every laser beam used in modern spectroscopy has certain spectral width too. 
According to the Maxwell-Boltzmann law for speeds of molecules in gas in local thermal equilibrium the line shape function associated with the Doppler broadening can be written as:

\begin{equation}
f^{Doppler}(\nu)=\frac{1}{\nu_0}\left(\frac{mc^2}{2\pi k_bT}\right)^{\frac{1}{2}}\exp\left(-\frac{mc^2}{2k_bT}\left(\frac{\nu-\nu_0}{\nu_0}\right)^2\right)
\label{eq:maxwell}
\end{equation}

where $m$ is the mass of the molecule and $\nu_0$ is the line centre (transition wavenumber). Function given in eq. \ref{eq:maxwell} define the \textit{Doppler line shape}.

\item \textbf{Lifetime Broadening} \\
Each quantum state possesses a characteristic lifetime, which is associated with the probability of a spontaneous transition to lower energy states. This probability is proportional to the spontaneous emission Einstein coefficient, which is proportional to the Einstein A-coefficient. Einstein A-coefficients have a direct relation to the absorption line strength and the Einstein B-coefficient for stimulated emission. From the time-energy formulation of the \textit{uncertainty principle} we know that every system undergoing finite-time $\tau$ interaction with the electromagnetic radiation has a natural uncertainty in its energy $\Delta E$. As a result of the uncertainty relation $\Delta E \cdot \tau \propto \hslash $, the energetic separation between the initial and the final energy level in the molecule has some inherent quantum uncertainty, which leads to statistical absorption in some frequency range. This natural radiative broadening is typically small and can be modelled with the Lorentzian function of full-width at half maximum ($FWHM\equiv \gamma$) of about $FWHM/$\cm $\approx \frac{10.6}{\tau /ps}$: 

\begin{equation}
f^{lifetime}(\nu)=\frac{1}{\pi}\frac{\gamma}{(\nu_0-\nu)^2+(\gamma)^2}
\label{eq:lorentz0}
\end{equation}
where $\gamma = \sum_{f} A_{if}$ is the sum of all Einstein A-coefficients, which contribute to the total spontaneous emission rate from state $i$.

\item \textbf{Pressure Broadening} \\
When pressure in the gas sample is increased the average time between molecular collisions decreases. Collisions between molecules often lead to \textit{collision induced emission}, which reduces the lifetime of energy levels and adds up to the overall lifetime broadening.:
\begin{equation}
f^{Pressure}(\nu)=\frac{1}{\pi}\frac{\Gamma}{(\nu_0-\nu)^2+\Gamma^2}
\label{eq:lorentz}
\end{equation}

where $\Gamma$ depends on the collision rate at given temperature, that is on pressure and molecular density of the sample.
At low pressures ($< 10$Torr \footnote{$1 Torr \approx 133.3 Pa$}) it is reasonable to neglect any effects of absorption from interacting, that is non-isolated, molecules. In other words, the situation when two molecules are so close to one another that they significantly affect each other's energy levels is rare and it does not visibly influence the measured spectrum.

There are a number of other factors contributing to the overall line shape, to name a few: power broadening, transit time broadening or speed-dependence in pressure broadening. These subtle effects become important when very high accuracy ($<5\%$) for integral transition intensities is needed.
\end{enumerate}
\subsection{Nuclear spin statistical weights}
\label{sec:spinstat}
The nuclear spin statistical weight $g_{ns}$ given in eq. \ref{eq:Int3} can be calculated with the aid of group-theoretic techniques when individual nuclear spins of atoms in a given isotopologue are known \cite{06BuJexx.method}. Having said that the total internal wavefunction of the molecule is a product of the ro-vibronic wavefunction and the nuclear spin part, the dynamics of nuclear spins is decoupled from the dynamics of the remaining degrees of freedom in the molecule.

If a molecule contains nuclei with non-zero spins, individual spin functions can be combined in a number of different ways for a given value of the total nuclear spin. This means that for every ro-vibronic state there exist many states with different values of the total nuclear spin as well as the projection of the total nuclear spin on the space-fixed $z$-axis. Because, to a very good approximation, all interactions involving the nuclear spin can be neglected, different values of the total nuclear spin do not affect the energy levels. In a such case, the only signature of existence of these \textit{hyperfine energy levels}, which are assumed degenerate, is of purely statistical origin. 

If one has at a disposal two initial states of identical energy connected to a final state, then statistically the probability for transition is two times the transition probability for a transition between a single initial state and the final state. In brief, the spin statistical weight gives the number of these degenerate nuclear spin states, which correspond to a given initial ro-vibronic molecular state. \textbf{More specifically $g_{ns}$ is the number of nuclear spin states, which when combined in a product with a ro-vibronic state of a given symmetry, produces the total internal state of the molecule (see eq. \ref{eq:wfint}) which has the symmetry allowed by the \textit{Bose-Einstein} or the \textit{Fermi-Dirac} statistics \cite{06BuJexx.method}.} This statement is based on the assumption that the nuclear spin does not interact with spins of electrons in the molecule or the rotational motion of the molecule, so that the total internal wavefunction of the molecule can be factorized into the nuclear spin part and the ro-vibronic part: $\Phi_{int}=\Phi_{nspin}\Phi_{rve}$. Neglecting nuclear spin interactions with other forms of motion in the molecule results in degenerate nuclear spin energy levels. The number of these degenerate nuclear spin states, which for a given ro-vibronic symmetry of the wavefunction generate a representation allowed by the Bose-Einstein or Fermi-Dirac statistics, constitutes the spin statistical weight. Hyperfine interactions, which are not considered here, remove this nuclear spin degeneracy, which causes splitting in energies of spin-rovibronic states.

To give an example of how the spin statistical weights can be determined we focus below on isotopologues of carbon dioxide.
If the molecule possesses identical fermions, such as the $^{17}$O nuclei  (spin $5/2$), then every even permutation of these atoms (which is a symmetry operation of this molecule), changes sign of the total internal wavefunction (Fermi-Dirac statistics). On the other hand, if two or more identical atoms in the molecule are bosons, as $^{18}$O is, the permutation of two $^{18}$O atoms leaves the total internal wavefunction of the molecule unchanged (Bose-Einstein statistics). These permutation symmetry requirements on the total internal wavefunction are considered fundamental, meaning that so far no experiment have shown violation of this rule. 

The total nuclear spins of naturally abundant isotopologues of carbon and oxygen atoms are: i($^{12}$C) = 0, i($^{13}$C) = 1/2, i($^{14}$C) = 3, i($^{16}$O) = 0, i($^{17}$O) = 5/2, i($^{18}$O) = 0. This means that the nuclei of $^{13}$C and  $^{17}$O atoms are fermions and the remaining nuclei are bosons. Apart from translational, time-reversal and rotational symmetries, which are irrelevant for the present discussion, two fundamental symmetries are related to spin statistical weights: permutation of identical nuclei and parity. The total internal wavefunction of the molecule can transform either symmetrically or anti-symmetrically with respect to the parity transformation: $E^*\Phi_{int}=\pm \Phi_{int}$. Odd permutation of identical nuclei in the molecule changes sign of the total internal wavefunction when the permuted nuclei are fermions and does not change sign for boson nuclei. Even permutation of identical nuclei preserves the sign of $\Phi_{int}$, hence always has character +1. 
In order to find spin statistical weights for isotopologues of CO$_2$, it is first necessary to choose a symmetry group in which the ro-vibronic, spin and total internal states will be classified. 
Molecular states of the symmetric isotopologues of CO$_2$ can be classified in the $D_{\infty h}(M)$ molecular symmetry group, which is isomorphic with the CNPI \textbf{$G_4$} group for this molecule. CNPI stands for \textit{complete nuclear permutation inversion group} and contains all possible permutations of identical nuclei and parity operation. Asymmetric isotopologues, on the other hand, have no permutation symmetry of identical nuclei and will be classified in the $C_{\infty v}(M)$ molecular symmetry group, which is isomorphic with the  \textbf{$G_2$} CNPI group. Let us choose two example isotopologues,$^{16}$O$^{12}$C$^{16}$O and $^{17}$O$^{12}$C$^{17}$O, for which the procedure of finding spin statistical weights will be presented.

Because the nucleus of $^{17}$O is a fermion with spin 5/2, an odd permutation of oxygen atoms in $^{17}$O$^{12}$C$^{17}$O will cause a sign change in the total internal molecular wavefunction, as stated by the Pauli Principle. 
Character table for the $D_{\infty h}(M)$ molecular symmetry group given in Table \ref{table:Dinfh} shows that only $\Sigma_u^+$ and $\Sigma_g^-$ irreducible representations of this group are allowed for the total internal wavefunction, because in these representations the (12) permutation of oxygen atoms has character -1. Ro-vibronic states of  $^{17}$O$^{12}$C$^{17}$O can transform as any of the four irreducible representations of $D_{\infty h}(M)$. Thus, the question is, which irreducible representations of the nuclear spin states can generate representations of the total internal wavefunctions allowed by the Fermi-Dirac statistics? The other question is: what is the degeneracy of these statistically allowed nuclear spin states? To answer both questions, first let us classify the nuclear spin functions $\Phi_{spin}$ in the $D_{\infty h}(M)$ group. 
\begin{table}[h]
\begin{center}
\setlength{\tabcolsep}{4pt}
\caption{Part of the character table for the $D_{\infty h}(M)$ molecular symmetry group used in determination of nuclear spin statistical weights for CO$_2$.}
\begin{tabular}{c r r r r}
\hline\hline
$D_{\infty h}(M)$ &	E & (12) & E* & (12)* \\ [0.3ex] 
\hline 
$\Sigma_g^+$	& 1	& 1	& 1	& 1 	 	\\
$\Sigma_u^+$	& 1	& -1	& 1	& -1 	 	\\
$\Sigma_g^-$	& 1	& -1	& -1	& 1 	 	\\
$\Sigma_u^-$	& 1	& 1	& -1	& -1 	 	\\
\hline\hline
\end{tabular}
\label{table:Dinfh}
\end{center}
\end{table}
The total number of product spin functions of atoms in $^{17}$O$^{12}$C$^{17}$O  is $(2i_{^{17}O}+1)^2=6^2=36$. For the nuclei of $^{17}$O there are 5 possible values of the projection of the total nuclear spin $i$ on the Z-axis of the LAB frame, which corresponds to respective degenerate nuclear spin states: $|\frac{5}{2},+\frac{5}{2}\rangle \equiv \delta_{+\frac{5}{2}}$, $|\frac{5}{2},+\frac{3}{2}\rangle \equiv \delta_{+\frac{3}{2}}$, $|\frac{5}{2},+\frac{1}{2}\rangle \equiv \delta_{+\frac{1}{2}}$, $|\frac{5}{2},-\frac{1}{2}\rangle \equiv \delta_{-\frac{1}{2}}$, $|\frac{5}{2},-\frac{3}{2}\rangle \equiv \delta_{-\frac{3}{2}}$, $|\frac{5}{2},-\frac{5}{2}\rangle \equiv \delta_{-\frac{5}{2}}$. For the  $^{12}$C nucleus there is only one spin state: $|0,0\rangle \equiv \gamma$. The product state of the three states of individual nuclear spins corresponds to states with values of the projection of the total nuclear spin on the Z-axis of the LAB frame in the range $m_I= -|2\cdot i_{^{17}O}+i_{^{12}C}|, ..., +|2\cdot i_{^{17}O}+i_{^{12}C}|= -5, -4, ... ,+4, +5$. Product functions of the three nuclear states form a reducible representation for each value of $m_I$ separately. 

\begin{table}[h]
\begin{center}
\setlength{\tabcolsep}{2pt}
\caption{Product nuclear spin states for $^{17}$O$^{12}$C$^{17}$O. $m_I$ is the projection of the total nuclear spin on the Z-axis of the LAB frame. Shown are only positive values of $m_I$. Negative $m_I$'s are generated in analogical way. The spin state $\gamma$ of the $^{12}$C nucleus is omitted. In the right column given are irreducible representations of the $D_{\infty h}(M)$ group generated by appropriate products of the nuclear spin functions.   }
\begin{tabular}{c l l}
\hline\hline
$m_I$ &	nuclear spin states & $\Gamma_{spin}^{m_{I}}$ \\ [0.3ex] 
\hline 
+$5$	& $\delta_{+\frac{5}{2}}\delta_{+\frac{5}{2}}$ & $\Sigma_g^+$ 	\\
+$4$	& $\delta_{+\frac{5}{2}}\delta_{+\frac{3}{2}}$, $\delta_{+\frac{3}{2}}\delta_{+\frac{5}{2}}$,   & $\Sigma_g^+ \oplus \Sigma_u^+$	\\
+$3$	& $\delta_{+\frac{5}{2}}\delta_{+\frac{1}{2}}$,  $\delta_{+\frac{1}{2}}\delta_{+\frac{5}{2}}$, $\delta_{+\frac{3}{2}}\delta_{+\frac{3}{2}}$ & $2\Sigma_g^+ \oplus \Sigma_u^+$	 \\
+$2$	& $\delta_{+\frac{5}{2}}\delta_{-\frac{1}{2}}$,  $\delta_{-\frac{1}{2}}\delta_{+\frac{5}{2}}$, $\delta_{+\frac{3}{2}}\delta_{+\frac{1}{2}}$ , $\delta_{+\frac{1}{2}}\delta_{+\frac{3}{2}}$ &  $2\Sigma_g^+ \oplus 2\Sigma_u^+$		\\
+$1$	& $\delta_{+\frac{5}{2}}\delta_{-\frac{3}{2}}$,  $\delta_{-\frac{3}{2}}\delta_{+\frac{5}{2}}$, $\delta_{+\frac{1}{2}}\delta_{+\frac{1}{2}}$ , $\delta_{+\frac{3}{2}}\delta_{-\frac{1}{2}}$, $\delta_{-\frac{1}{2}}\delta_{+\frac{3}{2}}$	&  $3\Sigma_g^+ \oplus 2\Sigma_u^+$	\\
$0$	& $\delta_{+\frac{5}{2}}\delta_{-\frac{5}{2}}$,  $\delta_{-\frac{5}{2}}\delta_{+\frac{5}{2}}$, $\delta_{+\frac{3}{2}}\delta_{-\frac{3}{2}}$ , $\delta_{-\frac{3}{2}}\delta_{+\frac{3}{2}}$, $\delta_{-\frac{1}{2}}\delta_{+\frac{1}{2}}$, $\delta_{+\frac{1}{2}}\delta_{-\frac{1}{2}}$ & $3\Sigma_g^+ \oplus 3\Sigma_u^+$		\\
-$1$	& $\delta_{-\frac{5}{2}}\delta_{+\frac{3}{2}}$,  $\delta_{+\frac{3}{2}}\delta_{-\frac{5}{2}}$, $\delta_{-\frac{1}{2}}\delta_{-\frac{1}{2}}$ , $\delta_{-\frac{3}{2}}\delta_{+\frac{1}{2}}$, $\delta_{+\frac{1}{2}}\delta_{-\frac{3}{2}}$& $3\Sigma_g^+ \oplus 2\Sigma_u^+$		\\
-$2$	& $\delta_{-\frac{5}{2}}\delta_{+\frac{1}{2}}$,  $\delta_{+\frac{1}{2}}\delta_{-\frac{5}{2}}$, $\delta_{-\frac{3}{2}}\delta_{-\frac{1}{2}}$ , $\delta_{-\frac{1}{2}}\delta_{-\frac{3}{2}}$	& $2\Sigma_g^+ \oplus 2\Sigma_u^+$\\
-$3$	& $\delta_{-\frac{5}{2}}\delta_{-\frac{1}{2}}$,  $\delta_{-\frac{1}{2}}\delta_{-\frac{5}{2}}$, $\delta_{-\frac{3}{2}}\delta_{-\frac{3}{2}}$ & $2\Sigma_g^+ \oplus \Sigma_u^+$  \\
-$4$	& $\delta_{-\frac{5}{2}}\delta_{-\frac{3}{2}}$, $\delta_{-\frac{3}{2}}\delta_{-\frac{5}{2}}$   & $\Sigma_g^+ \oplus \Sigma_u^+$	\\
-$5$	& $\delta_{-\frac{5}{2}}\delta_{-\frac{5}{2}}$ & $\Sigma_g^+$  	\\
\hline\hline
\end{tabular}
\label{table:spinstates}
\end{center}
\end{table}

The right column in Table~\ref{table:spinstates} can be calculated with the use of the formula for the number of irreducible representations of a group included in a given reducible representation: $a_i=\frac{1}{h}\sum_{R}\chi^{\Gamma_{spin}^{m_{I}}}\left[R\right]\cdot \chi^{\Gamma_{i}}\left[R\right]^ *$, where $h$ is the order of the molecular symmetry group and $R$ stands for symmetry operations in the group, $\chi^{\Gamma_{spin}^{m_{I}}}\left[R\right]$ is the character of the operation $R$ in the reducible representation $\Gamma_{spin}^{m_{I}}$ and $\chi^{\Gamma_{i}}\left[R\right]$ is the character of the operation $R$ in the irreducible representation $\Gamma_{i}$. Let us choose an irreducible representation of the ro-vibronic state, say $\Sigma^+_u$, and find all irreducible representations of nuclear spin states, which if combined with $\Sigma^+_u$ in a product, will give one of the two representations allowed by the Fermi-Dirac statistics:  $\Sigma_u^+$ and $\Sigma_g^-$. There are only two types of representations available for the nuclear spin states: $\Sigma_g^+$ and $\Sigma_u^+$.  From the character Table \ref{table:Dinfh} $\Sigma^+_u \otimes \Sigma^+_g = \Sigma^+_u$ and from Table \ref{table:spinstates} we can see that there are 21 in total  $\Sigma^+_g$'s which satisfy this equation. Thus, the spin statistical weight for the $\Sigma^+_u$ ro-vibronic symmetry species is $g_{ns}=21$. Weights for other ro-vibronic symmetry species can be calculated in a similar way. All spin statistical weights  for $^{17}$O$^{12}$C$^{17}$O are listed in Table \ref{table:spinweights}.

\begin{table}[h]
\begin{center}
\setlength{\tabcolsep}{4pt}
\caption{Spin statistical weights $g_{ns}$ for the $^{17}$O$^{12}$C$^{17}$O molecule classified in the $D_{\infty h}(M)$ molecular symmetry group. $\Gamma_{rve}$, $\Gamma_{spin}$ and $\Gamma_{int}$ are the irreducible representations of the ro-vibronic state, nuclear spin state and the total internal molecular state, respectively.  }
\begin{tabular}{c r r r }
\hline\hline
$\Gamma_{rve}$ &	$\Gamma_{spin}$ & $\Gamma_{int}$ & $g_{ns}$ \\ [0.3ex] 
\hline 
$\Sigma_g^+$	& $15\Sigma_u^+$	& $\Sigma_u^+$	& 15		 	\\
$\Sigma_u^+$	& $21\Sigma_g^+$	& $\Sigma_u^+$		& 21		 	\\
$\Sigma_g^-$	& $21\Sigma_g^+$& $\Sigma_g^-$		& 21		 	\\
$\Sigma_u^-$	& $15\Sigma_u^+$	& $\Sigma_g^-$	& 15		 	\\
\hline\hline
\end{tabular}
\label{table:spinweights}
\end{center}
\end{table}

For the $^{17}$O$^{13}$C$^{17}$O isotopologue, for which there are two available nuclear spin functions for the $^{13}$C atom, all weights are simply multiplied by 2, as the single carbon atom does not affects any properties associated with the permutation symmetry of identical nuclei, yet has two independent nuclear spin states, which need to be accounted for. This gives a state independent factor (2), and a state dependent factor: 15:21 (for $\Sigma_g^+$, $\Sigma_u^-$ and  $\Sigma_u^+$, $\Sigma_g^-$ representations respectively).    In the case of the main $^{16}$O$^{12}$C$^{16}$O isotopologue, for which all nuclei are bosons with spin 0, there exists only one nuclear spin state, which belongs to the $\Sigma_g^+$ representation. In such case, the total internal wavefunctions can be generated only from ro-vibronic states of  $\Sigma_u^+$ and $\Sigma_g^-$ symmetry. For these ro-vibronic representations, there is a single corresponding spin state, hence the appropriate spin statistical factors are $g_{ns}=1$. For the other two ro-vibronic symmetry species: $\Sigma_g^+$ and $\Sigma_u^-$, there are no available spin functions, which would generate representations of the total internal wavefunction allowed by the Bose-Einstein statistics, hence the spin statistical weights for these representations are $g_{ns}=0$. Ro-vibronic states, for which no nuclear spin function can be found to generate an allowed representation of the total internal wavefunction do not exist in nature and are called \textit{missing states}. The corresponding hypothetical energy levels are called \textit{missing levels}. In the case of $^{16}$O$^{12}$C$^{16}$O the observable energy levels correspond to ro-vibrational states which consist of sums of products $\Phi_{vib}\Phi_{rot}$, which have the total symmetry either  $\Sigma_u^+$ or $\Sigma_g^-$ (in the electronic ground state). With these requirements it can be shown that observable ro-vibrational states with a symmetric vibrational part ($n\nu_1$, the symmetric part of the irreducible representation for $n\nu_2$ vibration, even quanta in the $\nu_3$ mode) have $J$-even and $e$-Wang symmetry ( $\Sigma_g^-$ total symmetry). For odd number of quanta in the asymmetric stretching $\nu_3$ mode, the allowed by statistics energy levels have $J$-odd and $e$-Wang symmetry. The $f$-Wang symmetry is allowed only for levels in the degenerate bending manifold and only for $J$-even when the vibrational state has odd number of $\nu_2$ quanta and for $J$-odd when the vibrational state has even number of $\nu_2$ quanta. These rules for observable energy levels are straightforwardly derived from the character table for the $D_{\infty h}(M)$ group and from the knowledge of the classification of rotational and vibrational basis wavefunctions in this molecular symmetry group.

\subsection{Dipole moment surfaces}
\label{sec:tdms}
The electronic transition dipole moment surface (TDMS) defined in eq. \ref{eltrans0} remains a function of internal coordinates $(Q_1,...,Q_D)$ of the molecule, as long as the electronic wavefunctions $ \Phi_{elec,(i')}$ depend on the geometry of the molecule:
\begin{equation}
\mu^{i'i''}_{\sigma}(Q_1,...,Q_D)=\langle \Phi_{elec,(i')}\left| \hat{\mu}_{\sigma}^{mol}\right| \Phi_{elec,(i'')} \rangle _{elec.}
\label{eltrans}
\end{equation}
where $\sigma=-1,0,1$ labels the spherical tensor components of the electric dipole moment $\hat{\mu}_{\sigma}^{mol}$ and $i',i''$ denote electronic states. In computational practice the shape of the hypersurface given in eq. \ref{eltrans} is determined by solving the electronic \SE\ at a sufficient number of nuclear geometries 
($(Q^{(l)}_1,...,Q^{(l)}_D), l=1,2,...,N_{points}$, i.e. \textit{ab initio} points) and subsequent least-square fitting of a predefined functional form for  $\mu^{i'i''}_{\sigma}(Q_1,...,Q_D)$ to the computed points. Popular quantum chemistry packages offer the possibility of calculating the electronic transition dipole moment together with the electronic energy. Another, a stable way of finding the electronic transition moments is thorough calculating the first derivative of the electronic energy with respect to the static and uniform electric field, taken at zero-field value. This derivative can be reasonably accurately approximated with the numerical two-point finite-differences:
\begin{equation}
\mu^{i'i''}_{\sigma}(Q_1,...,Q_D)=E'(0)=\frac{E(\lambda)-E(-\lambda)}{2\lambda}+\mathcal{O}(\lambda^2)
\label{eq:two-point}
\end{equation}
where $E'(0)$ is the derivative of electronic energy with respect to electric field strength $\lambda$ taken at zero-field. 
Because in general the molecule have three non-vanishing components of the dipole moment vector, calculation of a single TDMS point requires solving six different electronic \SE s for the $x$,$y$ and $z$ components of the dipole moment vector and for the positive and negative orientation of the electric field. Previous research by Tennyson \etal\ \cite{jt613,jt475}  suggests that in general the
derivative method yields more reliable dipole moments than those obtained from the simple expectation value evaluation given in eq. \ref{eltrans}, perhaps due to cancellation of errors.

\begin{figure}[H]
\includegraphics[width=12cm]{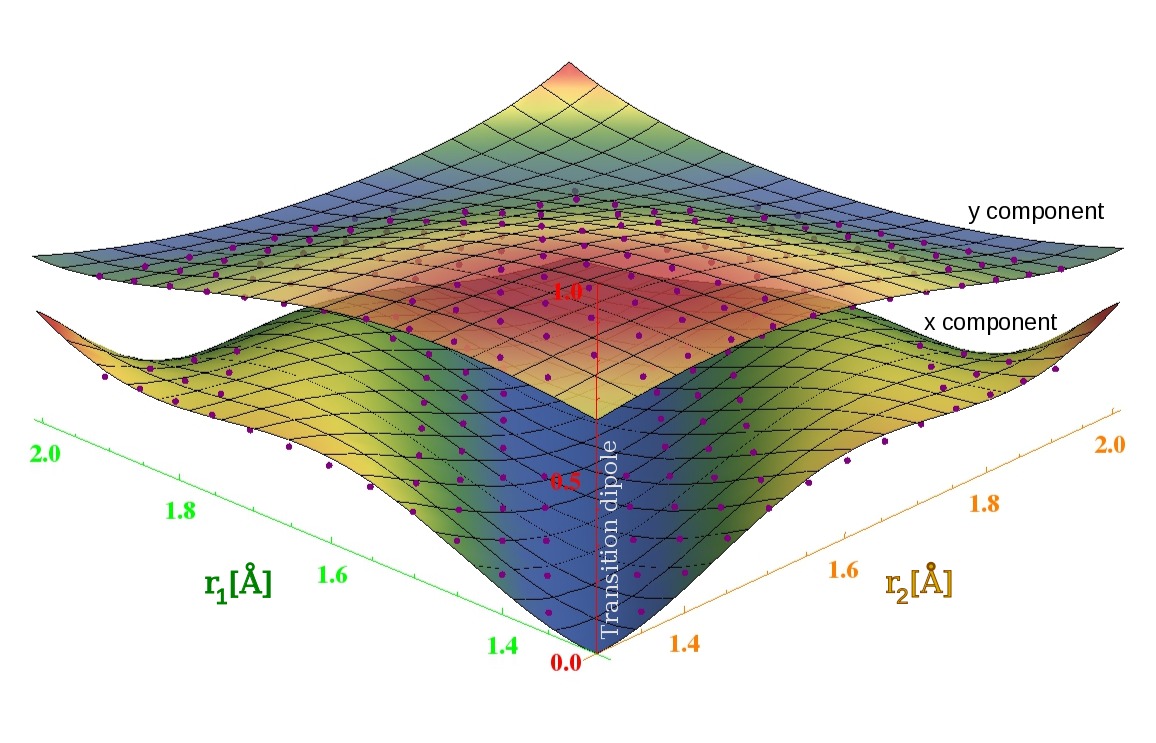}
\caption{Two components of the transition dipole moment function between  $\tilde{X}~^1A_1$ and  $\tilde{C}~^1B_2$ electronic states for the value of the angle between the S--O bonds $\gamma = 120.0 ^{\circ}$. The upper surface is the $y$-component and the lower surface is the $x$-component, which vanishes for C$_{2v}$ geometries. \textit{Ab initio} points are marked with red.}
\label{fig:DMS}
\end{figure}

With the \textit{ab initio} transition dipole moment points at hand the next step is to assume a functional form for the continuous TDMS function. In calculating transition intensities, often the number of grid points required to calculate the vibrational transition dipole moments (see eq. \ref{eq:vibtrans}) is greater than the number of \abinitio\ points obtained from quantum chemistry calculations. For this reason a form of interpolation/extrapolation of the calculated \abinitio\ transition dipole moment points is needed. The choice of the functional form for TDMS is a matter of experience and depends on the molecule. A Taylor-like expansion is a popular functional form used to represent the TDMS:
\begin{equation}
\mu^{i'i''}_{\alpha}(Q_1,...,Q_D)=\mu^{i'i''}_{\alpha}(Q^{ref}_1,...,Q^{ref}_D)+\sum_{i_1} f^{(1)}_{i_1}Q_{i_1}+\sum_{i_1,i_2} f^{(1)}_{i_1,i_2}Q_{i_1}Q_{i_2}+...
\label{eq:TDMSfit}
\end{equation}
where subsequent sums \textit{couple} together more and more internal coordinates  $Q_{i_c}$, or give higher powers of selected internal coordinates, providing better accuracy. The quality of the fit shown in eq. \ref{eq:TDMSfit} can be controlled with the \textit{root-mean-square deviation} (RMSD) between the \abinitio\ points and the values of the fitted function at these points. The RMSD is defined as: $RMSD = \sqrt{\frac{1}{N}\sum_{i=1}^N(\mu^{i'i'',fit}_{\alpha}(Q_{i})-\mu^{i'i''}_{\alpha}(Q_{i}))^2}$, where $Q_i$ stands for the $i$-th vector of the $(Q_1,...,Q_D)$ coordinates. An example TDMS between the $\tilde{X}~^1A_1$ and the $\tilde{C}~^1B_2$ electronic state of SO$_2$ is presented in Figure \ref{fig:DMS}.

\paragraph{Selection rules for vibrational transitions} 

Selection rules for rotational transitions were discussed in section \ref{par:selrules}. These rules follow directly from the rigorous treatment of the rotational motion with the complete symmetric-top basis set. The selection rules for isolated vibrational transitions are not rigorous, which means that in principle all vibrational transitions are allowed, yet some of them are very weak. 
Which transitions are allowed depends on the structure of the rotational-vibrational wavefunction and on the shape of the transition dipole moment surface.
A rule of thumb is that when the rotational-vibrational wavefunction becomes more accurate and the electronic transition dipole moment function becomes more strongly dependent on nuclear coordinates, then more and more transitions become allowed.

When the rotational-vibrational wavefunction accounts for more couplings between the vibrational degrees of freedom but also between the rotational and the vibrational degrees of freedom, and sometimes rotational vibrational and electronic degrees of freedom, the selection rules allow for more and more transitions, which is known as the\textit{ relaxation of selection rules}. 
On the other hand, if the wavefunction is an eigenfunction of a model system, which does not account for any couplings, then the selection rules allow for very limited set of transitions.
To show this, let us discuss a simple model Hamiltonian for the rotational-vibrational motion: the harmonic oscillator approximation for vibrations and rigid-rotor model for rotational motion. In this picture, the rotational-vibrational wavefunction takes the product form:
\begin{equation}
|\Psi_{rv}^h\rangle=|h\rangle|J,k,p\rangle
\label{eq:wfharmrigid}
\end{equation}
where $|h\rangle$ is the $h$-th Harmonic oscillator eigenfunction and $|J,k,p\rangle$ is the parity-adapted symmetric-top eigenfunction discussed in section \ref{sec:rotbasis}. For simplicity we chose one coordinate for the vibrational motion.
Selection rules for vibrational transitions originate from the \textit{vibrational transition dipole moment} given in eq. \ref{trans:transdip1}. In the present approximate model the vibrational transition dipole moment is given as
\begin{equation}
\langle h'| \mu^{i',i''}_{\sigma}(Q) |h''\rangle
\label{eq:vibtrans}
\end{equation}
In order to find out about which vibrational transitions are allowed and which are forbidden let us focus on the symmetry of the wavefunction alone. With the aid of basic tools of group theory it is possible to formulate the following rule:

\textit{The vibrational transition dipole moment between the vibrational states $\Psi^{h'}_{vib}$ and $\Psi^{h''}_{vib}$
can be non-zero only if the irreducible representation in which the total function $\Psi^{h'}_{vib}\mu^{i',i''}_{\sigma}(Q_1,...,Q_D)\Psi^{h''}_{vib}$ transforms under symmetry operations of the molecule contains the totally symmetric representation $A_1$.
}

If the above rule holds then the integral in eq. \ref{eq:vibtrans} does not vanish from symmetry reasons and the corresponding transition is considered vibrationally \textit{allowed}.
Any electronic transition dipole moment $\mu^{i'i''}_{\alpha}(Q_1,...,Q_D)$ function transforms in the molecular symmetry (MS) group of the molecule as the translational coordinate $T_{\alpha}$, which means that vibrational transition is allowed when $\Gamma ( T_{\alpha} ) \in \Gamma (\Psi^{h'}_{vib}) \otimes \Gamma (\Psi^{h'}_{vib}) $ for $\alpha = x,y$ or $z$. For further reading on molecular symmetry see ref. \cite{06BuJexx.method}.
For transitions in the same electronic state $i'=i''$  TDMS must change at least linearly with the nuclear coordinate, that is has to take the form
\begin{equation}
\mu^{i'i''}_{\alpha}(Q_1,...,Q_D)=\mu^{i'i''}_{\alpha}(Q^{ref}_1,...,Q^{ref}_D)+\sum_{j} f^{(1)}_{j}Q_{j}
\label{eq:TDMSfitharm}
\end{equation}
in order to observe vibrational transitions. The first term in eq. \ref{eq:TDMSfitharm} is the permanent dipole moment of the molecule. Because the electronic transition dipole moment must transform as one of the translational coordinates in the molecular symmetry group, and the constant scalar quantity $\mu^{i'i''}_{\alpha}(Q^{ref}_1,...,Q^{ref}_D)$ always transforms as the totally symmetric representation $A_1$, observation of purely rotational transitions is possible only if $\Gamma(T_{\alpha})=A_1$ for $\alpha=x,y$ or $z$. This fact can be intuitively understood as the interaction of the permanent dipole moment of the molecule with changing electric field of the incident radiation, which causes the dipole moment to rotate.

If we consider linear terms in $Q$ in eq. \ref{eq:TDMSfitharm}, then the integral $\langle \Psi^{h'}_{vib}|Q|\Psi^{h''}_{vib} \rangle$ is non-vanishing by symmetry when $\Gamma(Q)=\Gamma(T_{\alpha})$  for $\alpha=x,y$ or $z$. The coefficient $f^{(1)}_{j}$ standing next to $Q_j$ in eq. \ref{eq:TDMSfitharm} represents the change of the electronic dipole moment upon change in the nuclear geometry along the $Q_j$ coordinate, calculated at equilibrium. If this coefficient is non-zero and $\Gamma(Q)=\Gamma(T_{\alpha})$ then the vibrational transition is said to be \textit{infrared active}.

Two quantities contribute to the vibrational transition intensity from the theoretical perspective: electronic transition dipole moment function and the form of the vibrational wavefunction. If the harmonic oscillator eigenfunctions are used as vibrational wavefunctions $|\nu\rangle$ and we assume that the electronic TDMS expansion is truncated at linear terms as in eq. \ref{eq:TDMSfitharm}, then we can derive the matrix element of the vibrational TDMS explicitly:
\begin{equation}
\langle \nu '| Q |\nu''\rangle = \frac{1}{\sqrt{2}}\langle \nu '| a+a^{\dag} |\nu''\rangle = \sqrt{\nu''}\delta_{\nu'\nu''-1}+\sqrt{\nu''+1}\delta_{\nu'\nu''+1}
\label{eq:vibtransharm}
\end{equation}
where $Q$ stands for an arbitrary nuclear coordinate and $\nu$ labels harmonic oscillator states associated with this coordinate.
By analysing when the above expression vanishes we can see that the selection rule for the vibrational transitions is $\Delta \nu = \nu''-\nu'=\pm 1$. If anharmonic wavefunctions were used in eq. \ref{eq:vibtransharm}, the selection rule would relax. Anharmonic wavefunctions (eigenfunctions of an anharmonic Hamiltonian) can be viewed as linear combinations harmonic oscillator eigenfunctions. Therefore anharmoncity is the cause of \textit{overtone transitions}, which correspond to changes in the harmonic oscillator principal quantum number $\Delta v = \pm 1, \pm 2, ....$. When more than one internal coordinate is considered, the terms such as $Q_1^2Q_2$ give rise to so called \textit{combination bands}, that is in the harmonic oscillator picture $v_1=\pm 1, \pm 2$ and $v_2 = \pm 1$ in a single transition. Combination bands can be intuitively understood in the following way: the electromagnetic radiation activates a normal vibration $Q_i$ which is however coupled to some other coordinate $Q_j$ either in the form of the wavefunction or in the form of the electronic transition dipole moment. The former vibration transfers its energy to the latter, infrared active vibration. This general mechanism is responsible for intensity borrowing in vibronically forbidden transitions.
In addition to the relaxation of the selection rules caused by the structure of the wavefunction, higher the power of $Q_i$ in the expansion of the TDMS also allow for overtone transitions, meaning that selection rules also depend on the shape of the TDMS.

At this point it is important to take notice of a common problem with fitting of a functional form to \abinitio\ points. When the fit of the PES (or TDMS) to  \textit{ab initio} points is performed, the continuous functional form can take non-physical, sometimes negative values when it reaches beyond the regions in which the \textit{ab initio} points are computed. This rapidly brings up problems in the nuclear motion calculations, especially when a quadrature grid point is located in this extrapolation region which is poorly represented by the fit. For this reason, it is reasonable to put 'walls' to the \textit{ab initio} surface or better to enforce a reasonable asymptotic behaviour of the functional form used in the fit.

\section{UV rotational-vibrational-electronic spectrum of SO$_2$}
This section presents example calculations of rotational-vibrational-electronic spectrum for the  $\tilde{C}~^1B_2 \leftarrow \tilde{X}~^1A_1$ electronic in the SO$_2$ molecule. $B_2$ in $\tilde{C}~^1B_2$ denotes the irreducible representation according to which the electronic wavefunction transforms under operations in the molecular symmetry group of SO$_2$ (C$_{2v}(M)$ group), whereas $\tilde{C}$ is an additional label to distinguish between electronic states, in case they have identical symmetry.  The electronic transition discussed here is located in the 220--235~nm wavelength region, that is in the ultraviolet region of the electromagnetic spectrum. The aim of this section is to present a practical procedure for generating spectra of triatomic molecules from first principles of quantum mechanics. Therefore the specific electronic structure and nuclear motion methods and basis sets used to solve the rotational-vibrational-electronic \SE\ are here less important. The focus is on familiarizing the reader with the process of calculating the spectrum from beginning to end. Because only selected technical details are presented below, the reader should refer to electronic structure textbooks for further information about methods for solving the electronic \SE .

\subsection{Potential energy surfaces}
The potential energy surface for the $\tilde{C}~^1B_2$ electronic state (electronic excited state of SO$_2$) was generated from 3000 SO$_2$ molecular geometries in the S--O bond-length and the $\sphericalangle$OSO bond angle coordinates. Stretching (bond-length) coordinates were chosen in the range: $r_1,r_2 \in [1.2;1.9] $  \r{A}, with $0.05$ \r{A} increments. Angles between the $S-O$ bonds were sampled from 60$^{\circ}$ to 180$^{\circ}$with 5$^{\circ}$ increments. The choice of ranges for the internal coordinates is primarily dictated by the maximum energy up to which the PES is expected to be valid. It is important that \abinitio\ points cover molecular geometries which correspond to energies below the established cut-off value. Missing any significant regions in the configuration space of the nuclear coordinates leads to very inaccurate vibrational energy levels. 
For every discrete molecular geometry a separate run of the electronic structure calculations must be executed. For the $\tilde{C}~^1B_2$ of SO$_2$ the electronic \SE\ was  solved in the \textit{aug}-cc-pVTZ basis set. The 'cc-p', stands for 'correlation-consistent polarized' and the 'V' indicates that basis functions are for valence-only electrons whilst the core electrons are frozen. 'TZ' (triple-zeta) means that three primitive basis functions (usually gaussian functions)  are used to model the shape of a single orbital. See for instance \cite{Helgaker2000} for further reading about basis sets used in electronic structure theory. 
 
The method used to solve the electronic \SE\ was the explicitly correlated multi-reference internally contracted configuration interaction method with Davidson correction (ic-MRCI-F12+Q). MRCI methods can provide very accurate electronic energies, but due to their high computational cost, are usually limited to systems with less than 20 electrons. There are a number of alternative methods applicable to larger systems. For instance the coupled-clusters method (CC) is a common choice.  

\begin{figure}[H]
\includegraphics[width=10cm]{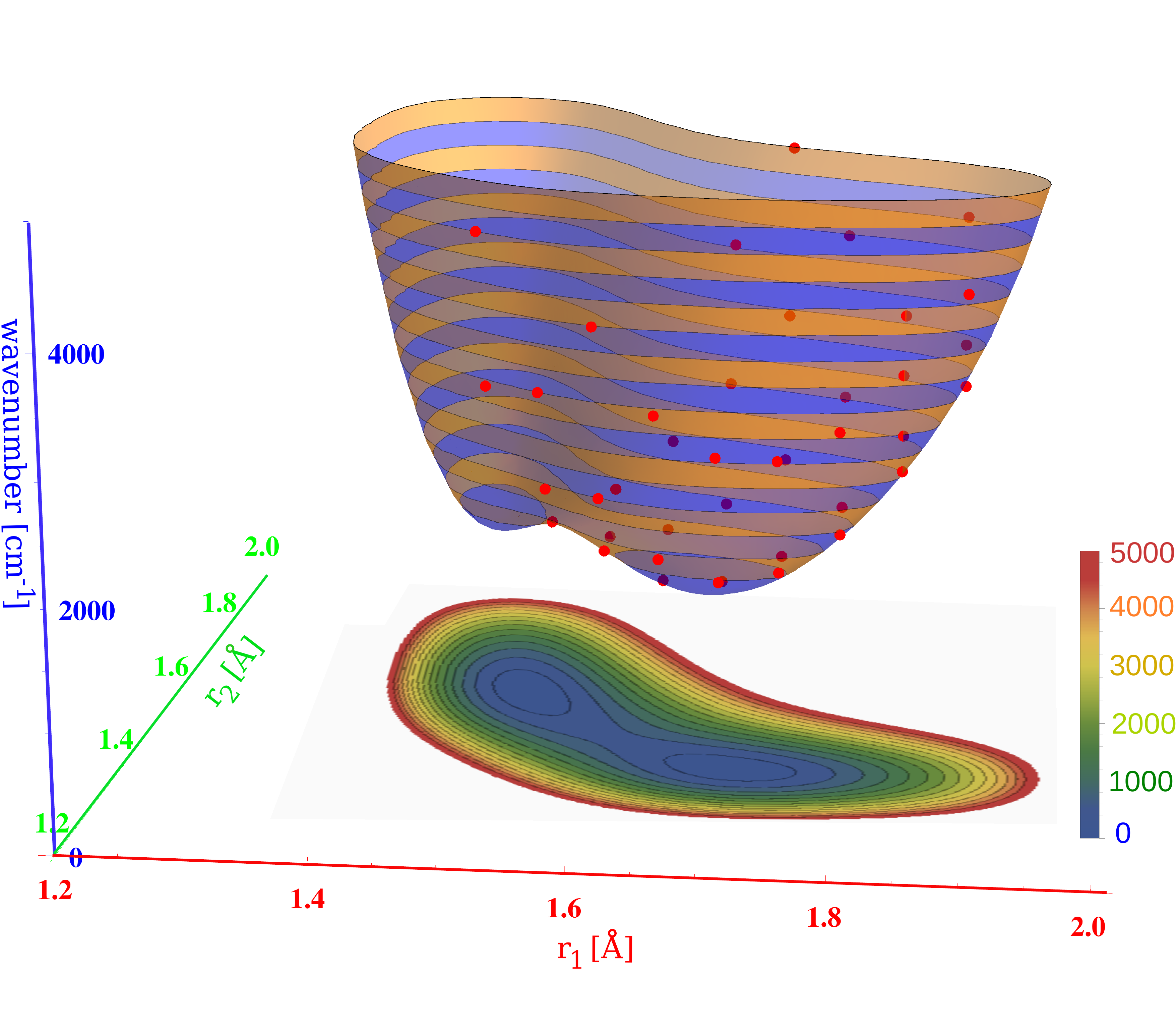}
\caption{Potential energy surface for the $\tilde{C}~^1B_2$ electronic state of SO$_2$ calculated at $\theta = 120.0 ^{\circ}$.  \textit{Ab initio} points are marked in red. Contour lines of equal energy are projected on the $r_1$-$r_2$ plane. Two local minima are visible in the $r_1$-$r_2$ plane, suggesting non-$C_{2v}$ equilibrium geometry.}
\label{fig:PESC2}
\end{figure}

The calculated \textit{ab initio} electronic energy points were fitted with the least-squares method to the functional form:

\begin{equation}
V\left(y_1,y_2,y_3\right)=\sum_{j,k,l}C_{jkl}y^j_1y^k_2y^l_3
\label{eq:PES}
\end{equation}
where $y_1=\frac{1}{2}(x_1 + x_2)$, $y_2=\frac{1}{2}(x_1 - x_2)$ and  $y_3=\theta-\theta_{eq}$. Here $x_1,x_2$ are Morse coordinates $x_1=1-e^{-a_1\left(r_1-r_1^{eq}\right)}$,   $x_2=1-e^{-a_2\left(r_2-r_2^{eq}\right)}$. In the least-squares method one determines the values of $C_{jkl}$ coefficients by solving a system of linear algebraic equations which ensure that the RMSD between the fit and the set of \abinitio\ points is minimal. Sometimes, it is convenient to impose a certain shape of the PES guided by the knowledge of the structure of the molecule. For instance, the functional form and coefficients $C_{jkl}$ for SO$_2$ were chosen to secure the correct shape of the PES at C$_{2v}$ geometries, that is when $r_1=r_2$. For a fixed angle $\theta=\theta_0$, the PES $V(r_1,r_2,\theta_0)$ has a saddle point when $r_1=r_2$, and two non-C$_{2v}$  minima in the $r_1$--$r_2$ plane, which are symmetry connected, as shown in Figure \ref{fig:PESC2}.  Non-C$_{2v}$  minima in the PES indicate that the equilibrium S--O bond lengths are not equal in this electronic state, which is schematically shown in Figure \ref{fig:so2embedding}. 

The art of fitting molecular surfaces (PES, TDMS, etc.) often requires a considerable level of patience topped with physical intuition about the possible shape of the surface. Often the fitting procedure starts with a preliminary fit to a small set of points with a low degree polynomial. In this way, approximate values for the leading coefficients can be found. Sometimes, the fitting procedure starts with fitting only 1-D cuts of the molecular surface to \abinitio\ points, followed by determining the coefficients which stand by the coupling terms. A common practice in the least-squares fitting procedure is to weight lower points lying at higher energies and weight higher points near the equilibrium geometry.

The quality of the fit can be controlled with root-mean square deviation (RMSD) between the fitted surface and \textit{ab initio} points: $RMSD = \sqrt{\frac{1}{N_{points}}\sum_{i=1}^{N_{points}}\left(V^{(fit)}(r_1^{(i)},r_1^{(i)},\theta^{(i)})-V^{(ab initio)}(r_1^{(i)},r_1^{(i)},\theta^{(i)})^2\right)}$.
For the construction of the $\tilde{C}~^1B_2$ state PES, a non-uniformly weighted fit to 623 \textit{ab initio} electronic energy points with energies below 5000 \cm\ was performed in the $\theta \in [90 ^{\circ};130 ^{\circ}]$ range. It gave 12 \cm\ RMSD between the fitted surface and the \textit{ab initio} points.

\begin{figure}[H]
\includegraphics[width=10cm]{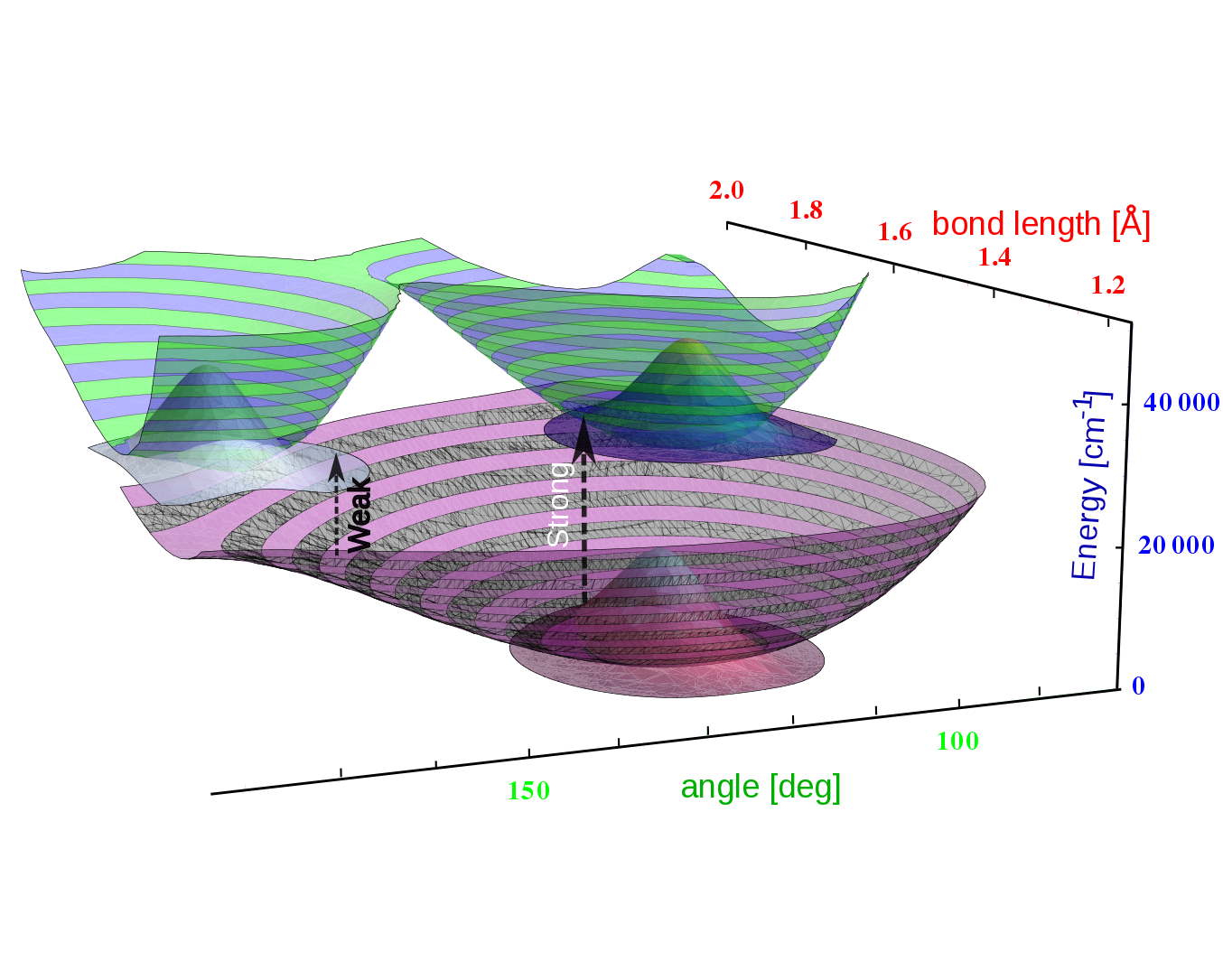}
\caption{Potential energy surfaces for $\tilde{X}~^1A_1$ electronic state (purple/grey) and  $\tilde{C}~^1B_2$ electronic state (green/blue) of SO$_2$. The other bond length is fixed at $r_2=1.7 $\AA . Wavefunctions for the vibrational ground state of each well are added, with arrows marking Franck-Condon vertical transitions from the electronic ground state.}
\label{fig:second_well}
\end{figure}

By inspecting Figure \ref{fig:PESC2}, we can conclude that the equilibrium geometry in the $\tilde{C}~^1B_2$ electronic state of SO$_2$ has unequal S--O bond-lengths. This suggests a non-$C_{2v}$ equilibrium structure. Indeed, the global equilibrium geometry possesses the low $C_s$ symmetry with $r^{eq}_1=1.640 $ \AA, $r^{eq}_2=1.496$ \AA, $\theta^{eq}=104.3^{\circ}$. There are two other local minima near $\theta=80 ^{\circ}$ and $\theta=165 ^{\circ}$. The latter is a C$_{2v}$ symmetric minimum with energy 400 \cm\ above the global minimum. This second potential well, located near linearity of SO$_2$, as displayed in Figure \ref{fig:second_well}, generates an independent set of vibrational energy levels. A separate fit for the second well gave RMSD$ = 17$ \cm .

For the electronic ground $\tilde{X}~^1A_1$ state of SO$_2$ a highly accurate potential energy surface of Huang \etal\ \cite{Schwenke2014} was used in solving the nuclear motion problem. This is a semi-empirical PES based on CCSD(T)/cc-pVQZ-DK calculations and a refinement to the experimental energy levels in the J=0--80 range. The RMSD of the fit to \textit{ab initio} points was 0.21 \cm\ below 30 000 \cm\ and the RMSD with respect to experimental ro-vibrational levels was 0.013 \cm. Note that in general a low RMSD of the fit to \textit{ab initio} points does not imply high accuracy of the surface. Depending on the electronic structure method used, electronic basis set, and other subtle factors, even a perfectly fitted PES can generate quite inaccurate vibrational energy levels. Only comparison to absolute ro-vibrational energy levels can validate the accuracy of the PES.

\begin{figure}[H]
\includegraphics[width=9cm]{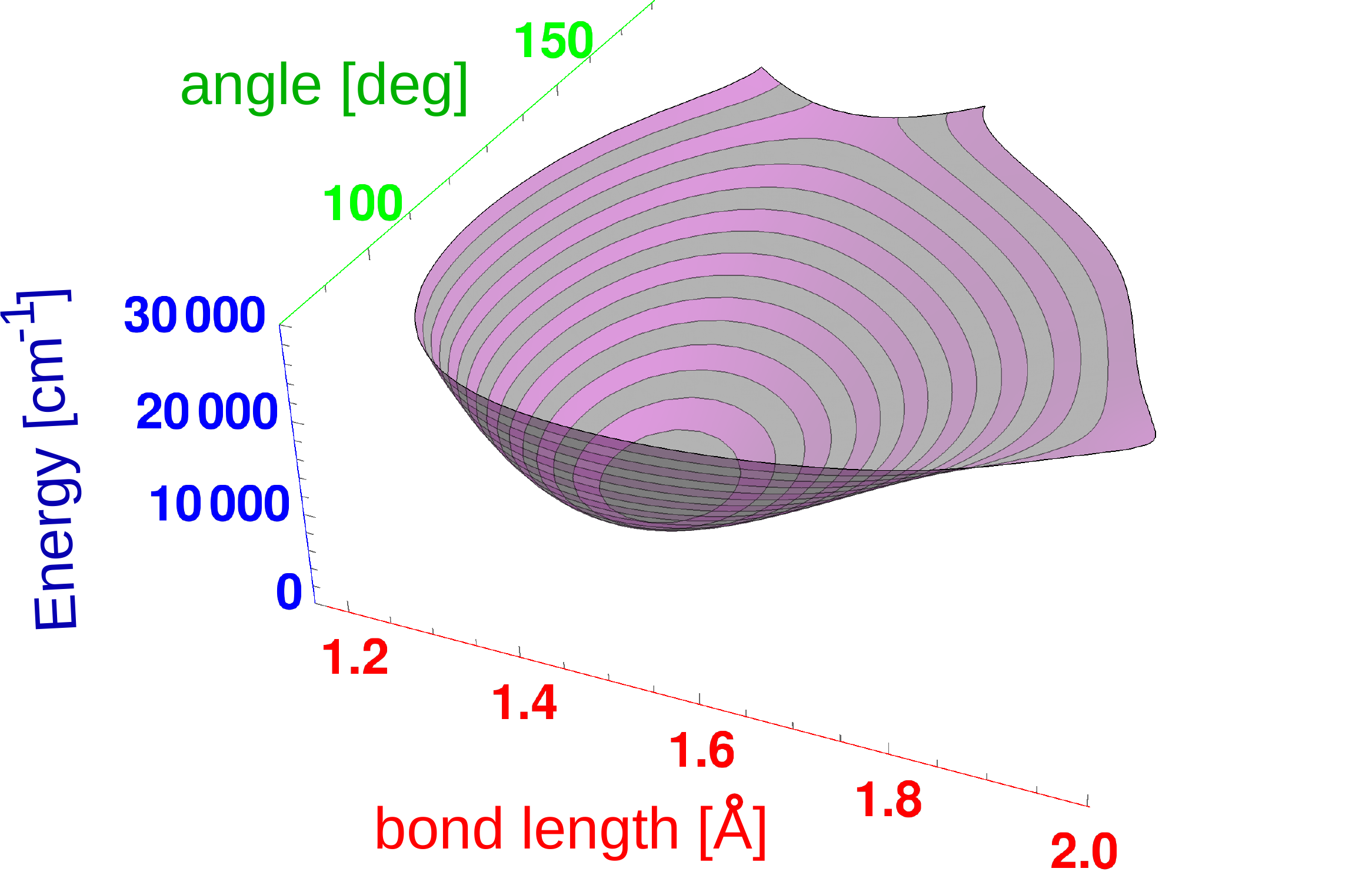}
\caption{Potential energy surface for the $\tilde{X}~^1A_1$ electronic state of SO$_2$ calculated at $\theta = 120.0^{\circ}$.}
\label{fig:PESA1}
\end{figure}

The equilibrium  geometry of the electronic ground state  $r^{eq}_1=1.431 $\AA, $r^{eq}_2=1.431 $\AA, $\theta^{eq}=119.32 ^{\circ} $ corresponds to C$_{2v}$ symmetry, as displayed in Figure \ref{fig:PESA1}.

What is the reason for fitting a predefined functional form to \textit{ab initio} points? First of all, whenever the electronic structure calculations are expensive (e.g. when the molecule has many electrons), it is computationally impractical to generate many accurate electronic energies for a molecule. Secondly, with the increasing number of atoms in the molecule, the number of internal coordinates also increases ($3N_{atoms}-6$), which means that the dimensionality of the potential energy surface also increases. If one takes a 5 atom molecule, such as CH$_4$, then constructing the 9-dimensional PES with only three \abinitio\ points per dimension, would require calculating the total of about 20 000 solutions to the electronic \SE\ at different geometries. Including just 4 points per dimension in the PES would require about 260 000 \abinitio\ points.

Finally, popular methods for solving the rotational-vibrational \SE, such as the finite basis representation or discrete variable representation methods discussed in the previous chapter, require values of the PES at quadrature grid points. Sometimes the number of quadrature grid points is much too high for calculating the electronic energy at every grid point. For this reason a form of interpolation is usually necessary and fitting of the PES provides exactly that.

\subsection{Nuclear motion calculations}
\label{sec:so2nuc}
With available potential energy surfaces for the $\tilde{X}~^1A_1$ and $\tilde{C}~^1B_2$ electronic states the next step is to solve the rotational-vibrational \SE\ for each of these electronic states separately, which is equivalent to assuming the Born-Oppenheimer approximation. As a result we separately obtain ro-vibrational wavefunctions and energy levels for the  $\tilde{X}~^1A_1$ and the $\tilde{C}~^1B_2$ electronic states. Because the total angular momentum is a conserved quantity it is possible to solve separately the rotational-vibrational \SE\ for each value of the total angular momentum quantum $J$.
The effective equations for the vibrational wavefunctions, as given in eq. \ref{eq:Heff0}, are solved in the next step. A Morse-like oscillator basis set (as described in section \ref{sec:theorybasis}) is suitable for modelling the vibrational wavefunction along the S--O stretching coordinates. Associated Legendre functions discussed in section \ref{sec:theorybasis} are used to describe the bending motion.
Because the vibrational Morse basis functions depend on three parameters: width $\omega_0$, depth $D_e$ and the position of the minimum $r_e$, the values of these parameters will correlate with how good the basis function is for a given PES.  A common practice is to optimize the parameters of the vibrational basis functions in order to reduce the number of functions needed to accurately represent the wavefunction. This procedure is nothing else but adjusting the shape of the basis functions, within their range of flexibility, to maximally mimic the exact wavefunctions. Because the shape of the PES in both electronic states is different, the parameters of the optimal basis set for one electronic state will be different for another electronic state.

A general recipe for finding reasonable values of basis set parameters (such as the dissociation energy $D_e$, the equilibrium bond length $r_e$ and width $\omega_0$ in the Morse-like oscillator basis functions) is to run multiple nuclear motion calculations (solve the vibrational, $J=0$ \SE ) with increasing basis set size and analyse for which set of basis function parameters the number of basis functions required to achieve certain convergence of the calculated energy levels is the lowest.

For SO$_2$, the set of optimized parameters for the vibrational basis functions and other parameters of the nuclear motion calculations are listed in Table \ref{table:nuc}.

 \begin{figure}[H]
 \begin{center}
 \includegraphics[width=6cm]{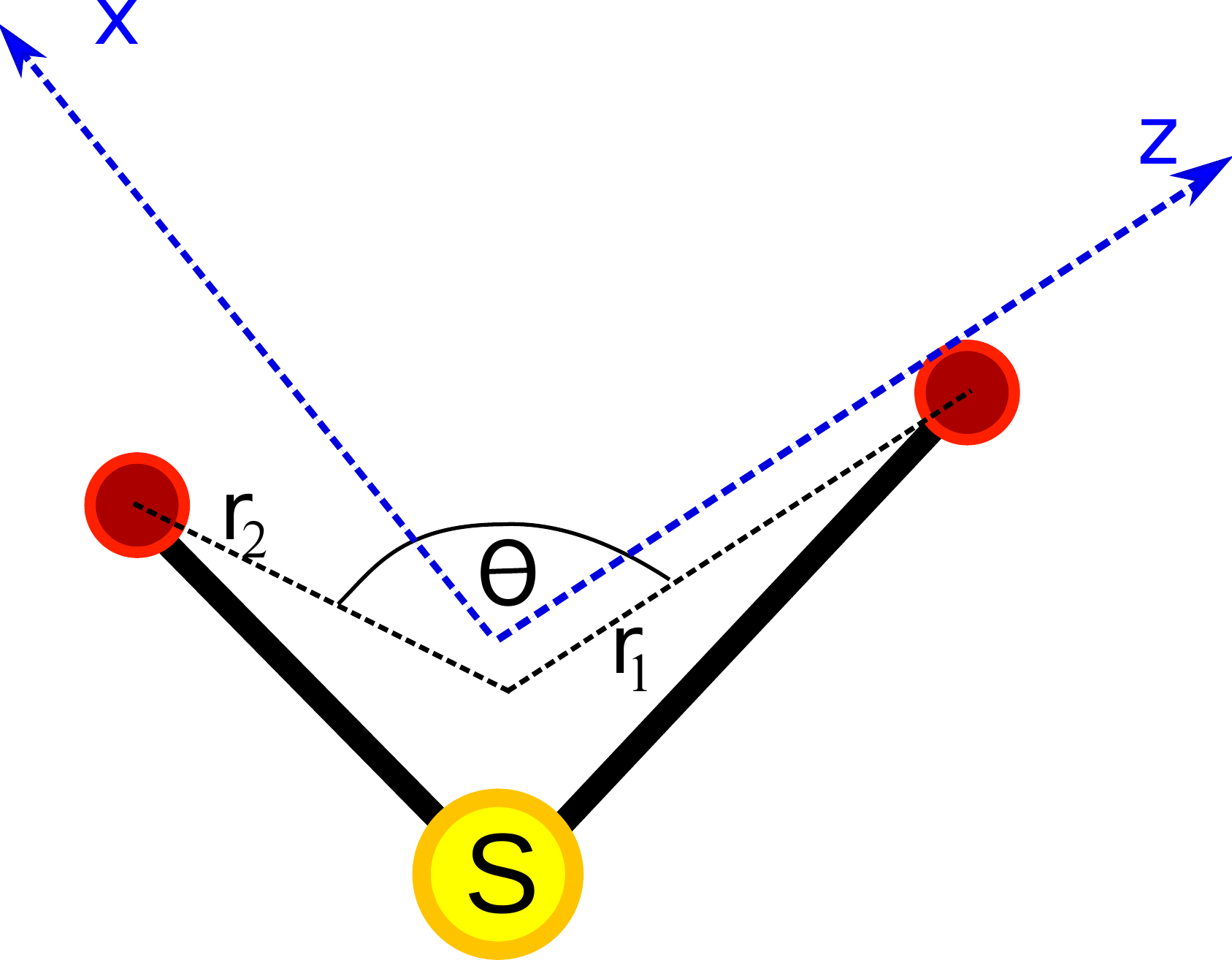}
 \caption{Schematic picture of the molecule-fixed frame embedding used in nuclear motion calculations for SO$_2$. $r_1$ and $r_2$ are Radau stretching coordinates.}
  \label{fig:so2embedding}
\end{center}
\end{figure}

\begin{table}[H]
\begin{center}
\setlength{\tabcolsep}{5pt}
\small
\vspace{0.05cm}
\begin{tabular}{l l}
\hline\hline
Coordinates:	& Radau  (r$_1$,r$_2$,$\theta$) \\ [0.3ex] 
BF frame embedding:	& $z$-axis along r$_1$ \\
\hline \hline 
\textbf{Basis set}	& identical for both el. states  \\
\hline 
'Stretching' r$_1$,r$_2$:	& 90 Morse-like oscillator functions  \\ [0.3ex] 
Morse-like oscillator basis parameters: & $r_e=2.9\; a_0$, $D_e=0.30\; E_h$,  $\alpha=0.012\; E_h$ \\ [0.3ex] 
Number of DVR points for stretching:				& 90 (Gauss-Laguerre quadrature)  \\ [0.3ex] 
						
Bending $\theta$:	&  60 (Associated Legendre Polynomials)  \\ [0.3ex] 
		Number of DVR points for bending:							& 60 (Gauss-Legendre quadrature)\\ [0.3ex] 
Rotations	&  Complete basis set of symmetric-top wavefunctions  \\ [0.3ex] 				
Truncated Hamiltonian size in the first step: & 1000 \\
Truncated Hamiltonian size in the second step: & 500 \\
Computation time$^a$ ($J=0$) & 10 min. \\
Scaling with J (computation time)  &  $\sim J$ (first step), $\sim J^2$ (second step)\\
\hline
\textbf{Intensity calculations }& \\
Common DVR grid for both electronic states & \\
Computation time ($J''=0 \rightarrow J'=1$) & 15 min. \\
Scaling with $J''$ (computation time)  & $\sim J^2$ \\
\hline 

\hline\hline
\end{tabular}
\caption{Summary of the parameters of the nuclear motion calculations for the  $\tilde{X}~^1A_1$ and the $\tilde{C}~^1B_2$ electronic states of SO$_2$. In the table given are: the type of coordinates used, the type of molecule-fixed frame embedding, basis set parameters and some technical details of the computation. The two-step DVR procedure described in Chapter 2 was here used. $^a$ Test computations were performed on a stationary PC with Intel(R) Core(TM) i5-2500@3.30~GHz processor and 8 GB of RAM. }
\label{table:nuc}
\end{center}
\end{table}

For calculations with $J > 0$ one must choose an embedding of the molecule-fixed coordinate frame.
Identical embeddings (see Figure \ref{fig:so2embedding}) of the molecule-fixed coordinate frame were chosen for both electronic states. Such strategy of identical embeddings allow to keep identical definitions of the Euler angles in both electronic states and avoid potential complications with  the use of different rotational basis sets the two electronic states. The $z$-axis of the molecule-fixed frame was chosen along one of the Radau coordinates ($r_1$), which nearly overlaps with one of the S--O bonds ('bond embedding', see eq.~\ref{eq:bond} and Figure~\ref{fig:embeddings}). The nuclear motion calculations with the DVR technique described in section \ref{sec:dvr} were performed for the $\tilde{X}~^1A_1$ and the $\tilde{C}~^1B_2$ electronic states of SO$_2$ separately with identical Radau coordinates and identical DVR grids. The detailed numbers are given in Table \ref{table:nuc}. A Gauss-Hermite quadrature was used in the DVR method for the stretching coordinates $r_1,r_2$ and a  Gauss-Legendre quadrature was used for the bending coordinate.

\subsection{Transition dipole moment surface and transition intensities calculations}

The transition dipole moment surface between the  $\tilde{X}~^1A_1$ and  $\tilde{C}~^1B_2$ electronic states was calculated as the expectation value of the electric dipole moment operator. A fit to the functional form given in eq. (\ref{eq:PES}) was performed with 1852 \textit{ab initio} points in the [85$^{\circ}$:140$^{\circ}$] angle range. For the triatomic SO$_2$ molecule only two components of the dipole moment vector are non-zero.
The RMSD of the fit for the $x$-component of the surface ($x$-axis chosen to bisect the angle between S--O bonds) was 0.03 a.u.\footnote{a.u. means 'atomic units' and for the dipole moment is 1 Debye = 3.33564$\cdot 10^{-30}$ Coulomb$\cdot$m and  1 Debye = 0.393456 a.u}, and the RMSD for the $z$-component of the surface was 0.02 a.u. 

\begin{figure}[H]
\begin{center}
  \includegraphics[width=10cm]{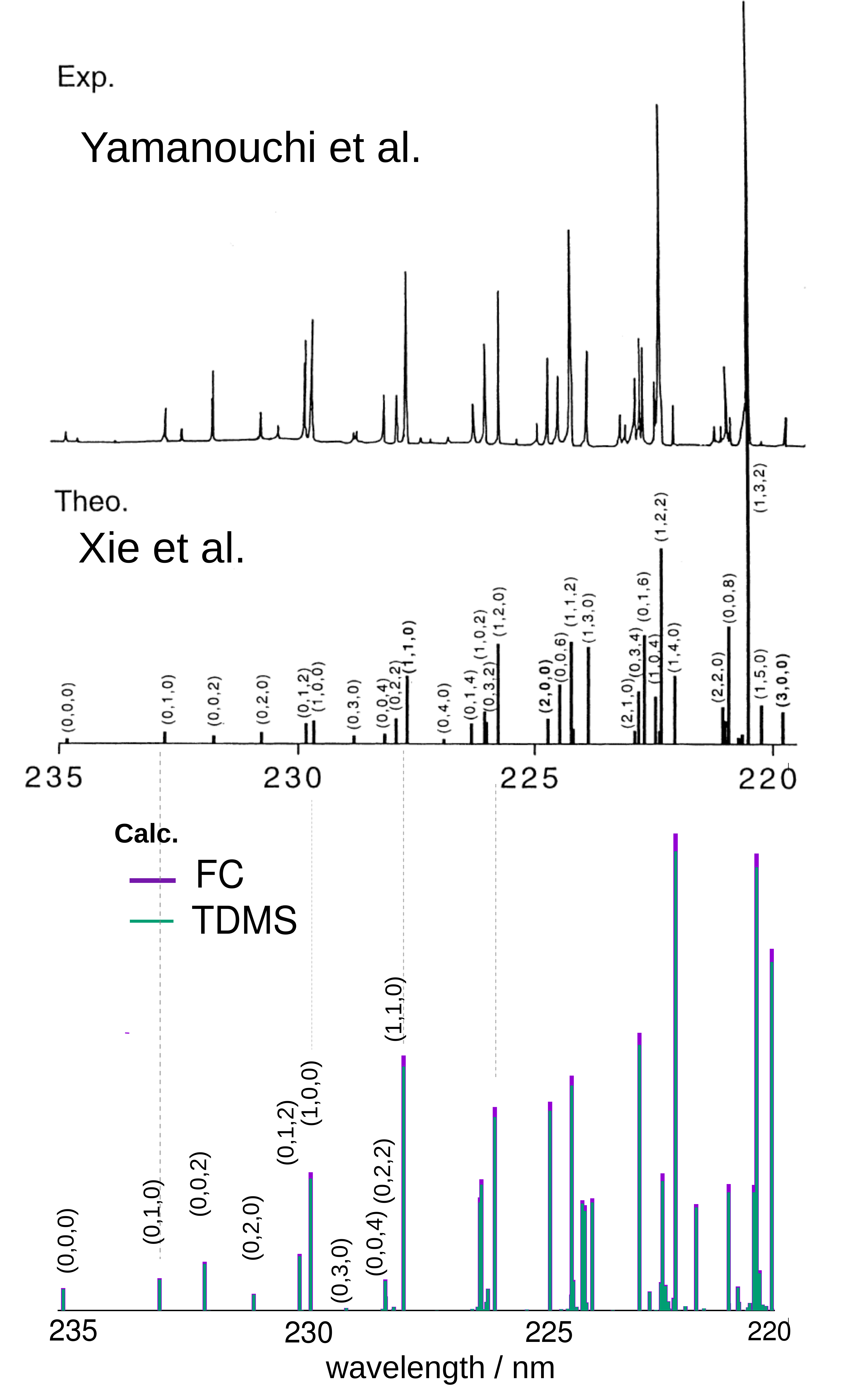}
\caption{Vibronic spectrum for $\tilde{X}~^1A_1 \rightarrow \tilde{C}~^1B_2 $ electronic transition of SO$_2$ calculated in the Frank-Condon approximation (green) and with full transition dipole moment surface(magenta), which accounts for the change in the electronic transition moment with internal coordinates of the molecule. In brackets are given approximate vibrational quantum numbers ($\nu_1,\nu_2,\nu_3$) of the final state.}
\label{fig:FCspect}
\end{center}
\end{figure}

At the equilibrium geometry of the electronic ground state (that is C$_{2v}$ geometry) the $z$-component of the transition dipole moment vanishes, as shown in Figure \ref{fig:DMS}. The transition dipole moment depends relatively weakly on the nuclear coordinates, nevertheless in high-accuracy calculations of transition intensities this dependence must be included in the model. Comparison of the Frank-Condon spectrum (see eq. \ref{trans:FC}) with the spectrum calculated with the full TDMS is depicted in Figure \ref{fig:FCspect}. Here the non-Frank-Condon effects account for about 10\% of the total transition intensity, meaning that the FC approximation is not sufficient for high-accuracy calculations. However the relatively smooth shape of the TDMS displayed in Figure \ref{fig:DMS} justifies expansion of the TDMS function only to low order terms in eq. \ref{eq:TDMSfit}. 

Combination of the calculated rotational-vibrational transition frequencies (details of the method are outlined in section \ref{sec:so2nuc}) with transition intensities and quantum numbers forms a \textit{line list}. There are several online databases archiving high-accuracy line lists for small molecules: HITRAN (www.hitran.org), ExoMol (www.exomol.com), CDMS (cdms.astro.uni-koeln.de) and other.

\section{Infrared rotational-vibrational spectra of CO$_2$ isotopologues}
\label{sec:co2}
In this section we discuss some aspects of calculating rotational-vibrational spectra on the example of the CO$_2$ molecule.

Accurate infrared line lists for carbon dioxide are of high importance in atmospheric sciences. Carbon dioxide is an inert gas in the Earth's atmosphere, therefore it can be relatively easily traced for studying the atmospheric circulation. Monitoring the concentration of carbon dioxide in the Earth's atmosphere remains a priority task for a constantly growing number of government funded projects.  This greenhouse gas has been a fingerprint of anthropogenic activity since the industrial revolution in the early 19th century, by which time its atmospheric concentration is estimated to have risen from 280~ppm to over 380~ppm \cite{Sabine2004}. A 20\%  increase in the CO$_2$ atmospheric concentration has been observed over the past half-century - the most dramatic change in human history \cite{12BaAlMi.CO2}. Mapping the circulation of the CO$_2$ gas in the troposphere 
is clearly vital to understanding and hopefully controlling the CO$_2$ content and hence the climate change \cite{Emmert2012,Masarie1995}. Several space missions are dedicated to explicitly monitor the atmospheric CO$_2$ molar fraction in high geographic resolution: GOSAT \cite{GOSAT}, ASCENDS \cite{ASCENDS}, AIRS \cite{AIRS}, CarbonSat \cite{CarbonSat} and recently launched NASA's OCO-2 mission \cite{OCO-2,11WuWeTo.CO2,16TaChDe.CO2,16WuWeOs.CO2}. Remote sensing measurements are cross-compared with ground based projects, such as Total Carbon Column Observing Network (TCCON) \cite{TCCON,Chevallier2011} or Network for Detection of Atmospheric Composition Change
(NDACC) \cite{NDACC}, to look at the overall CO$_2$ concentration and its time variation, but more importantly to pinpoint where CO$_2$ is being produced (sources) and where it is going (sinks). Future missions, such as ARIEL and NASA's JWST \cite{JWST1}, are designed to probe atmospheres of exoplanets, many of which are believed to have carbon dioxide as its major component.  

All the projects mentioned above require reference line strengths and line shapes for selected infrared transitions in CO$_2$. Theoretical calculations has been very successful in providing reference transition intensities at high accuracy, sufficient for the remote sensing purposes. In section \ref{sec:co2mu} we focus on analysis of an atmospherically important absorption band in CO$_2$ - the \textit{2 $\mu$m band}. Before that however, we need to introduce a nomenculature used in characterizing energy levels and transitions in CO$_2$. This nomenculature is quite general throughout molecular spectroscopy.

\subsection{Approximate vibrational quantum numbers}
As discussed in section \ref{sec:strategies} the exact quantum numbers for a general triatomic molecule in the Born-Oppenheimer approximation are: $J$ - the total angular momentum and $p$ - parity. However from the diagonalization of the Hamiltonian matrix one obtains eigenvalues per $J$ and $p$. These eigenvalues correspond to vibrational energy levels.

Normal mode analysis performed on CO$_2$ defines the \textit{normal coordinates}, which are linear combinations of the Cartesian coordinates of atoms. When the electronic potential energy surface is calculated along these normal coordinates and is subsequently truncated at the quadratic term in the normal coordinate, then we can find the values of the \textit{Harmonic frequencies}. These are approximately given as $\tilde{v}_1 = 1285$\cm (symmetric stretching), $\tilde{v}_2 = 667$\cm (bending), $\tilde{v}_3 = 2349$\cm (asymmetric stretching). Note that the accuracy of the harmonic frequencies depends on the level of electronic structure theory used. One needs at least two electronic energies at different geometries along the normal coordinate in order to calculate the value of the harmonic frequency.

If for a non-linear triatomic molecule the internal degrees of freedom behaved as the ideal harmonic oscillators and the vibrations were not coupled to rotations (rigid rotor model for rotations), then there would be three exact vibrational quantum numbers: $v_1$, $v_2$, and $v_3$.  The $v_1$, $v_2$, and $v_3$ numbers label consecutive solutions to 1-D harmonic oscillator Hamiltonians associated with internal (normal) coordinates in the molecule. Because non-linear triaotmic molecules are asymmetric tops, the $k$ quantum number is not an exact, or a \textit{good} quantum number.

Assignment of approximate quantum numbers is often based on exactly solvable quantum-mechanical models. When these models are not ideal representations of the state of the molecule, the associated quantum numbers are no longer good reflections of the true state of the molecule. 
When the model used to define approximate quantum numbers is poor, a pair of states can become a nearly half/half mixture of states with two different model quantum numbers, in which case it is said that the zero-order (model) states are interacting strongly, or are in (near)resonance. As a result the individual zero-order states can no longer be unambiguously traced and the approximate quantum numbers of the zero-order model loose their validity. For this reason, the assignment of non-exact quantum numbers is to some extent arbitrary: choosing a more accurate zero-order model, which captures a considerable portion of interactions in the system, would make the corresponding quantum numbers to be good approximations for the true state of the molecule. Sometimes, when the energy levels interact strongly through higher degree polynomial terms in the PES (Fermi resonances), there is an additional quantum number $n$, which enumerates states within the set of Fermi-interacting zero-order (harmonic oscillator) states. Resonance interactions of rotational-vibrational energy levels are discussed in detail in Chapter 4.

\begin{figure}[H]
\includegraphics[width=12cm]{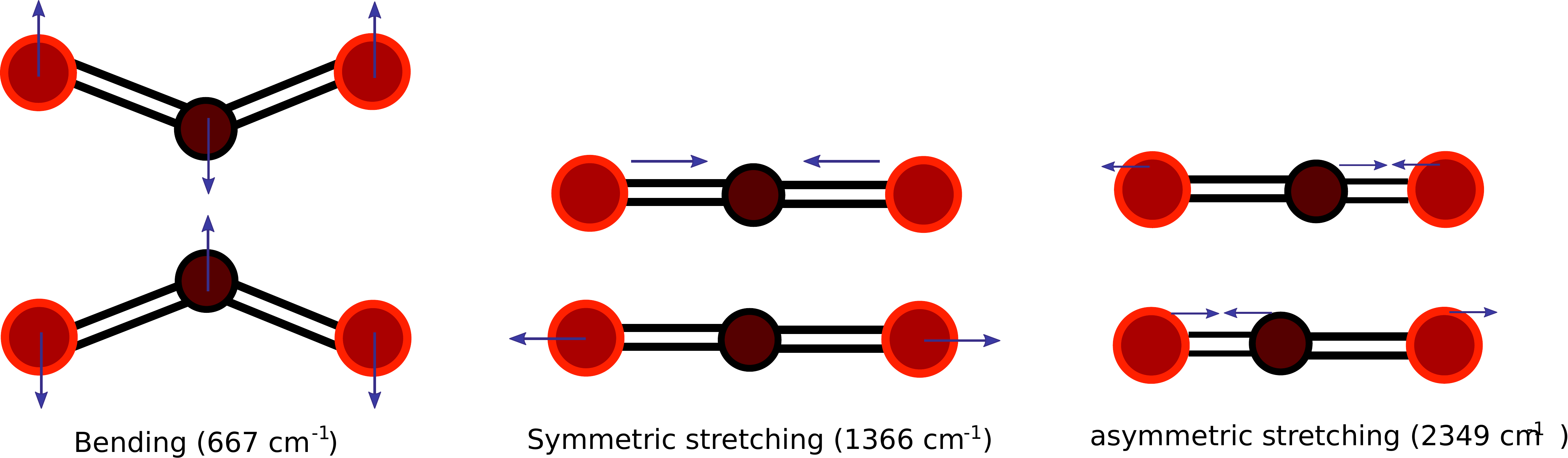}
\caption{Representation of normal vibrations in the carbon dioxide molecule.  }
\label{fig:vibCO2}
\end{figure}

For carbon dioxide, which is a linear molecule (the global minimum in the PES is located at linear geometry), there are only two rotational degrees of freedom, because one of the moments of inertia of the molecule is always zero, regardless of the choice of the molecule-fixed coordinate frame. As a result there are three vibrational quantum numbers: $v_1$, $v_2$, and $v_3$, labelling excitations in the symmetric stretching, bending and the asymmetric stretching normal mode (see \cite{Wilson1955} for further discussion of normal vibrations). The normal vibrations for CO$_2$ are displayed in Figure \ref{fig:vibCO2}. The bending mode is doubly degenerate. This is because the CO$_2$ molecule can bend in two perpendicular directions - motions which are energetically equivalent. For this reason the 2-D harmonic oscillator model is needed to describe the bending motion of the molecule. This 2-D bending motion in the Cartesian coordinates becomes in fact a combination of a radial harmonic motion and a circular motion. This circular motion introduces a form of the \textit{vibrational angular momentum}. As a result the $v_2$ quantum number for the doubly-degenerate bending mode of CO$_2$ becomes $v_2, l$ where the first number labels the amplitude of the bending motion whereas $l$ labels the vibrational angular momentum associated with the internal rotational motion of bent CO$_2$ around the centre-of-mass. In other words, with no additional angular momentum from the motion of electrons ($\Lambda = 0$ states), the vibrational quantum number $l$ is $k$, that is the projection of the total angular momentum on the molecule-fixed z-axis. 

To summarize, for CO$_2$ we have 5 approximate quantum numbers describing the vibrational state:   $v_1 v_2 l v_3 n$.

\subsection{Example results for main isotopologue of CO$_2$}
\label{sec:co2mu}
\paragraph{The 2$\mu$m absorption band.}

\begin{figure}[H]
\includegraphics[width=12cm]{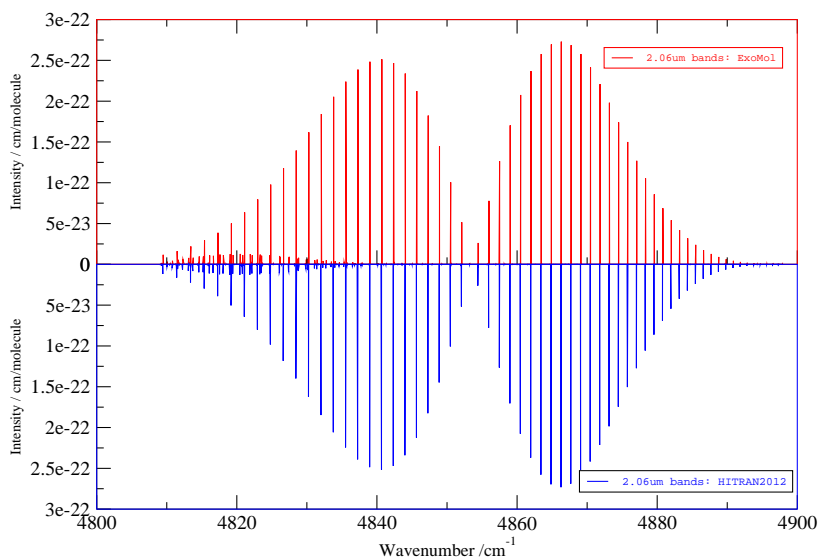}
\caption{Example rotational-vibrational spectrum for the main isotopologue of CO$_2$ calculated with the DVR3D program in Ref. \cite{DVR3D}. The main absorption in this region corresponds to combination vibrational absorption band: $v_1=2, v2=0, l=0, v_3=1, n=2 \leftarrow v_1=0, v_2=0 l=0, v_3=0, n=1$. Individual rotational lines divide into P and R branch. The calculated spectrum is compared to empirical spectrum from the HITRAN 2012 database.}
\label{fig:206}
\end{figure}

In Figure \ref{fig:206} an example rotational-vibrational spectrum is presented for the main isotopologue of CO$_2$ calculated with the DVR3D program in Ref. \cite{DVR3D}. The DVR3D code uses the Sutcliffe-Tennyson Hamiltonian described in Chapter 2 and the DVR technique outlined in section \ref{sec:dvr} with wavefunction anzatz given in eq. \ref{eq:rvwf}. In Figure \ref{fig:206} only the wavelength region near 2$\mu$m is presented. The wavelength is related to the wavenumber (in \cm ) by the relation: $\lambda = \frac{100}{\tilde{v}}$. The main absorption in the 2$\mu$m  region corresponds to the combination vibrational absorption band: $v_1=2, v2=0, l=0, v_3=1, n=2 \leftarrow v_1=0, v_2=0 l=0, v_3=0, n=1$. Individual rotational absorption lines visible in Figure  \ref{fig:206} correspond to P and R rotational branches. The P rotational branch corresponds to transition with lowering of the $J$ quantum number by one, whereas the $R$ rotational branch corresponds to transition with raising  of the $J$ quantum number by one. The calculated spectrum is compared to empirical spectrum from the HITRAN 2012 database (bottom panel). In DVR3D calculations, a high quality potential energy surface of Huang \etal\ \cite{Schwenke2012} and an \abinitio\ calculated dipole moment surface \cite{jt613} was used to generate the CO$_2$ spectrum presented in Figure \ref{fig:206}. 

Comparisons between line lists can be done in the frequency domain, but comparing transition intensities sheds light directly on the accuracy of the underlying TDMS. The latter is important in atmospheric science and in quantitative spectroscopy of molecules.
A popular measure for quantifying the deviation between transition intensities from two different sources is given as:
\begin{equation}
S=\left(\frac{I_{1}}{I_{2}}-1\right)\cdot 100 \%
\label{eq:asymmetric}
\end{equation}
where $I_{1}, I_2$ stand for line intensities, given in cm/molecule, from sources $1$ and $2$, respectively. This measure is adequate for small deviations but poorly illustrates highly discrepant intensities, due to its asymmetric functional form. Sometimes it is more instructive to use a symmetrized measure:

\begin{equation}
S_{sym}=\frac{1}{2}\left(\frac{I_{1}}{I_{2}}-\frac{I_{2}}{I_{1}}\right)\cdot 100 \%
\label{eq:symmetric}
\end{equation} 
which is good for large deviations, but yields far from intuitive numbers near $0\%$ deviation. In comparisons, the two measures given in eqs. \ref{eq:asymmetric}-\ref{eq:symmetric} can be used interchangeably, depending on the span of intensity deviations. 

\begin{figure}[H]
\begin{center}
\includegraphics[width=10cm]{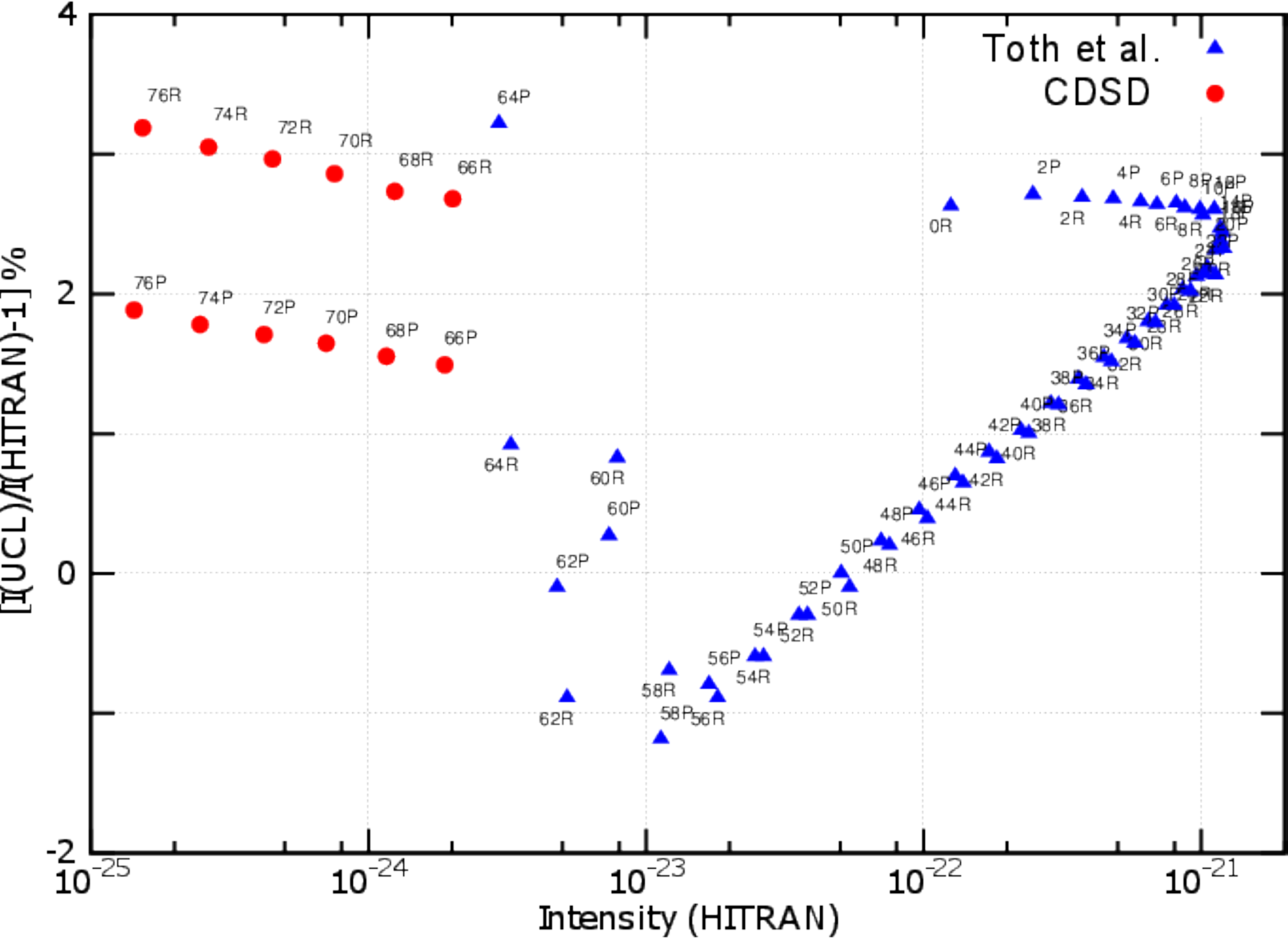}
\caption{HITRAN 2012 vs. UCL line intensities comparison for the 20012 -- 00001 band in the main isotopologue of CO$_2$. Two HITRAN data sources are marked with red (CDSD-296: semi-empirical calculations \cite{11TaPe.CO2}) and blue (Toth \etal\ - experimental  \cite{08ToBrMi.CO2}) circles.}
\label{fig:Int20012}
\end{center}
\end{figure}

Intensities comparison for the discussed here 2$\mu$m absorption band is presented in Figure \ref{fig:Int20012}. Results of DVR3D calculations (marked as UCL) are compared to intensities taken from the HITRAN 2012 database. In the figure a shorthand notation for transitions is used:  20012 -- 00001, means that the initial state is the ro-vibrational ground state 00001 ($v_1=0, v_2=0 l=0, v_3=0, n=1$) and the final state is 20012 ($v_1=2, v2=0, l=0, v_3=1, n=2$). Individual points in  Figure \ref{fig:Int20012}  stand for rotational-vibrational transitions marked with the name of the branch (P,R) and the initial rotational quantum number ($J$). The HITRAN 2012 database uses two different data sources for the 20012 -- 00001 band, showing discontinuity in the transition intensity pattern. This issue of empirical databases is not present in \abinitio\ calculations. This consistency in accuracy is one attractive feature of theoretical calculations.

In the calculation of the transition intensities shown in Figure \ref{fig:Int20012} the Hamiltonian from eq. \ref{eq:ham1} was used for CO$_2$. The exact KEO used in the calculations (within the Born-Oppenheimer approximation) means that the rotational motion of the molecule is described exactly. Because the wavefunctions are converged with respect to the basis set size, one can expect that the accuracy in all rotational transitions will be constant. Only the PES and the DMS are the sources of  inaccuracy in vibrational transition intensities (systematic shifts in intensity of the whole vibrational absorption band). The errors in the former introduces inaccuracies in the vibrational wavefunctions whereas errors in the DMS introduce errors directly in the linestrength. Similar accuracy should be expected in all rotational transitions within a vibrational band. An exception to this rule is when some rotational-vibrational energy levels are involved in a strong resonance interaction, in which case the accuracy in the energy levels and consequently in the transition intensities strongly depends on the accuracy of the wavefunction. These resonance interactions are discussed in Chapter 4.

To summarize, the process of generation of theoretical line list can be the following: the potential energy surface in an analytical form serves as an input function in the ro-vibrational calculations, as displayed in Figure \ref{fig:scheme}. Although the nuclear motinon calculations most often take a pivotal role for triatomic molecules, an efficient and recommended approach is to divide the problem into two steps: 1) solve the Coriolis-decoupled problem (see the Hamiltonian given in eq. \ref{eq:ham1}); 2) use the rotational-vibrational basis from step 1) to find the ro-vibrational energy levels and ro-vibrational wavefunctions (see eq. \ref{eq:anzatzrv}). This is called the \textit{contraction scheme}. The wavefunctions are next passed into a computer program which combines them with the the transition dipole moment surface. Calculated line strengths are further converted into transition intensities (in cm/molecule units) according to eq. \ref{eq:Int}. Combined: lower ro-vibrational energy levels $E_i$, transition wavenumbers ($\tilde{\nu}_{if}$), transition intensities $I_{if}$ and quantum numbers ($J p v_1 v_2 l v_3 n$) form a line list.

\paragraph{Isotopologues and purely rotational transitions.}

Isotopologues of carbon dioxide are species which have either $^{12}$O replaced with $^{13}$O or $^{14}$O, or have one or two oxygen atoms $^{16}O$ replaced with $^{17}$O or $^{18}$O. On account of the permutation symmetry of identical nuclei in symmetric isotopolouges of carbon dioxide, such as $^{16}$O$^{12}$C$^{16}$O (626) or $^{18}$O$^{13}$C$^{18}$O (838), the permanent electric dipole moment is zero in these molecules. For this reason in symmetric isotopologues no purely rotational transitions are observed. This can be shown by taking the TDMS given in eq. \ref{eq:TDMSfitharm} and computing the transition dipole matrix element between identical vibrational states and two distinct rotational states:

\begin{equation}
\begin{split}
\langle J,K,p | \langle \phi_{vib,h}| \mu^{i'i'}_{\alpha}(Q_1,...,Q_D)| \phi_{vib,h}\rangle | J',K',p'\rangle =\mu^{i'i'}_{\alpha}(Q^{ref}_1,...,Q^{ref}_D)\langle J,K,p | J',K',p'\rangle\langle \phi_{vib,h}| \phi_{vib,h}\rangle+ \\
\sum_{j} f^{(1)}_{j}\langle \phi_{vib,h}|Q_{j}| \phi_{vib,h}\rangle\langle J,K,p | J',K',p'\rangle+
\end{split}
\label{eq:purerot}
\end{equation}
From the above equation it is seen that whenever the permanent electric dipole moment is zero $\mu^{i'i'}_{\alpha}(Q^{ref}_1,...,Q^{ref}_D)=0$ the first term vanishes and all terms with linear powers of internal coordinates $Q_j$ also vanish due to symmetry. The integral $\langle \phi_{vib,h}|Q_{j}| \phi_{vib,h}\rangle$ vanishes because the product of identical vibrational wavefunctions $\phi_{vib,h}$ generates the fully symmetric $A_1$ irreducible representation, whereas the internal coordinate $Q_j$ never transforms as the fully symmetric representation, meaning that the total symmetry of the integrated function is not $A_1$. Integrals over functions which do not transform fully symmetrically in the molecular symmetry group are always zero. This is known as the \textit{vanishing integral rule}.

\begin{figure}[H]
\begin{center}
\includegraphics[width=10cm]{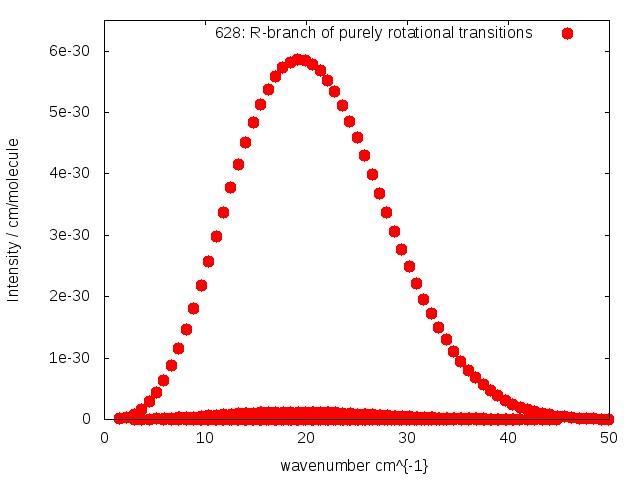}
\caption{A fragment of the rotational-vibrational spectrum of the 628 carbon dioxide isotopologue. The strongest visible peak corresponds to the R branch of purely rotational transitions in the vibrational ground state. Weak diffuse absorption peak near 20\cm\ corresponds to hot vibrational bands. Source: the HITRAN 2016 database.  }
\label{fig:purerot}
\end{center}
\end{figure}

Non-symmetric isotopologues of carbon dioxide have a very small non-zero permanent dipole moment, which allows for the existence of purely rotational spectrum, that is rotational transitions without any accompanying vibrational transitions (see eq. \ref{eq:TDMSfitharm}). When the oxygen atom in $^{16}$O$^{12}$C$^{16}$O is replaced with   $^{17}$O or $^{18}$O, the C$_{2v}$ symmetry of the molecule is lowered to C$_s$, meaning that the only feasible symmetry operations for asymmetric isotopologues of CO$_2$ are the parity operation $E^*$ and the identity operation $E$, which together constitute the C$_s$ group. Molecules belonging to the C$_s$ point symmetry group have a non-vanishing dipole moment. A purely rotational spectrum for $^{16}$O$^{12}$C$^{18}$O (628) isotopologue is displayed in Figure \ref{fig:purerot}. The purely rotational transitions are very weak, with peak intensity of just above $5\cdot 10^{-30}$ cm/molecule. In computational practice, even though identical Born-Oppenheimer PES is used to calculate rotational-vibrational energy levels of different isotopologues of CO$_2$, due to different masses in asymmetric isotopologues, such as 628, which enter the nuclear KEO, the rotational-vibrational wavefunctions differ among isotopologues. Because transition intensities depend on the vibrational transition dipole moment, the symmetry of vibrational wavefunctions is broken for asymmetric isotopologues and the permanent dipole moment is non-zero. Despite all isotopologues of CO$_2$ having identical electronic structure, different nuclear masses, generate a notably distinct dynamics among isotopologues.
\bibliographystyle{plain}
\bibliography{References}
\chapter{Resonance interactions of rotational-vibrational energy levels}

This chapter focuses specifically on the concept of a resonance interaction between ro-vibrational energy levels. Resonance interactions between quantum-mechanical states are an important part of the quantum theory of molecules. The phenomenon of interacting energy levels is responsible for the transition intensity borrowing, the alternations in transition intensities, they cause the normally forbidden transitions to be allowed. A general theoretical formulation of theory of resonances given below is exemplified   on the CO$_2$ molecule in sections \ref{sec:resfermi} and \ref{sec:rescoriol}. Finally, a simple theory for transition intensity borrowing, which  accompanies resonance interactions, is developed. This theory serves in section \ref{sec:theoryscatter} to formulate a theoretical descriptor, which serves as a quantitative measure to the sensitivity of line intensities for small perturbations in the ro-vibrational wavefunction. Further on, this new descriptor will be used in a qualitative discussion of reliability of calculated line intensities in CO$_2$. 

\section{What is a resonance interaction?}
\label{sec:resonances}
A resonance interaction between two energy levels occurs when three conditions are satisfied: a) the levels have similar (identical) energies; b) quantum states associated with the energy levels have identical symmetry; c) an interaction operator exists, which mixes the interacting states. 

In the variational nuclear motion calculations, the \SE\ is represented by an eigenvalue problem. The ro-vibrational Hamiltonian, which is represented by a square Hermitian matrix is diagonalized to obtain energy levels and wavefunctions. The values of individual matrix elements depend on the choice of the variational basis functions. The choice of the basis set determines which matrix elements of the Hamiltonian are large, which are small and which vanish by symmetry conditions. The primitive rotational-vibrational basis typically is a product of the vibrational and rotational basis function:

\begin{equation}
\Phi_{rv}^0=\Phi_{rot}\Phi_{vib}
\label{eq:zeroorder}
\end{equation}
in which  $\Phi_{rot}$ is the symmetric-top (rigid-rotor) eigenfunction $|J,k\rangle$ and $\Phi_{vib}$ is either an eigenfunction of the 3-D harmonic oscillator or the 3-D Morse oscillator.  In triatomic molecules, such as CO$_2$, $\Phi_{vib}$ can be written in the  \textit{bra-ket} form as: $|\nu_1,\nu_2,l,\nu_3 \rangle =|\nu_1 \rangle |\nu_2,l \rangle |\nu_3 \rangle$, where $\nu_1,\nu_2,\nu_3$ are quantum numbers characterizing the vibrational zero-order basis state. $l$ is the vibrational angular momentum quantum number, originating from the degenerate bending motion in a linear triatomic molecule, as discussed in section \ref{sec:co2}. The full ro-vibrational wavefunction is a linear combination of basis functions:

\begin{equation}
\Phi^{(n)}_{rv}= \sum_{\nu_1,\nu_2,l,\nu_3,k} C^{(n)}_{\nu_1,\nu_2,l,\nu_3,k}|J,k,M\rangle |\nu_1,\nu_2,l,\nu_3 \rangle
\label{eq:reswf}
\end{equation}
The quantum numbers $\nu_1,\nu_2,l,\nu_3$ are good symmetry labels only for the harmonic oscillator Hamiltonian.  Similarly, the $k$ quantum number is a good quantum number for the symmetric-top rigid rotor Hamiltonian, that is also for molecules at linear geometries.  Any small deviations from these model systems cause the $\nu_1,\nu_2,l,\nu_3$ and $k$ quantum numbers to be only \textit{near quantum numbers} \cite{06BuJexx.method}. For large deviations from the harmonic oscillator model the vibrational labelling given in eq. \ref{eq:reswf} harmonic oscillator starts loosing its sense. Similarly, strong rotation-vibration interactions can cause $k$ to be no longer even a near-good quantum number. Such situations occur predominantly when two zero-order states are involved in a resonance interaction. We call basis functions $\Phi_{rv}^0$ given in eq. \ref{eq:zeroorder} zero-order states. Let us analyse resonance interactions between ro-vibrational zero-order states on a simple example.
 
\begin{figure}[H]
\begin{center}
 \includegraphics[width=10cm]{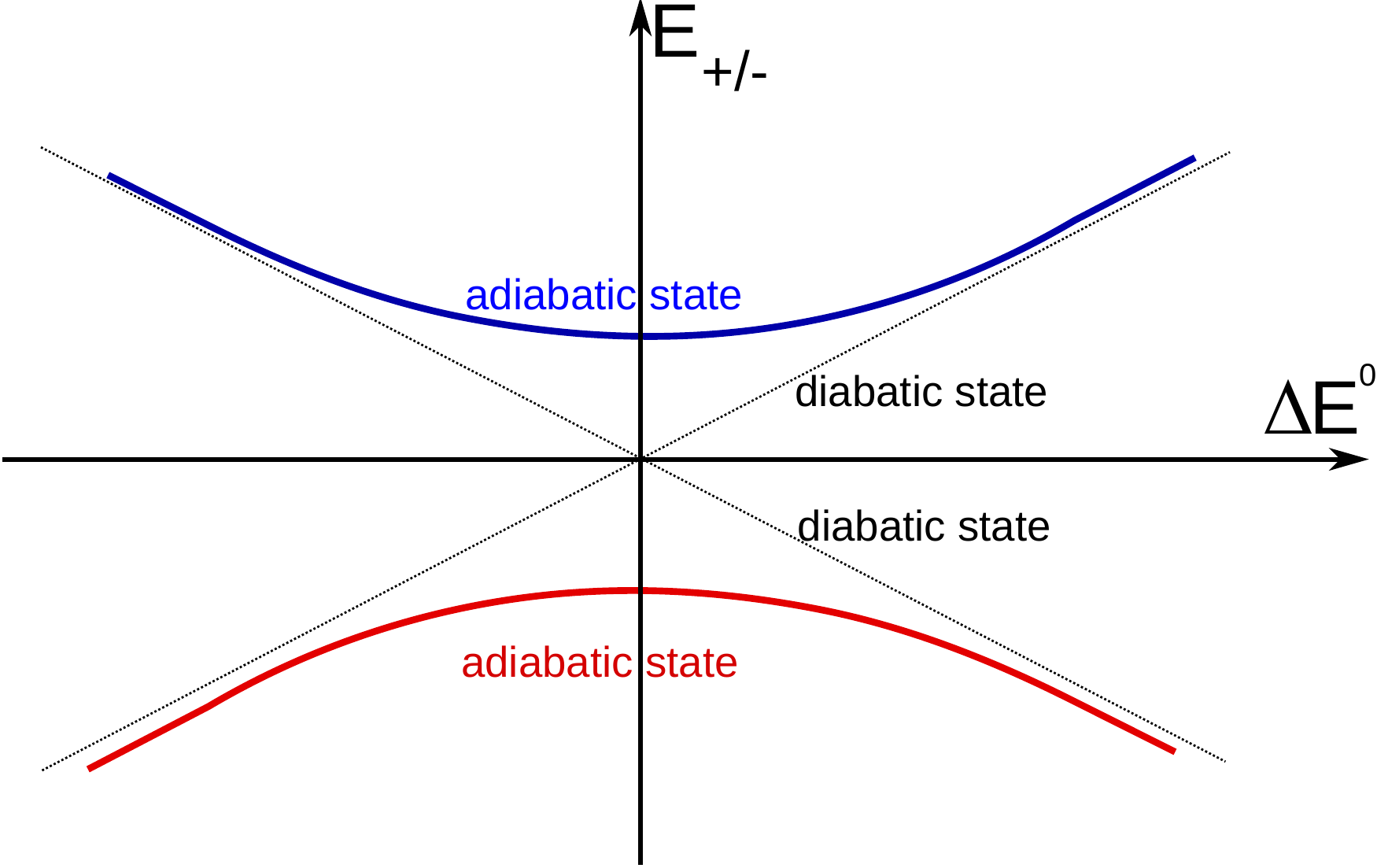}
 \caption{Schematic illustration of an avoided crossing of two states with the same symmetry. The dashed black lines represent energies of bare (diabatic) non-interacting states, whereas the blue and red thick curves are adiabatic states. In the $x$-axis given is the energetic separation of the diabatic states $\Delta E^0$.    }
 \label{fig:avoidedcrossing}
\end{center}
\end{figure}

Consider a generic two-level system  with zero-order basis states having energies $E_S^0$ and $E_W^0$ \textit{perturbed by an interaction} $C$. Off-diagonal elements are sometimes called couplings or resonance interaction terms, whereas diagonal elements are called unperturbed or diabatic energies. The Hamiltonian matrix for the two-level system is written as:
\begin{equation}
 \begin{pmatrix} E_S^0 & C \\
                 C & E_W^0\\
 \end{pmatrix}
 \label{eq:adiabaticresonance}
\end{equation}
The zeroth-order unperturbed basis is called the \textit{diabatic basis} \cite{Baer2002}.
Diagonalisation of this matrix gives \textit{adiabatic} energy levels, which have an avoided crossing, as depicted in Figure \ref{fig:avoidedcrossing}. Adiabatic energies are eigenvalues of the matrix given in eq. \ref{eq:adiabaticresonance}:

\begin{equation}
E_{\pm}=\frac{1}{2}\left( E_S^0 + E_W^0\right) \pm  \frac{1}{2}\sqrt{4|C|^2+\Delta E^0}
\label{eq:adiabaticeigen}
\end{equation}
where $\Delta E^0=E^0_W-E^0_S$ is the separation of \textit{diabatic} energy levels. The new adiabatic states are mixtures of diabatic states:
\begin{equation}
 \begin{pmatrix}|\tilde{W}\rangle \\
                |\tilde{S}\rangle \\
 \end{pmatrix} =\begin{pmatrix} c_1 & -c_2 \\
                c_2 & c_1\\
 \end{pmatrix} \begin{pmatrix}|W\rangle \\
                |S\rangle \\
 \end{pmatrix} 
\label{eq:adiabaticstates}
\end{equation}
where mixing coefficients are given as: $c_1=\left(\frac{\sqrt{4|C|^2+\Delta E^0}+\Delta E^0}{2\sqrt{4|C|^2+\Delta E^0}}\right)^{\frac{1}{2}}$ and $c_2=\left(\frac{\sqrt{4|C|^2+\Delta E^0}-\Delta E^0}{2\sqrt{4|C|^2+\Delta E^0}}\right)^{\frac{1}{2}}$. In the limit of no interaction ($C=0$) the two diabatic states remain unperturbed. In the limit of the exact resonance ($\Delta E^0=0$) we get 50\% / 50\% mixtures (in-phase and anti-phase) of the two diabatic states.

Thus adiabatic levels exhibit an avoided crossing at the point where the diabatic levels intersect. Near the resonance point, that is near the crossing of the diabatic energy levels, the majority of the standard variational and perturbative formulas fail to accurately reproduce empirical results. 
In the following paragraphs, we are going to analyse the resonantly interacting energy levels and discuss what consequences such interactions carry for the transition intensity measurements. 

Knowing that the interaction operator $\hat{C}$ can mix different ro-vibrational states, which manifests itself in a non-zero off diagonal matrix element $C$ in eq. \ref{eq:adiabaticresonance}, the next question to ask is: what is the necessary condition for a resonance interaction in terms of  symmetry of states? 

The interaction operator, as a form of energy, must have the symmetry of the ro-vibrational Hamiltonian, which is totally symmetric under operations of the molecular symmetry group. For this reason, the vanishing integral rule introduced in section \ref{sec:co2mu} allows  only for rotation-vibration interaction of diabatic states with the same ro-vibrational symmetry: $\Gamma^1_{rv}=\Gamma^2_{rv}$, where $\Gamma^i_{rv}=\Gamma^i_{v}\otimes \Gamma^i_{r}$. In other words, a coupled ro-vibrational wavefunction, that is the one which is not a single product of a rotational and a vibrational state must have identical symmetries of its component terms. This is because the whole ro-vibrational wavefunction must transform according to one of the irreducible representations of the molecular symmetry group and cannot be a sum of terms with different symmetry (different irreducible representations). 

For further discussion, the harmonic oscillator basis is chosen for vibrations and the symmetric-top basis for rotations. We are going to classify resonance interactions, by the type of the perturbing operator $\hat{C}$.

\section{Types of resonance interactions}

There are two basic types of perturbations of ro-vibrational energy levels caused by the interaction with other ro-vibrational energy levels: \textit{Fermi-type interactions} and \textit{Coriolis-type interactions}. The former type of the resonance interaction occurs when two vibrational levels have the same symmetry, $\Gamma^1_{v}=\Gamma^2_{v}$, meaning that this is a purely vibrational anharmonic effect. Fermi-resonance leads to shifting of ro-vibrational energy levels of the whole vibrational band involved in the interaction. The Coriolis-type interaction on the other hand is $J$-specific and depends both on the symmetry of the vibrational and the rotational part of the wavefunction. As we will show further on, the Coriolis-type resonance is possible between ro-vibrational states with differ in the $k$ quantum number by $\Delta k =\pm 1,\pm 2$, which for a linear triatomic molecule means also  $\Delta l =\pm 1,\pm 2$ ($l$ is the vibrational angular momentum quantum number). The  $\Delta l = \pm 2$ interaction is often called the \textit{l-type resonance} or \textit{l-doubling resonance}. 

For future discussion the concept of the \textit{polyad number} \cite{06BuJexx.method} must be introduced. On the example of CO$_2$, the polyad number is defined as $P=2\nu_1+\nu_2+3\nu_3$, where
$\nu_1, \nu_2, \nu_3$ are the vibrational quantum numbers of the symmetric
stretching, bending and the asymmetric stretching, respectively. The weights standing by the vibrational quantum numbers in the polyad number formula are associated with the relative energetics of vibrational fundamental frequencies: $\tilde{\nu}_3\approx 3\tilde{\nu}_2 \approx 2\tilde{\nu}_1$. The majority of resonance interactions occur between states with the same polyad number, which is referred to as \textit{intra-polyad resonances}. However sometimes an \textit{inter-polyad anharmonic interaction} can occur, which will be distinguished from the intra-polyad interactions.

\subsection{Example: Fermi-type resonance in CO$_2$}
\label{sec:resfermi}

Anharmonic terms in the PES, that is terms such as $Q_iQ_j, Q_j^3, Q_i^2Q_j^2$, etc. but not $Q_j^2$ (which are \textit{harmonic terms}) cause Fermi-interaction of vibrational levels of the same symmetry, as shown in eq. \ref{eq:adiabaticstates}.  An example of a Fermi-resonance interaction is mixing between the $\nu_1\nu_2l\nu_3n$ =  10001 and $\nu_1\nu_2l\nu_3n$ = 02001 energy levels in CO$_2$. Both states have identical $\Sigma_g^+$ symmetry in the $D_{\infty h}$ group and the perturbing operators are the cubic $Q_j^3$ and quartic terms $Q_j^4$ in the PES.  Here $n$ is defined as an index labeling vibrational states which are subject to Fermi-mixing with some other vibrational state. A fundamental law of quantum mechanics, the so called \textit{no cloning} theorem, ensures that whenever two zero-order energy levels interact, two new (perturbed) levels must appear in their place. For this reason if two states are involved in the Fermi-interaction, then the $n$ number takes values $n=1,2$ and enumerates \textit{Fermi components}. 

Interactions of energy levels carry consequences for transition intensities.
If one of the energy levels involved in a weak transition in the harmonic approximation is in Fermi resonance with an energy level which participates in a strong transition, then an intensity borrowing occurs, which is due to mixing of the \textit{dark state} with the \textit{bright state}. This borrowing can significantly increase the transition intensity of the weak (dark) vibrational band, regardless of $J$ value. As a result, intensity of the whole band is shifted by a constant number.

\subsection{Example: Coriolis-type resonance in CO$_2$}
\label{sec:rescoriol}
Coriolis-type interactions are usually associated with operators, which are products of vibrational coordinates, vibrational momentum and angular momentum operators, e.g. $\zeta Q_ip_iJ_j$. Here $\zeta$ is the scalar Coriolis-constant \cite{Watson1968,06BuJexx.method}.
To show this for a general triatomic molecule let us come back to eq. \ref{eq:Kvr} from Chapter 2:
\begin{equation}
\begin{split}
\hat{K}_{VR}(r_1,r_2,\gamma,\phi,\theta,\chi)=\frac{1}{2}\left[M_{xx}\hat{J}_x^2+M_{yy}\hat{J}_y^2+M_{zz}\hat{J}_z^2+M_{xz}\left(\hat{J}_x\hat{J}_z+\hat{J}_z\hat{J}_x\right)\right]+\\
+\frac{1}{i}\left[\left(\frac{1-a}{\mu_1r_1^2}-\frac{a}{\mu_2r_2^2}\right)\left(\frac{\partial}{\partial\gamma}+\frac{\cot}{2}\right)+\frac{2a-1}{\mu_{12}r_1r_2}\left(\cos\gamma\frac{\partial}{\partial\gamma}+\frac{1}{2\sin\gamma}\right)\right.+\\
\left.+\frac{\sin\gamma}{\mu_{12}}\left(\frac{a}{r_2}\frac{\partial}{\partial r_1}-\frac{(1-a)}{r_1}\frac{\partial}{\partial r_2} \right)\right]\hat{J}_y
\end{split}
\label{eq:Kvr2}
\end{equation}

In eq. \ref{eq:Kvr2} giving the rotational-vibrational interaction terms in the kinetic energy operator of the nuclei in the Born-Oppenheimer approximation we can identify terms of general form $\zeta f(Q_i)p_jJ_k$ and $\zeta f(Q_i)J_j$, where $p_j$ stands for derivative operator with respect to some internal coordinate ($Q_j = r_1,r_2$ or $\gamma$). Such terms generate matrix elements which are off-diagonal in the $k$ quantum number, as shown in eq. \ref{eq:Kvibvr}. Only states with $\Delta k =\pm 1,\pm 2$ are mixed (they are connected by off-diagonal matrix elements of the Hamiltonian):

\begin{equation}
\hat{K}_{VR}(r_1,r_2,\gamma)=\delta_{k'k\pm 2}\frac{1}{4}C^{\pm}_{Jk\pm 1}C^{\pm}_{Jk}b_-+\delta_{k'k\pm 1}\frac{1}{2}C^{\pm}_{Jk}\lambda^{\pm}+\delta_{k'k}\frac{1}{2}\left(b_+(J(J+1)-k^2)+b_0k^2\right)
\label{eq:Kvibvr2}
\end{equation}

In operators of type $Q_ip_jJ_k$, $Q_i$ and $p_i$ have the same symmetry (derivative $d/dx$ carries the same symmetry as $x$). As a result the product $Q_ip_j$ transforms as the totally symmetric representation $A_1$ (as any product of identical irreducible representations). The vanishing rule for integrals (matrix elements of $Q_ip_jJ_k$) states that the product of irreducible representations of two interacting states must contain the irreducible representation of the $j$-th component of the total angular momentum operator $\vec{J}$. Only then the integrated function transforms as the totally symmetric representation $A_1$ in the molecular symmetry group and the integral can be non-vanishing.

\begin{figure}[H]
\begin{center}
 \includegraphics[width=10cm]{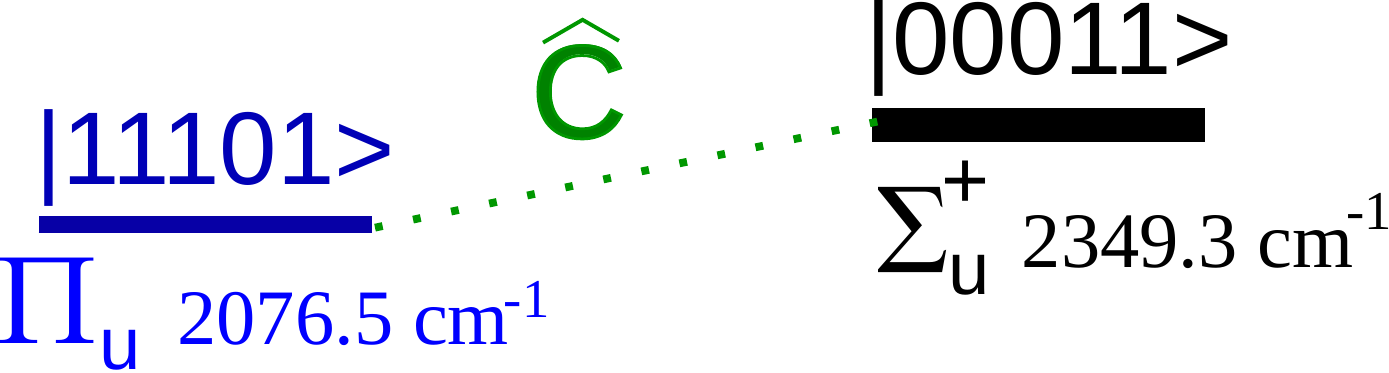}
 \caption{Schematic illustration of a resonance interaction between two energy levels in CO$_2$ through operator the Coriolis operator $\hat{C}$. $\Pi_u$ is the vibrational symmetry of the $|11101\rangle$ state and  $\Sigma_u^+$ is the vibrational symmetry of the $|00011\rangle$ state. Energies of respective states are given in wavenumbers.  }
 \label{fig:co2borrowing}
 \end{center}
\end{figure}

As an example of the Coriolis-type resonance let us consider two energy levels in CO$_2$: 11101 ($\nu_1=1, \nu_2=1, l=1, \nu_3=0$) and 00011 ($\nu_1=0, \nu_2=0, l=0, \nu_3=1$). Figure \ref{fig:co2borrowing} displays these energy levels with their vibrational symmetries and energies. It is clear that the 11101 and 00011 energy levels have different vibrational symmetries, thus cannot interact by a pure vibrational-anharmonic mechanism (Fermi resonance). However, when we list possible symmetries of ro-vibrational states generated from these vibrational states, there are combinations of $J$ and $k$ quantum numbers for which ro-vibrational energy levels in the  11101 and 00011 manifolds have the same symmetry. For example, when the symmetry of the 11101 energy level ($\Pi_u$) is combined with the symmetry of $k$-odd rotational wavefunction $\Pi_g$ it produces the following sum of irreducible representations:  $\Pi_u\times\Pi_g=\Sigma_g^+\oplus \Sigma_g^-\oplus \Delta_g$. The symbols of irreducible representations used here correspond to the $D_{\infty h}(M)$ molecular symmetry group.  At this point it is convenient and sufficient to move into the finite $C_{2v}(M)$ group with the classification of ro-vibrational states. In this group the $D_{\infty h}$'s $\Pi_u$ state correlates with $A_1\oplus B_1$ symmetry and the $\Sigma_u^+$ correlates with the $B_2$ symmetry. 

One may ask a question: why did we move into the $C_{2v}(M)$ group? First of all, it is subjectively more convenient to operate with a simple four element group, which contains no degenerate representations. Secondly, selection rules for the interaction matrix elements derived in the less general $C_{2v}(M)$ group will also hold in the more general $C_s$ group. The implication in the opposite direction is not always true. 

The symmetry of rotational wavefunctions $|J,k\rangle$ in the $C_{2v}(M)$ group is $A_1\oplus B_1$ for even values of $k$  and $A_2\oplus B_2$ for odd values of $k$. Only states with the same $J$ quantum number can be mixed by Coriolis-resonance interactions. This is because $J$ is a good quantum number labelling the Hamiltonian matrix.  By combining the rotational and vibrational symmetries we obtain for the 11101 vibrational state two possible ro-vibrational symmetries: $A_1\oplus B_1$ for even values of $k$ and  $A_2\oplus B_2$ for odd values of $k$. The 00011 vibrational state also generates two possible ro-vibrational symmetries: $A_2\oplus B_2$ for even values of $k$ and  $A_1\oplus B_1$ for odd values of $k$. We can therefore see, that only states with opposite parity of the $k$ quantum number can be connected via a Coriolis interaction. Indeed, shown in  Figure \ref{fig:symrho3} in section \ref{sec:sensitivity} the intensity of the R branch of the 11101 -- 00001 band is affected by $J$-localized resonance interaction between the 11101 and the 00011 state. Although the whole  11101 -- 00001  band is subject to Coriolis interaction with the 00011 state, only near the particular value of $J$ this interaction grows rapidly to give a noticeable alternation in transition intensity pattern. This deviation of the pattern in rotational transition intensities means that the typical the intensity pattern which is good for the rigid-rotor model is no longer observed. This is what we call deviation from the 'natural pattern'.

\section{Ro-vibrational transition intensity borrowing}
\label{sec:borrowing}
In this section we provide a theoretical description of the phenomenon of transition intensity borrowing caused by resonance interactions of ro-vibrational energy levels. In the case of transitions in which one or both energy levels are involved in a form of a resonance interaction, the zero-order states can mix significantly, which leads to strong dependence of the transition intensity on the admixture of the perturbing state. We will derive a measure of 'reliability' of the calculated transition intensities. Because the accuracy of variational wavefunctions and energy levels is determined by the accuracy of the underlying potential energy surface (PES), the term 'reliability' means 'the accuracy of the PES used'. In this sense, we are aiming at finding a measure of how sensitive are the variational matrix elements (e.g. the transition dipole moment) to small changes in the PES. As we will see further on, this measure formally depends on the energetic separation of states involved in the interaction $\Delta E_{ij}=E_i-E_j$ (not the states between which the transition occurs) and on the deviation of the PES from a reference PES: $\Delta V(r_1,r_2,\gamma)= V(r_1,r_2,\gamma)-V^0(r_1,r_2,\gamma) $. With these assumptions, the term 'reliability' refers to the accuracy of the most accurate PES considered, rather then to the exact (experimental) energy levels. In other words, we assume the most accurate PES (reference PES $V^0$) as the exact model for reality. 
Of course this does not mean that we are here providing a method for estimating the uncertainty for matrix elements of operators, which can be later compared to experimental values. Rather we aim to give a procedure for stating how sensitive these matrix elements are when a different PES is used from the reference PES. The reference PES may generate energy levels with some systematic shift with respect to experiment, and so the sensitivity measure can not be directly related to the absolute error of the calculated matrix elements. Nevertheless, such multiple-PES-based procedure can deliver information about how strong are the resonance interactions of energy levels, provided that we have control over the deviation between the reference PES and the working PES. Having such a measure of the strength of resonance interaction, it can be for example possible to estimate which variationally calculated transition intensities can be trusted and which of them are very sensitive to the quality of the PES (accuracy of the wavefunction).  

\subsection{Matrix elements between interacting states}
It has been already shown in section \ref{sec:resonances}, that the resonance between energy levels is possible between states with the same ro-vibrational symmetry. Coriolis-interaction operators are proportional to the components of the total angular momentum operators $\hat{J}_x,\hat{J}_y,\hat{J}_z$, which in turn can be expressed by the  \textit{molecule-fixed angular momentum ladder operators}  $\hat{J}_{m}^{\pm}:=\hat{J}_{x}\pm i\hat{J}_{y}$. As shown by eq. \ref{eq:Kvr2} the Coriolis operator has the symmetry of the total ro-vibrational Hamiltonian. 

A Coriolis operator can be a product of the linear momentum and position operators, which mix different vibrational zero-order states (see eq. \ref{eq:vibtransharm}), but it can also contain squares of the total angular momentum operator, which mix of states with $\Delta k = \pm 1,\pm 2$ by means of the equation:

\begin{equation}
\hat{J}_{m}^{\pm}|J,k\rangle=\hslash \sqrt{J(J+1)-k(k\mp 1)}|J,k\mp 1\rangle
\end{equation} 
where $k$ is the projection of the total angular momentum on the $z$-axis of the molecule-fixed frame.

The other necessary condition for a resonance is non-symmetry based, but related to the energetics of the two interacting energy levels. The strength of the anharmonic and ro-vibrational interactions strongly depend on the energetic separation of the zero-order energy levels. Quantitatively, the non-vanishing by symmetry interaction matrix element $C$ from eq. (\ref{eq:adiabaticresonance}) can be rewritten in the following form:

\begin{equation}
C_{ij}=\langle \Phi_i^0 | \hat{C} | \Phi_j^0 \rangle = \frac{\langle \Phi_i^0 | [\hat{C} ,\hat{H}^0_{rv}] | \Phi_j^0 \rangle}{E_i^0-E_j^0}
\label{eq:resonance}
\end{equation}
which reveals singularity at $E_i=E_j$. Here the zero-order states $\Phi_i^0$ are eigenstates of the Coriolis-decoupled ro-vibrational Hamiltonian: $ H^0_{rv}\Phi_i^0=E_i^0\Phi_i^0$. Matrix elements of the commutator $ [\hat{C} ,\hat{H}^0_{rv}]$ are finite and non-zero, thus the coupling matrix element approaches infinity when the energy levels become degenerate. In reality, ideal accidental degeneracies of two zero-order energy levels are never observed, hence the interaction matrix elements are always finite, although they can take very large values and be very sensitive to small changes in the accuracy of the wavefunction. Because the denominator in eq. \ref{eq:resonance} is very small, any small perturbation $\delta \Phi_i^0$ to the wavefunction can cause a significant change to the value of the interaction term $C_{ij}$. As a consequence, variational wavefunctions need to be calculated with a very high accuracy, to properly reproduce the values of the interaction matrix elements near resonances. This is obviously one of the drawbacks of the variational methodology.

\begin{figure}[H]
\begin{center}
 \includegraphics[width=10cm]{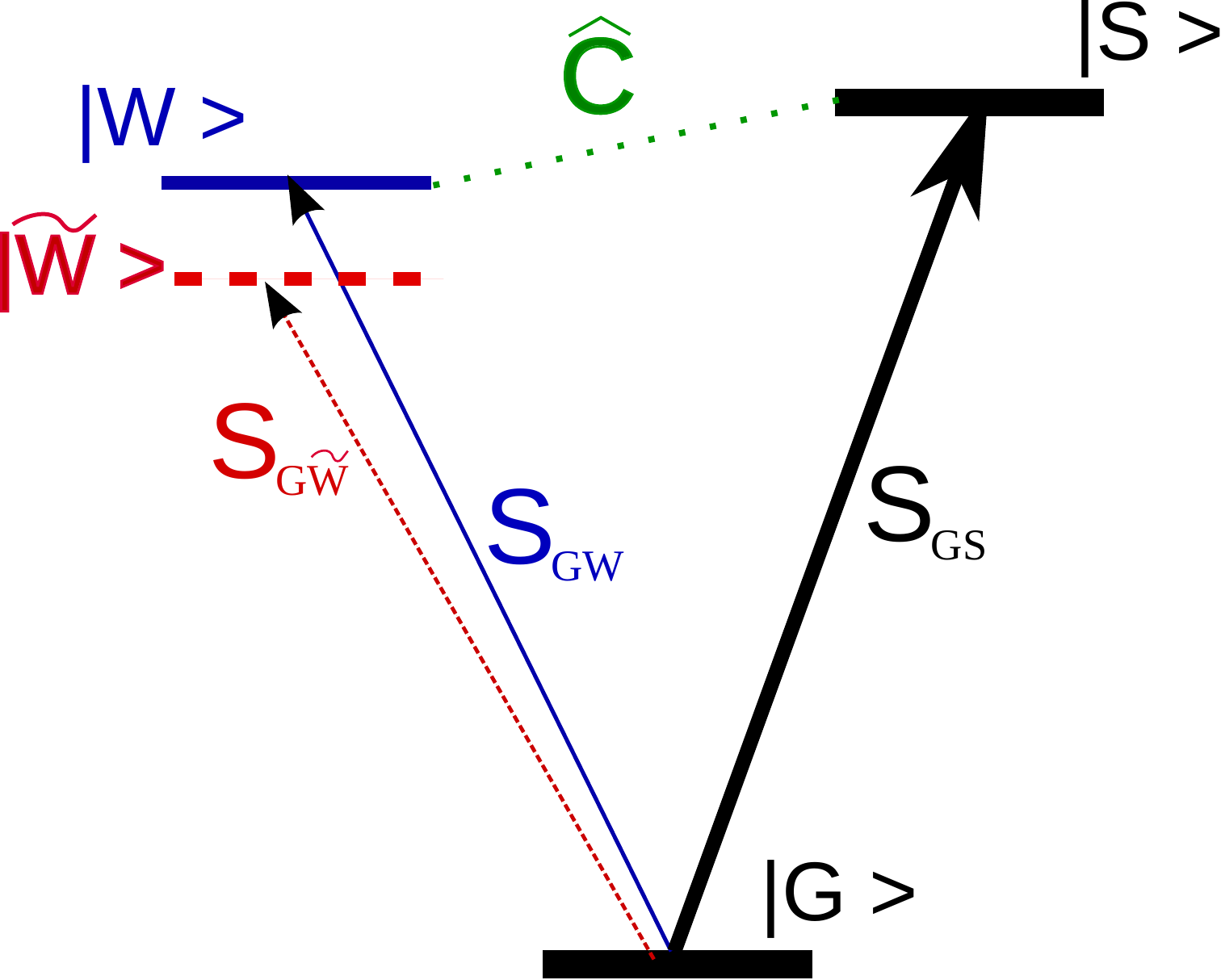}
 \caption{Schematic illustration of intensity borrowing caused by a resonance interaction between two energy levels through operator $\hat{C}$. The  $S\leftarrow G$ transition is from the ro-vibrational ground state $|G\rangle$ and the associated line strength is $S_{GS}$. $W\leftarrow G$ transition is from the ro-vibrational ground state $|G\rangle$ and the associated line strength is $S_{GW}$. The interaction between $|W\rangle$ and $|S\rangle$ energy levels leads to intensity borrowing of the $W\leftarrow G$ transition, which results in an altered transition line strength $S_{G\tilde{W}}$.  }
 \label{fig:intensityborrowing}
 \end{center}
\end{figure}

Now, the question is: near a resonance, when $E_i^0-E_j^0$ is small, how does a small change in the wavefunction $\delta \Phi_i^0$ affect the matrix element of the interaction operator $C_{ij}$? In answering this question, we shall first follow an informal and somewhat more intuitive picture and then compare this result to a more formal approach of perturbation theory.

\paragraph{Infinitesimal perturbation approach. }
Assume that the full ro-vibrational Hamiltonian given by $\hat{H}^0_{rv}=\hat{T}+\hat{C}+\hat{V}^0$, with the reference PES $\hat{V}^0$, generates a set of \textit{reference }eigenfunctions $\{|\psi^0_i\rangle\}_{i=1,2,...}$. In this eigenbasis the Coriolis operator $\hat{C}$ is diagonal, because $\hat{H}^0_{rv}|\psi^0_i\rangle=E^0_i|\psi^0_i\rangle$ and $\hat{C}$ is part of $\hat{H}^0_{rv}$. The full ro-vibrational Hamiltonian  $\hat{H}_{rv}=\hat{T}+\hat{C}+\hat{V}$, with the second test-PES $\hat{V}$, generates a set of slightly different \textit{test} eigenfunctions $\{|\psi_i\rangle\}_{i=1,2,...}$, which correspond to slightly different eigenvalues:  $\hat{H}_{rv}|\psi_i\rangle=E_i|\psi_i\rangle$. In both cases the kinetic energy operator is the same. Nevertheless, a change in the PES ($\Delta \hat{V} = \hat{V}^0-\hat{V}$) will result in a change to the wavefunction, so that $\hat{H}^0_{rv}|\psi_i\rangle\neq E_i^0|\psi_i\rangle$.
The variational wavefunction is expanded as a linear combination of basis functions:
\begin{equation}
|\psi^0_i\rangle=\sum_{j}d_{ij}^0|\Phi^0_j\rangle
\end{equation}
Basis sets are identical in both cases (reference and test); the only difference is in the variational expansion coefficients. 
\begin{equation}
|\psi_i\rangle=\sum_{j}d_{ij}|\Phi^0_j\rangle
\end{equation}
Matrix elements of the Coriolis operator $\hat{C}$ in the reference eigenbasis can be written as:
\begin{equation}
\langle \psi_i^0 | \hat{C} | \psi_j^0 \rangle=C^0_{ij}\delta_{ij}
\end{equation}
This matrix is diagonal and the expectation values of the Coriolis operator in this basis are considered as known reference values based on the highest quality PES $\hat{V}^0$. Now assume analogous matrix elements in the test eigenbasis:
\begin{equation}
\langle \psi_i| \hat{C} | \psi_j \rangle=C_{ij}\delta_{ij}
\label{eq:coriolis1}
\end{equation}
This matrix is also diagonal, but the expectation values are different with respect to the reference case. After expanding matrix elements from eq. \ref{eq:coriolis1} in the variational basis we get:

\begin{equation}
C_{ij}= \sum_{k,l}d^*_{ik}d_{jl}\langle \Phi^0_k| \hat{C} | \Phi^0_l \rangle
\label{eq:coriolis2}
\end{equation}
Note that the linear combination  coefficients are different for the test and reference wavefunctions, because different PESs were used to calculate these wavefunctions. Now, the change in the values of the matrix elements of the Coriolis operator in the test eigenbasis $C_{ij}$ with respect to the reference values $C^0_{ij}$  comes mainly from non-vanishing elements $\langle \Phi^0_k| \hat{C} | \Phi^0_l \rangle$, for which the zero-order basis states are energetically close. All other terms in the sum in eq. \ref{eq:coriolis2} will contribute negligibly if the PES distortion is small. For this reason, we may focus on a particular matrix element which significantly contributes to the change in value of $C^0_{ij}$ upon switching to the test PES. Because the new wavefunctions $|\psi_i\rangle$ (test eigenfunctions) have slightly different expansion coefficients $d_{ij}$ than in the reference eigenfucntions $|\psi^0_i\rangle$, one can think of it as if the basis for the test set of functions has been slightly changed (we incorporate the part of the value of the new expansion coefficient, $d_{ij}=d_{ij}^0+\delta d_{ij}$, into the new basis function). Thus we may write: $| \Phi_j \rangle:=| \Phi^0_j \rangle+ |\delta \Phi_j \rangle$. Working with variations in expansion coefficients or variations of the basis functions is equivalent, but for more compact notation it was decided to vary the basis function. 
As a consequence, in the test basis set, the Coriolis interaction matrix element, for which resonance interaction occurs is given as

\begin{equation}
\langle \Phi_k | \hat{C} | \Phi_l \rangle = \langle \Phi_k^0 | \hat{C} | \Phi_l^0 \rangle + \langle \delta\Phi_k| \hat{C} | \Phi_l^0 \rangle + \langle \Phi_k^0 | \hat{C} |  \delta\Phi_l \rangle + \langle \delta\Phi_k | \hat{C} |  \delta\Phi_l \rangle  
\label{eq:resonance2}
\end{equation}

The last term is of order of $(\delta \Phi_k^0)^2$ and can be neglected. The perturbed interaction matrix elements are represented as a sum of the original interaction matrix elements $\langle \Phi_k^0 | \hat{C} | \Phi_l^0 \rangle$ plus correction terms of type $ \langle\delta\Phi_i | \hat{C} | \Phi_j^0 \rangle$. The latter can be further written as follows:

\begin{equation}
\langle \delta\Phi_i^0 | \hat{C} | \Phi_j^0 \rangle = \frac{\langle \delta\Phi_i^0 | [\hat{C},\hat{H}^0_{rv}] | \Phi_j^0 \rangle}{E_i^0\prime+(E_i^0-E_j^0)}
\label{eq:resonance3}
\end{equation}
where $E_i^{0 \prime}=E_i^0+\delta E_i$ and $E_i^0-E_j^0:= \Delta E^0$. This correction term contributes to the change in value of the matrix elements of the Coriolis interaction operators. With a constant, known distortion in the PES $\Delta \hat{V}$, the change in the value of the energy level is assumed to be known and small: $\delta E_i$. Then one may write

\begin{equation}
\delta E_i\langle \delta\Phi_i^0 | \hat{C} | \Phi_j^0 \rangle = \frac{\langle \delta\Phi_i^0 | [\hat{C},\hat{H}^0_{rv}] | \Phi_j^0 \rangle}{1+\frac{\Delta E^0}{\delta E_i}}
\label{eq:resonance4}
\end{equation}
As soon as $\delta E_i$ is non-zero, the magnitude of the correction to the interaction matrix element depends on $\Delta E^0$ as $\frac{const}{1+\frac{\Delta E^0}{\delta E_i}}$, hence for $\Delta E^0 = 0$ takes the maximal value. 

To summarize, we have shown above that separation of zero-order non-interacting energy levels  $\Delta E^0$ determines the magnitude of the resonance-induced variation in the Coriolis interaction matrix elements. 

The present discussion of resonance interactions is primarily dedicated to application in estimating the reliability of calculated transition intensities. Henceforth, the key question to find answer to is: \textit{how do resonance interactions of energy levels influence intensities of ro-vibrational transitions?}

We are going to assume a situation, when the upper energy level involved in a transition accidentally near-crosses another energy level and the ro-vibrational symmetries of these two levels are identical. Such situation is schematically depicted in Figure \ref{fig:intensityborrowing}, where the weak transition from state $|G\rangle$ to $|W\rangle$ is accompanied by the strong transition from state $|G\rangle$ to $|S\rangle$. The energy levels associated with states $|W\rangle$ and $|S\rangle$ are nearly degenerate. Also, ro-vibrational symmetries of both states are identical (they have appropriate vibrational symmetries, equal $J$ quantum numbers and the approximate quantum number $k$ differing by 1 or 2). We are interested in calculating the transition line strength:
\begin{equation}
T_{GW}=|S_{GW}|^2=|\langle G | \hat{\mu}(Q)|W\rangle|^2
\label{eq:LS}
\end{equation}
where G and W labels ro-vibrational states between which the transition occurs. $\hat{\mu}(Q) $ is the transition dipole moment surface (TDMS), which depends on nuclear coordinates denoted as $Q$. 

\paragraph{Perturbation theory approach. }
Until now we have been operating on 'small' distortions of wavefunctions and surfaces in the configuration space. A more formal way to approach the problem of sensitivity of transition intensities is by the use of perturbation theory. 
In the same way as the \textit{Herzberg-Teller effect}  \cite{06BuJexx.method} is responsible for intensity borrowing in vibronic transitions, the present formulation explains the intensity borrowing in terms of resonance interactions of ro-vibrational energy levels. 
Let us employ the interaction scheme from Figure \ref{fig:intensityborrowing}. Assume that the $|W\rangle = |\phi^W_{vib}\rangle|\phi^W_{rot}\rangle$ state is perturbed by the $|S\rangle= |\phi^S_{vib}\rangle|\phi^S_{rot}\rangle$ state via an interaction described by operator $\hat{C}$.  In the first-order perturbation theory the  $|W\rangle$ state is given by:

\begin{equation}
|W^{(1)}\rangle=|\phi^{W,(1)}_{vib}\rangle|\phi^{W,(1)}_{rot}\rangle =|\phi^W_{vib}\rangle|\phi^W_{rot}\rangle + \sum_I \frac{ \langle\phi^W_{vib}|\langle\phi^W_{rot} | \hat{C} |\phi^I_{rot}\rangle|\phi^I_{vib} \rangle}{E_I^0-E_W^0}|\phi^I_{vib}\rangle|\phi^I_{rot}\rangle
\label{eq:borrowing}
\end{equation}
If the symmetry of the $|S\rangle$ state is appropriate, that is $\Gamma^W_{vib}\otimes \Gamma^W_{rot}=\Gamma^S_{vib}\otimes \Gamma^S_{rot}$, then the matrix element $\langle\phi^W_{vib}|\langle\phi^W_{rot} | \hat{C} |\phi^S_{rot}\rangle|\phi^S_{vib}\rangle $ is non-vanishing. Contributions to the perturbed wavefunction from other energy levels can be small if $|S\rangle$  and  $|W\rangle$ states are well isolated from other states (as in the case of 00011 and 11101 states in CO$_2$, see section \ref{sec:rescoriol}). Then the sum in eq. \ref{eq:borrowing} has only two terms.  The perturbed wavefunction can be inserted in the expression for the transition dipole moment:

\begin{equation}
\langle G| \hat{\mu}(Q) |W^{(1)}\rangle
\label{eq:borrowingtransition}
\end{equation}
where $|G\rangle$ represents the lower energy level (the ground state) in Figure \ref{fig:intensityborrowing}. Expanding the \textit{bra-}state in the above equation according to eq. \ref{eq:borrowing} gives:

\begin{equation}
\langle G| \hat{\mu}(Q)|W^{(1)}\rangle= \langle G| \hat{\mu}(Q) |W\rangle + \frac{ \langle W | \hat{C} |S \rangle}{E_W^0-E_S^0}\langle G| \hat{\mu}(Q) |S \rangle = S_{GW}+ \lambda_{SW}S_{GS}
\label{eq:borrowingtransition1}
\end{equation}
The magnitude of the admixture of the $|S\rangle$ state to the $|W\rangle$ state resulting in intensity borrowing by the $|W\rangle$ state depends on the value of the $\lambda_{GW}= \frac{ \langle W | \hat{C} |S \rangle}{E_W^0-E_S^0}$ parameter, which in turn depends on the energetic separation of the two energy levels. Conversely, the intensity of the $S\leftarrow G$ transition is not significantly altered by the interaction with the $|W\rangle$ state because the intensity of the  $W\leftarrow G$ transition is low (low value of $S_{GW}$). 

\subsection{Quantitative measures for the strength of resonance interactions of energy levels}
\label{sec:theoryscatter}
The intensity borrowing effect quantified by eq. \ref{eq:borrowingtransition1} gives an estimate for the strength of resonance interaction between respective ro-vibrational energy levels. This knowledge, in turn, can be utilized in quantifying how reliable are the calculated transition intensities. Transition intensities insensitive to small changes in the PES are considered reliable, because no resonance interaction affects the energy levels involved in a transition. On the other hand, one can expect that when a resonance interaction occurs between energy levels, then by means of eq. \ref{eq:borrowingtransition1}, the transition intensity will be sensitive to even small change in the shape of the ro-vibrational wavefunction, and consequently to even small changes in the PES.

Below, we introduce a descriptor which can serve as a measure of reliability of calculated transition intensities.
Let PES1 be a reference PES for which the line strength for the $W\leftarrow G$
transition is given by $S_{GW}$. The second PES2 generates the ro-vibrational wavefunction $|\psi\rangle$, which is very similar to the wavefunction $|\psi^0\rangle$ calculated with the PES1 . This variation in the wavefunction causes the change in the line strength according to eq. \ref{eq:borrowingtransition1}. The ratio of intensities calculated with PES1 and PES2 and with identical DMS, which we will be calling \textit{scatter factor}, is given by
\begin{equation}
\rho_{GW} = \frac{|S_{GW}+ \lambda_{GW}S_{GS}|^2}{|S_{GW}|^2}\circeq\left|1+\lambda_{SW}\frac{S_{GS}}{S_{GW}}\right|^2
\label{eq:definitionscatter}
\end{equation}
where the last equality holds only when the expression in the bracket is real (a special case of \textit{Schwartz inequality}). 
Eq. \ref{eq:definitionscatter} provides a direct relation between the strength of the interaction of two energy levels $\lambda_{SW}$ and the sensitivity of the transition intensity to the PES change. It is assumed that the change in the PES is small and controlled, meaning both surfaces are of similar quality, so that $\Delta PES (Q) = PES1(Q)-PES2(Q)$ is small for all $Q$'s. The value of the scatter factor should rapidly grow for transitions involving energy levels affected by resonance interactions. Typically, the energetic condition for a resonance and the $\Delta J = 0 $ selection rule lead to J-localized resonances. Such resonances occur when two energy levels, with the same ro-vibrational symmetry and the same $J$ become energetically close. This observation cues into the idea of a method for detection of resonance interactions of energy levels with \textit{ab initio} calculations. 
If one calculates the ratio of intensities for the same transition, but calculated with slightly different PESs, it can be possible to elucidate information about the strength of interactions of energy levels involved in the transition with other energy level. 

The idea of using the scatter factor was originally introduced by Lodi and Tennyson \cite{jt522,jt509} for water to capture accidental resonances which were not fully characterized by the underlying PES.
Under these circumstances calculations with different procedures should give markedly different results. It should be stressed here that the procedure does not yield an uncertainty as such, it rather establishes which transition intensities are correctly 
characterized by the calculation and hence have an uncertainty reflecting the underlying TDMS, and which are not, in which case the
predictions were deemed as unreliable and alternative sources of intensity information is recommended.

In other words, trustworthy lines should be stable under minor PES/DMS
modifications.  One problem with this strategy is that if the
alternate PES (or DMS) differs too much from the reference (best) PES then large
intensity variations can be found which do not reflect problems with
the best calculation, but rather inaccuracy of the lower quality PES.

\section{Reliability analysis of calculated transition intensities}
\label{sec:sensitivity}
In calculations of transition intensities, the dominant source of uncertainty in the absence of resonance interactions is given by the {\it ab initio} DMS. As shown in section \ref{sec:resonances}, the nuclear motion wavefunctions give a secondary but, under certain circumstances, important contribution to the uncertainties.
Variational nuclear motion programs yield very highly converged wavefunctions
and in situations where the PES is precise the 
intensities show little sensitivity to the details of how they are calculated.

For example, ro-vibrational wavefunctions of triatomics, such as CO$_2$, calculated using Radau coordinates give intensities very similar (typically to within 0.1~\%) to those computed using Jacobi coordinates and different basis sets.

As discussed in section \ref{sec:borrowing}, where the wavefunctions do play an important role
is in capturing the interaction between different ro-vibrational states. Such resonance interactions can lead to intensity stealing
and, particularly for so-called \textit{dark states}, huge changes in transition intensities (cf. Figure \ref{fig:co2borrowing}).

Here we are going to show an example sensitivity analysis of transition intensities to the accuracy of the PES (ro-vibrational wavefunctions). With the use of the scatter factor introduced in eq.\ref{eq:definitionscatter}, we are going to show a method of detecting plausible resonances between ro-vibrational energy levels, through analysis of the sensitivity in transition intensities to small changes in the PES/DMS. 
   
In practice the the Lodi-Tennyson strategy \cite{jt522} for calculating the scatter factor given in eq. \ref{eq:definitionscatter} consists of doing ro-vibrational calculations with different PESs and DMSs. {A general scheme for calculating the scatter factor $\rho$ is shown in Figure \ref{fig:scatterscheme}, where transition intensities calculated with several sets of PES/DMS combinations are compared. 

\begin{figure}[H]
\begin{center}
 \includegraphics[width=10cm]{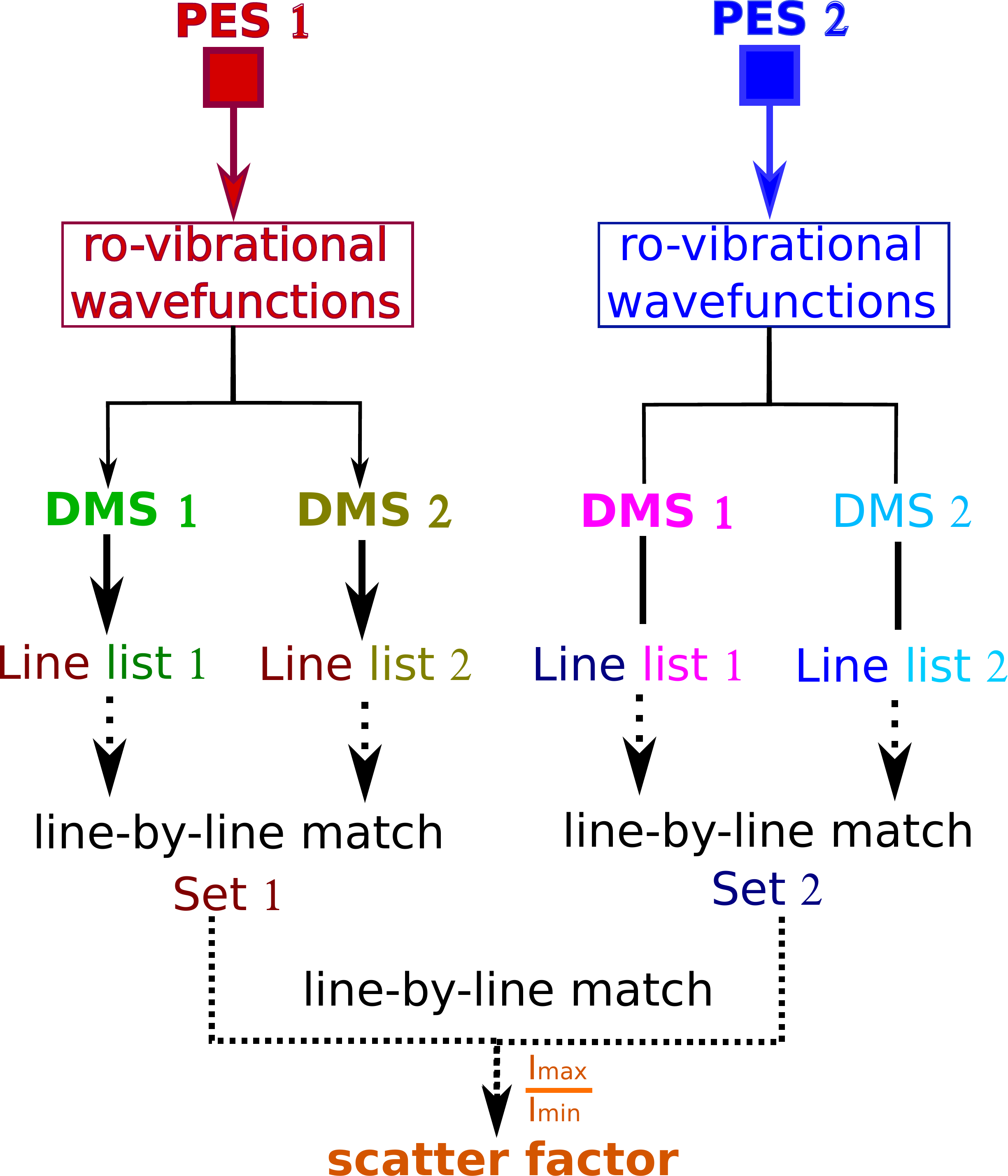}
 \caption{A general scheme for calculating the scatter factor $\rho$.}
 \label{fig:scatterscheme}
\end{center}
\end{figure}
Practically, the procedure for finding the scatter factor is as follows: two sets of ro-vibrational wavefunctions are produced, with two different PESs (PES1 and PES2). For each set of ro-vibrational wavefunctions transition intensities are then calculated with two different DMSs (DMS1 and DMS2). This gives four line lists: (PES1,DMS1), (PES1,DMS2), (PES2,DMS1) and (PES2,DMS2). In the next step, transition lines are matched between the four line lists in a two-step algorithm. First, a straightforward match between lines calculated with the same PES is made. This generates two sets of line lists, which contain two transition intensities for each matched transition line.  In the second stage, a match between the two sets of lines from stage 1 is made. The most efficient way of doing so is by prior matching of energy levels through available quantum numbers $J$ and $e/f$ as well as energetic proximity criteria. Usually not all lines between the two sets from stage one can be unambiguously matched. The percentage of matched lines strongly depends on the difference in quality of the two PESs. Having all 4 line lists matched line-by-line,  for each 'matched' line, the ratio of strongest to weakest transition intensity is calculated, yielding a scatter factor $\rho$, as schematically depicted in Figure \ref{fig:scatterexample}. 

\begin{figure}[H]
\begin{center}
 \includegraphics[width=10cm]{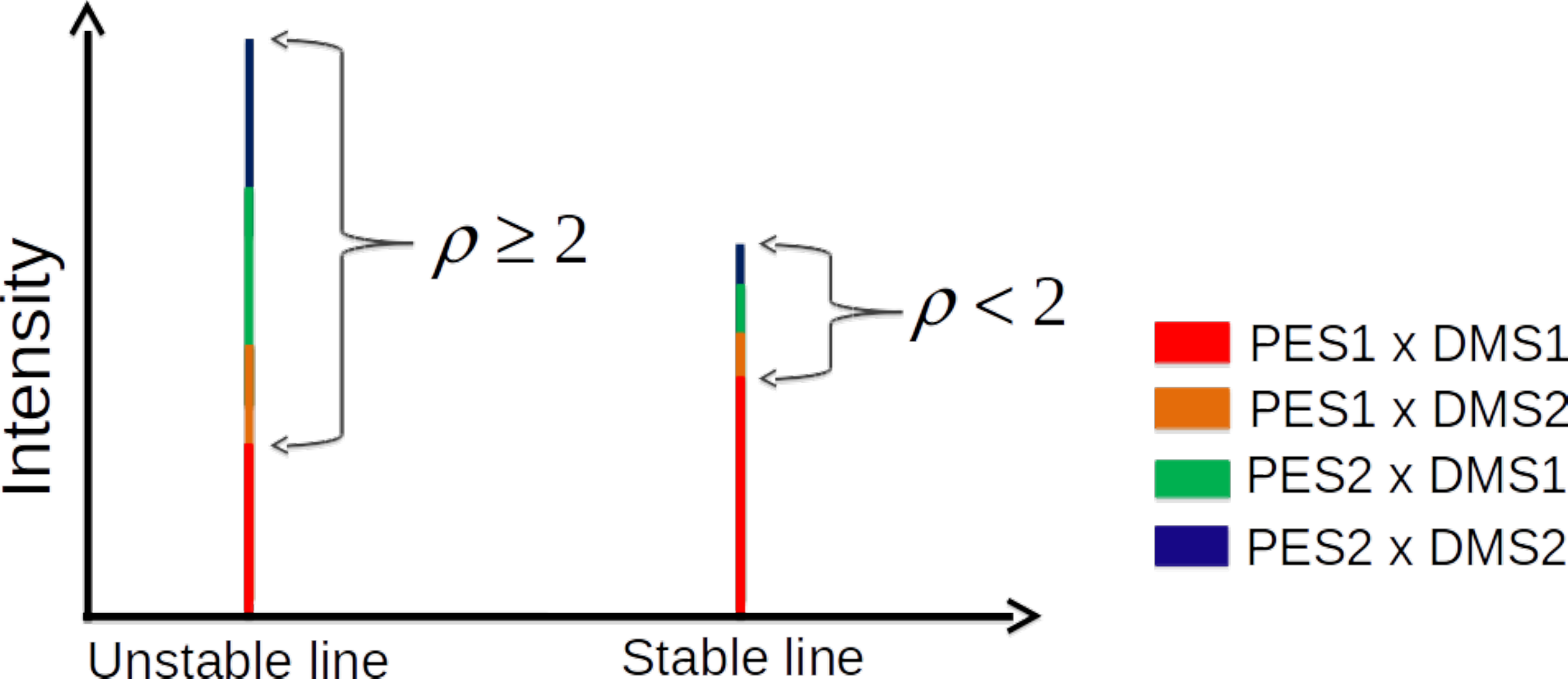}
 \caption{Schematic illustration of the concept of the scatter factor $\rho$. Two situations are given. Left: an unstable line; right: a stable line. Here, the critical value for the scatter factor, dividing stable lines from unstable lines, was chosen to be $\rho_{crit}=2$.}
 \label{fig:scatterexample}
 \end{center}
\end{figure}
The magnitude of the scatter factor, determined by the ratio of the strongest to the weakest transition intensity for a given line informs about the sensitivity of this transition line to minor PES and DMS changes. The division between stable and unstable absorption lines is somewhat arbitraty and can be guided by the statistics of the scatter factor, by for instance choosing only 10\% of lines with highest values of $\rho$ as unstable. Such statistics is depicted in Figure \ref{fig:626rhostat}.

\begin{figure}[H]
\begin{center}
\includegraphics[width=12cm]{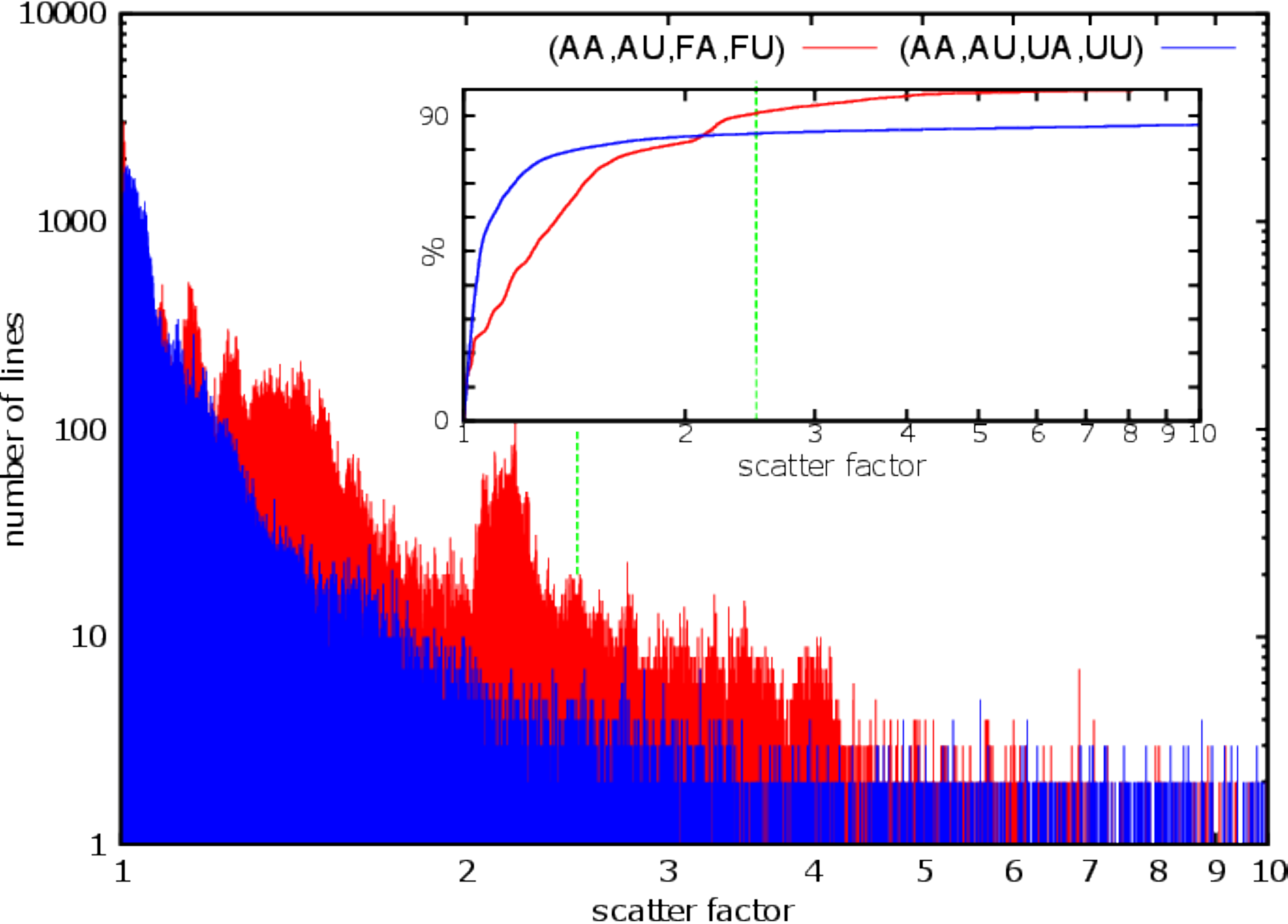}
\caption{Scatter factor, $\rho$, statistics for two sets of PES-DMS combination ((AA,AU,FA,FU) and (AA,AU,UA,UU)) for the main isotopologue of CO$_2$. Inset: cumulative distribution function for $\rho$. See text for further details.}
\label{fig:626rhostat}
\end{center}
\end{figure}

Figure \ref{fig:626rhostat} suggests $\rho=2.5$  is a reasonable value for the critical value of the scatter factor. The plateau of the cumulative distribution function for (AA,AU,FA and FU combinations of PES/DMS) is reached at $\rho\approx4$, at which around 99\% of all lines having a smaller value of the scatter factor. This potentially determines another critical value, separating 'intermediate' and  'unstable' lines. For CO$_2$ the ro-vibrational transitions (details of calculations are given in section \ref{sec:co2} can be divided into three classes of lines: stable, intermediate and unstable; following established arbitrary limits on $\rho$ for a line to be considered stable ($1.0 \leq \rho<2.5$), intermediate ($2.5 \leq \rho<4.0$) and unstable ($\rho \geq 4.0$).

The energetic landscape of the scatter factor can be visualized with the use of maps, as displayed in Figure \ref{fig:626map}, which present the values of the scatter factor as a function of the lower and upper energy level involved in a transition.

\begin{figure}[H]
  \includegraphics[width=10cm,angle=-90]{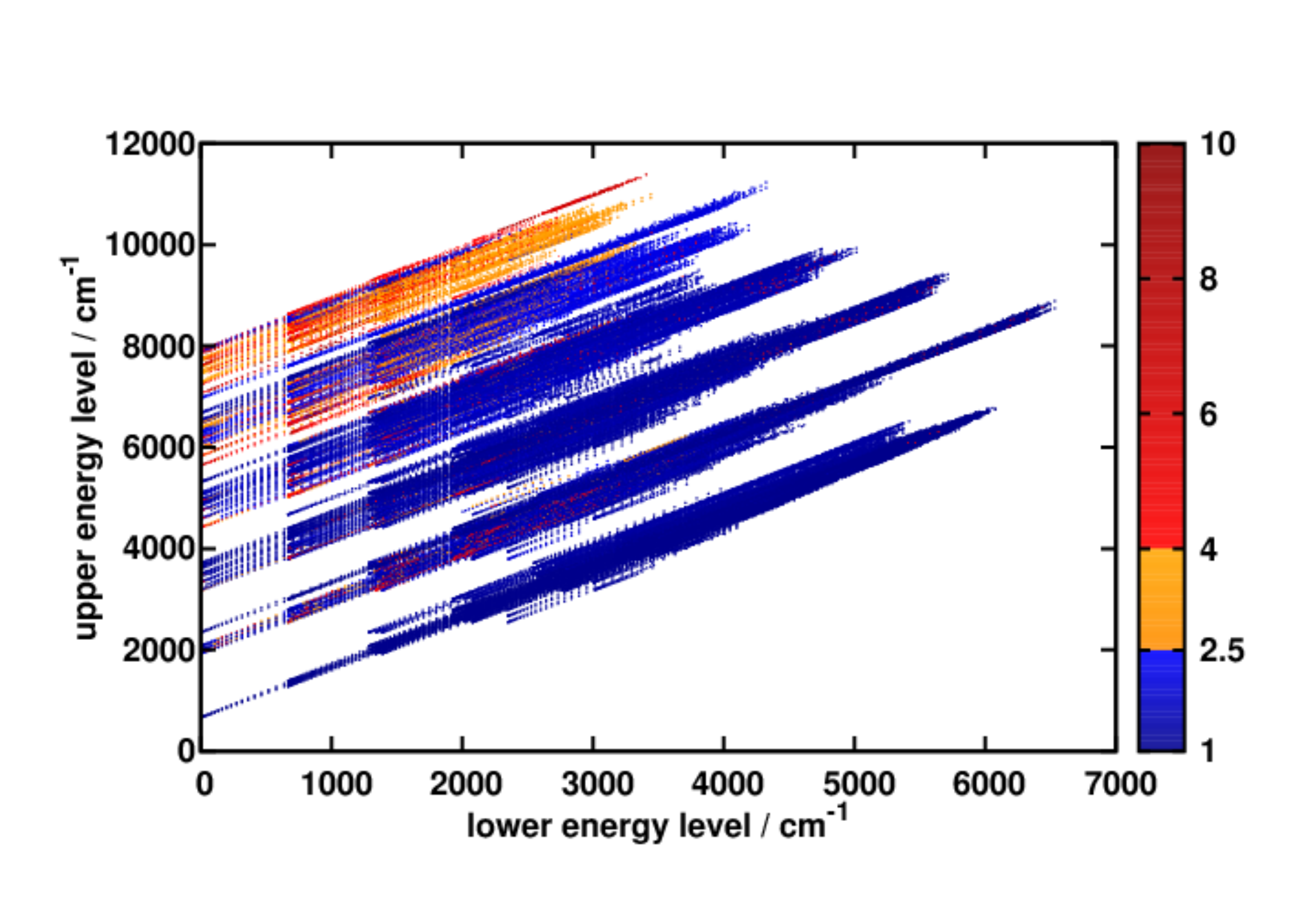}
\caption{Scatter factor map for the main isotopologue (626) as a function of lower and upper energy level for transitions stronger than $10^{-30}$ cm/molecule. 
The color code represents the values of scatter factor, $\rho$. Three regions of 
line stability were determined:  blue-stable, orange-intermediate and red-unstable. Only transitions below 8000\cm\ are shown.  See text for further details.}
\centering
\label{fig:626map}
\end{figure}
The fundamental bands are easily identified as straight lines originating at 0
\cm\ lower energy in Figure \ref{fig:626map}. The lowest hot bands originate at around 668 \cm, that is at the wavenumber of the first excitation in the bending vibration of CO$_2$, complicating the whole picture. A general conclusion from  Figure 
\ref{fig:626map} is that the higher energy of a level involved in a
transition, the higher tendency for the transition to be unstable.  The color coding in the figure
divides scatter factor space into 3 regions of increasing instability, marked blue,
orange and red, respectively. The blue region is considered to be stable
and corresponding intensities are reliable. The orange region is intermediate
between stable and unstable, 
hence transitions marked orange need careful consideration.
The red region contains highly unstable lines whose computed line intensities
should not be trusted.  There are
a few super-unstable transitions ($\rho > 10$) which are not shown on the plots;
these lines are usually
associated with a strong resonance interaction with some other  energetically-close level. Analysis of scatter factors for individual bands can yield insight. For instance, by zooming in an energetic region of interest, as done in Figure \ref{fig:map828}, it is straightforward to determine entirely unstable bands or single transitions which are affected by a resonance.

\begin{figure}[H]
  \includegraphics[width=14cm]{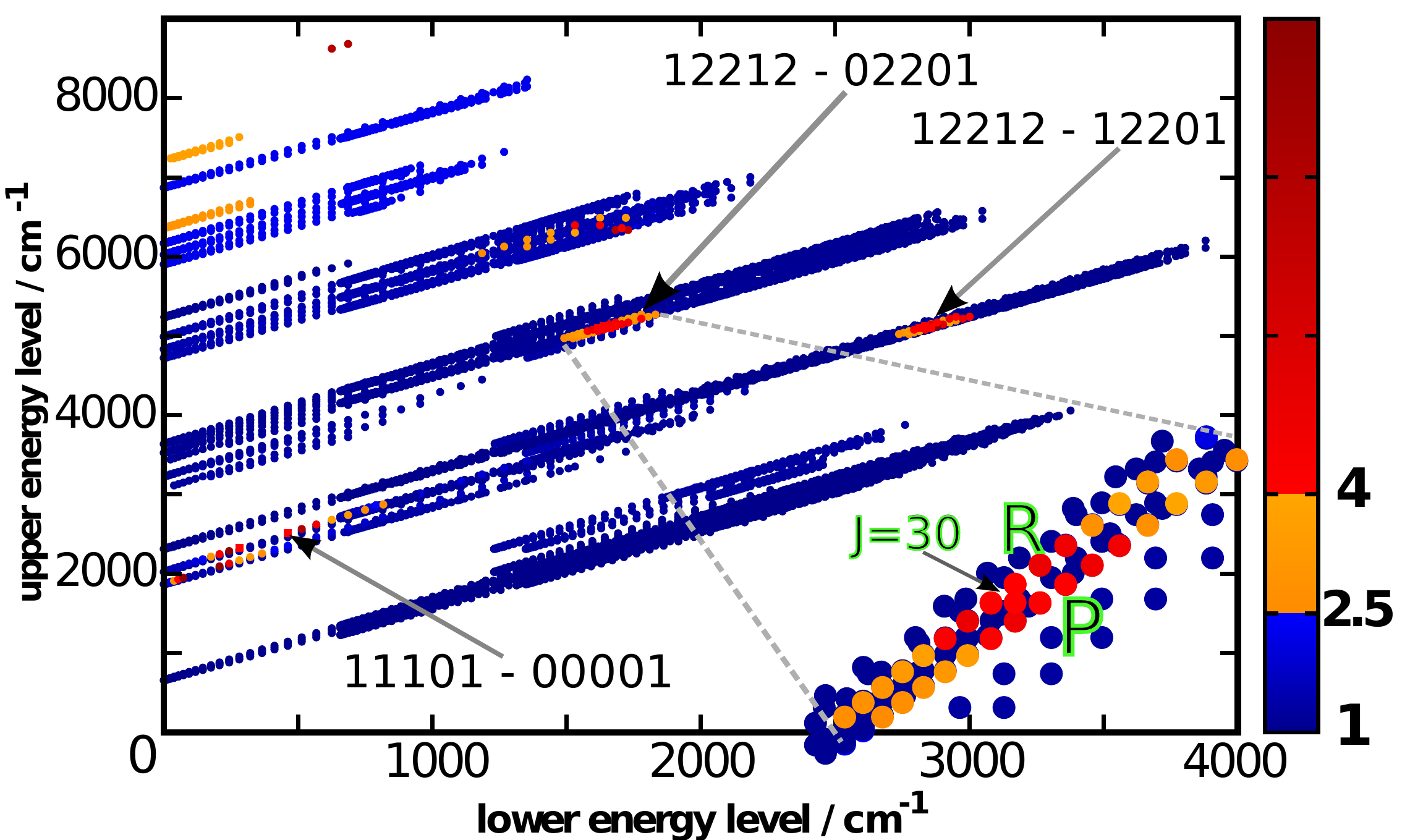}
\caption{Scatter factor map for the 828 isotopologue. Colour coding denotes respective classification of lines: blue  stands for stable lines, orange for intermediate lines and red for unstable lines. The arrows indicate selected bands for which a $J$-localized peak in the scatter factor is observed. The  zoomed inset in right bottom corner shows the peak region of the scatter factor for the 12212 -- 02201 band. Both P and R branches are affected by the interaction around $J=30$. }
\centering
\label{fig:map828}
\end{figure}
Figure \ref{fig:map828} illustrates the general trend of
decreasing stability of lines with increasing energy of states
involved in a transition. This has been already observed for the main
626 isotopologue in Figure \ref{fig:626map}. 
In general, the scatter factor pattern does not change significantly over different symmetric isotopologues, which means that resonance interactions are mostly common for all symmetric isotopologues. This is because changing  nuclear masses in symmetric isotopologues of CO$_2$ shifts vibrational energy levels by a few wavenumbers, and resonance interactions of vibrational energy levels present for the main isotopologue persist for other isotopologues too. For asymmetric isotopologues, a qualitatively different situation with broken symmetry of identical nuclei leads to appearance of some new resonances and disappearance of others, which is captured by scatter factor maps displayed in Figure \ref{fig:asymmaps}.
Sporadic red points localized in small energetic areas are indicative of $J$-localized resonances,
while long chains of unstable points suggest instability of whole
bands. The latter effect can be associated with combination of Fermi-type resonance and limited accuracy of the Fitted PES, especially for higher energies.

\begin{figure}[H]
  \includegraphics[width=14cm]{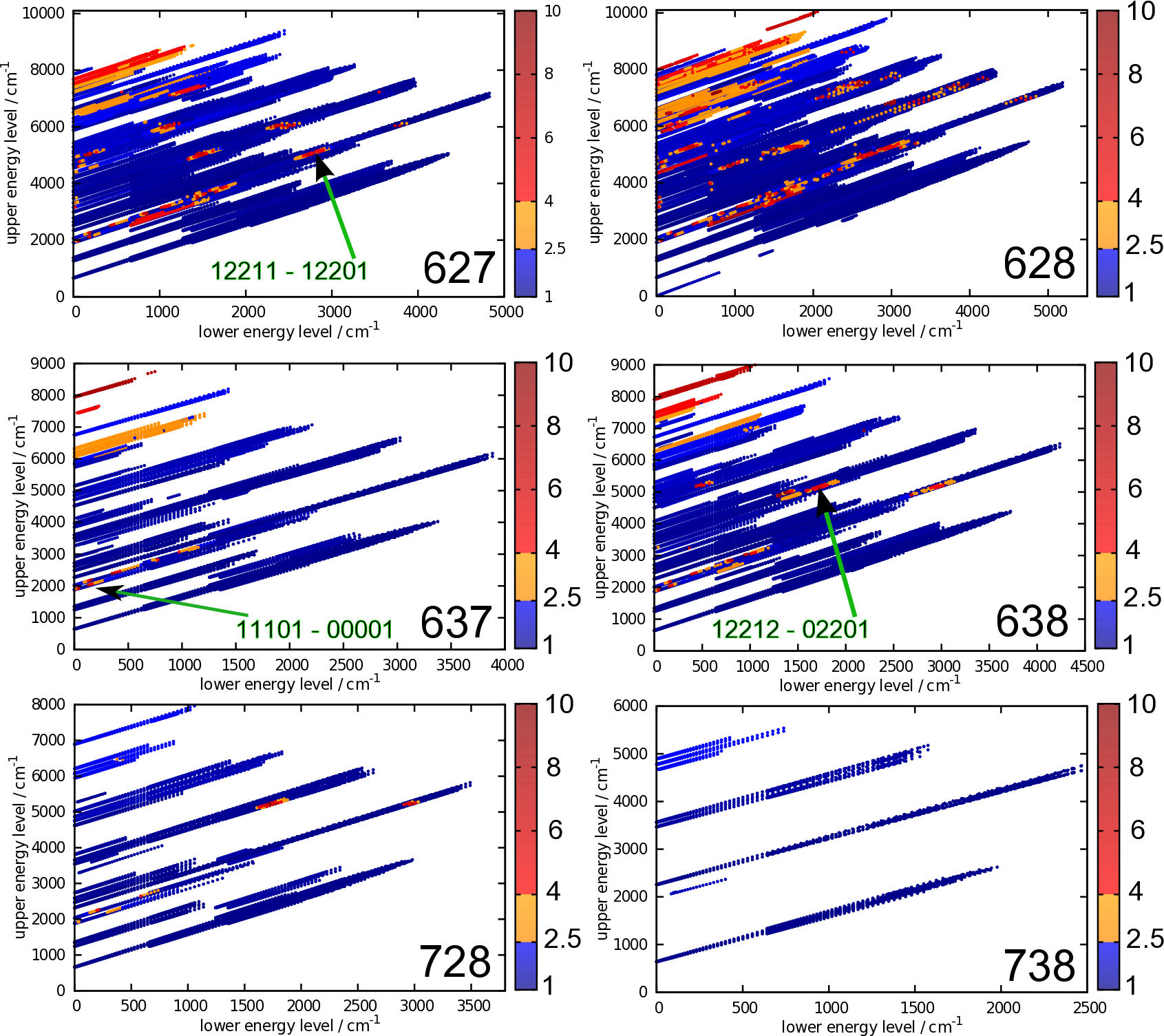}
\caption{Scatter factor maps for all six asymmetric isotopologues of CO$_2$. Colour coding classifies transitions as: stable(blue), intermediate(orange) and red(unstable). The arrows indicate examples of bands involved in a resonance interactions.}
\centering
\label{fig:asymmaps}
\end{figure}
Often it is instructive to give a more detailed insight into
resonances by plotting scatter factor as a function of $m$ quantum
number for each band separately within a given polyad number change
($\Delta P$). $m$ is the rotationally-derived quantum number defined as equal to -$J$(lower energy level) for the P branch, $J$(lower energy level) for the Q branch, and $J$(lower energy level)+1 for the R branch, and $J$ is the rotational quantum number. The polyad
number for carbon dioxide is defined as $P=2\nu_1+\nu_2+3\nu_3$, where
$\nu_1, \nu_2, \nu_3$ are the vibrational quantum numbers of symmetric
stretching, bending and asymmetric stretching, respectively. Figure \ref{fig:symrho1} displays scatter factor analysis of several bands with $\Delta P = 3$ in the 828 isotopologue of CO$_2$.

\begin{figure}[H]
\begin{center}
  \includegraphics[width=14cm]{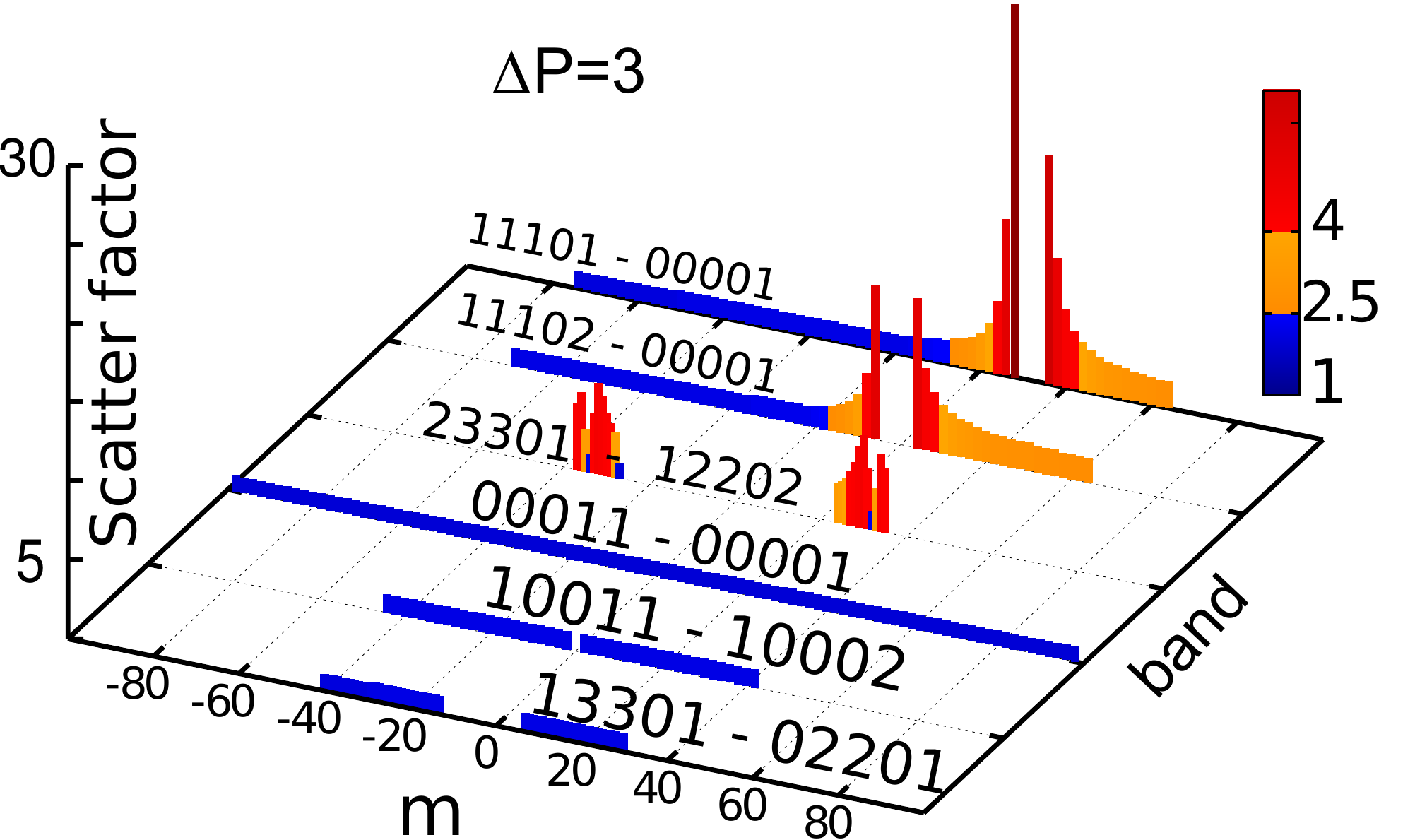}
\caption{Scatter factor distribution for selected bands of 828 with polyad change $\Delta P = 3$. Colour code denotes classification of transition as stable (blue), orange (intermediate) or unstable (red), measured by the scatter factor.}
\label{fig:symrho1}
\end{center}
\end{figure}

For $\Delta P = 3$ three unstable bands were found: 23301
-- 12202 , 11101 -- 00001 and 11102 -- 00001, as shown in Figure \ref{fig:symrho1}. The first of the three
bands contain transitions for which upper energy levels (localized
around a particular $J$ value) become energetically close to
rotational states of some other vibrational state; in this case to
levels from the 12212 state. This may lead to a strong resonance
interaction between states. In the case of the last two bands, an
intensity borrowing mechanism from the strong asymmetric stretching
fundamental is responsible for the instability of line intensities
around a particular $J$. Bands
with higher polyad change number ($\Delta P =5,7,9,11$) are in general
less stable, following uniform distribution of the scatter factor.

Resonances occur when ro-vibrational energy levels of two or more states cross 
or nearly cross in the vicinity of a single $J$ value. A prominent example of near 
crossing situation is the 11101 -- 00001 vibrational band in CO$_2$ (626 isotopologue), which is perturbed by the 00011 
state (intrapolyad interaction). Because the 00011 -- 00001 fundamental is very 
strong and the perturbed band is relatively weak, significant intensity stealing 
is observed. This case is depicted in Figure \ref{fig:symrho2}, where empirically derived transition intensities (taken from the HITRAN2012 database) are compared with theoretically calculated intensities (marked UCL). In Figure \ref{fig:symrho2} colour coding quantifies the stability of the transition intensity. A $J$-localized resonance is visible around $m=+36$, clearly 
correlating with both high instability of lines (marked by red points) and large 
deviations of UCL line intensities from HITRAN2012 line intensities.  For this reason, the theoretical transition intensities near  $m=36$ are unreliable. Present example establishes a procedure for assessing the reliability of theoretically calculated transition intensities. Intensities with high scatter factor value are likely to be inaccurate.

\begin{figure}[H]
\begin{center}
  \includegraphics[width=14cm]{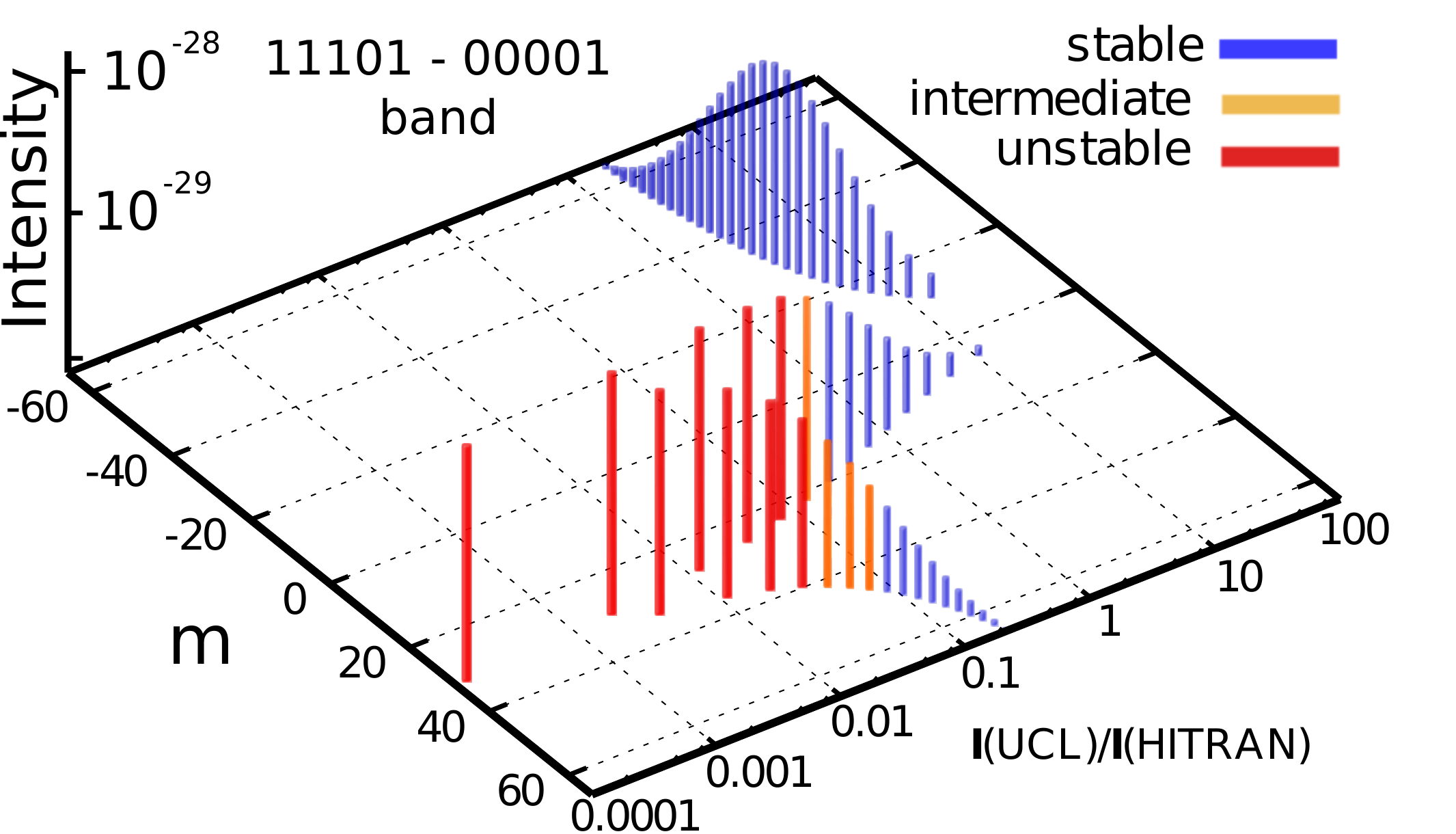}
\caption{Relative intensities (calculated/empirical) plotted against HITRAN2012 line intensities for the 
11101--00001 band for the 828 isotopologue. This is an example of a band 
involved in resonant Coriolis interaction. Blue, orange and red points denote 
stable, intermediate and unstable lines, respectively.}
\label{fig:symrho2}
\end{center}
\end{figure}
This quasi-singularity in line intensity occurs due to the Coriolis interaction with 
the strong absorption 00011--00001 band (asymmetric stretching), which equally perturbs $P$ and $R$ branches of the 
11101--00001 band, and manifests itself by intensity borrowing, which in turn 
leads to the strengthening of the P-branch and to suppression of the R-branch. 

A view of the 636 isotopologue in Figure \ref{fig:symrho3} supports the thesis that resonance interactions may affect only selected rotational branches. Here the scatter factor for the $P$,$Q$ and $R$ branches of the  11101 -- 00001 band in 636 is plotted as a function of $m$.  Only the R branch is affected by intensity borrowing. Similar picture emerges from Figure \ref{fig:626_rho11102}, where the scatter factor for P,Q and R branches of the  11102 -- 00001 band in 626 is plotted as a function of the upper energy level; showing energetic localization of the resonance. Analogical behaviour is observed for resonance-affected bands in other CO$_2$ isotopologues.

Thus, to summarize this part, the scatter factor analysis is capable of detecting resonance interactions of ro-vibrational energy levels, which are branch-specific and $J$-specific. 

\begin{figure}[H]
\begin{center}
  \includegraphics[width=10cm]{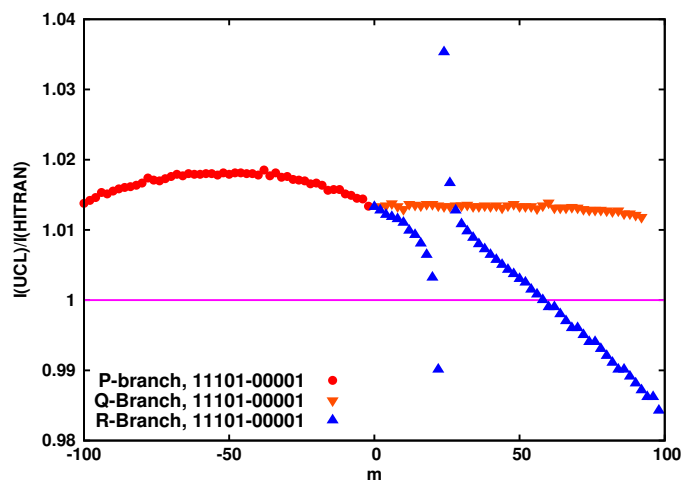}
\caption{Relative intensities  plotted against HITRAN2012 line 
intensities for 11101--00001 band for the 636 isotopologue. This is an example 
of a band involved in resonant Coriolis interaction.}
\label{fig:symrho3}
\end{center}
\end{figure}

\begin{figure}[H]
\begin{center}
 \includegraphics[width=10cm]{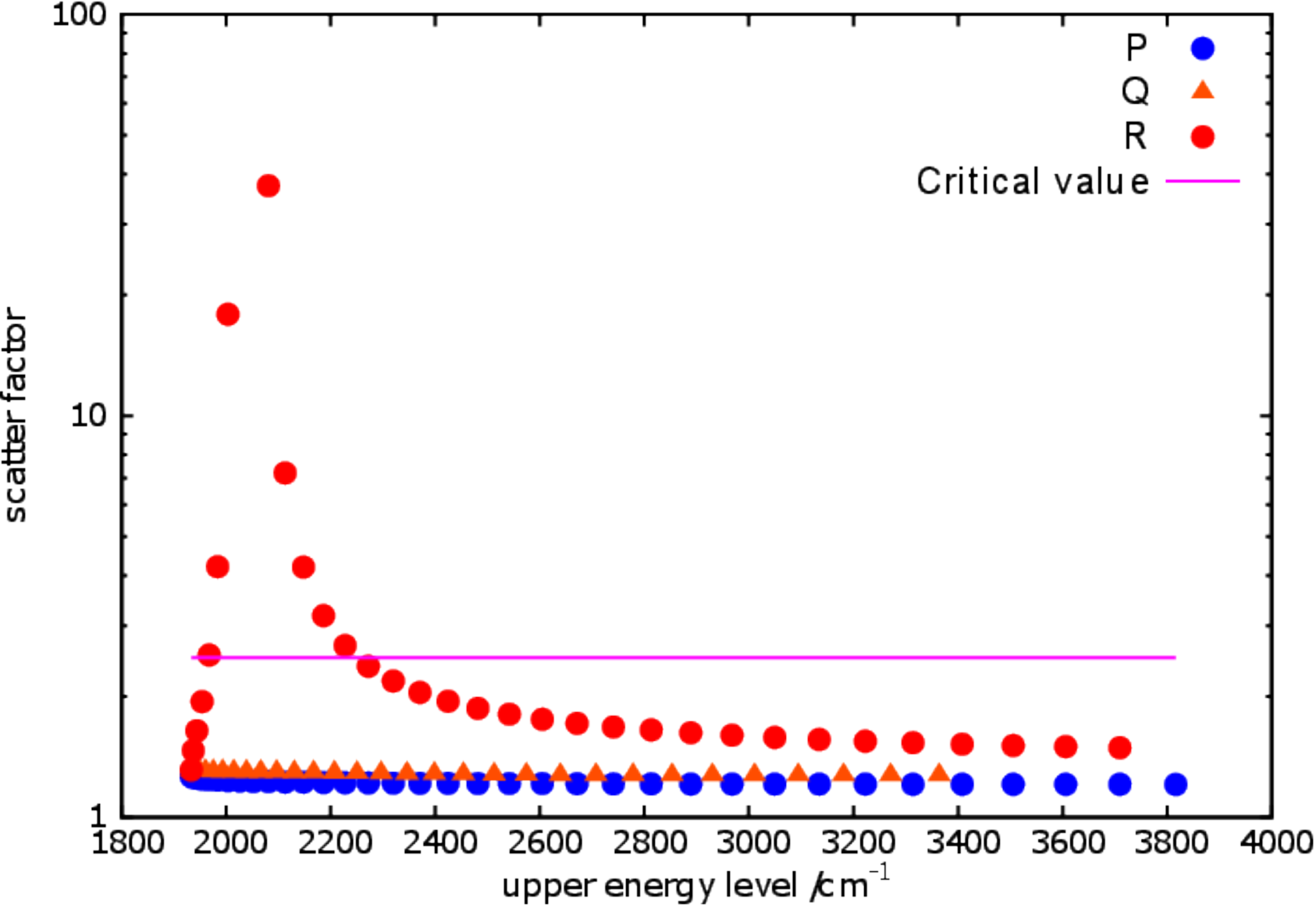}
\caption{Scatter factor as a function of upper energy level for the 
11102 -- 00001 band in 626. The purple line denotes critical value of
the scatter factor ($\rho=2.5$). Different colouring was used for the P, Q and R branches. }
\label{fig:626_rho11102}
\end{center}
\end{figure}

Another interesting example, this time the the intrapolyad interaction, is the pair: 23301 (perturber) and 12212 -- 02201 (perturbed band), for which the intensities scheme is depicted in Figure \ref{fig:symrho4}.

\begin{figure}[H]
\begin{center}
  \includegraphics[width=12cm]{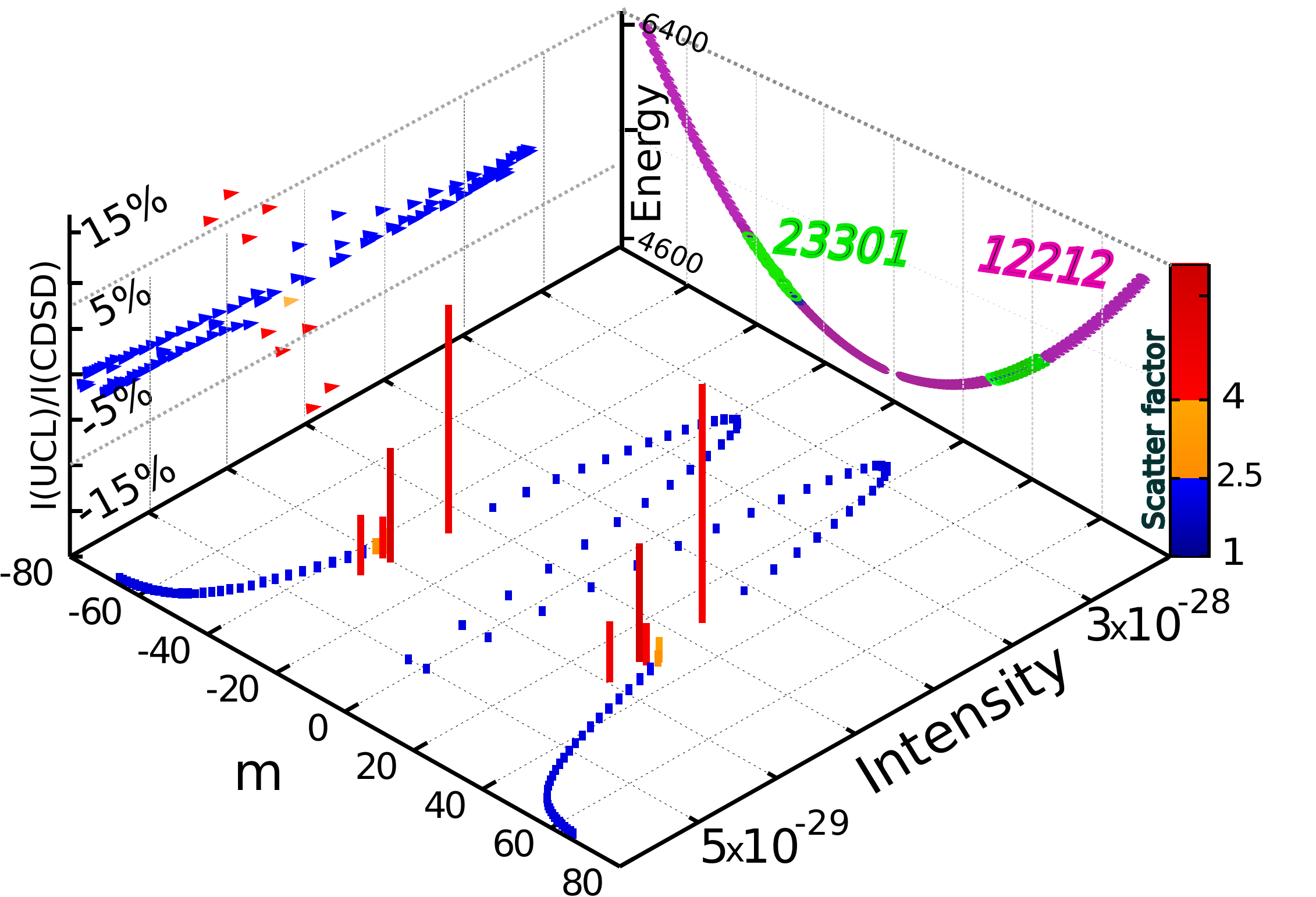}
\caption{Multidimensional graph characterising the 12212 -- 02201 band of 
$^{12}$C$^{18}$O$_2$. The base plane depicts $m$ dependence of line intensities 
with bar height and color code measuring the value of the scatter factor. The 
far right plane represents $m$ dependence of energy levels of the perturbed 
state (12212) and perturber (23301), which happen to nearly overlap around 
$m=\pm 36$. Left plane gives intensity ratios of lines taken from the present line 
list and semi-empirical CDSD-296 database \cite{15TaPeGa.CO2}. }
\label{fig:symrho4}
\end{center}
\end{figure}
Figure \ref{fig:symrho4} shows perfect correlation between line stability 
measured by the scatter factor and agreement with  semi-empirical 'CDSD-296' \cite{15TaPeGa.CO2} line intensities, 
where large discrepancies surround the region of elevated scatter factor (marked 
with red filled triangles in Figure \ref{fig:symrho4}).

One would expect that at least some of the large deviations in line
intensities (see Figures \ref{fig:symrho3} and
\ref{fig:symrho4}) can be assigned to the influence of a
resonance. Indeed, the correlation between high deviations in
intensity and high scatter factor values is strikingly pronounced.  Therefore we may consider the scatter factors used as a
legitimate measure of reliability of a theoretical line list.

An effective reproduction of the experimental line intensities for the 'resonance bands' is a challenge in the variational calculations. The transition dipole moment is very sensitive to small inaccuracies of the ro-vibrational wavefunction and requires an almost perfect reproduction of the PES in this region, which is currently beyond the reach of the variational methodology (electronic structure calculations). For the time being, the best that can be done is to identify these sensitive transitions and use other sources of transition intensities, such as the Effective Hamiltonian calculations, which have proven to be very successful for reproduction of resonance affected bands \cite{14KaCaMo.CO2, 15JaBoLy.CO2}. However, the main drawback of the Effective Hamiltonians is a necessity of a very detailed semi-empirical parametrisation of these resonance affected bands, which requires a lot of experimental data. In atmospheric science, where accurate and reliable transition intensities are necessary to retrieve the concentrations of the measured chemical compounds, a list of ro-vibrational transitions potentially affected by the resonance interactions is very valuable.

Table \ref{table:resonances} gathers information about vibrational bands perturbed by a resonance interaction with other vibrational state for the $^{16}$O$^{12}$C$^{18}$O (628) isotopologue.  Data on other isotopologues can be found in refs. \cite{15ZaTePo.CO2,17ZaTePo.CO2,Zak2017} also available as a part of the ExoMol database (www.exomol.com). 

\begin{table}[H]
\setlength{\tabcolsep}{2pt}
\caption{List of selected $^{16}$O$^{12}$C$^{18}$O vibrational bands  perturbed by a resonance interaction. The columns give: vibrational quantum numbers of the perturbed band, vibrational assignment of the perturbing state, type of interaction: Inter-polyad or Coriolis, band centre, total band strength, the total number of lines in the band in UCL line list, the number of stable lines, 
the number of intermediate lines, median of the scatter factor in the band $\tilde{\rho}$ , maximum scatter factor in the   band $\rho_{max}$, minimum scatter factor in the band $\rho_{min}$ and instability classification: J-localized(branch) or diffuse.}
\vspace{0.2cm}
\footnotesize
\begin{tabular}{l l l l l r r r r r r l}
\hline\hline
Vibrational band			 & 	Perturber  	& Type		  & Centre & Strength & Total & Stable & Inter.&$\tilde{\rho}$ & $\rho_{max}$ & $\rho_{min}$&  Stability   \\ [0.3ex] 
\hline 
 \\
11111 -- 00001			& 31104 		& Inter-pol. 	&  4346.974 		 & 3.88E-27		& 154		&153   	& 1  & 1.2  & 3.1  &  1.1 & J-local 	\\
31112 -- 01101			& 51105 		& Inter-pol. 	&  6263.825 		 & 1.02E-25	& 332		& 312   	& 4  & 2.2  & 422.5  &  2.2 & J-local 	\\
11101 --  00001			& 	 00011	&  Coriolis 	&  2050.068	 & 1.69E-23 & 277	& 261  	& 9  & 1.2  &  2124.0  &  1.1 & J-local(R)	\\
11102 --  00001			& 	 00011	&  Coriolis 	&  1902.447	 & 2.95E-24 & 266	& 251 	& 9  & 1.2  &  2599.0  &  1.2 & J-local(R)	\\
12212 --  00001		& 	 23301	&  Coriolis 	&  4838.085	 & 1.52E-26 & 148	& 138 	& 0  & 1.5  &  15.2  &  1.4 & J-local	\\
23301 --  00001	& 	 12212	&  Coriolis 	&  4825.853	 & 1.46E-27 & 61	& 24	& 15   & 1.1  &  8.0  &  1.1 & sensitive	\\
21112 -- 01101		& 41105 		& Inter-pol. 	&  4894.770	 & 9.37E-24 & 448	& 422  	& 2  & 1.5  &  7843.0  &  1.4 & J-local	\\
21102 -- 00001 & 	 10012	&  Coriolis 	&  3281.717	 & 3.76E-25 & 239	& 138 	& 0  & 1.2  &  115.2  &  1.2 & sensitive	\\
21111 -- 01101 & 	 41104	&  Inter-pol.	&  5063.241	 & 2.92E-24 & 410	& 363 	& 0    & 1.4  &  3.1$\times$10$^{5}$   &  1.4 & sensitive	\\
40014 -- 00001 & 	 60007	&  Inter-pol. 	&  7338.180	 & 2.95E-26 & 134	& 0 	& 123    & 3.6  &  3480.0  &  3.6 & J-local	\\
31113 -- 01101 & 	 42202	&  Coriolis 	&  6098.911	 & 1.56E-25 & 345	& 326	& 0    & 2.3  &  7.6$\times$10$^{5}$  &  2.2 & sensitive	\\
22212 -- 22202			& 25501 		& Cor.+l-type 	&  2262.766		 & 3.07E-27 & 227	& 207   	& 0  & 1.0  & 3444.0  &  1.0 & J-local 	\\
30003 -- 00001			& 14402 		& Anh.+l-type 	&  3855.968	 & 1.43E-24 & 162	& 158   	& 0  & 1.2  & 1.1$\times$10$^{7}$  &  1.2 & J-local 	\\
30013 -- 00001			& 50006 		& Inter-pol. 	&  6127.111	 & 2.24E-24 & 165	& 160   	& 0  & 2.3  & 4.6$\times$10$^{5}$  &  1.3 & J-local 	\\
41113 -- 01101			& 61106 		& Inter-pol. 	&  7459.917	 & 2.45E-27 & 199	& 5  	& 8  & 4.1  & 114.2  &  1.7 & sensitive	\\
05521 -- 00001			& 33314 		& Inter-pol. 	&  7851.812	 & 3.98E-29 & 14	& 0  	& 0  & 102.2  &  1436.0  &  94.5 & sensitive	\\
\hline\hline
\end{tabular}
\label{table:resonances}
\end{table}
\bibliographystyle{plain}
\bibliography{References}

\chapter{Rotational-vibrational-electronic transitions}

Among a number of ways of describing jointly the electronic and the nuclear motion in polyatomic molecules, a formulation given by Sutcliffe and Tennyson \cite{jt96,Sutcliffe2007,DVR3D,jt46} stands out for the following reasons: a) it uses an elegant mathematical framework to derive an exact kinetic energy operator; b) the Hamiltonian is written in an permutationally invariant form; c) the centre-of-mass motion is separated from the internal coordinates; d) it has been succesfully applied to calculate highly-accurate rotational-vibrational spectra of triatomic molecules. A non-Born-Oppenheimer rotational-vibrational-electronic methodology still awaits an implementation. Therefore, in this chapter we introduce the general spin-rotational-vibrational-electronic Hamiltonian for a polyatomic molecule given in the framework of Sutcliffe and Tennyson.

A number of programs for solving the (spin-)ro-vibronic Schr{\"o}dinger equation are available, such as RENNER \cite{02OdHiJe.method,RENNER07,08MeOdJe.method} by Odaka \etal\ , which is designed for linear \textit{Renner-type} triatomic molecules, or more general variational codes for solving the triatomic spin-ro-vibronic problem based on MORBID by Jensen \etal\ \cite{95JeBrKr.method}  which uses an approximate kinetic energy operator for nuclei, RVIB3 \cite{00CaHaPu.method} by Carter, Handy \etal\ is designed only for semi-rigid triatomic molecules with three or less interacting electronic states. 
In the light of limitations of each of these computer codes, it seems natural to search for a more general code, which does not carry any significant inherent approximation. No such program has been yet developed, to the best of author's knowledge.

For diatomic molecules, the DUO computer code of Yurchenko
and co-workers \cite{jt609} solves the fully coupled ro-vibronic \SE\ using an exact kinetic energy operator.  A possible analogue of DUO for triatomic (and other polyatomic) systems is discussed in this chapter.

\section{General rotational-vibrational-electronic Hamiltonian for a triatomic molecule}

The general space-fixed nuclear Hamiltonian given in eq. \ref{eq:Htcoord} can be extended to take into account simultaneously electronic and nuclear degrees of freedom. With such an extension, following Sutcliffe \cite{Sutcliffe2007} the body-fixed \textit{molecular} Hamiltonian can be written as:
\begin{equation}
\hat{H}(\Omega, q,\xi)= \hat{K}(\Omega, q,\xi)+\hat{H}_{el.}(\xi;q)
\label{eq:generalHam}
\end{equation}
where  $\Omega \equiv (\phi,\theta,\chi)$ denotes the three \textit{Euler angles} describing the rotational motion, $q$ stands for a general vector $(q_1,q_2,...,q_{3N_{nuc}-6})$ of vibrational coordinates. $\xi$ is a vector of $3N_e$ electronic position coordinates. The reader is referred to  \cite{Sutcliffe1993} for a detailed discussion and derivation of eq. \ref{eq:generalHam}.

The KEO given in eq. \ref{eq:generalHam} can be represented by following contributions
\begin{equation}
\begin{split}
\hat{K}(\Omega, q,\xi)= \hat{K}_V(q)+\hat{K}_{SRV}(\Omega, q)+\hat{K}_{SRVE}(\Omega, q,\xi)
\end{split}
\label{eq:eHbody}
\end{equation}

where the KEO was split into the vibrational part:

\begin{equation}
\begin{split}
\hat{K}_V(q)= -\frac{1}{2}\left[\sum_{\mu,\nu=1}^{3N_{nuc}-6}G_{\mu\nu}\frac{\partial^2}{\partial q_{\mu}\partial q_{\nu}}+\sum_{\mu=1}^{3N_{nuc}-6}\tau_{\mu}\frac{\partial}{\partial q_{\mu}}\right]
\end{split}
\label{eq:eKV}
\end{equation}
the spin-ro-vibrational part:

\begin{equation}
\begin{split}
\hat{K}_{SRV}(\Omega, q)= \frac{1}{2}\left[\sum_{\alpha,\beta}M_{\alpha,\beta}\hat{N}_{\alpha}\hat{N}_{\beta}+\sum_{\alpha}\lambda_{\alpha}\hat{N}_{\alpha}\right]
\end{split}
\label{eq:eKSRV}
\end{equation}
and the spin-ro-vibronic part:

\begin{equation}
\begin{split}
\hat{K}_{SRV}(\Omega, q)= \frac{1}{2}\left[\sum_{\alpha,\beta}M_{\alpha,\beta}\hat{L}_{\alpha}\hat{L}_{\beta}+\sum_{\alpha}\left(\lambda_{\alpha}\hat{L}_{\alpha}+2\left(M\hat{L}\right)_{\alpha}\right)\hat{N}_{\alpha}\right]
\end{split}
\label{eq:eKSRVE}
\end{equation}
The quantities appearing in the above equations are defined as in eq. \ref{eq:Hbody}, \ref{eq:Kv1},  \ref{eq:Kv2} and  \ref{eq:Kvr}.

The electronic Hamiltonian $\hat{H}_{el.}(\xi;q)$ from eq.  \ref{eq:generalHam} is a sum of the following terms:
\begin{equation}
\hat{H}_{el.}(\xi;q)= \hat{T}_e(\xi)+V_{ne}(\xi;q)+V_{ee}(\xi)+V_{nn}(q)
\end{equation}
where $\hat{T}_e(\xi)$ denotes the kinetic energy operator for electrons, $V_{ne}(\xi;q)$ is the electron-nuclei attraction potential energy, $V_{ee}(\xi)$ is the electron-electron repulsion potential energy and $V_{nn}(q)$ is the nucleus-nucleus repulsion potential energy.  

Here, we operate in the Hund's case (b) \cite{10BrCaxx.method}, with the rotational energy operator proportional to $\hat{N}-\hat{L}$, that is the difference between the ro-vibronic angular momentum and the electronic angular momentum. This way of coupling the nuclear and electronic angular momentum is suitable when the spin-orbit interaction is small compared to rotational energy of the molecule. In such a non-relativistic scheme it is a custom to include terms, which account for the 'spin-orbit' and 'spin-rotation' interactions. These operators are usually parametrized empirically. Here all nuclear-spin interactions are neglected. 
With these spin-relativistic add-ons the total Hamiltonian can be written as:
\begin{equation}
\hat{H}(\Omega, q,\xi)= \hat{K}(\Omega, q,\xi)+\hat{H}_{el.}(\xi;q)+\hat{H}_{SO}(q,\xi)+\hat{H}_{SR}(q)
\label{eq:generalHam2}
\end{equation}
The spin-orbit Hamiltonian can be written in a general form as:
\begin{equation}
\hat{H}_{SO}(q,\xi)=\sum_{\mu\nu}A^{SO}_{\mu\nu}(q)\hat{L}_{\mu}(\xi)\hat{S}_{\nu}
\label{eq:eSO}
\end{equation}

and the spin-rotation Hamiltonian is given by:
\begin{equation}
\hat{H}_{SR}(q)=\sum_{\mu\nu}\gamma^{SR}_{\mu\nu}(q)\hat{N}_{\mu}\hat{S}_{\nu}
\label{eq:eSR}
\end{equation}

The spin-orbit coupling and spin-rotation coupling operators are often approximated by effective operators \cite{10BrCaxx.method}: 
$\hat{H}_{SR}(q)\approx \gamma^{SR}(q)\sum_{\mu}\hat{N}_{\mu}\hat{S}_{\mu}$, $\hat{H}_{SO}(q)\approx A^{SO}(q)\sum_{\mu}\hat{L}_{\mu}\hat{S}_{\mu}$. Such an approximation suggest neglecting the directionality of the spin-orbit and spin-rotation interactions in the molecule. 

In equations \ref{eq:eKV}-\ref{eq:eKSRVE} defining the ro-vibronic kinetic energy operator $M_{\alpha\beta}$ denotes the generalized inverse moment of inertia introduced and explained by Sutcliffe in \cite{Sutcliffe2007}. $\lambda_{\alpha}=-i\left(v_{\alpha}+2\sum_{\mu=1}^{3N_{nuc}-6}W_{\mu\alpha}\frac{\partial}{\partial q_{\mu}}\right)$ and quantities $v_{\alpha}$ and $W_{\mu\alpha}$ depend on vibrational coordinates only and are defined in ref. \cite{Sutcliffe2007}. $\lambda_{\alpha}$ is related to Coriolis-couplings and cannot be eliminated by a choice of embedding. We do not specify the \textit{molecule-fixed} embedding of the coordinate frame at this point, but the form of $\lambda_{\alpha}$ depends on the choice of the molecule-fixed frame embedding. In the vibrational part of the kinetic energy operator $K_V(q)$ the tensor $G_{\mu\nu}$  and vector $\tau_{\mu}$ are functions of coordinates, independent of embedding, and are defined in ref. \cite{Sutcliffe2007}. All internal-coordinate dependent quantities presented above can be calculated with a little use of algebra when a choice of particular internal coordinates in made. Here we only specify that the internal coordinates are orthogonal ($\mu_{ij}=0$, see eq. \ref{eq:Hbody}).

With the total Hamiltonian defined, the next step is to choose a spin-ro-vibronic basis set.

\subsection{Spin-ro-vibronic basis set}
We shall follow the standard approach to the vibronic-coupling problem, where a Born-Huang (BH) \cite{Baer2002} type expansion, discussed in Chapter 1, is assumed for the ro-vibronic wavefunction.
This approach however is not the only one \cite{Baer2002,Mukherjee2017,Min2014} available, and perhaps is not the best one. Unfortunately, to the best of author's knowledge, better representations of the ro-vibronic wavefunction are still lacking. 
The problem with the BH representation to the wavefunction is that when a finite number electronic states is considered, non-removable singularities appear in the ro-vibronic Hamiltonian, in regions of the configuration space where two adiabatic electronic states intersect. Approximate diabatization schemes are possible, which eliminate the singularities in the non-adiabatic coupling terms (NACT), as discussed in Chapter 1. Yet, an open challenge is to find an alternative procedure for defining the ro-vibronic matrix elements in terms of some new, neither adiabatic nor diabatic representation. 

In analogy with the Born-Huang expansion the spin-rovibronic wavefunction the wavefunction is expanded as a sum of products of an 'electronic wavefunction' and a 'spin-ro-vibrational wavefunction':

\begin{equation}
|\Psi(\Omega,q,\xi )\rangle = \sum_{i}|\psi_{el.,i}(\xi ;q)\rangle |\Phi_{srv,i}(\Omega,q)\rangle
\label{eq:bornhuang}
\end{equation}

Effectively the expansion in eq. \ref{eq:bornhuang} is truncated at few electronic states, which  couple substantially. Coupled states must lie energetically close and have appropriate  symmetry, as discussed in Chapter 3. The electronic part of the wavefunction is obtained from electronic structure calculations, in a solution to the eigenvalue problem for the electronic Hamiltonian $\hat{H}_{el.}(\xi ;q)|\psi_{el.,i}(\xi ;q)\rangle=V_i(q)|\psi_{el.,i}(\xi ;q)\rangle$. The electronic basis forms a complete orthonormal set of functions with the standard scalar product: $\langle \psi_{el.,i'}(\xi  ;q)|\psi_{el.,i}(\xi ;q)\rangle= \int \psi^*_{el.,i'}(\xi ;q)\psi_{el.,i}(\xi ;q) d \xi = \delta_{i'i}$. For the primitive rotational basis a complete set of symmetric-top Hamiltonian eigenvectors  $|N,k,M\rangle $ is used. Here $N$ stands for the ro-vibronic angular momentum quantum number, $k$ is the projection of the total angular momentum $J$ on the molecule-fixed $z$-axis and $M$ is  the projection of the total angular momentum $J$ on the space-fixed $Z$-axis. A set of commuting observables for this eigenbasis is $\hat{N}^2, \hat{N}_z, \hat{J}^2, \hat{J}_z, \hat{S}^2$, where $\hat{N}$ is the ro-vibronic angular momentum operator, $\hat{J}$ is the total angular momentum operator and $\hat{S}$ is the electron-spin angular momentum operator. These operators correspond to the following eigenvalue problems\footnote{written in atomic units}

\begin{equation}
\begin{split}
\hat{N}^2|N,k,M\rangle=N(N+1)|N,k,M\rangle \\
\hat{N}_z|N,k,M\rangle=k|N,k,M\rangle \\
\hat{J}^2|N,k,M\rangle=J(J+1)|N,k,M\rangle \\
\hat{J}_z|N,k,M\rangle=M_J|N,k,M\rangle \\
\hat{S}^2|N,k,M\rangle=S(S+1)|N,k,M\rangle \\
\end{split}
\label{eq:roteigenval}
\end{equation}
This basis can be further symmetry-adapted with the use of the parity transformation $\hat{E}^*$:

\begin{equation}
\begin{split}
|N,K,M,p\rangle = \frac{1}{\sqrt{2}}\left[|N,k,M\rangle + (-1)^{N+k+p}|N,-k,M\rangle\right], \qquad K=|k|>0 \\
|N,K,M,p\rangle = |N,0,M\rangle,  \qquad K=0
\end{split}
\end{equation}
where $p=0,1$ denotes the parity quantum number of the rotational wavefunction. 
This rotational basis is then coupled in a symmetry-adapted way with the spin basis \cite{06BuJexx.method}:

\begin{equation}
\begin{split}
|\Psi_{spinrot}^{J,K,S,p}\rangle = \sum_{N=|J-S|}^{J+S}\sum_{M=-N}^{N}\sum_{M_s=-S}^{S}(-1)^{N-S+M_J}\sqrt{2J+1}\tj{N}{S}{J}{M}{M_s}{-M_J}\times \\
\times |S,M_s\rangle|N,K,M,p\rangle
\end{split}
\label{eq:srbasis}
\end{equation}
where $|S,M_s\rangle$ is the standard  spin  basis: $\hat{S}^2 |S,M_s\rangle = S(S+1) |S,M_s\rangle,  \hat{S}_z |S,M_s\rangle = M_s |S,M_s\rangle$ and $ \langle S',M'_s | S,M_s\rangle = \delta_{S'S}\delta_{M'_s,M_s}$. The total spin quantum number $S$ is associated with a given isolated electronic state. Such coupled representation to the spin-electronic-rotational wavefunction guarantees that it is an eigenfunction of 
$\hat{N}^2, \hat{N}_z, \hat{J}^2, \hat{J}_z, \hat{S}^2$ operators. 
Note that the energy of the molecule in free-space is independent of any space-fixed defined quantum number: $M, M_J,M_s$. 

Stepping down from the general picture, we narrowing our focus on triatomic molecules only. With the electronic and the spin-rotational basis defined above, we are ready to write the total spin-ro-vibronic wavefunction function:

\begin{equation}
\begin{split}
|\Psi^{J,p,h}\rangle = \sum_{S}\sum_{N=|J-S|}^{J+S}\sum_{M=-N}^{N}\sum_{M_s=-S}^{S}\sum_{K=p}^{N}\sum_{m,n,j}^{vib.states}\sum_{i}^{el.states}(-1)^{N-S+M_J}\sqrt{2J+1}\tj{N}{S}{J}{M}{M_s}{-M_J}\times \\
\times C^{J,p,h}_{i,N,S,K,m,n,j}|i\rangle|S,M_s\rangle|N,K,M,p\rangle|m^{(i)}\rangle|n^{(i)}\rangle|jK^{(i)}\rangle
\end{split}
\label{eq:srvebasis}
\end{equation}
where $|i\rangle$ is a shorthand notation for an $i$-th eigenfunction of the electronic Hamiltonian $H_{el.}(\xi;q)$, $|S,M_s\rangle$ is the standard spin basis function defined above,  $|N,K,M,p\rangle $ is the parity-adapted symmetric-top wavefunction, $|m^{(i)}\rangle$, $|n^{(i)}\rangle$ are vibrational basis functions for the $r_1$ and $r_2$ stretching, respectively and $|jK^{(i)}\rangle$ is a basis function for the bending motion ($\gamma$ coordinate, see for instance eq. \ref{eq:Hbody}). Note that the vibrational basis in general depends on the electronic state $i$ as indicated with the superscript $(i)$. Summation over the $S$ quantum number accounts for mixing of electronic states with different spin multiplicity through spin-orbit coupling or spin-rotation coupling.  Additionally the bending vibrational basis state is assumed to depend on $K$, which as discussed in chapter 1, brings the benefit of eliminating spurious singularities in the Hamiltonian, which result from the vanishing Jacobian of the space-fixed to molecule-fixed transformation. Such regularization of the Hamiltonian is possible when an associated Legendre polynomial basis is used $|jK\rangle = P^{(K)}_{j}(\cos\theta)$. The superscript in the basis set notation is reserved for good quantum numbers which in the present case are $J$ and $p$, while $h$ enumerates the final spin-rovibronic energy states.  

Because rotational basis set $\{|N,k,M\rangle\}_{k=-N,-N+1,...,N-1,N}$ used here is complete, the  rotational degrees of freedom of the molecule can be integrated out in an exact, formal way. The matrix of the total Hamiltonian in this basis can be evaluated analytically. Similarly, it is possible to formally integrate out all electronic and spin degrees of freedom yielding an effective vibrational Hamiltonian in the following form:
\begin{equation}
\begin{split}
\hat{H}(q)= \hat{K}_V(q)+\hat{K}_{SRV}(q)+\hat{K}_{SRVE}(q)+\hat{H}_{el.}(q)+\hat{H}_{SO}(q)+\hat{H}_{SR}(q)
\end{split}
\label{eq:eHbodyeff}
\end{equation}

where

\begin{equation}
\begin{split}
2\hat{K}_V(q)= \delta_{K'K}\delta_{s's}\left[\delta_{i'i}\hat{K}_V(q)+\tilde{G}^{i'i}+\tilde{\tau}^{i'i}\right]
\end{split}
\label{eq:eKVeff}
\end{equation}

\begin{equation}
\begin{split}
2\hat{K}_{SRV}(q)= \delta_{K'K}\delta_{s's}\left[\delta_{i'i}\left(2b\left(N(N+1)-K^2\right)-K^2b_0+k\lambda_0\right)+\hat{\Lambda}_0^{i'i}\right]+\\
\delta_{K'K+1}\delta_{s's}\frac{1}{2}C^+_{NK}\left[\delta_{i'i}\left(b_{+1}(2K+1)+\lambda_{+}\right)+\hat{\Lambda}_+^{i'i}\right]+\\
+\delta_{K'K-1}\delta_{s's}\frac{1}{2}C^-_{NK}\left[\delta_{i'i}\left(b_{-1}(2K-1)+\lambda_{-}\right)+\hat{\Lambda}_-^{i'i}\right]+\\
+\frac{1}{2}\delta_{K'K+2}\delta_{s's}\delta_{i'i}b_{+2}C^+_{NK}C^+_{NK+1}+\\
+\frac{1}{2}\delta_{K'K-2}\delta_{s's}\delta_{i'i}b_{-2}C^-_{NK}C^-_{NK-1}
\end{split}
\label{eq:eKSRVeff}
\end{equation}

\begin{equation}
\begin{split}
2\hat{K}_{SRVE}(q)= \delta_{K'K}\delta_{s's}\left[\sum_{\alpha\beta}M_{\alpha\beta}L_{\alpha\beta}^{i'i}+\sum_{\alpha}\lambda_{\alpha}L_{\alpha}^{i'i}-K\langle i'|(M\hat{L})_z | i \rangle + \hat{\mathcal{O}}^{i'i}\right]+\\
+\delta_{K'K+1}C_{NK}^+\langle i'|(M\hat{L})_+ | i \rangle+\delta_{K'K-1}C_{NK}^+\langle i'|(M\hat{L})_- | i \rangle
\end{split}
\label{eq:eKSRVEeff}
\end{equation}

\begin{equation}
\begin{split}
\hat{H}_{el.}(q)=\delta_{s's}\delta_{K'K}\delta_{i'i}V_{i'i}(q)
\end{split}
\label{eq:eHeleff}
\end{equation}

\begin{equation}
\begin{split}
\hat{H}_{SO}(q)=A^{SO}(q)\left[\frac{1}{2}\left[\delta_{s's+1}\delta_{K'K}C_{SM_S}^+L_{+}^{i'i}+\delta_{s's-1}\delta_{K'K}C_{SM_S}^-L_{-}^{i'i}\right]+\delta_{s's}\delta_{K'K}L_z^{i'i}\right]
\end{split}
\label{eq:eHSOeff}
\end{equation}

\begin{equation}
\begin{split}
\hat{H}_{SR}(q)=\gamma^{SR}(q)\left[\frac{1}{2}\left[\delta_{s's+1}\delta_{K'K+1}C_{SM_S}^+C_{NK}^+ + \delta_{s's-1}\delta_{K'K-1}C_{SM_S}^-C_{NK}^-\right]+\delta_{s's}\delta_{K'K}K\right]
\end{split}
\label{eq:eHSReff}
\end{equation}
where $C_{NK}^{\pm}=\left(N(N+1)\mp K(K+1)\right)^{\frac{1}{2}}$. Here, for clarity of presentation the integration was performed with the use of the primitive spin-rotational-electronic basis: $|i\rangle|s,M_s\rangle|N,K,M\rangle$. Note that the $M$ symbol appearing in eq. \ref{eq:eKSRVEeff} is a function of internal coorinates, not the $M$ rotational quantum number.

All operators appearing in eq. \ref{eq:eHbodyeff} are formally labeled by the quantum numbers, which define the spin-electronic-rotational basis: $J,K,i,S,p$. We omit the space-fixed related quantum numbers as irrelevant to the dynamics in free-space. Each operator written above is defined for a given set of good quantum numbers $J$ and $p$, which are assumed implicit and are not denoted for the clarity of presentation. Thus, there are three sets of indices, which define the Hamiltonian hyper-matrix: $K'K$, $i'i$ and $S'S$. The quantities appearing in  equations \ref{eq:eKVeff} - \ref{eq:eHSReff}  for the effective operators are as follows:

\begin{equation}
\begin{split}
\tilde{G}^{i'i} = -\frac{1}{2}\sum_{\mu,\nu =1}^{3N_{nuc.}-6}\left[G_{\mu\nu}\beta^{i'i}_{\mu\nu}+\tau_{\mu}\alpha^{i'i}_{\mu}\right] \\
\beta^{i'i}_{\mu \nu} = \langle i'| \frac{\partial^2}{\partial q_{\mu} \partial q_{\nu}}| i \rangle,  \qquad \alpha^{i'i}_{\mu} = \langle i'| \frac{\partial}{\partial q_{\mu}}| i \rangle \\
\tilde{\tau}^{i'i} = -\frac{1}{2}\sum_{\mu,\nu =1}^{3N_{nuc.}-6}G_{\mu\nu}\left(\alpha^{i'i}_{\mu} \frac{\partial}{\partial q_{\mu}}+\alpha^{i'i}_{\nu} \frac{\partial}{\partial q_{\nu}}\right) \\
\hat{\Lambda}_{\gamma}^{i'i} = -2i\sum_{\mu =1}^{3N_{nuc}-6}W_{\mu \gamma} \alpha^{i'i}_{\mu}  \quad \gamma = x,y,z \\
\hat{\Lambda}_0^{i'i} = \hat{\Lambda}_z^{i'i}, \quad \hat{\Lambda}_{\pm}^{i'i} = \hat{\Lambda}_x^{i'i} \mp i  \hat{\Lambda}_y^{i'i} \\
L^{i'i}_{\alpha \beta} = \langle i' | \hat{L}_{\alpha}\hat{L}_{\beta} | i \rangle \\
L^{i'i}_{\alpha} = \langle i' | \hat{L}_{\alpha}| i \rangle \\
\mathcal{O}^{i'i} =  -2i\sum_{\mu =1}^{3N_{nuc}-6} \langle i' | (W\hat{L})_{\mu}\frac{\partial}{\partial q_{\mu}}|i \rangle \\
V_{i'i}(q) = \langle i' | \hat{H}_{el.}(\xi ; q) | i \rangle
\end{split}
\label{eq:quantities}
\end{equation}
and all $\pm$ sub-scripted operators appearing in eqs. \ref{eq:eKVeff} -- \ref{eq:eHSReff} are defined as $A_{\pm}=A_x\mp i A_y$. There are essentially two types of operators, which act on the vibrational basis: functions of vibrational coordinates and differential operators. In the above Hamiltonian there are two types of differential operators enclosed in $\tilde{\tau}^{i'i}$ and $\lambda_{\alpha}$. The former one is associated with derivative couplings of different electronic states (NACT) and the latter is associated with the Coriolis-coupling of ro-vibrational states.  

To summarize, the fully coupled spin-ro-vibronic procedure outlined in the present section requires the following input functions for solving the molecular time-independent Schr{\"o}dinger equation:

\begin{enumerate}
\item $V_{i}(q)$ -  potential energy surfaces (PES) for each adiabatic electronic state of interest. 
\item $\alpha_{\mu}^{i'i}(q)$ - non-adiabatic coupling matrix elements (NACT) between electronic states $i$ and $i'$ and for vibrational coordinate $\mu$. 
\item $\beta_{\mu \nu}^{i'i}(q)$ - diagonal Born-Oppenheimer correction (DBOC) terms
\item $A^{SO}(q)$ - spin-orbit coupling surface(s). Note that in general the spin-orbit coupling operator is proportional to a tensor quantity $A_{i'i}^{SO}(q)$ which mixes different spin states. 
\item $L_{\alpha}^{i'i}(q), L_{\alpha\beta}^{i'i}(q)$ - $L$-coupling and  $L^2$-coupling surfaces, respectively. These terms represent coupling of electronic angular momentum to the vibrational coordinates of nuclei.
\item $\mathcal{O}^{i'i}(q)$ - $L$-vibronic coupling surfaces.
\item $\gamma^{SR}(q)$ -spin-rotation coupling surface(s). Note that in general the spin-rotation coupling operator is proportional to a tensor quantity $\gamma_{i'i}^{SR}(q)$ which mixes different spin states. 
\end{enumerate}

\section{Spin-ro-vibronic transition intensities}
With the wavefunction given in eq. \ref{eq:srvebasis} and the procedure described in eqs. \ref{trans:transdip}-\ref{trans:strength2}, it is straightforward to derive an expression for the spin-ro-vibronic transition line strength:
\begin{equation}
\begin{split}
S_{if}=\frac{1}{4}\left(2J''+1\right)\left(2J'+1\right)\times \\
\times \left|\sum_{S'',S'} (-1)^{S'}\delta_{S''S'} \sum_{N'=|J'-S'|}^{J'+S'}\sum_{N''=|J''-S''|}^{J''+S''}\sj{N'}{J'}{S'}{J''}{N''}{1} \left[(-1)^{N''+N'+1}+(-1)^{p''+p'}\right] \right. \times \\  \left. \sum_{i'',i'}\sum_{\sigma=-1}^{+1} \sum_{\substack{K'=p' \\ K''=p''}}^{N',N''}(-1)^{K''} \tj{1}{N'}{N''}{\sigma}{K'}{K''}\sum_{\textbf{n}',\textbf{n}''} c_{\textbf{n}',N',K',i',S'}^{J',h',p'} c_{\textbf{n}'',N'',K'',i'',S''}^{J'',h'',p''}   M_{\textbf{n}',\textbf{n}'',K',K''}^{\sigma,i',i''} \right|^2
\end{split}
\label{trans:strengthspinrovib}
\end{equation}
where $c_{\textbf{n}'',N'',K'',i'',S''}^{J'',h'',p''} $ are variational coefficients defined in eq. \ref{eq:srvebasis}, $ M_{\textbf{n}',\textbf{n}'',K',K''}^{\sigma,i',i''} $ is defined in eq. \ref{trans:transdip1} and $\sj{N'}{J'}{S'}{J''}{N''}{1}$ is the 6-j symbol (see Book by Bunker and Jensen \cite{06BuJexx.method}, section 14.1.5). 
\section{Ro-vibronic transition intensity borrowing}
One of the consequences of the Born-Huang representation to the ro-vibronic wavefunction is that different Born-Oppenheimer electronic states can become significantly mixed under certain circumstances. Nuclear vibrational and electronic degrees of freedom in the molecule can interact in a similar way as the nuclear rotational and vibrational degrees couple to each other. Such coupling carries consequences to transition intensities. The formulation of the rotational-vibrational transition intensity borrowing presented in section \ref{sec:borrowing} can be quite straightforwardly generalized to rotational-vibrational-electronic transitions. 

Let us begin with a simple example of two internal states of the molecule, $\psi_l$ and $\psi_u$, between which a transition occurs. The lower state is completely separable into nuclear spin part, ro-vibrational part and electronic part (Born-Oppenheimer approximation) $\psi_l = \Phi_{ns,l}|J,p;l\rangle\Phi_l (\mathbf{q};\mathbf{Q})$, but the upper state is a mixture of two electronic states $|\Phi_a\rangle$, $|\Phi_b\rangle$: $\psi_u = \Phi_{ns,u}\left(|J,p;a\rangle\Phi_a (\mathbf{q};\mathbf{Q}) + |J,p;b\rangle\Phi_b (\mathbf{q};\mathbf{Q}) \right) $ . Consider a situation in which the transition  $\Phi_l \rightarrow \Phi_a$ is electronically allowed ($\Gamma(\Phi_l)\otimes \Gamma(\Phi_a) \supset \Gamma(\hat{\mu}_{elec.})$), but the transition  $\Phi_l \rightarrow \Phi_b$ is electronically forbidden ($\Gamma(\Phi_l)\otimes \Gamma(\Phi_b) \not\supset \Gamma(\hat{\mu}_{elec.})$). In the Born-Oppenheimer approximation there will be zero transition intensity between the $\Phi_l $ and  $ \Phi_b$ electronic states. 
We say that states $\Phi_l $ and  $ \Phi_b$ are not connected by the electric dipole moment operator $\hat{\mu}_{elec.}$. It means that the product of the irreducible representations of both electronic states and the dipole moment operator does not contain the totally symmetric representation.

However, similarly as the rotation-vibration coupling (discussed in Chapter 3) can relax selection rules, the electronic-vibrational coupling can relax selection rules for electronic transitions. Even when a transition is electronically forbidden it can be allowed in a more general vibronic framework. Taking into account vibrations of atoms in the molecule, it is possible that one or more of such vibrations can assist in relaxation of the electronic selection rules. The mechanism of this relaxation relies on a more general selection rule, which states that a transition is \textit{vibronically} allowed when the product of \textit{vibronic} symmetries of the wavefunctions and the dipole moment operator contain the totally symmetric representation:
\begin{eqnarray}
\Gamma(\psi_{l}) \otimes \Gamma(\psi_{u}) \supset \Gamma(\hat{\mu}_{elec.})
\label{eq:el_forbidden2}
\end{eqnarray}

As discussed in Chapter 1, electronic states can be coupled through non-adiabatic coupling terms (NACT):
\begin{equation}
\vec{\mathbf{F}}^{\alpha}_{ab}=\frac{\langle \Phi_a|( \vec{\nabla}_{\alpha}\hat{H}_{el})|\Phi_b\rangle}{E_b(\mathbf{Q})-E_a(\mathbf{Q})}
\label{eq:BOcommut2}
\end{equation}
For our two-state model the system of vibronic equations can be written as:
\begin{equation}
\begin{pmatrix}
\hat{K}_{nuc}(\Theta,\mathbf{Q})+E_a(\mathbf{Q})-E & \vec{\mathbf{F}}_{ab}\cdot  \vec{\nabla} \\
\vec{\mathbf{F}}_{ba}\cdot \vec{\nabla} & \hat{K}_{nuc}(\Theta,\mathbf{Q})+E_b(\mathbf{Q})-E  \\
\end{pmatrix}\begin{pmatrix}
|J,p;a\rangle\\
|J,p;b\rangle \\
\end{pmatrix}=0
\label{vibmatrix}
\end{equation}
where $|J,p;a\rangle,|J,p;b\rangle$ are ro-vibrational wavefunctions, which satisfy the normalisation condition: $|\langle J,p;a |J,p;a\rangle|^2+|\langle J,p;b |J,p;b\rangle|^2=1$.
The form of the above system of equations suggests that only if $\vec{\mathbf{F}}^{\alpha}_{ab}$ is non-zero, there will be mixing of Born-Oppenheimer electronic wavefunctions. Determining the coefficients in eq. \ref{vibmatrix} is in general not an easy task and approximations must be made. A standard approach to this problem is to switch to a \textit{crude adiabatic representation} of the vibronic wavefunction: 
\begin{equation}
\psi_u = \Phi_{ns,u}\left(|J,p;a\rangle\Phi_a (\mathbf{q};\mathbf{Q}_0) + |J,p;b\rangle\Phi_b (\mathbf{q};\mathbf{Q}_0) \right) 
\label{eq:diabatic}
\end{equation}
where $Q_0$ is the equilibrium nuclear geometry in electronic state $\Phi_a$. Electronic wavefunctions satisfy the \SE : $\hat{H}_{el}(\mathbf{q};\mathbf{Q}_0)\Phi_a (\mathbf{q};\mathbf{Q}_0)=E_a(\mathbf{Q}_0)\Phi_a (\mathbf{q};\mathbf{Q}_0)$. Putting anzatz given in eq. \ref{eq:diabatic} into the \SE\ results in a set of ro-vibronic equations in which the kinetic energy part is diagonal and the off-diagonal elements are given as:
$\mathbf{U}_{ab}=\langle \Phi_a (\mathbf{q};\mathbf{Q})|\hat{H}_{el}(\mathbf{q};\mathbf{Q})-\hat{H}_{el}(\mathbf{q};\mathbf{Q}_0)|\Phi_b (\mathbf{q};\mathbf{Q})\rangle$
these matrix elements mix electronic states $a$ and $b$. Such approach is only good when we consider only small displacements from the equilibrium geometry of the molecule. Then the vibronic coupling elements (electronic Hamitlonian) can be represented by low-order Taylor expansion:

\begin{equation}
\hat{H}_{el}(\mathbf{q};\mathbf{Q}) = \hat{H}_{el}(\mathbf{q};\mathbf{Q}_0)+\frac{\partial \hat{H}_{el}(\mathbf{q};\mathbf{Q})}{\partial \mathbf{Q}}|_{\mathbf{Q}=\mathbf{Q}_0}\cdot \mathbf{Q}+o(\mathbf{Q}^2)
\end{equation}
so that $\mathbf{U}_{ab}=\langle \Phi_a (\mathbf{q};\mathbf{Q})|\frac{\partial \hat{H}_{el}(\mathbf{q};\mathbf{Q})}{\partial \mathbf{Q}}|_{\mathbf{Q}=\mathbf{Q}_0}|\Phi_b (\mathbf{q};\mathbf{Q})\rangle\mathbf{Q}$. We can use perturbation theory to assess the coefficient standing by the  $\Phi_a $ electronic state in the vibronic wavefunction. Because $\mathbf{Q}$ is assumed small, $U_{ab}$ can also be considered small. We can expect only small admixture of the $|\phi_a\rangle$ state in $|\psi_e\rangle$, so that $|\langle J,p;a |J,p;a\rangle|^2 \ll |\langle J,p;b |J,p;b\rangle|^2$. Zeroth-order perturbed wavefunction is given by:

\begin{equation}
|\psi_u\rangle=|\phi_{b}\rangle|J,h;b\rangle + \sum_n \frac{ \langle \phi_{a}| \frac{\partial \hat{H}_{el}(\mathbf{q};\mathbf{Q})}{\partial \mathbf{Q}_n}|_{\mathbf{Q}_n=\mathbf{Q}_0} |\phi_{b}\rangle}{E_b(\mathbf{\mathbf{Q}})-E_a(\mathbf{\mathbf{Q}})}\langle J,h,b|\mathbf{Q}_n|J,h;a\rangle|\phi_{a}\rangle|J,h;a\rangle
\label{eq:perturbed}
\end{equation}
where summation over $\mathbf{n}$ is carried over vibrational coordinates. Note analogy to eq. \ref{eq:borrowing}. The symmetry of $\mathbf{Q}_n$ is the same as the symmetry of $ \frac{\partial \hat{H}_{el}(\mathbf{q};\mathbf{Q})}{\partial \mathbf{Q}_n}$. The mixing of the two electronic states is possible through a vibration  $\mathbf{Q}_n$ of symmetry such that $\Gamma(\mathbf{Q}_n)\otimes \Gamma(\phi_a)\otimes \Gamma(\phi_a) \supset A_1$.

The consequences of vibronic coupling are visible in alterations of transition intensities between electronically allowed states, but most of all in appearing of weak electronically forbidden transitions. The transition dipole moment between an isolated vibronic state and a mixed vibronic state can be written as
\begin{equation}
\mu_{le}=\langle \psi_l  |\hat{\mu}_{elec.}|\psi_e\rangle 
\label{transdipmix}
\end{equation}
The square of the transition dipole moment, which is proportional to the absorption line strength, is given by
\begin{equation}
S_{le}=|\langle \psi_l  |\hat{\mu}_{elec.}|\psi_e\rangle |^2=S_{la}
\label{linestrengthvibr}
\end{equation}
because $\langle \psi_l  |\hat{\mu}_{elec.}|\psi_b\rangle=0$, the only contribution to the line strength is from the admixture of state $|\Phi_a\rangle $, which is dipole-connected with state $|\Phi_l\rangle $.
It is easily seen that due to the admixture of the $a$ electronic state  through the vibronic coupling mechanism, the Born-Oppenheimer forbidden transition gains intensity. We say that electronic state $b$ borrows intensity from electronic state $a$.

\section{Axis switching effects}
Not only does the equilibrium geometry of the molecule change upon the electronic transition,  but also an additional rotation of the molecule-fixed coordinate system accompanies such transitions \cite{Hougen1965}. The former effect can be directly attributed to the difference in shapes of the potential energy surfaces for the two electronic states that causes the vibrational basis set optimized for the electronic excited state to be no longer optimal for the electronic ground state. In the terminology of normal modes it means that normal coordinates in the electronic excited states are rotated (leading to the so-called Duschinsky effect \cite{Meier2015}) with respect to normal coordinates in the electronic ground state.  The effect of rotation of the molecule-fixed coordinate system affects the Euler angles, causing rotation of the rotational basis set.  
Although in many systems these artifacts of the electronic transition are marginal, they sometimes significantly soften rotational selection rules, allowing for appearance of whole vibrational "forbidden" bands, as observed in HCN \cite{Jonas1992} and SiHD \cite{Kokkin2016}. For example, in the HCN molecule, for the ($\pi^* \leftarrow \pi$) electronic transition only $\Delta K=\pm 1$ sub-bands are allowed by rotational selection rules. However stimulated-emission-pumping (SEP) experiments \cite{Yang1990} observed weak $\Delta K=0$ transitions to levels with non-zero vibrational angular momentum ($l=1$). This type of transition is forbidden by rotational selection rules and has been convincingly attributed to non-rigidity of the molecule during the transition between linear $\tilde{X}$ electronic ground state and bent $\tilde{A}$ electronic state \cite{Hougen1965,Jonas1992}.
In calculations, the magnitude of the axis-switching effect depends on the choice of the molecule-fixed frame, the choice of coordinates and the basis set. Axis-switching is strongly pronounced in the \textit{Eckart frame}, the molecule-fixed coordinate system which needs to be rotated when changing the electronic state, in order to satisfy the conditions of the minimal rotational-vibrational coupling in both states separately. 

The axis-switching effect suggests that the rotational basis functions should be labelled with the quantum numbers for electronic states too. However, the completeness of the rotational basis guarantees that the rotational part is accounted for exactly regardless of the electronic states. Therefore, rotational states in the electronic excited state, which are nominally functions of rotated Euler angles can be modeled with the un-rotated rotational basis of the electronic ground state (or vice-versa). An appropriately large vibrational basis set can also eliminate any inaccuracies resulting from the \textit{Duschinsky effect}, meaning that the vibrational basis is nearly complete hence does not depend on the electronic state. As a result, with appropriate choice of the rotational and vibrational basis, it is possible to use identical basis set for both (all) electronic states of interest. Of course situations where the geometry displacement upon electronic excitation in large, sometimes render the task of finding a universal vibrational basis set difficult. 
\bibliographystyle{plain}
\bibliography{References}

\end{document}